%% file: fl3-all.tex
\def\ftmagnification{1200}
\def\spacingNumerator{5}
\def\spacingDenominator{4}
\input jfvmacros

\chapno=\cHintroIII
      

{\nopagenumbers
\multiply\baselineskip by \spacingDenominator\divide \baselineskip by\spacingNumerator

\null\vskip3truecm

%
%
\centerline{\tafontt A Two Dimensional Fermi Liquid }
\vskip0.1in
\centerline{\tbfontt Part 3: The Fermi Surface}

\vskip0.75in
\centerline{Joel Feldman{\parindent=.15in\footnote{$^{*}$}{Research supported 
in part by the
 Natural Sciences and Engineering Research Council of Canada and the Forschungsinstitut f\"ur Mathematik, ETH Z\"urich}}}
\centerline{Department of Mathematics}
\centerline{University of British Columbia}
\centerline{Vancouver, B.C. }
\centerline{CANADA\ \   V6T 1Z2}
\centerline{feldman@math.ubc.ca}
\centerline{http:/\hskip-3pt/www.math.ubc.ca/\squiggle
feldman/}
\vskip0.3in
\centerline{Horst Kn\"orrer, Eugene Trubowitz}
\centerline{Mathematik}
\centerline{ETH-Zentrum}
\centerline{CH-8092 Z\"urich}
\centerline{SWITZERLAND}
\centerline{knoerrer@math.ethz.ch, trub@math.ethz.ch}
\centerline{http:/\hskip-3pt/www.math.ethz.ch/\squiggle
knoerrer/}

\vskip0.75in
\noindent
%
{\bf Abstract.\ \ \ } 
We show that the particle number density derived from the  
thermodynamic Green's function at temperature zero 
constructed in the second part of this series 
has a jump across the Fermi curve, a basic property of a  
Fermi liquid. We further show that the two particle 
thermodynamic Green's function at temperature zero has  
the regularity behavior expected in a Fermi liquid.

\vfill
\eject


\titleb{Table of Contents}
\halign{\hfill#\ &\hfill#\ &#\hfill&\ p\ \hfil#&\ p\ \hfil#\cr
\noalign{\vskip0.05in}
\S XI&\omit Introduction                          \span&\:\pgNPXI&\omit\cr
\noalign{\vskip0.05in}
\S XII&\omit Momentum Green's Functions              \span&\:\pgNPXII\cr
\noalign{\vskip0.05in}
\S XIII&\omit Momentum Space Norms                   \span&\:\pgNPXIII\cr
\noalign{\vskip0.05in}
\S XIV&\omit Ladders with External Momenta             \span&\:\pgNPXIV\cr
\noalign{\vskip0.05in}
\S XV&\omit Recursion Step for Momentum Green's Functions \span&\:\pgNPXV\cr
&&More Input and Output Data                           &\omit&\:\pgNPXVa\cr
&&Integrating Out a Scale                             &\omit&\:\pgNPXVb\cr
&&Sector Refinement, ReWick ordering and Renormalization  &\omit&\:\pgNPXVc\cr
\noalign{\vskip0.05in}
{\bf Appendices}\span\cr
\noalign{\vskip0.05in}
\S C&\omit H\"older Continuity of Limits             \span&\:\pgNPC\cr
\noalign{\vskip0.05in}
\S D&\omit Another Description of Particle--Hole Ladders  \span&\:\pgNPD\cr
\noalign{\vskip0.05in}
 &\omit References                                    \span&\:\pgNPIIIref \cr
\noalign{\vskip0.05in}
 &\omit Notation                                      \span&\:\pgNPIIInot \cr
}
\vfill\eject
\multiply\baselineskip by \spacingNumerator\divide \baselineskip by\spacingDenominator}
\pageno=1


\chap{Introduction}\PG\pgNPXI

This paper, together with [FKTf1] and [FKTf2] provides a construction of a
two dimensional Fermi liquid at temperature zero. This paper contains
Sections \CHintroIII\ through \CHtildestep\ and 
Appendices \APappHoelder\ and \APappPhladders.  
Sections \CHintroI\ through \CHrenmap\ and Appendix \APappModelComp\ are 
in [FKTf1] and Sections \CHintroII\ through \CHrecurs\ and Appendix
\APappRewick\  are in [FKTf2]. Notation tables are provided at the end
of the paper. The main goal
of this part is the proof of the existence of a Fermi surface,
stated in Theorem  \theoremNPmainthII. The proof of this theorem
follows  Lemma \:\lemTNPabstractjump. We assume for the rest of this paper that
the interaction $V$ satisfies the reality condition (\eqnNPreal) and is
bar/unbar exchange invariant in the sense of (\eqnNPphexchange). The latter is not essential\footnote{$^{(1)}$}{See footnote (1) in \S\CHtildestep.}. It is used only for notational convenience at intermediate stages of the proof.

\vfill\eject
 
\chap{ Momentum Green's Functions}\PG\pgNPXII

Recall that the momentum distribution function $n(\k)$ of Theorem 
\theoremNPmainthII\ is expressed in terms of the Fourier transform $\check G_2(k)$ of the two point Green's function $G_2(x,y)$. 
In Theorem \theoremNPinduction, we defined a generating functional 
$\cG^\rg_j(K)$. Following the statement of Theorem \theoremNPinduction,
we constructed
$G_2(x,y)$ as the limit of functions $2G^\rg_{j,2}({\sst(x,1), (y,0)})$,
 where 
$\int d\eta_1 d\eta_2\ G^\rg_{j,2}(\eta_1, \eta_2)\ \phi(\eta_1)\phi(\eta_2)
=2\int dx dy\ G^\rg_{j,2}({\sst(x,1), (y,0)})\ \phi({\sst x,1})\phi({\sst y,0})$
is the part of $\cG^\rg_j(K)\big|_{K=0}$ that is homogeneous of degree two.
In \S\CHtildestep, we shall prove the following decomposition of the Fourier transforms,
$2\check G^\rg_{j,2}(k)$, of these functions.

\theorem{\STM\theoremNPtildeinduction}{
Let $\aleph<\aleph'<\sfrac{2}{3}$. If the constants $\bar\al,\ \bar\la>0$ of Theorem \theoremNPinduction\ are big, respectively small, enough, the data 
of Theorem  \theoremNPinduction\ may be chosen such that 
$2\check G^\rg_{j,2}(k)$
has the decomposition 
$$
2\check G^\rg_{j,2}(k)=  C^{(\le j)}_{u_j(0)}(k)+ \sfrac{1}{[ik_0-e(\k)]^2}
\sum_{i=2}^{j}\smsum_{\ell=i}^{j}q^{(i,\ell)}(k)
$$
Here $u_j(0)$ is the sectorized
function $u_j\big((\xi_1,s_1),(\xi_2,s_2);K\big)\big|_{K=0}$
and $q^{(i,\ell)}(k)$, $\ell\ge i\ge 2$  is a family of  functions 
with $q^{(i,\ell)}(k)$  vanishing when $k$ is in the 
$(i+2)^{\rm nd}$ neighbourhood and and when $\k$ is not in the support
of $U(\k)$ and
obeying, for each multiindex $\de=(\de_0,\bde)$ with $|\de|\le 2$,
$$
\sup_k\big|\rD^\de q^{(i,\ell)}(k) \big|
\le2\la_0^{1-2\upsilon}\sfrac{\fl_\ell}{M^\ell}M^{\aleph'(\ell-i)}
M^{\de_0 i}M^{|\bde|\ell}
\EQN\eqnNPqderiv$$
Furthermore $q^{(i,\ell)}(-k_0,\k)=\overline{q^{(i,\ell)}(k_0,\k)}$.

}
\vskip.1in
For the  rest of this section, we deduce consequences of
Theorem \theoremNPtildeinduction. In Lemma \:\lemNPgtwoamp, we describe
regularity properties of $\lim\limits_{j\rightarrow\infty}\check G^\rg_{j,2}(k)$ and in Lemma  \:\lemNPgtwoft\ we show that Fourier transforms
commute with the limit $j\rightarrow\infty$. From this we derive properties 
of the proper self--energy and use them to prove that there is a jump
in the momentum distribution function $n(\k)$, thus showing that 
Theorem \theoremNPtildeinduction\ implies Theorem \theoremNPmainthII.
 
\lemma{\STM\lemNPgtwoamp}{ 
\Item i)
The sequence of functions 
$ \check u_j(k;0)$ converges uniformly to a function 
$ P(k)$ that vanishes at $k_0=0$ and obeys
$$\eqalign{
\big| P(k)\big|
&\le\la_0^{1-2\upsilon}\min\{\big|k_0\big|,1\big\}\cr
\big|\nabla P(k)\big|
&\le \la_0^{1-2\upsilon}\cr
\big|\nabla P(k)-\nabla P(k')\big|
&\le \la_0^{1-2\upsilon} |k-k'|^{1\over 2}\cr
\big| P(k)- \check u_j(k;0)\big|
&\le\la_0^{1-2\upsilon}\fl_j\min\{\big|k_0\big|,1\big\}\cr
}$$
\Item ii)
The sequence of functions 
$Q_j(k)=\sum_{i=2}^{j}\smsum_{\ell=i}^{j}q^{(i,\ell)}(k)$ 
converges uniformly to a function 
$Q(k)$ that vanishes unless $\k$ is in the support of $U(\k)$ and obeys
$$\eqalign{
\big| Q(k)\big|
&\le\la_0^{1-3\upsilon}\min\{\big|ik_0-e(\k)\big|^{3\over 2},1\big\}\cr
\big|\sfrac{\partial Q}{\partial k_0}(k)\big|
&\le \la_0^{1-3\upsilon}\min\{\big|ik_0-e(\k)\big|^{1\over 2},1\big\}\cr
\big| Q(k)- Q(k')\big|,\ 
\big|\sfrac{\partial Q}{\partial k_0}(k)
-\sfrac{\partial Q}{\partial k_0}(k')\big|
&\le \la_0^{1-3\upsilon} |k-k'|^{1\over 2}\cr
\big| Q(k)- Q_j(k)\big|
&\le\la_0^{1-3\upsilon}\fl_j\min\{\big|ik_0-e(\k)\big|,1\big\}\cr
}$$
\Item iii) $P(-k_0,\k)=\overline{P(k_0,\k)}$ and $Q(-k_0,\k)=\overline{Q(k_0,\k)}$.

}
\prf  i)
By Lemma \lemOSNormMom\ of [FKTo3] and (\eqnNPpbound)
$$
\sup_k\big|\rD^\de \check p^{(i)}(k) \big|
\le 4\la_0^{1-\upsilon}
\sfrac{\fl_i}{M^i}M^{i|\de|}
\EQN\eqnNPpderiv$$
for all $|\de|\le 2$.
Consequently, the sequence of functions 
$ \check u_j(k;0)=\smsum\limits_{i=2}^{j-1}\check p^{(i)}(k)$ 
converges uniformly to
$ 
 P(k)=\smsum\limits_{i=2}^{\infty}\check p^{(i)}(k)
$
and, by Lemma \lemNPhoelder, with $\al=\aleph$ and $\be=1-\aleph$,
$$\eqalign{
|P(k)|,\ \big|\nabla P(k)\big|
&\le \const\la_0^{1-\upsilon} \le \la_0^{1-2\upsilon}\cr
\big|\nabla P(k)-\nabla P(k')\big|
&\le \const \la_0^{1-\upsilon} |k-k'|^{\aleph}
\le \la_0^{1-2\upsilon} |k-k'|^{1\over 2}\cr
}$$
if $\la_0$ is small enough. Since each $\check p^{(i)}(k)$ vanishes at
$k=0$, the same is true for $P(k)$ and $|P(k)|\le \la_0^{1-2\upsilon}|k_0|$.
Furthermore,
$$\eqalign{
\big|P(k)- \smsum_{i=2}^{j-1}\check p^{(i)}(k)\big|
&\le \smsum\limits_{i=j}^{\infty}\big|\check p^{(i)}(k)\big|
\le \smsum\limits_{i=j}^{\infty}4\la_0^{1-\upsilon}\fl_i \min\{\big|k_0|,1\big\}
\le \const\la_0^{1-\upsilon} \fl_j \min\{\big|k_0|,1\big\}\cr
&\le\la_0^{1-2\upsilon}\fl_j\min\{\big|k_0|,1\big\}\cr
}$$

\Item ii)
As $\big|q^{(i,\ell)}(k) \big|
\le2\la_0^{1-2\upsilon}\sfrac{\fl_\ell}{M^\ell}M^{\aleph'(\ell-i)}$ and
$$
\sum_{i=2}^\infty\smsum_{\ell=i}^\infty\sfrac{\fl_\ell}{M^\ell}
               M^{\aleph'(\ell-i)}
=\sum_{i=2}^\infty M^{-\aleph' i}
        \sfrac{1}{1-M^{-(1+\aleph-\aleph')}}M^{-(1+\aleph-\aleph')i}
<\infty
$$
 the sequence of functions 
$Q_j(k)={\dst\sum\limits_{2\le i\le\ell\le j}}q^{(i,\ell)}(k)$ converges uniformly to
$$ 
 Q(k)={\dst\sum\limits_{2\le i\le\ell}}q^{(i,\ell)}(k)
$$
By Lemma \lemNPhoelder, with $j=\ell-i$, 
$C_0=\la_0^{1-2\upsilon}\sfrac{\fl_i}{M^i}$, $C_1=\la_0^{1-2\upsilon}\fl_i$,
 $\al=1-\aleph'+\aleph$ and $\be=\aleph'-\aleph$
$$\eqalign{
\big| \smsum_{\ell=i}^\infty q^{(i,\ell)}(k) 
        - \smsum_{\ell=i}^\infty q^{(i,\ell)}(k')\big|
&\le \const \la_0^{1-2\upsilon} \fl_i|k-k'|^{1-\aleph'+\aleph}
\cr
}$$
and by Lemma \lemNPhoelder, with $j=\ell-i$, 
$C_0=\la_0^{1-2\upsilon}\fl_i$, $C_1=\la_0^{1-2\upsilon}\fl_iM^i$,
 $\al=1-\aleph'+\aleph$ and $\be=\aleph'-\aleph$
$$\eqalign{
\big| \smsum_{\ell=i}^\infty \sfrac{\partial q^{(i,\ell)}}{\partial k_0}(k) 
- \smsum_{\ell=i}^\infty \sfrac{\partial q^{(i,\ell)}}{\partial k_0}(k')\big|
&\le \const \la_0^{1-2\upsilon} \fl_iM^{(1-\aleph'+\aleph)i}
                    |k-k'|^{1-\aleph'+\aleph}
\cr
}$$
Pick any $\half<\aleph''<\aleph$ and 
set $\ga=\sfrac{\aleph''}{1-\aleph'+\aleph}$. Note that, since $1-\aleph'>0$,
$0<\ga<1$.
Taking the $\ga^{\rm th}$ power of this bound and multiplying by the
$(1-\ga)^{\rm th}$ power of the bound
$$\eqalign{
\big| \smsum_{\ell=i}^\infty \sfrac{\partial q^{(i,\ell)}}{\partial k_0}(k) 
- \smsum_{\ell=i}^\infty \sfrac{\partial q^{(i,\ell)}}{\partial k_0}(k')\big|
&\le 2\sup_k\Big|\sfrac{\partial q^{(i,\ell)}}{\partial k_0}(k)\Big|
\le 4\smsum_{\ell=i}^\infty\la_0^{1-2\upsilon}\fl_i\sfrac{1}{M^{(1-\aleph'+\aleph)(\ell-i)}}\cr
&\le \const \la_0^{1-2\upsilon} \fl_i
\cr
}$$
gives
$$\eqalign{
\big| \smsum_{\ell=i}^\infty \sfrac{\partial q^{(i,\ell)}}{\partial k_0}(k) 
- \smsum_{\ell=i}^\infty \sfrac{\partial q^{(i,\ell)}}{\partial k_0}(k')\big|
&\le  \const \la_0^{1-2\upsilon} \fl_iM^{\aleph''i}|k-k'|^{\aleph''}
\cr
}$$
Hence
$$\deqalign{
\big| Q(k)- Q(k')\big|
&\le \const\la_0^{1-2\upsilon} |k-k'|^{1-\aleph'+\aleph}
&\le \la_0^{1-3\upsilon} |k-k'|^{5\over 6}\cr
\big|\sfrac{\partial Q}{\partial k_0}(k)
-\sfrac{\partial Q}{\partial k_0}(k')\big|
&\le \const\la_0^{1-2\upsilon} |k-k'|^{\aleph''}
&\le \la_0^{1-3\upsilon} |k-k'|^{1\over 2}\cr
}\EQN\eqnTNPQkbnd$$
if $\la_0$ is small enough. 

By hypothesis, every $q^{(i,\ell)}(k)$ vanishes on the 
$(i+2)^{\rm nd}$ neighbourhood.
Hence, for $k$ in the support of $\nu^{(m)}(k)$, $q^{(i,\ell)}(k)$ vanishes
when $i+2\le m$ and 
$$\eqalign{
\big| Q(k)\big|
&\le \smsum_{i=m-2}^{\infty}\ \smsum_{\ell=i}^\infty\big|q^{(i,\ell)}(k)\big|
\le 2\la_0^{1-2\upsilon}\smsum_{i=m-2}^{\infty}\ \smsum_{\ell=i}^\infty
          \sfrac{\fl_\ell}{M^\ell}M^{\aleph'(\ell-i)}
\le \const\la_0^{1-2\upsilon}\sfrac{\fl_m}{M^m}\cr
&\le\const\la_0^{1-2\upsilon}\min\big\{\big|ik_0-e(\k)\big|^{1+\aleph},1\big\}
\le\la_0^{1-3\upsilon}\min\big\{\big|ik_0-e(\k)\big|^{3\over 2},1\big\}\cr
\big| \sfrac{\partial Q}{\partial k_0}(k)\big|
&\le \smsum_{i=m-2}^{\infty}\ \smsum_{\ell=i}^\infty
             \big|\sfrac{\partial q^{(i,\ell)}}{\partial k_0}(k)\big|
\le 2\la_0^{1-2\upsilon}\smsum_{i=m-2}^{\infty}\ \smsum_{\ell=i}^\infty\fl_i 
                           M^{-(1-\aleph'+\aleph)(\ell-i)}
\le \const\la_0^{1-2\upsilon}\fl_m\cr
&\le\const\la_0^{1-2\upsilon}\min\big\{\big|ik_0-e(\k)\big|^\aleph,1\big\}
\le\la_0^{1-3\upsilon}\min\big\{\big|ik_0-e(\k)\big|^{1\over 2},1\big\}\cr
}\EQN\eqnTNPQkbndb$$
In general
$$
Q(k)- Q_j(k)=\smsum_{i=2}^{j}\smsum_{\ell=j+1}^\infty q^{(i,\ell)}(k)
+\smsum_{i=j+1}^{\infty}\smsum_{\ell=i}^\infty q^{(i,\ell)}(k)
$$
So, for $k$ in the support of $\nu^{(m)}(k)$,
$$\eqalign{
\big|Q(k)- Q_j(k)\big|
&\le \sum_{m-2\le i\le j}\ \smsum_{\ell=j+1}^\infty 
2\la_0^{1-2\upsilon}\sfrac{\fl_\ell}{M^\ell}M^{\aleph'(\ell-i)}
+\hskip-8pt\sum_{i=\max\{m-2,j+1\}}^{\infty}\ \smsum_{\ell=i}^\infty
2\la_0^{1-2\upsilon}\sfrac{\fl_\ell}{M^\ell}M^{\aleph'(\ell-i)}\cr
&\le \const\la_0^{1-2\upsilon}\Big\{ \sum_{m-2\le i\le j}
\sfrac{\fl_j}{M^j}M^{\aleph'(j-i)}
+\sum_{i=\max\{m-2,j+1\}}^{\infty}\sfrac{\fl_i}{M^i}\Big\}\cr
&\le \const\la_0^{1-2\upsilon}\Big\{ 
\sfrac{\fl_{\max\{m,j\}}}{M^{\max\{m,j\}}}M^{\aleph'({\max\{m,j\}}-m)}
+\sfrac{\fl_{\max\{m,j\}}}{M^{\max\{m,j\}}}\Big\}\cr
&\le \const\la_0^{1-2\upsilon}\sfrac{\fl_j}{M^m}
\Big\{M^{(\aleph'-1)({\max\{m,j\}}-m)}+1\Big\}\cr
&\le \const\la_0^{1-2\upsilon}\sfrac{\fl_j}{M^m}\cr
&\le\la_0^{1-3\upsilon}\fl_j\min\big\{\big|ik_0-e(\k)\big|,1\big\}\cr
}$$

\Item iii) That $\check u_j(-k_0,\k;0)=\overline{\check u_j(k_0,\k;0)}$
and $q^{(i,\ell)}(-k_0,\k)=\overline{q^{(i,\ell)}(k_0,\k)}$ are
consequences of Theorems \theoremNPinduction\ and \theoremNPtildeinduction,
respectively.

\endproof

\lemma{\STM\lemNPgtwoft}{ 
$$\eqalign{
G_{2}\big((0,0,\uparrow),(x_0,\x,\uparrow)\big)
&= \int \sfrac{dk_0}{2\pi} \sfrac{d^d\k}{(2\pi)^d}\ 
e^{-i(-k_0x_0 +\k\cdot\x)}\Big\{
\sfrac{U(\k)}{ik_0-e(\k)-P(k)} 
+\sfrac{Q(k)}{[ik_0-e(\k)]^2}\Big\}\cr
}$$
(with the value of $G_{2}\big((0,0,\uparrow),(x_0,\x,\uparrow)\big)$
at $x_0=0$ defined through the limit $x_0\rightarrow 0+$)
and the Fourier transform of $G_{2}\big((0,0,\uparrow),(x_0,\x,\uparrow)\big)$ 
is $\sfrac{U(\k)}{ik_0-e(\k)-P(k)} 
+\sfrac{Q(k)}{[ik_0-e(\k)]^2}$, which is continuous at all points 
$(k_0,\k)$ for which $ik_0-e(\k)\ne 0$.

}
\prf
By the definitions of $Q_j$ and $\check u(k)$ (Definition \defNPsectrepr.iv)
$$
2G^\rg_{j,2}\big((0,0,\uparrow,1),(x_0,\x,\uparrow,0)\big)
= \int \sfrac{dk_0}{2\pi} \sfrac{d^d\k}{(2\pi)^d}\ 
e^{-i(-k_0x_0 +\k\cdot\x)}\,\Big[
\sfrac{\nu^{(\le j)}(k)}{ik_0-e(\k)-\check u_j(k;0)}
+\sfrac{Q_j(k)}{[ik_0-e(\k)]^2}\Big]
$$
Hence
$$\eqalign{
&\int \sfrac{dk_0}{2\pi} \sfrac{d^d\k}{(2\pi)^d}\ 
e^{-i(-k_0x_0 +\k\cdot\x)}\Big\{
\sfrac{U(\k)}{ik_0-e(\k)-P(k)} 
+\sfrac{Q(k)}{[ik_0-e(\k)]^2}\Big\}
-2G^\rg_{j,2}\big((0,0,\uparrow,1),(x_0,\x,\uparrow,0)\big)\cr
&\hskip.1in= \int\!\! \sfrac{dk_0}{2\pi} \sfrac{d^d\k}{(2\pi)^d}\ 
e^{-i(-k_0x_0 +\k\cdot\x)}
\Big[\sfrac{U(\k)-\nu^{(\le j)}(k)}{ik_0-e(\k)-P(k)}
+\sfrac{\nu^{(\le j)}(k)[P(k)-\hat u_j(k;0)]}
{[ik_0-e(\k)-\hat u_j(k;0)][ik_0-e(\k)-P(k)]}
+\sfrac{Q(k)-Q_j(k)}{[ik_0-e(\k)]^2}\Big]\cr
}$$
Let $\tilde U(\k)$ be the characteristic function of the support of $U(\k)$.
By Lemma \lemNPgtwoamp, all three of
$$\eqalign{
\Big|\sfrac{Q(k)-Q_j(k)}{[ik_0-e(\k)]^2}\Big|
&\le \sfrac{\fl_j\min\{|ik_0-e(\k)|,1\}}{|ik_0-e(\k)|^2}\tilde U(\k)\cr
\Big|\sfrac{U(\k)-\nu^{(\le j)}(k)}{ik_0-e(\k)-P(k)}\Big|
&=\Big|\sfrac{\nu^{(> j)}(k)}{ik_0-e(\k)-P(k)}\Big|
\le 2\sfrac{\nu^{(> j)}(k)}{|ik_0-e(\k)|}\cr
\Big|\sfrac{\nu^{(\le j)}(k)[P(k)-\hat u_j(k;0)]}
{[ik_0-e(\k)-\hat u_j(k;0)][ik_0-e(\k)-P(k)]}\Big|
&\le 4\sfrac{\fl_j\min\{|k_0|,1\} }{|ik_0-e(\k)|^2}U(\k)
}$$
converge to zero in $L^1(\bbbr^{d+1})$ as $j\rightarrow\infty$.
Consequently, $2G^\rg_{j,2}\big((0,0,\uparrow,1),(x_0,\x,\uparrow,0)\big)$ 
converges uniformly to $\int \sfrac{dk_0}{2\pi} \sfrac{d^d\k}{(2\pi)^d}\ 
e^{-i(-k_0x_0 +\k\cdot\x)}
\Big\{\sfrac{U(\k)}{ik_0-e(\k)-P(k)} 
+\sfrac{Q(k)}{[ik_0-e(\k)]^2}\Big\}$ as $j$
tends to infinity. In the proof of Theorem \theoremNPmainthI, we defined
$G_2({\sst (0,0,\uparrow),(x_0,\x,\uparrow)})$ as the pointwise limit
of the $2G^\rg_{j,2}({\sst (0,0,\uparrow,1),(x_0,\x,\uparrow,0)})$'s. Hence
$$
G_{2}\big((0,0,\uparrow),(x_0,\x,\uparrow)\big)
= \int \sfrac{dk_0}{2\pi} \sfrac{d^d\k}{(2\pi)^d}\ 
e^{-i(-k_0x_0 +\k\cdot\x)}\Big\{
\sfrac{U(\k)}{ik_0-e(\k)-P(k)} 
+\sfrac{Q(k)}{[ik_0-e(\k)]^2}\Big\} 
$$

Write
$$
G_{2}\big((0,0,\uparrow),(x_0,\x,\uparrow)\big)
=a(x_0,\x)+b(x_0,\x)+c(x_0,\x)
$$
with
$$\eqalign{
a(x_0,\x)
&=\int \sfrac{dk_0}{2\pi} \sfrac{d^d\k}{(2\pi)^d}\ 
e^{-i(-k_0x_0 +\k\cdot\x)}\,\sfrac{U(\k)}{i(1-w(\k))k_0-e(\k)}\cr
b(x_0,\x)
&=\int \sfrac{dk_0}{2\pi} \sfrac{d^d\k}{(2\pi)^d}\ 
e^{-i(-k_0x_0 +\k\cdot\x)}\,\sfrac{U(\k)[P(k)-iw(\k)k_0]}{[i(1-w(\k))k_0-e(\k)][ik_0-e(\k)-P(k)]}\cr
c(x_0,\x)
&=\int \sfrac{dk_0}{2\pi} \sfrac{d^d\k}{(2\pi)^d}\ 
e^{-i(-k_0x_0 +\k\cdot\x)}\,\sfrac{Q(\k)}{[ik_0-e(\k)]^2}\cr
w(\k)&=\sfrac{1}{i}\sfrac{\partial P}{\partial k_0}(0,\k)\cr
}$$
By repeated use of Lemma \lemNPgtwoamp,
$$\deqalign{
\Big|\sfrac{Q(\k)}{[ik_0-e(\k)]^2}\Big|
&\le \la_0^{1-3\upsilon}\sfrac{\min\{|ik_0-e(\k)|^{3\over 2},1\}}
                        {|ik_0-e(\k)|^2}\tilde U(\k)
&\in L^2(\bbbr^{d+1})\cr
\Big|\sfrac{U(\k)[P(k)-iw(\k)k_0]}{[i(1-w(\k))k_0-e(\k)][ik_0-e(\k)-P(k)]}\Big|
&\le 8\la_0^{1-2\upsilon}\sfrac{U(\k)\min\{|k_0|^{3/2},|k_0|\}}
                        {|ik_0-e(\k)|^2}
&\in L^2(\bbbr^{d+1})\cr
}$$
so the Fourier transform of $b(x_0,\x)+c(x_0,\x)$ exists and equals 
$$\sfrac{U(\k)[P(k)-iw(\k)k_0]}{[i(1-w(\k))k_0-e(\k)][ik_0-e(\k)-P(k)]}
+\sfrac{Q(\k)}{[ik_0-e(\k)]^2}$$
Observe that 
$$\eqalign{
a(x_0,\x)
&=\sfrac{1}{1-w(\k)}\int  \sfrac{d^d\k}{(2\pi)^d}\ 
e^{-i\k\cdot\x}\,U(\k)e^{x_0e(\k)/(1-w(\k))}\chi(\k,x_0)\cr
}$$
where
$$
\chi(\k,x_0)=\cases{1& if $e(\k)<0$ and $x_0\ge 0$\cr
                                         -1& if $e(\k)>0$ and $x_0< 0$\cr       
                                         0& otherwise\cr}
$$
As $\sfrac{1}{1-w(\k)}U(\k)e^{x_0e(\k)/(1-w(\k))}$ is in $L^2(\bbbr^d)$, 
the spatial Fourier transform $\int d^d\x\ e^{i\k\cdot\x}\,a(x_0,\x)$ exists 
and equals $\sfrac{1}{1-w(\k)}U(\k)e^{x_0e(\k)/(1-w(\k))}\chi(\k,x_0)$. 
A direct computation now shows that the temporal Fourier transform
$\sfrac{1}{1-w(\k)}\int dx_0\ e^{-ik_0x_0}\ U(\k)e^{x_0e(\k)/(1-w(\k))}
\chi(\k,x_0)$
 exists and equals $\sfrac{U(\k)}{i(1-w(\k))k_0-e(\k)}$. Thus
the Fourier transform of $G_{2}\big((0,0,\uparrow),(x_0,\x,\uparrow)\big)$ 
exists and equals 
$\sfrac{U(\k)}{ik_0-e(\k)-P(k)}+\sfrac{Q(k)}{[ik_0-e(\k)]^2}$, which,
by Lemma \lemNPgtwoamp, is continuous except when $ik_0-e(\k)\ne 0$.

\endproof

\lemma{\STM\lemTNPabstractjump}{
Let $S(k_0,\k)$ be a function that obeys
\item{$\bullet$} $S(k_0,\k)$ and $\sfrac{\partial S}{\partial k_0}(k_0,\k)$
are continuous in $(k_0,\k)$ and there are $C,\veps>0$ such that
$$
\big|\sfrac{\partial S}{\partial k_0}(k_0,\k)
   -\sfrac{\partial S}{\partial k_0}(0,\k)\big|\le C|k_0|^\veps
$$
\item{$\bullet$} $S(0,\k)$ and $\sfrac{1}{i}\,\sfrac{\partial S}{\partial k_0}(0,\k)$
are real. 
\item{$\bullet$} $\big|S(k_0,\k)\big|$, 
$\Big|\sfrac{\partial S}{\partial k_0}(k_0,\k)\Big|\le\half$
and $\big|S(0,\k)\big|\le\half |e(\k)|$

\noindent 
Define
$$
N(\k,\tau)=\int\sfrac{dk_0}{2\pi}\ 
         \sfrac{e^{i k_0\tau}}{i k_0-e(\k)-S(k_0,\k)}
\qquad
N(\k)=\lim_{\tau\rightarrow 0+}N(\k,\tau)
$$
Then
$N(\k)$ is continuous except on the Fermi surface $F$.
If $\bar\k\in F$, then $\lim\limits_{\k\rightarrow\bar\k\atop e(\k)>0}N(\k)$ and 
$\lim\limits_{\k\rightarrow\bar\k\atop e(\k)<0}N(\k)$ exist and obey
$$
\lim_{\k\rightarrow\bar\k\atop e(\k)<0}N(\k)
-\lim_{\k\rightarrow\bar\k\atop e(\k)>0}N(\k)
=\big[1-\sfrac{1}{i}\,\sfrac{\partial S}{\partial k_0}(0,\bar\k)\big]^{-1}
$$
}
\prf
Define
$$\eqalign{
A(\k)&=1-\sfrac{1}{i}\,\sfrac{\partial S}{\partial k_0}(0,\k)\cr
E(\k)&=e(\k)+ S(0,\k)\cr
R(k_0,\k)&= S(k_0,\k)- S(0,\k)
-\sfrac{\partial S}{\partial k_0}(0,\k)\,k_0\cr
}$$
and observe that
$$\eqalign{
i k_0-e(\k)- S(k_0,\k)
&=i k_0-\sfrac{\partial S}{\partial k_0}(0,\k)\,k_0
-e(\k)- S(0,\k)-R(k_0,\k)\cr
&=i A(\k)k_0-E(\k)-R(k_0,\k)\cr
}$$
Hence, for any $\et>0$,
$$
N(\k,\tau)=I_1(\k,\tau)+I_2(\k,\tau)+I_3(\k,\tau)-I'_3(\k,\tau)+I_4(\k,\tau)
$$
with
$$\eqalign{
I_1(\k,\tau)&=\int_{|k_0|<\et}\sfrac{dk_0}{2\pi}\ 
         \sfrac{e^{i k_0\tau}}{i A(\k)k_0-E(\k)}\cr
I_2(\k,\tau)&=\int_{|k_0|<\et}\sfrac{dk_0}{2\pi}\ \Big[
         \sfrac{e^{i k_0\tau}}
         {i A(\k)k_0-E(\k)-R(k_0,\k)}
         -\sfrac{e^{i k_0\tau}}
         {i A(\k)k_0-E(\k)}\Big]\cr
I_3(\k,\tau)&=\int_{\bbbr}\sfrac{dk_0}{2\pi}\ 
         \sfrac{e^{i k_0\tau}}{i k_0-e(\k)}\cr
I'_3(\k,\tau)&=\int_{|k_0|<\et}\sfrac{dk_0}{2\pi}\ 
         \sfrac{e^{i k_0\tau}}{i k_0-e(\k)}\cr
I_4(\k,\tau)&=\int_{|k_0|\ge\et}\sfrac{dk_0}{2\pi}\ \Big[
         \sfrac{e^{i k_0\tau}}
         {i k_0-e(\k)- S(k_0,\k)}
         -\sfrac{e^{i k_0\tau}}
         {i k_0-e(\k)}\Big]\cr
}$$
We shall later fix some small $\et$.

\noindent{\it Control of $I_1$:\ \ }
Let $\ep>0$. On the set $D_\ep=\set{(\tau,\k)}{\tau\in\bbbr,\ |e(\k)|>\ep}$, the integrand
$\sfrac{e^{i k_0\tau}}{i A(\k)k_0-E(\k)}$ is continuous in $(\tau,\k)$,
for each fixed $k_0$, and is uniformly bounded by $\sfrac{2}{\ep}$. Hence,
by the Lebesgue dominated convergence theorem, $I_1(\tau,\k)$ is continuous 
on $D_\ep$ and obeys
$$\eqalign{
\lim_{\tau\rightarrow 0}I_1(\tau,\k)
&=\int_{|k_0|<\et}\sfrac{dk_0}{2\pi}\ 
         \sfrac{1}{i A(\k)k_0-E(\k)}\cr
&=-\int_{|k_0|<\et}\sfrac{dk_0}{2\pi}\ 
         \sfrac{i A(\k)k_0+E(\k)}{A(\k)^2k_0^2+E(\k)^2}\cr
&=-\int_{|k_0|<\et}\sfrac{dk_0}{2\pi}\ 
         \sfrac{E(\k)}{A(\k)^2k_0^2+E(\k)^2}\cr
&=-\sfrac{\sgn\, E(\k)}{A(\k)}\int_{|k_0'|<\et\big|{A(\k)\over E(\k)}\big|}\sfrac{dk_0'}{2\pi}\ 
         \sfrac{1}{{k_0'}^2+1}\cr
&=-\sfrac{\sgn\, e(\k)}{\pi A(\k)}
\tan^{-1}\big(\et\big|\sfrac{A(\k)}{ E(\k)}\big|\big)\cr
}$$ 
Since
$$
\lim_{e(\k)\rightarrow 0\atop e(\k)\ne 0}
\tan^{-1}\big(\et\big|\sfrac{A(\k)}{ E(\k)}\big|\big)
=\sfrac{\pi}{2}
$$
we have
$$
\lim_{\k\rightarrow\bar\k\atop e(\k)<0}\lim_{\tau\rightarrow 0}I_1(\tau,\k)
-\lim_{\k\rightarrow\bar\k\atop e(\k)>0}\lim_{\tau\rightarrow 0}I_1(\tau,\k)
=1/A(\bar\k)
\EQN\eqnLPone$$

\noindent{\it Control of $I_2$:\ \ }
\noindent
Since $|i A(\k)k_0-E(\k)|\ge\half|k_0|$ and, 
for some $\tilde k_0$ between $0$ and $k_0$,
$$
\big|R(k_0,\k)\big|
=\big|k_0\big[\sfrac{\partial S}{\partial k_0}(\tilde k_0,\k)
-\sfrac{\partial S}{\partial k_0}(0,\k)\big]\big|\le C|k_0|^{1+\veps}
$$ 
and
$$\eqalign{
\big|i A(\k)k_0-E(\k)&-R(k_0,\k)\big|
\ge \big|A(\k)k_0-\Im R(k_0,\k)\big|
= \big|A(\k)k_0-\Im [R(k_0,\k)-R(0,\k)]\big|\cr
&= \big|k_0\big[A(\k)-\Im \sfrac{\partial R}{\partial k_0}(\tilde k_0,\k)\big]\big|
= \big|k_0\big[A(\k)-\Im \big\{\sfrac{\partial S}{\partial k_0}(\tilde k_0,\k)
-\sfrac{\partial S}{\partial k_0}(0,\k)\big\}\big]\big|\cr
&\ge\sfrac{1}{4}|k_0|
}$$
if $\et$ is small enough, the integrand
$$
\Big[
         \sfrac{e^{i k_0\tau}}
         {i A(\k)k_0-E(\k)-R(k_0,\k)}
         -\sfrac{e^{i k_0\tau}}
         {i A(\k)k_0-E(\k)}\Big]
=\sfrac{e^{i k_0\tau}R(k_0,\k)}
         {[i A(\k)k_0-E(\k)][i A(\k)k_0-E(\k)-R(k_0,\k)]}
$$
is uniformly bounded in magnitude, on the domain of integration,
 by the integrable function $\sfrac{8C}{|k_0|^{1-\veps}}$.
Since the integrand is, for each fixed $k_0\ne 0$, continuous in $(\tau,\k)$,
the Lebesgue dominated convergence theorem implies that  $I_2(\tau,\k)$ is 
continuous $(\tau,\k)$ and, in particular,
$$
\lim_{\k\rightarrow\bar\k\atop e(\k)<0}\lim_{\tau\rightarrow 0}I_2(\tau,\k)
-\lim_{\k\rightarrow\bar\k\atop e(\k)>0}\lim_{\tau\rightarrow 0}I_2(\tau,\k)
=0
\EQN\eqnLPtwo$$

\noindent{\it Control of $I_3$ and $I'_3$:\ \ }
\noindent
By residues, for all $\tau>0$ and $e(\k)\ne 0$,
$$
I_3(\tau,\k)=\cases{0 &if $e(\k)>0$\cr
                    e^{e(\k)\tau} &if $e(\k)<0$\cr}
$$
Hence $I_3(\tau,\k)$ is continuous in $(\tau,\k)$ for all $\k$ with 
$e(\k)\ne0$, and obeys
$$
\lim_{\k\rightarrow\bar\k\atop e(\k)<0}\lim_{\tau\rightarrow 0+}I_3(\tau,\k)
-\lim_{\k\rightarrow\bar\k\atop e(\k)>0}\lim_{\tau\rightarrow 0+}I_3(\tau,\k)
=1
\EQN\eqnLPthree$$
As in the ``Control of $I_1$'', but with $A(\k)=1$ and $E(\k)=e(\k)$,
$I'_3(\tau,\k)$ is continuous in $(\tau,\k)$ for all $\k$ with $e(\k)\ne 0$, 
and obeys
$$
\lim_{\k\rightarrow\bar\k\atop e(\k)<0}\lim_{\tau\rightarrow 0}I'_3(\tau,\k)
-\lim_{\k\rightarrow\bar\k\atop e(\k)>0}\lim_{\tau\rightarrow 0}I'_3(\tau,\k)
=1
\EQN\eqnLPthreep$$
\noindent{\it Control of $I_4$:\ \ }
Since $\big| S(k_0,\k)\big|\le\half$, $|i k_0-e(\k)|\ge |k_0|$ and,
for some $\tilde k_0$ between $0$ and $k_0$,
$$\eqalign{
\big|i k_0-e(\k)- S(k_0,\k)\big|
&\ge \big|k_0-\Im  S(k_0,\k)\big|
= \big|k_0-\Im [ S(k_0,\k)- S(0,\k)]\big|\cr
&= \big|k_0\big[1-\Im \sfrac{\partial  S}{\partial k_0}(\tilde k_0,\k)\big]\big|
\ge\half|k_0|
}$$
the integrand 
$$
\Big[ \sfrac{e^{i k_0\tau}}{i k_0-e(\k)- S(k_0,\k)}
         -\sfrac{e^{i k_0\tau}}{i k_0-e(\k)}\Big]
=\sfrac{e^{i k_0\tau} S(k_0,\k)}
         {[i k_0-e(\k)][i k_0-e(\k)- S(k_0,\k)]}
$$
is uniformly bounded in magnitude, on the domain of integration,
 by the integrable function $\sfrac{1}{k_0^2}$.
Since the integrand is, for each fixed $k_0$, continuous in $(\tau,\k)$,
the Lebesgue dominated convergence theorem implies that  $I_4(\tau,\k)$ is 
continuous $(\tau,\k)$ and, in particular,
$$
\lim_{\k\rightarrow\bar\k\atop e(\k)<0}\lim_{\tau\rightarrow 0}I_4(\tau,\k)
-\lim_{\k\rightarrow\bar\k\atop e(\k)>0}\lim_{\tau\rightarrow 0}I_4(\tau,\k)
=0
\EQN\eqnLPfour$$

\vskip.3cm
Combining (\eqnLPone--\eqnLPfour) gives the desired jump.

\endproof

\proof{ of Theorem \theoremNPmainthII\ from Theorem \theoremNPtildeinduction}
Define
$$\meqalign{
N_1(\k,\tau)&=\int\sfrac{dk_0}{2\pi}\ e^{i k_0\tau}
\sfrac{U(\k)}{ik_0-e(\k)-P(k)} 
&&
N_1(\k)&=\lim_{\tau\rightarrow 0+}N_1(\k,\tau)\cr
N_2(\k,\tau)&=\int\sfrac{dk_0}{2\pi}\ e^{i k_0\tau}
\sfrac{Q(k)}{[ik_0-e(\k)]^2}
&&
N_2(\k)&=\lim_{\tau\rightarrow 0+}N_2(\k,\tau)\cr
}$$
Then, by Lemma \lemNPgtwoft,
$$
n(\k)=N_1(\k)+N_2(\k)
$$
By Lemma \lemNPgtwoamp.ii, the integrand 
$e^{i k_0\tau}\sfrac{Q(k)}{[ik_0-e(\k)]^2}$ is continuous in $\tau$ and
$\k$, except possibly when $ik_0-e(\k)=0$, and is uniformly bounded
by 
$$
\Big|e^{i k_0\tau}\sfrac{Q(k)}{[ik_0-e(\k)]^2} \Big|
\le \la_0^{1-3\upsilon}\sfrac{\min\{|ik_0-e(\k)|^{3\over 2},1\}}
                        {|ik_0-e(\k)|^2}
\le  \la_0^{1-3\upsilon}\min\Big\{\sfrac{1}{k_0^2},\sfrac{1}{\sqrt{|k_0|}}\Big\}
\in L^1(\bbbr)
$$
Hence, by the Lebesgue dominated convergence theorem,
$N_2(\k,\tau)$ is continuous. Hence, so is $N_2(\k)$.
In Lemma \lemNPgtwoamp, parts (i) and (iii), we showed that $P(k)$ satisfies the
conditions imposed on the function $ S(k)$ in Lemma  \lemTNPabstractjump.
So, by Lemma  \lemTNPabstractjump, with $ S(k)$ replaced by $P(k)$, 
$N_1(\k)$ is continuous except on $F$ and if $\bar\k\in F$, then 
$\lim\limits_{\k\rightarrow\bar\k\atop e(\k)>0}N_1(\k)$ and 
$\lim\limits_{\k\rightarrow\bar\k\atop e(\k)<0}N_1(\k)$ exist and obey
$$
\lim_{\k\rightarrow\bar\k\atop e(\k)<0}N_1(\k)
-\lim_{\k\rightarrow\bar\k\atop e(\k)>0}N_1(\k)
=\big[1-\sfrac{1}{i}\,\sfrac{\partial P}{\partial k_0}(0,\bar\k)\big]^{-1}
$$
Since $N_2(\k)$ is continuous, $n(\k)$ is continuous except on $F$
and if $\bar\k\in F$, then 
$\lim\limits_{\k\rightarrow\bar\k\atop e(\k)>0}n(\k)$ and 
$\lim\limits_{\k\rightarrow\bar\k\atop e(\k)<0}n(\k)$ exist and obey
$$
\lim_{\k\rightarrow\bar\k\atop e(\k)<0}n(\k)
-\lim_{\k\rightarrow\bar\k\atop e(\k)>0}n(\k)
=\big[1-\sfrac{1}{i}\,\sfrac{\partial P}{\partial k_0}(0,\bar\k)\big]^{-1}
$$
The remaining conclusions of Theorem \theoremNPmainthII\ were proven in 
Lemma \lemNPgtwoft.
\endproof

If $\k$ is such that $U(\k)=1$, then
$$
\sfrac{U(\k)}{ik_0-e(\k)-P(k)} 
+\sfrac{Q(k)}{[ik_0-e(\k)]^2}
=\sfrac{1}{ik_0-e(\k)- \Si(k)}
$$
with
$$
\Si(k)
=\frac{P(k)+Q(k)-Q(k)\sfrac{P(k)}{ik_0-e(\k)}}
{1+\sfrac{Q(k)}{ik_0-e(\k)}-\sfrac{P(k)}{ik_0-e(\k)}\sfrac{Q(k)}{ik_0-e(\k)}}
$$
$\Si(k)$ is usually called the proper self--energy. For the sake of completeness, we summarize the properties of $\Si(k)$ proven in this paper and its relationship to the function $P(k)$ used above.

\lemma{\STM\lemNPpsebnd}{ Let $\half<\aleph''<\aleph<\sfrac{2}{3}$. Then
$$\eqalign{
\big|\Si(k)-P(k)\big|
&\le\la_0^{1-4\upsilon}\min\{\big|ik_0-e(\k)\big|^{3\over 2},1\big\}\cr
\big|\Si(k)\big|,\big|\sfrac{\partial\Si}{\partial k_0}(k)\big|
&\le \la_0^{1-4\upsilon}\cr
\big|\Si(k)-\Si(k')\big|,\ 
\big|\sfrac{\partial\Si}{\partial k_0}(k)
-\sfrac{\partial\Si}{\partial k_0}(k')\big|
& \le\const\la_0^{1-2\upsilon}|k-k'|^{\aleph''}\le \la_0^{1-4\upsilon} |k-k'|^{1\over 2}\cr
}$$

}
\prf 
Writing $E(k)=ik_0-e(\k)$ and differentiating
$$\eqalign{
\Si(k)
&=\sfrac{E(k)^2P(k)+E(k)^2Q(k)-E(k)P(k)Q(k)}
{E(k)^2+E(k)Q(k)-P(k)Q(k)}
}$$
with respect to $k_0$ yields
$$\eqalign{
\sfrac{\partial \Si}{\partial k_0}(k)
&=\frac{2iEP+E^2\sfrac{\partial P}{\partial k_0}
+2iEQ+E^2\sfrac{\partial Q}{\partial k_0}
-iPQ-E\sfrac{\partial P}{\partial k_0}Q
-EP\sfrac{\partial Q}{\partial k_0}}
{E(k)^2+E(k)Q(k)-P(k)Q(k)}
\cr
&\hskip.5in
-\frac{[E^2P+E^2Q-EPQ][2iE+iQ+E\sfrac{\partial Q}{\partial k_0}
-\sfrac{\partial P}{\partial k_0}Q-P\sfrac{\partial Q}{\partial k_0}]}
{[E(k)^2+E(k)Q(k)-P(k)Q(k)]^2}\cr
}$$
Implementing the cancellation (of terms of the form $E^nP$)
$$
\sfrac{2iEP}{E^2+EQ-PQ}
-\sfrac{[E^2P+E^2Q-EQP]2iE}{[E^2+EQ-PQ]^2}
=-\sfrac{2iE^3Q+2iEP^2Q-4iE^2PQ}{[E^2+EQ-PQ]^2}
$$
we have
$$\eqalign{
\sfrac{\partial \Si}{\partial k_0}(k)
&=\frac{E^2\sfrac{\partial P}{\partial k_0}
+2iEQ+E^2\sfrac{\partial Q}{\partial k_0}
-iPQ-E\sfrac{\partial P}{\partial k_0}Q
-EP\sfrac{\partial Q}{\partial k_0}}
{E(k)^2+E(k)Q(k)-P(k)Q(k)}
\cr
&-\frac{[E^2P+E^2Q-EPQ][iQ+E\sfrac{\partial Q}{\partial k_0}
-\sfrac{\partial P}{\partial k_0}Q-P\sfrac{\partial Q}{\partial k_0}]
+2iE^3Q+2iEP^2Q-4iE^2PQ}
{[E(k)^2+E(k)Q(k)-P(k)Q(k)]^2}\cr
}$$
Observe that, if we weight $E$, $P$ and $Q$ with degree one and 
we weight $\sfrac{\partial P}{\partial k_0}$ and 
$\sfrac{\partial Q}{\partial k_0}$ with degree zero, then both the
numerator and denominator of the first term are homogeneous of degree two
and both the numerator and denominator of the second term are homogeneous of 
degree four. This sort of homogeneity information can be used in place
of the explicit formula for $\sfrac{\partial \Si}{\partial k_0}(k)$.

Define, for $m=0,1,2$,
$$\deqalign{
\tilde Q^{(m)}(k)&=\sfrac{(ik_0)^m Q(k)}{[ik_0-e(\k)]^{m+1}}
&=\sum_{2\le j\le\ell}\sfrac{(ik_0)^mq^{(j,\ell)}(k)}{[ik_0-e(\k)]^{m+1}}
&=\sum_{2\le j\le\ell}\tilde q^{(j,\ell,m)}(k)\cr
\tilde Q^{(m)}_0(k)&=\sfrac{ (ik_0)^m}{[ik_0-e(\k)]^m}
\sfrac{\partial Q}{\partial k_0}
&=\sum_{2\le j\le\ell}\sfrac{(ik_0)^m}{[ik_0-e(\k)]^m}
\sfrac{\partial q^{(j,\ell)}}{\partial k_0}
&=\sum_{2\le j\le\ell}\tilde q^{(j,\ell,m)}_0(k)\cr
\tilde P(k)&=\sfrac{ P(k)}{ik_0}
&=\sum_{j=2}^\infty\sfrac{\check p^{(j)}(k)}{ik_0}
&=\sum_{j=2}^\infty\tilde p^{(j)}(k)\cr
}$$
with $\tilde q^{(j,\ell,m)}(k)=\sfrac{(ik_0)^mq^{(j,\ell)}(k)}{[ik_0-e(\k)]^{m+1}}$
and $\tilde q^{(j,\ell,m)}_0(k)=\sfrac{(ik_0)^m}{[ik_0-e(\k)]^m}
\sfrac{\partial q^{(j,\ell)}}{\partial k_0}$
 vanishing on the $(j+2)^{\rm nd}$ neighbourhood and obeying
$$\eqalign{
\sup_k\big|\rD^\de \tilde q^{(j,\ell,m)}(k) \big|,
\sup_k\big|\rD^\de \tilde q^{(j,\ell,m)}_0(k) \big|
&\le \const\la_0^{1-2\upsilon}\sfrac{\fl_\ell}{M^\ell}M^{\aleph'(\ell-j)}
M^jM^{j\de_0}M^{\ell|\bde|}\cr
&\le 
\const\la_0^{1-2\upsilon}\fl_jM^{-(1+\aleph-\aleph')(\ell-j)}M^{\ell|\de|}\cr
}\EQN\eqnTNPtildeqbnds$$
for $|\de|\le 1$ and 
$\tilde p^{(j)}(k)=\sfrac{\check p^{(j)}(k)}{ik_0}
=-i\int_0^1\sfrac{\partial\check p^{(j)}}{\partial\, k_0}(tk_0,\k)\,dt$ obeying
$$
\sup_k\big|\rD^\de \tilde p^{(j)}(k) \big|
\le \const \la_0^{1-\upsilon}\fl_jM^{j|\de|}
$$
for all $|\de|\le 1$.
 The right hand side of (\eqnTNPtildeqbnds) coincides with the bounds on
$\sfrac{\partial q^{(i,\ell)}}{\partial k_0}$ that were used in the derivation 
of the second bounds of (\eqnTNPQkbnd) and (\eqnTNPQkbndb). 
Hence, for $m=0,1,2$,
$$
|\tilde Q^{(m)}(k)|, |\tilde Q^{(m)}_0(k)|\le\const\la_0^{1-2\upsilon}
                         \min\{|ik_0-e(\k)|^\aleph,1\}\qquad
 |\tilde P(k)|\le\la_0^{1-3\upsilon}
\EQN\eqnTNPtildeQbnds$$
and
$$
\big|\tilde Q^{(m)}(k)-\tilde Q^{(m)}(k')\big|, 
\big|\tilde Q^{(m)}_0(k)-\tilde Q^{(m)}_0(k')\big|, 
\big|\tilde P(k)-\tilde P(k')\big|
\le\const\la_0^{1-2\upsilon}|k-k'|^{\aleph''}
\EQN\eqnTNPtildeQmQbnds$$
The bound on $\tilde P$ is a direct application of Lemma \lemNPhoelder\ 
with $\al=\aleph$ and $\be=1-\aleph$.
In terms of these new functions,
$$\eqalign{
\Si(k)
&=\frac{P(k)+Q(k)-P(k)\tilde Q^{(0)}(k)}
{1+\tilde Q^{(0)}(k)-\tilde P(k)\tilde Q^{(1)}(k)}\cr
\Si(k)&-P(k)
=\frac{Q(k)-2P(k)\tilde Q^{(0)}(k)+P(k)\tilde P(k)\tilde Q^{(1)}(k)}
{1+\tilde Q^{(0)}(k)-\tilde P(k)\tilde Q^{(1)}(k)}\cr
\sfrac{\partial \Si}{\partial k_0}(k)
&=\frac{\sfrac{\partial P}{\partial k_0}
+2i\tilde Q^{(0)}+\tilde Q^{(0)}_0
-i\tilde P\tilde Q^{(1)}-\sfrac{\partial P}{\partial k_0}\tilde Q^{(0)}
-\tilde P\tilde Q^{(1)}_0}
{1+\tilde Q^{(0)}-\tilde P \tilde Q^{(1)}}
\cr
&-\frac{\tilde P[i\tilde Q^{(1)}+\tilde Q^{(1)}_0
-\sfrac{\partial P}{\partial k_0}\tilde Q^{(1)}-\tilde P\tilde Q^{(2)}_0]}
{[1+\tilde Q^{(0)}-\tilde P \tilde Q^{(1)}]^2}\cr
&-\frac{[\tilde Q^{(0)}-\tilde P\tilde Q^{(1)}][i\tilde Q^{(0)}
+\tilde Q^{(0)}_0
-\sfrac{\partial P}{\partial k_0}\tilde Q^{(0)}
-\tilde P\tilde Q^{(1)}_0]
+2i\tilde Q^{(0)}+2i\tilde P^2\tilde Q^{(2)}-4i\tilde P\tilde Q^{(1)}}
{[1+\tilde Q^{(0)}-\tilde P \tilde Q^{(1)}]^2}\cr
}\EQN\eqnNPexplicitdsigmadk$$
Both $\Si(k)$ and $\sfrac{\partial \Si}{\partial k_0}(k)$ are rational
functions in the variables $P$, $\tilde P$, $\sfrac{\partial P}{\partial k_0}$,
$Q$, $\tilde Q^{(0)}$, $\tilde Q^{(1)}$, $\tilde Q^{(2)}$, $\tilde Q^{(0)}_0$, 
$\tilde Q^{(1)}_0$ and $\tilde Q^{(2)}_0$. As all of these variables are
bounded in magnitude by $\la_0^{1-3\upsilon}$, the numerators contain 
no constant terms and the denominators are bounded away from zero,
$$
\big|\Si(k)\big|,\big|\sfrac{\partial \Si}{\partial k_0}(k)\big|
\le\const \la_0^{1-3\upsilon}
\le \la_0^{1-4\upsilon}
$$
Each term in the numerator of $\Si(k)-P(k)$ contains either a factor 
of $Q(k)$ or a factor of $P(k)\tilde Q^{(m)}(k)$ so that, using the bound
on $Q(k)$ in (\eqnTNPQkbndb),  
$$
\big|\Si(k)-P(k)\big|
\le\const \la_0^{1-2\upsilon} \min\{|ik_0-e(\k)|^{1+\aleph},1\}
\le \la_0^{1-4\upsilon}\min\{|ik_0-e(\k)|^{3\over 2},1\}
$$
Applying a Taylor expansion with degree one remainder and with expansion point
$x_0=P(k)$, $y_0=\tilde P(k)$, $\cdots$ , $z_0=\tilde Q^{(2)}_0(k)$
and evaluation point  $x=P(k')$, $y=\tilde P(k')$, $\cdots$ , 
$z=\tilde Q^{(2)}_0(k')$,
$$\eqalign{
\big|\Si(k)-\Si(k')\big|
&\le \const\max\Big\{|P(k)-P(k')|, |\tilde P(k)-\tilde P(k')|,
\cdots, |\tilde Q^{(2)}_0(k)-\tilde Q^{(2)}_0(k')| \Big\}\cr
&\le \const\la_0^{1-2\upsilon}|k-k'|^{\aleph''}\cr
\big|\sfrac{\partial \Si}{\partial k_0}(k)
       -\sfrac{\partial \Si}{\partial k_0}(k')\big|
&\le \const\max\Big\{|P(k)-P(k')|, |\tilde P(k)-\tilde P(k')|,
\cdots, |\tilde Q^{(2)}_0(k)-\tilde Q^{(2)}_0(k')| \Big\}\cr
&\le  \const\la_0^{1-2\upsilon}|k-k'|^{\aleph''}
}$$

\endproof

\remark{\STM\remNPpseregularity}{
We have proven, in Lemma \lemNPpsebnd, only very limited regularity properties
of the proper self--energy $\Si(k)$. It is proven in [FST3], that, to all finite
orders of perturbation theory, the proper self--energy is $C^{2-\veps}$
for every $\veps>0$. While we prove the convergence of perturbation theory
in the course of proving Theorems \theoremNPmainthI\ and \theoremNPmainthII,
the convergence proof does not give sufficient control over derivatives
to allow us to rigorously conclude that $\Si$ is  $C^{2-\veps}$.
 
}

The amputation used in the input and output data of \S\CHtildestep\ 
multiplies all external legs by $A(k)=ik_0-e(\k)$. The following lemma
bounds the functions  used to change to the amputation in which one
multiplies by $ik_0-e(\k)-\Si(k)$, the inverse of the physical two--point
function, instead.

\lemma{\STM\lemNPampcorrect}{
\Item i) Let $A_1(k)=\nu^{(\le i)}(k)\sfrac{ik_0-e(\k)-P(k)}{ik_0-e(\k)}$. Then
$$
\|A_1(k)\tnorm\ \le \const\cb_i
$$
\Item ii) Let $A_2(k)=\sfrac{ik_0-e(\k)-\Si(k)}{ik_0-e(\k)-P(k)}$. Then
$$
\big|A_2(k)\big|\le 2\qquad
\big|A_2(k)-A_2(k')\big|\le 
\la_0^{1-3\upsilon}|k-k'|^{1\over 2}
$$

}
\prf 
Since $\check p^{(j)}(k)$ is supported on the $j^{\rm th}$ neighbourhood,
$$\eqalign{
A_1(k)&=\nu^{(\le i)}(k)-\nu^{(\le i)}(k)\sfrac{P(k)}{ik_0-e(\k)}\cr
&=\nu^{(\le i)}(k)-\sfrac{\nu^{(\le i)}(k)}{ik_0-e(\k)}\smsum_{j=2}^{i+1}\check
p^{(j)}(k)\cr
}$$
Since $\cb_j\le 1+\const \sfrac{M^j}{M^i}\cb_i$ for all $j\le i+1$,
$$\eqalign{
\Big\|\smsum_{j=2}^{i+1}\check p^{(j)}(k)\TNorm\ 
&\le \smsum_{j=2}^{i+1}2\la_0^{1-\upsilon}\sfrac{\fl_j}{M^j}\cb_j\cr
&\le \const\la_0^{1-\upsilon}\big[1+\sfrac{1}{M^i}\cb_i\big]\cr
}$$
Also $\big\|\nu^{(\le i)}(k)\Tnorm\le\const\cb_i$ and 
$\big\|\sfrac{\nu^{(\le i)}(k)}{ik_0-e(\k)}\Tnorm\le \const M^i\cb_i$, 
so that
$$
\sup_k\Big|\sfrac{\partial^\de\hfill}{\partial k^\de}\nu^{(\le i)}(k)\Big|
\le\const M^{i|\de|}
$$
and
$$
\sup_k\Big|\sfrac{\partial^\de\hfill}{\partial k^\de}
\sfrac{\nu^{(\le i)}(k)}{ik_0-e(\k)}\Big|
\sup_k\Big|\sfrac{\partial^{\de'}\hfill}{\partial k^{\de'}}
\smsum_{j=2}^{i+1}\check p^{(j)}(k)\Big|
\le\const M^{i(|\de|+1)}\la_0^{1-\upsilon}M^{i(|\de'|-1)}
\le\const \la_0^{1-\upsilon}M^{i|\de+\de'|}
$$
provided $\de'\ne 0$.
This handles all contributions to $\|A_1(k)\tnorm\ $ except those for which no
derivatives act on $P(k)$. For those contributions, we write $P(k)=ik_0\tilde
P(k)$ and use that $|\tilde P(k)|\le\la_0^{1-3\upsilon}$  so that
$$
\sup_k\Big|P(k)\sfrac{\partial^\de\hfill}{\partial k^\de}
\sfrac{\nu^{(\le i)}(k)}{ik_0-e(\k)}\Big|
\le \la_0^{1-3\upsilon}\sup_k\Big|k_0\sfrac{\partial^\de\hfill}{\partial k^\de}
\sfrac{\nu^{(\le i)}(k)}{ik_0-e(\k)}\Big|
\le\const \la_0^{1-3\upsilon}M^{i|\de|}
$$

Now we move on to
$$
A_2(k)=1-\sfrac{\Si(k)-P(k)}{ik_0-e(\k)-P(k)}
$$
In the notation of the proof of Lemma  \lemNPpsebnd,
$$\eqalign{
A_4(k)=\sfrac{\Si(k)-P(k)}{ik_0-e(\k)-P(k)}
&=\sfrac{Q(k)-2P(k)\tilde Q^{(0)}(k)+P(k)\tilde P(k)\tilde Q^{(1)}(k)}
{[1+\tilde Q^{(0)}(k)-\tilde P(k)\tilde Q^{(1)}(k)][E(k)-P(k)]}\cr
&=\sfrac{\tilde Q^{(0)}(k)-2\tilde P(k)\tilde Q^{(1)}(k)+\tilde P(k)^2\tilde Q^{(2)}(k)}
{[1+\tilde Q^{(0)}(k)-\tilde P(k)\tilde Q^{(1)}(k)]
[1-\tilde K_0(k)\tilde P(k)]}\cr
&=\sfrac{A_3(k)}{1-\tilde K_0(k)\tilde P(k)}
}$$
where $\tilde K_0(k)=\sfrac{ik_0}{E(k)}$ and
$
A_3(k)=\sfrac{\tilde Q^{(0)}(k)-2\tilde P(k)\tilde Q^{(1)}(k)+\tilde P(k)^2\tilde Q^{(2)}(k)}
{1+\tilde Q^{(0)}(k)-\tilde P(k)\tilde Q^{(1)}(k)}
$.
By the  bounds
(\eqnTNPtildeQbnds) and (\eqnTNPtildeQmQbnds) on $\tilde Q^{(m)}$
and $\tilde P$
$$
|A_3(k)|\le \const\la_0^{1-2\upsilon}
                         \min\{|ik_0-e(\k)|^\aleph,1\}\qquad
\big|A_3(k)-A_3(k')\big|\le \const\la_0^{1-2\upsilon}|k-k'|^{\aleph''}
\EQN\eqnTNPAthreebnds$$
To bound $\big|A_3(k)-A_3(k')\big|$, we used a Taylor expansion argument as 
in Lemma \lemNPpsebnd.

Let $\ell$ be such that $\sfrac{1}{M^\ell}\le |k-k'|\le\sfrac{M}{M^\ell}$.
If either $|E(k)|\le \sfrac{1}{M^\ell}$ or $|E(k')|\le \sfrac{1}{M^\ell}$,
then both $|E(k)|,|E(k')|\le \sfrac{\const}{M^\ell}$ and, by 
(\eqnTNPtildeQbnds) and (\eqnTNPAthreebnds),
$$
|A_4(k)|, |A_4(k')|\le\const\la_0^{1-2\upsilon}\sfrac{1}{M^{\aleph\ell}}
\le\const\la_0^{1-2\upsilon}|k-k'|^\aleph
$$
If both $|E(k)|\ge \sfrac{1}{M^\ell}$ and $|E(k')|\ge \sfrac{1}{M^\ell}$,
then 
$$
\max\{|E(k)|,|E(k')|\}\le\const\min\{|E(k)|,|E(k')|\}
$$ 
and
$$
\big|\tilde K_0(k)-\tilde K_0(k')\big|
\le \min\Big\{2,\const \sfrac{|k-k'|}{\min\{|E(k)|,|E(k')|\}}\Big\}
\le\const \Big(\sfrac{|k-k'|}{\max\{|E(k)|,|E(k')|\}}\Big)^\aleph
$$
Hence, by (\eqnTNPAthreebnds), (\eqnTNPtildeQbnds)  and (\eqnTNPtildeQmQbnds)
$$\eqalign{
\big|A_4(k)-A_4(k')\big|
&=\Big|
\sfrac{A_3(k)-A_3(k')}{1-\tilde K_0(k')\tilde P(k')}
+A_3(k)\sfrac{\tilde K_0(k)\tilde P(k)-\tilde K_0(k')\tilde P(k')}
{[1-\tilde K_0(k)\tilde P(k)][1-\tilde K_0(k')\tilde P(k')]}
\Big|\cr
&\le \const\la_0^{1-2\upsilon}|k-k'|^{\aleph''}
+\const\la_0^{1-2\upsilon}|E(k)|^\aleph
\big|\tilde K_0(k)-\tilde K_0(k')\big|\,\big|\tilde P(k)\big|\cr
&\hskip1in+\const\la_0^{1-2\upsilon}\big|\tilde K_0(k')\big|
\big|\tilde P(k)-\tilde P(k')\big|\cr
&\le \const\la_0^{1-2\upsilon}|k-k'|^{\aleph''}
+\const\la_0^{1-2\upsilon}|k-k'|^\aleph\cr
}$$
The desired bounds on $A_2$ follow.
\endproof

\vfill\eject

\chap{ Momentum Space Norms}\PG\pgNPXIII

Theorem \theoremNPtildeinduction\ is concerned with the Fourier transforms of the two point functions $G^\rg_{j,2}(\eta_1, \eta_2)$. In the proof of Theorem
\theoremNPinduction, $G^\rg_{j,2}$ is built up recursively from contributions to effective interactions with two, one or no external fields. To control
$\check G^\rg_{j,2}$, we control the partial Fourier transforms, with 
respect to the external variables, of these contributions. In this chapter,
we provide the notation to do this. In particular, we specify norms for functions that depend on ``external'' momentum and ``internal'' 
sectorized position variables.

\definition{\STM\defNPtildefourtrans (Partial Fourier transforms)}{
Let $f({\sst \eta_1,\cdots,\eta_m;\,\xi_1,\cdots,\xi_n})$
be a translation invariant function on $\cB^m \times \cB^n$. 
If $n\ge 1$, the partial Fourier transform $f^\sim$ is defined,
using the notation of Definition \defNPfourtrans, by
$$
f^\sim({\sst\check\eta_1,\cdots,\check\eta_m;\,\xi_1,\cdots,\xi_n }) 
= \int \Big(\smprod_{i=1}^m E_+(\check\eta_i,\eta_i)\,
d\eta_i\Big)\  f({\sst \eta_1,\cdots,\eta_m;\,\xi_1,\cdots,\xi_n })
$$
or, equivalently, by
$$
f({\sst \eta_1,\cdots,\eta_m;\,\xi_1,\cdots,\xi_n })
= \int \Big(\smprod_{i=1}^m E_-(\check\eta_i,\eta_i)\,
\sfrac{d\check\eta_i}{(2\pi)^{d+1}}\Big)\, f^\sim({\sst
\check\eta_1,\cdots,\check\eta_m;\,\xi_1,\cdots,\xi_n }) 
$$ 
If $n=0$, we set $f^\sim =\check f$. 
}
\definition{\STM\defNPphicheckfourtrans}{
If $\phi(\et)$ is a Grassmann field, we set for,
 $\check \eta=(k,\si,a) \in \check \cB$,
$$
\check \phi(\check \eta)
=\int d\et\ E_-(\check\et,\et)\phi(\et) 
=  \int dx_0d^d\x\ e^{-(-1)^a\imath<k,x>_-}\phi(x_0,\x,\si,a)
$$

}

\remark{\STM\remNPphicheckfourtrans}{
A translation invariant sectorized Grassmann function $\cW$ can be uniquely written in the form
$$\eqalign{
\cW(\phi,\psi) = \smsum_{m,n}  
\int {\sst d\eta_1\cdots d\eta_m\,d\xi_1\cdots d\xi_n}\ & 
W_{m,n}({\sst \eta_1,\cdots, \eta_m\,;\,\xi_1,\cdots ,\xi_n})
\ \phi({\sst \eta_1})\cdots \phi({\sst \eta_m})\
\psi({\sst \xi_1})\cdots \psi({\sst \xi_n\,})\cr
}$$
with $W_{m,n}$ antisymmetric separately in the $\eta$ and in the $\xi$ 
variables. Then, under the Fourier transform Definitions \defNPtildefourtrans\ 
and \defNPphicheckfourtrans,
$$\eqalign{
\cW(\phi,\psi) &=
 \smsum_m 
\int \smprod_{i=1}^m \sfrac{d\check\eta_i}{(2\pi)^{d+1}}\  
W_{m,0}^\sim({\sst\check \eta_1,\cdots,\check \eta_m})\
{\sst (2\pi)^{d+1}}
 \de({\sst\check \eta_1+\cdots+\check \eta_m})\ 
 \check\phi({\sst \check\eta_1})\cdots \check\phi({\sst \check \eta_m}) \cr
&\hskip.25in+ \smsum_{m,n \atop n \ge 1} 
\int \smprod_{i=1}^m \sfrac{d\check\eta_i}{(2\pi)^{d+1}}
\smprod_{i=1}^n {\sst d\xi_i}\  
W_{m,n}^\sim({\sst\check \eta_1,\cdots,\check \eta_m\,;\,\xi_1,\cdots ,\xi_n})
\  \check\phi({\sst \check\eta_1})\cdots \check\phi({\sst \check \eta_m})\
\psi({\sst \xi_1})\cdots \psi({\sst \xi_n\,})\cr
}$$

}

\definition{\STM\defNPtransinv}{
A function $f({\sst \check\eta_1,\cdots,\check\eta_m;\,\xi_1,\cdots,\xi_n })$ on $\check\cB^m\times\cB^n$, with $n\ge 1$,
 is said to be translation invariant if 
$$
f({\sst \check\eta_1,\cdots,\check\eta_m;\,\xi_1+t,\cdots,\xi_n+t })
= e^{\imath \<\check\eta_1+\cdots+\check\eta_m,t\>_-}\, 
f({\sst \check\eta_1,\cdots,\check\eta_m;\,\xi_1,\cdots,\xi_n })
$$
for all $t\in\bbbr \times \bbbr^d$. A distribution 
$f({\sst \check\eta_1,\cdots,\check\eta_m})$ on $\check\cB^m$
 is said to be translation invariant if it is supported on
$\check\eta_1+\cdots+\check\eta_m=0$. Recall, from just before
Definition \defNPfourtrans, that if 
$\check\et = (k,\si,a), \,\check\et' = (k',\si',a') \in \check \cB$, then
$\ 
\check\et+\check\et' = (-1)^a\,k + (-1)^{a'}\,k'\ \in\bbbr \times \bbbr^d
$.

}

\noindent The partial Fourier transform of a translation invariant function 
on $\cB^m \times \cB^n$ is a translation invariant function 
on $\check\cB^m \times \cB^n$.

\definition{\STM\defNPdiffdecay (Differential--decay operators)}{
Let $m,n\ge 0$. If $n\ge 1$, 
let $f$ be a function on $\check \cB^m \times \cB^n$. If $n=0$, let
$f$ be a function on 
$$
\check \cB_m= \set{(\check \eta_1,\cdots,\check \eta_m)\in \check \cB^m}
{\check \eta_1+\cdots+\check \eta_m=0} 
$$
\Item{i)} 
For $1\le j \le m$ and a multiindex $\de$ set
$$\eqalign{
&\rD^\de_j
f\,({\sst(p_1,\tau_1,b_1),\cdots,(p_m,\tau_m,b_m);\,\xi_1,\cdots,\xi_n}) \cr
& \hskip 1.5cm= 
\big[\imath(-1)^{b_j}\big]^{\de_0}
\smprod_{\ell=1}^d\big[-\imath(-1)^{b_j}\big]^{\de_\ell}\ 
\sfrac{\partial^{\de_0}\hfill}{\partial p_{j,0}^{\de_0}}\,
\sfrac{\partial^{\de_1}\hfill}{\partial \p_{j,1}^{\de_1}} \cdots
\sfrac{\partial^{\de_d}\hfill}{\partial \p_{j,d}^{\de_d}}\,
f({\sst(p_1,\tau_1,b_1),\cdots,(p_m,\tau_m,b_m);\,\xi_1,\cdots,\xi_n})
}$$

\Item{ii)}
Let $1\le i\ne j\le m+n$ and $\de$ a multiindex. Set
$$\deqalign{
\rD_{i;j}^\de f & = (\rD_i-\rD_j)^\de\,f  
           &{\rm if}\ 1\le i <j \le m \cr
\rD_{i;j}^\de f & = (\rD_i-\xi_{j-m})^\de\,f   
           &{\rm if}\ 1\le i  \le m,\ m+1\le j \le m+n \cr
\rD_{i;j}^\de f & = (\xi_i-\rD_{j-m})^\de\,f   
           &{\rm if}\ m+1\le i  \le m+n,\ 1\le j \le m \cr
\rD_{i;j}^\de f & = (\xi_{i-m}-\xi_{j-m})^\de\,f \ = \cD_{i-m,j-m}^\de f  \hskip 1cm 
           &{\rm if}\  m+1\le i<j \le m+n \cr
}$$
The decay operator $\cD_{i,j}$ was defined in Definition \defNPdecayop.

\Item{iii)}
A differential--decay operator (dd--operator) of type $(m,n)$, with 
$m+n\ge 2$, is an operator $\rD$
of the form
$$
\rD = \rD^{\de^{(1)}}_{i_1;j_1}\, \cdots  \rD^{\de^{(r)}}_{i_r;j_r}
$$
with $1\le i_\ell\ne j_\ell\le m+n$ for all $1\le\ell\le r$. 
A dd--operator of type $(1,0)$
is an operator of the form
$\ 
\rD = \rD^{\de^{(1)}}_{1}\, \cdots  \rD^{\de^{(r)}}_{1}
.$
The total order of $\rD$ is $\de(\rD) =\de^{(1)}+\cdots+\de^{(r)}$.

}

\goodbreak
\remark{\STM\remNPdiffdecay}{

\Item{i)} For a translation invariant function $\varphi$ on $\cB^m \times \cB^n$
$$
\rD_{i;j}(\varphi^\sim) = (\cD_{i,j}\varphi)^\sim
$$
In particular, Leibniz's rule also applies for differential--decay operators.

\Item{ii)} Let $f$ be a translation invariant function on 
$\check\cB^m \times \cB$. Then, for $\xi =(x_0,\x,\si,a) \in \cB$,
$$
f(\check\eta_1,\cdots,\check\eta_m;\xi)  = 
 e^{\imath \<\check\eta_1+\cdots+\check\eta_m,(x_0,\x)\>_-}\,
    f(\check\eta_1,\cdots,\check\eta_m;(0,\si,a))  
$$
Consequently, for $1\le i\le m$ and a multiindex $\de$
$$
\rD_{i;m+1}^\de f({\sst \check\eta_1,\cdots,\check\eta_m;\xi})
= e^{\imath \<\check\eta_1+\cdots+\check\eta_m,(x_0,\x)\>_-}\,
    \rD_i^\de f(\check\eta_1,\cdots,\check\eta_m;(0,\si,a))  
$$

}

\definition{\STM\defNPdiffdecaynorm}{
For  a function $f$ on $\check\cB_m$, set
$$
\| f\tnorm 
= \smsum\limits_{\de \in \bbbn_0\times\bbbn_0^2} \sfrac{1}{\de !}\ 
\max\limits_{\rD\ {\rm dd-operator} 
\atop{\rm with\ } \de(\rD)=\de}\
\sup\limits_{\check\eta_1,\cdots,\check\eta_m \in \check \cB}
\big| \rD f\,
({\sst \check\eta_1,\cdots,\check\eta_m}) \big|\ t^\de
$$
Let $f$ be a function on $\check\cB^m\times \cB^n$ with $n\ge 1$. Set
$$
\| f\tnorm 
= \smsum_{\de\in \bbbn_0\times\bbbn_0^2} \sfrac{1}{\de!}
\max_{\rD\, {\rm dd-operator} \atop{\rm with\ } \de(\rD) =\de}\
\sup_{\check\eta_1,\cdots,\check\eta_m \in \check \cB}
\TN \rD\, f\,({\sst\check\eta_1,\cdots,\check\eta_m; \xi_1,\cdots,\xi_n})
\TN_{1,\infty}
\ t^\de
$$
The norm $\tn\,\cdot\,\tn_{1,\infty}$ of Definition \defNPSymmNorm\ 
refers to the variables ${\sst \xi_1,\cdots,\xi_n}$. That is,
$$
\TN \rD\, f\,({\sst\check\eta_1,\cdots,\check\eta_m; \xi_1,\cdots,\xi_n})
\TN_{1,\infty}
=\max\limits_{1\le j_0 \le n}\ 
\sup\limits_{\xi_{j_0} \in \cB}\  
\int \prod\limits_{j=1,\cdots, n \atop j\ne j_0} d\xi_j\, 
| \rD\, f\,({\sst\check\eta_1,\cdots,\check\eta_m; \xi_1,\cdots,\xi_n}) |
$$
}

\remark{\STM\remNPdiffdecaynorm}{
In the case $m=0$ the norm $\| \,\cdot\,\|_{1,\infty}$ of Definition \defNPSymmNorm\ and the norm $\| \,\cdot\,\tnorm$ of Definition 
\defNPdiffdecaynorm\ agree.
}

\definition{\STM\defNPampGreen}{We amputate a Grassmann function by 
applying the Fourier transform $\hat A$, in the sense of Notation
\notNPfourierTI, of $A(k)=ik_0-e(\k)$  to its 
external arguments. Precisely, if $\cW(\phi,\psi)$ is a Grassmann function,
then
$$
\cW^a(\phi,\psi)=\cW(\hat A\phi,\psi)
$$
where
$$
(\hat A\phi)(\xi)=\int d\xi'\ \hat A(\xi,\xi')\phi(\xi')
$$

}

\remark{\STM\remNPampGreen}{
If $C(\xi,\xi')$ is the covariance associated to  $C(k)$
in the sense of (\eqnNPcovFT) and (\eqnNPantisymmCov) and $J$ is the particle
hole swap operator of (\eqnNPjdef), then,
by parts (i) and (ii) of Lemma \lemOSjhat\ of [FKTo2],
$$
\int d\et d\et'\ C(\xi,\et)J(\et,\et')\hat A(\et',\xi)=\hat E(\xi,\xi')
$$
where
$\hat E$ is the Fourier transform of $\big(ik_0-e(\k)\big)C(k)$ in the sense
of Notation \notNPfourierTI.
}

Integrating out the first few scales is controlled much as in Theorem
\thmNPfirststep.

\theorem{\STM\thmNPTfirststep}{ 
There are ($M$ and $j_0$--dependent) constants
$\bar\la,\ \mu$ and $\be_0$ such that, for all 
$\la_0<\bar\la$ and $\be_0\le\be\le \sfrac{1}{\la_0^{\upsilon/5}}$, 
the following holds:

\noindent
Let $X \in \fN_{d+1}$ with $X_\0<\mu$, $\de e\in\cE$ with
$ \|\de\hat e\|_{1,\infty}\le X$ and
$$
\cV(\psi) = \int_{\cB^4} {\sst d\xi_1 \cdots d\xi_4} \,V({\sst \xi_1,\cdots,\xi_4})
\,\psi({\sst \xi_1})\cdots \psi({\sst \xi_1})
$$
with an antisymmetric function $V$ fulfilling 
$$
\|V\|_{1,\infty} \le \la_0 \,\fe_0(X)
$$
Write
$$\eqalign{
&\tilde\Om_{C_{-\de e}^{(\le j_0)}}\big(\cV{\sst(\psi)}\big) (\phi,\psi) 
= \cV(\psi) +\half\phi JC^{(\le j_0)}_{-\de e}J\phi\cr 
& \hskip 1.5cm+ \smsum_{m,n\ge 0\atop m+n\ {\rm even}} 
\int_{\cB^{m+n}}\hskip-15pt {\sst d\et_1\cdots d\et_m\ d\xi_1\cdots d\xi_n}\ 
W_{m,n}({\sst\et_1\cdots \et_m,\xi_1,\cdots,\xi_n};\de e)\,
\phi{\sst(\et_1)}\cdots\phi{\sst(\et_m)}\,\psi{\sst(\xi_1)}\cdots{\sst\psi(\xi_n)}
}$$
with kernels $W_{m,n}$ that are separately antisymmetric under 
permutations of their $\et$ and $\xi$ arguments. Then
$$
\smsum_{m+n \ge 2 \atop m+n\ {\rm even}} \be^{m+n} \,\tilde\rho_{m;n}\,
\| W^{a\sim}_{m,n}(\de e)\tnorm \le  \be^4\la_0^\upsilon\,\fe_0(X) 
$$
and
$$
\smsum_{m+n \ge 2 \atop m+n\ {\rm even}} \be^{m+n} \,\tilde\rho_{m;n}\,
\big\|\sfrac{d\hfill}{ds} W^{a\sim}_{m,n}(\de e+s\de e')\big|_{s=0}\Tnorm 
\le  \be^4\la_0^\upsilon\,\fe_0(X)\ \|\de\hat e'\|_{1,\infty}
$$
where 
$$
\tilde\rho_{m;n}
=\sfrac{\la_0^{m\upsilon/7}}{\la_0^{(1-\upsilon)\max\{m+n-2,2\}/2}}
$$
and $\upsilon$ was fixed in Definition \defNPrhomn.

}

\prf 
Apply Theorem \thmOSTfirststep\ of [FKTo2] with $\rho_{m,n}=\tilde\rho_{m;n}$,
 $\veps=\abcst\,\be^4\la_0^\upsilon\le\abcst\la_0^{\up/5}$ 
and $\veps'=\la_0^{\up/7}$.
Observe that, $\tilde\rho_{m;n}$ is $\la_0^{m\up/7}$ times the $\rho_{m;n}$
of Remark \remOSthmV.iii of [FKTo2]. So the hypotheses of 
Theorem \thmOSTfirststep\ concerning $\rho_{m;n}$  are fulfilled. Choosing
$\bar\la$ small enough ensures that the hypotheses 
$\veps,\veps'\le\veps_0$ of Theorem \thmOSTfirststep\ are satisfied. 

\endproof

\noindent
In the course of exhibiting control over amputated Green's functions 
in momentum space, we must deal with kernels having external arguments
in momentum space and internal arguments sectorized and in position
space.

\definition{\STM\defNPsectdiffdecaynorm (Momentum Space Norms)}{
Let $p$ be a natural number and $\Si$ a sectorization.

\Item{(i)} 
For a function $f$ on $\check\cB_m$ we define
$$
\v f\tv_{p,\Si} 
= \cases{ \| f\tnorm & if $ p=m-1$, $m=2,4$ \cr
\ \ 0  & otherwise \cr
}$$

\Item{(ii)}
For a translation invariant function $f$ on 
$\check\cB^m \times (\cB\times \Si)^n$ with $n\ge 1$, 
we set $\v f\tv_{p,\Si} \ =\ 0 \ $ when $p>m+n$ or $p<m$, and
$$
\v f\tv_{p,\Si} 
= \smsum_{\de\in \bbbn_0\times\bbbn_0^2} 
\sup_{{1\le i_1<\cdots<i_{p-m}\le n \atop s_{i_1},\cdots,s_{i_{p-m}}\in\Si}
\atop \check\eta_1,\cdots,\check\eta_m \in \check \cB} 
\smsum_{s_i \in \Si \ {\rm for}\atop i\ne i_1,\cdots i_{p-m}} \hskip-5pt
\sfrac{1}{\de!}
\max_{\rD\, {\rm dd-operator} \atop{\rm with\ } \de(\rD) =\de} \TN\rD f\,({\sst\check\eta_1,\cdots,\check\eta_m;(\xi_1,s_1),\cdots,(\xi_n,s_n)})
\TN_{1,\infty}\ t^\de
$$
when $m\le p \le m+n\,$.  The norm
$\tn\,\cdot\,\tn_{1,\infty}$ of Definition \defNPSymmNorm\ refers to the variables 
${\sst \xi_1,\cdots,\xi_n}$. 

}

\remark{\STM\remNPsecdiffdecaynorm}{
In the case $m=0$ the norm $\v \,\cdot\,\v_{p,\Si}$ of Definition
\defNPsectnorm\ and the norm $\v \,\cdot\,\tv_{p,\Si}$ of Definition 
\defNPsectdiffdecaynorm\ agree.
}

\definition{\STM\defNPsectcheckcF }{
Let $m,n \ge 0$  and $\Si$ a sectorization.
For $n\ge 1$, denote by  $\check\cF_m(n;\Si)$ the space
of all translation invariant, complex valued functions 
$\ 
f({\sst\check\eta_1,\cdots,\check\eta_m;\,(\xi_1,s_1),\cdots,(\xi_n,s_n)} )
\ $
on $\check\cB^m \times  \big( \cB \times\Si \big)^n$ whose Fourier transform
$\check f({\sst\check\eta_1,\cdots,\check\eta_m;
\,(\check\xi_1,s_1),\cdots,(\check\xi_n,s_n)} )$
vanishes unless 
$ k_i\in \tilde s_i$ for all $1\le j\le n$. 
Here, $\check\xi_i=(k_i,\si_i,a_i)$.
Also, let $\check\cF_m(0;\Si)$  be the space of all momentum conserving,
complex valued functions 
$\ 
f({\sst\check\eta_1,\cdots,\check\eta_m} )
\ $
on $\check\cB^m $.
}

We now provide the analogue of Definition \defNPsectnorm.ii for the 
$\v\ \ \tv$--norms.
Let $j\ge 2$ and let $\Si_j$ be the sectorization of scale $j$ and length
$\fl_j = \sfrac{1}{M^{\aleph j}}$ fixed just before Definition \defNPresector.

\definition{\STM\defNPmomscalednorms}{
\Item{i)}
For $f\in \check\cF_m(n;\Si_j)$ set
$$\eqalign{
\v f \tv_j &= 
\tilde\rho_{m;n}\cases{
\v f \tv  _{1,\Si_j} +\v f \tv  _{2,\Si_j} 
+ \sfrac{1}{\fl_j}\,\v f \tv_{3,\Si_j}
+ \sfrac{1}{\fl_j}\,\v f \tv_{4,\Si_j}
+ \sfrac{1}{\fl_j^2}\,\v f \tv_{5,\Si_j}
+ \sfrac{1}{\fl_j^2}\,\v f \tv_{6,\Si_j}& if $m\ne 0$\cr
\noalign{\vskip.1in}
\v f \tv  _{1,\Si_j} + \sfrac{1}{\fl_j}\,\v f \tv_{3,\Si_j}
+ \sfrac{1}{\fl_j^2}\,\v f \tv_{5,\Si_j}& if $m= 0$\cr
}\cr
\v f \tv_{p,\Si_j,\tilde\rho} &= \tilde\rho_{m;n} \v f \tv_{p,\Si_j}
}$$
\Item{ii)}
An even sectorized Grassmann function $w$ can be uniquely written in the form
$$\eqalign{
w(\phi,\psi) = \smsum_{m,n} 
\int {\sst d\eta_1\cdots d\eta_m\,d\xi_1\cdots d\xi_n}\ & 
w_{m,n}({\sst \eta_1,\cdots, \eta_m\,(\xi_1,s_1),\cdots ,(\xi_n,s_n)})\cr
& \hskip 2cm \phi({\sst \eta_1})\cdots \phi({\sst \eta_m})\
\psi({\sst (\xi_1,s_1)})\cdots \psi({\sst (\xi_n,s_n)\,})\cr
}$$
with $w_{m,n}$ antisymmetric separately in the $\eta$ and in the $\xi$ variables.
Set,  for $\al >0$ and $X\in \fN_{d+1}$,
$$
N_j^\sim(w,\al,\,X)
=\sfrac{M^{2j}}{\fl_j}\,\fe_j(X) 
\smsum_{m,n\ge 0}\,
\al^{m+n}\,\big(\sfrac{\fl_j\,\IB}{M^j}\big)^{(m+n)/2} \,
\v w^\sim_{m,n}\tv_j 
$$
where $w_{m,n}^\sim$ is the partial Fourier transform of $w_{m,n}$ of 
Definition \defNPtildefourtrans\ and
$\IB=4\max\{4\IB_3,\IB_4\}$ with $\IB_3,\ \IB_4$ being 
the constants of Proposition \propOSmomcontrintboundsectors\ of [FKTo3]. 

}

\remark{\STM\remNPmomscalednorms}{
In particular, for the ``pure $\phi$'' part of $w$,
$$
N_j^\sim(w(\phi,0) ,\al,X)
=\fe_j(X) \bigg[\sfrac{\al^2\IB}{\la_0^{1-9\upsilon/7}}
 M^j \v w^\sim_{2,0}\tv_{1,\Si_j}
 + \sfrac{\al^4\IB^2}{\la_0^{1-11\upsilon/7}} \v w^\sim_{4,0}\tv_{3,\Si_j}
\bigg]
$$
}

The sectorized version of Theorem \thmNPTfirststep\ is

\theorem{\STM\thmNPTsetupinduction }{
For $K\in \fK_{j_0}$, set 
$$
u(K)= - \big[K_{\rm ext}\big]_{\Si_{j_0}} \in \cF_0(2,\Si_{j_0})
$$
where $K_{\rm ext}$ was defined in Definition \defOSzerosectorext\ of [FKTo4]. 
Then there exist constants $\bar\al,\bar\la>0$ such that for all 
$0\le \la_0 <\bar\la$, $\bar\al<\al<\sfrac{1}{\la_0^{\upsilon/10}}$ and all
$$
K\in\fK_{j_0}
 \qquad
\|V\|_{1,\infty}\le \la_0\fe_{j_0}\big(\|K\|_{1,\Si_{j_0}}\big)
$$
the $\Si_{j_0}$--sectorized representative $w(\phi,\psi;K)$ of
$
\tilde\Om_{C_{u(K)}^{(\le j_0)}}\big(\cV{\sst(\psi)}\big) (\phi,\psi) 
 -\half\phi JC^{(\le j_0)}_{u(K)}J\phi
$
constructed in Theorem \thmNPsetupinduction\ may be chosen to obey
$$\eqalign{
N^\sim_{j_0}\big(w^a(K),\al,\|K\|_{1,\Si_{j_0}}\big) 
&\le \const\,\al^4\la_0^\upsilon\,\fe_{j_0}\big(\|K\|_{1,\Si_{j_0}}\big) \cr
N^\sim_{j_0}\big(\sfrac{d\hfill}{ds}w^a(K+sK')\big|_{s=0}
,\al,\,\| K\|_{1,\Si_{j_0}}\big) 
   &\le \const\,\al^4\la_0^\upsilon\,\fe_{j_0}\big(\|K\|_{1,\Si_{j_0}}\big) \,  \|K'\|_{1,\Si_{j_0}}
\cr
}$$
for all $ K'$. 
Furthermore $w^a(\phi,\psi,K)-w^a(0,\psi,K)$ vanishes unless 
$\hat\nu^{(\le j_0)}\phi$ is nonzero.
}

\prf
Write
$$\eqalign{
&\tilde\Om_{C_{u(K)}^{(\le j_0)}}\big(\cV{\sst(\psi)}\big) (\phi,\psi) 
= \cV(\psi) +\half\phi JC^{(\le j_0)}_{u(K)}J\phi\cr 
& \hskip 1.5cm+ \smsum_{m,n\ge 0\atop m+n\ {\rm even}} 
\int_{\cB^{m+n}}\hskip-15pt {\sst d\et_1\cdots d\et_m\ d\xi_1\cdots d\xi_n}\ 
W_{m,n}({\sst\et_1,\cdots,\et_m,\xi_1,\cdots,\xi_n})\,
\phi{\sst(\et_1)}\cdots\phi{\sst(\et_m)}\,
\psi{\sst(\xi_1)}\cdots{\sst\psi(\xi_n)}
}$$
and set, as in Theorem \thmNPsetupinduction,
$$
w_{m,n}=\cases{\big(W_{m,n}\big)_{\Si_{j_0}}& if $(m,n)\ne(0,4)$\cr
\noalign{\vskip0.05in}
                \big(W_{0,4}+V\big)_{\Si_{j_0}}& if $(m,n)=(0,4)$\cr}
$$
using the sectorization $f_\Si$ of Definition \defOScreateSectoriz\ of [FKTo4]. 
By Proposition \propOScreateSectoriz\ of [FKTo4]
$$\eqalign{
N^\sim_{j_0}\big(&w^a(K),\al,\|K\|_{1,\Si_{j_0}}\big) 
=\sfrac{M^{2{j_0}}}{\fl_{j_0}}\,\fe_{j_0}\big(\|K\|_{1,\Si_{j_0}}\big) 
\smsum_{m,n\ge 0}\,
\al^{m+n}\,\big(\sfrac{\fl_{j_0}\,\IB}{M^{j_0}}\big)^{(m+n)/2} \,\v w^{a\sim}_{m,n}\tv_{j_0} 
\cr
&\le\const\,\cb_{j_0}\,\fe_{j_0}\big(\|K\|_{1,\Si_{j_0}}\big) \Big[
\sfrac{\al^4}{\la_0^{1-\upsilon}}\,\| V\|_{1,\infty}+
\smsum_{m,n\ge 0}\,(\const \al)^{m+n}\,\tilde\rho_{m;n}\,
\| W^{a\sim}_{m,n}\tnorm \ \Big]
}$$
By hypothesis
$$
\sfrac{\al^4}{\la_0^{1-\upsilon}}\,\| V\|_{1,\infty}
\le \al^4\la_0^{\upsilon}\fe_{j_0}\big(\|K\|_{1,\Si_{j_0}}\big)
$$
and by Theorem \thmNPTfirststep, with  $\de e=-\check u$, $X=\const \|K\|_{1,\Si_{j_0}}$ and $\be=\const\al$,
$$
\smsum_{m,n\ge 0}\,(\const \al)^{m+n}\,\tilde\rho_{m;n}\,
\| W^{a\sim}_{m,n}\tnorm\ 
\le\be^4\la_0^{\upsilon}\fe_0(X)
\le\const\al^4\la_0^\upsilon\, \fe_{j_0}\big(\|K\|_{1,\Si_{j_0}}\big)
$$
Therefore, by Corollary \corOSappMonoidIV.ii of [FKTo1],
$$\eqalign{
N^\sim_{j_0}\big(w^a(K),\al,\|K\|_{1,\Si_{j_0}}\big) 
&\le\const\,\al^4\la_0^\upsilon\,\cb_{j_0}\,\fe_{j_0}\big(\|K\|_{1,\Si_{j_0}}\big)^2\cr
&\le\const\,\al^4\la_0^\upsilon\,\fe_{j_0}\big(\|K\|_{1,\Si_{j_0}}\big)\cr
}$$
The proof of the bound on $N^\sim_{j_0}\big(\sfrac{d\hfill}{ds}w^a(K+sK')\big|_{s=0}
,\al,\,\| K\|_{1,\Si_{j_0}}\big)$ is similar.

By Lemma \lemOStworengrpmaps\ of [FKTo2],
$$
\tilde\Om_{C_{u(K)}^{(\le j_0)}}\big(\cV{\sst(\psi)}\big) (\phi,\psi) 
 -\half\phi JC^{(\le j_0)}_{u(K)}J\phi
=\Om_{C_{u(K)}^{(\le j_0)}}\big(\cV{\sst(\psi)}\big) 
(\phi,\psi+C_{u(K)}^{(\le j_0)}J\phi) 
$$
Since $\Om_{C_{u(K)}^{(\le j_0)}}\big(\cV{\sst(\psi)}\big) (\phi,\psi) $
is independent of $\phi$ and since $C_{u(K)}^{(\le j_0)}J\phi$
vanishes unless $\hat\nu^{(\le j_0)}\phi$ is nonzero, 
$w^a(\phi,\psi,K)-w^a(0,\psi,K)$ vanishes unless 
$\hat\nu^{(\le j_0)}\phi$ is nonzero.
\endproof

\vfill\eject

\chap{Ladders with External Momenta}\PG\pgNPXIV

In the proof of Theorem \theoremNPinduction, ladders and iterated particle hole ladders needed special attention. Due to the ``external improvement"
of Lemma \lemOSsectextimpr\ of [FKTo3], we needed to consider only ladders all of whose ``ends" correspond to $\psi$ fields and are integrated out at a later scale. This is not the case here. We consider ladders some of whose ``ends" correspond to $\psi$ fields and have sectorized position space arguments $(\xi,s)\in\cB\times \Si_j$, and some of whose ends correspond to $\phi$ fields and have momentum space arguments $\check \eta \in \check\cB$. To do this, we extend the definitions and estimates of ladders and iterated particle hole ladders from \S \CHphladders. 

The following Definition extends Definition \NPsomespaces.

\definition{\STM\defmoredisjointunions}{
Let $\Si$ be a sectorization. 
\Item i) Define
$$\eqalign{
\check\cB^\updownarrow&=\set{(k_0,\k,\si)}
            {k_0\in\bbbr,\ \k\in\bbbr^2,\ \si\in\{\uparrow,\downarrow\} }\cr
}$$
and the disjoint unions
$$\eqalign{
\fY^\updownarrow_\Si
&=\check \cB^\updownarrow\dunion(\cB^\updownarrow\times\Si)\cr
\fX_\Si&=\check \cB\dunion(\cB\times\Si)\cr
}$$
\Item ii) Let $z\in\fX_\Si$. Then we define its undirected part
 $u(z)\in\fY^\updownarrow_\Si$ and its creation/annihilation index $b(z)\in\{0,1\}$ by
$$\eqalign{
u(z)&=\cases{(k,\si) & if $z=(k,\si,b)\in\check\cB$\cr
            (x,\si,s) & if $z=(x,\si,b,s)\in\cB\times\Si$\cr}\cr
b(z)&=\cases{b & if $z=(k,\si,b)\in\check\cB$\cr
            b & if $z=(x,\si,b,s)\in\cB\times\Si$\cr}\cr
}$$
\Item iii) Let $z'\in\fY^\updownarrow_\Si$ and $b\in\{0,1\}$. Then we define 
$\iota_b(z')$ as the unique point $z\in\fX_\Si$ with $u(z)=z'$ and $b(z)=b$.
}

With this notation, Definition \defParicleHoleDecomp\ and Lemma \lemunordord.ii
about particle--particle and particle--hole reductions and values carry over almost verbatim.

\vskip.1in
\goodbreak
\definition{\STM\deftildeParicleHoleDecomp}{
\Item{ i)} 
Let $f$ be a four legged kernel over $\fX_\Si$. When $f$ is a rung, its 
particle--particle reduction is the four legged kernel over $\fY^\updownarrow_\Si$ given by
$$
f^{\rm pp}({\sst z'_1, z'_2, z'_3, z'_4}) = 
f({\sst \iota_0(z'_1),\iota_0(z'_2),\iota_1(z'_3),\iota_1(z'_4)}) = 
\figplace{PPvertex}{-.2in}{-.05in}
$$
and its particle--hole reduction is
$$
f^{\rm ph}({\sst z'_1, z'_2, z'_3, z'_4}) = 
f({\sst \iota_0(z'_1),\iota_1(z'_2),\iota_1(z'_3),\iota_0(z'_4)}) = 
\figplace{PHvertex}{-.2in}{-.05in}
$$
When $f$ is a bubble propagator, the corresponding reductions are
$$\eqalign{
{^{\rm pp}\!}f({\sst z'_1, z'_2, z'_3, z'_4}) &= 
f({\sst \iota_1(z'_1),\iota_1(z'_2),\iota_0(z'_3),\iota_0(z'_4)})\cr 
{^{\rm ph}\!}f({\sst z'_1, z'_2, z'_3, z'_4}) &= 
f({\sst \iota_1(z'_1),\iota_0(z'_2),\iota_0(z'_3),\iota_1(z'_4)})\cr 
}$$

\Item{ ii)} 
Let $f'$ be a four legged kernel over $\fY^\updownarrow_\Si$. Its particle--particle value
is the four legged kernel over $\fX_\Si$ given by
$$\eqalign{
V_{\rm pp}(f')({\sst z_1, z_2, z_3, z_4}) 
&=\de_{b(z_1) , 0}
  \de_{b(z_2) , 0}
  \de_{b(z_3) , 1}
  \de_{b(z_4) , 1}
\, f'({\sst u(z_1), u(z_2), u(z_3), u(z_4)}) \cr
&+\de_{b(z_1) , 1}
  \de_{b(z_2) , 1}
  \de_{b(z_3) , 0}
  \de_{b(z_4) , 0}
\, f'({\sst u(z_3), u(z_4), u(z_1), u(z_2)}) \cr  
}$$
and its particle--hole value is
$$\eqalign{
V_{\rm ph}(f')({\sst z_1, z_2, z_3, z_4}) 
&=\de_{b(z_1) , 0}
  \de_{b(z_2) , 1}
  \de_{b(z_3) , 1}
  \de_{b(z_4) , 0}
\, f'({\sst u(z_1), u(z_2), u(z_3), u(z_4)}) \cr
&+\de_{b(z_1) , 1}
  \de_{b(z_2) , 0}
  \de_{b(z_3) , 0}
  \de_{b(z_4) , 1}
\, f'({\sst u(z_2), u(z_1), u(z_4), u(z_3)}) \cr  
&-\de_{b(z_1) , 1}
  \de_{b(z_2) , 0}
  \de_{b(z_3) , 1}
  \de_{b(z_4) , 0}
\, f'({\sst u(z_2), u(z_1), u(z_3), u(z_4)}) \cr  
&-\de_{b(z_1) , 0}
  \de_{b(z_2) , 1}
  \de_{b(z_3) , 0}
  \de_{b(z_4) , 1}
\, f'({\sst u(z_1), u(z_2), u(z_4), u(z_3)}) \cr  
}$$
}

\lemma{\STM\lemtildeunordord}{
Let $f$ be an antisymmetric, particle number preserving, four
legged  kernel over $\fX_\Si$. Then
$$
f = V_{\rm pp}(f^{\rm pp}) + V_{\rm ph}(f^{\rm ph})
$$
}

\vskip .3cm
In \S \CHtildenewsectors\ we developed norms for functions on 
$\check\cB^m \times (\cB \times \Si)^n$, since we usually write Grassmann monomials in a way such that all $\phi$ fields stand before all $\psi$ fields.
However the ``ends" of ladders have a natural ordering, and $\phi$ and $\psi$ fields may be arbitrarily distributed among them. Therefore we extend some of the notation of \S\CHtildenewsectors\ to this situation
and repeat the detailed Definition of \S\CHsecmomnorm\ of [FKTo3].
 
\definition{\STM\defNPdisjointOrd}{
Set $\fX_0 = \check \cB$ and $\fX_1 = \cB \times \Si$. 
Let $\vec i = (i_1,\cdots,i_n) \in \{0,1\}^n$. 

\Item i) The inclusions of $\fX_{i_j},\ j=1,\cdots,n,\ $ in $\fX_\Si$ induce an inclusion of 
$\fX_{i_1} \times\cdots\times \fX_{i_n}$ in $\fX_\Si^n$. We identify 
$\fX_{i_1} \times\cdots\times \fX_{i_n}$ with its image in $\fX_\Si^n$.

\Item ii) Set $m(\vec i) = n-(i_1+\cdots +i_n)$. Clearly, $m(\vec i)$ is the number of copies of $ \check \cB$ in $\fX_{i_1} \times\cdots\times \fX_{i_n}$.

\Item iii)
If $f$ is a function on $ \fX_{i_1} \times\cdots\times \fX_{i_n}$, then 
$\ord f$ is the function on 
$\check \cB^{m(\vec i)}\times (\cB \times \Si)^{n-m(\vec i)}$ 
obtained from $f$ by shifting all of the $\check B$ arguments before all of the
$\cB \times \Si$ arguments, while preserving the relative order of the 
$\check B$ arguments and the 
relative order of the $\cB \times \Si$ arguments and multiplying by the sign of the permutation that implements the reordering of the arguments. That is, 
$\ord f(x_1,\cdots,x_n)= \sgn \pi f(x_{\pi(1)},\cdots,x_{\pi(n)})$
where the permutation $\pi\in S_n$ is determined by
$\ 
\pi(j)<\pi(j')\hbox{ if } i_{j}<i_{j'}\hbox{ or } i_{j}=i_{j'}\ j<j'.
\ $
}

\remark{\STM\remNPbigdisjointunion}{
Using the identification of Definition \defNPdisjointOrd.i, 
$$
\fX_\Si^n\ =\ 
\bigcup_{i_1,\cdots,i_n \in \{0,1\}}\kern-2.8em\cdot\kern2.8em \fX_{i_1} \times\cdots\times \fX_{i_n}
$$
where, on the right hand side we have a disjoint union. If $f$ is a function on
$\fX_\Si^n$ and $\vec i = (i_1,\cdots,i_n) \in \{0,1\}^n$, we denote by $f\big|_{\vec i}$ the restriction of $f$ to 
$\fX_{i_1} \times\cdots\times \fX_{i_n}$.
}
 
\definition{\STM\defNPdisjointfnspaces}{

\Item i) We denote by $\check \cF_{n;\Si}$ the set of functions on $\fX_\Si^n$
with the property that for each $\vec i = (i_1,\cdots,i_n) \in \{0,1\}^n$
with $m(\vec i) <n$
$$
\ord \big(f\big|_{\vec i} \big) \in \check \cF_{m(\vec i)}(n-m(\vec i);\Si)
$$
and such there is a function $g$ on $\check \cB_n$ such that
$$
f\big|_{(0,\cdots, 0)}(\check\eta_1,\cdots,\check\eta_n) = 
(2\pi)^3\de(\check\eta_1+\cdots+\check\eta_n)\,
g(\check\eta_1,\cdots,\check\eta_n)
$$

\Item ii)
For a function $f\in \check \cF_{n;\Si}$ and a natural number $p$ we set
$$\eqalign{
\v f\tv_{p,\Si} &=  \v g \tv_{p,\Si} +
\sum_{\vec i \in \{0,1\}^n \atop m(\vec i) \ne 0} 
\V \ord \big(f\big|_{\vec i} \big) \tV_{p,\Si}\cr 
\v f\tv_{p,\Si,\tilde\rho} &=  \v g \tv_{p,\Si,\tilde\rho} +
\sum_{\vec i \in \{0,1\}^n \atop m(\vec i) \ne 0} 
\V \ord \big(f\big|_{\vec i} \big) \tV_{p,\Si,\tilde\rho}\cr 
}$$
where $g$ is the function on $\check \cB_n$ such that
$$
f\big|_{(0,\cdots, 0)}(\check\eta_1,\cdots,\check\eta_n) = 
(2\pi)^3\de(\check\eta_1+\cdots+\check\eta_n)\,
g(\check\eta_1,\cdots,\check\eta_n)
$$
}

We extend Definition \deNPdefsectbubbleprop\ of ladders to the case that the rungs are all defined over $\fX_\Si=\check \cB\dunion(\cB\times\Si)$ and the bubble propagators are defined over $\cB$. 

\definition{\STM\deftildeladders}{ Let $\Si$ be a sectorization.

\Item i) Let $P$ be a bubble propagator over $\cB$ and $r$
a rung over $\fX_\Si$.
We set
$$
(r\bullet P)({\sst y_1,y_2;x_3,x_4}) 
=  \smsum_{s'_1,s'_2\in \Si} \int_{\cB\times \cB} {\sst dx'_1 dx'_2}\ 
r({\sst y_1,y_2,(x'_1,s'_1),(x'_2,s'_2)})\ 
P({\sst x'_1,x'_2;x_3,x_4}) 
$$
$(r\bullet P)$ is a function on $\fX_\Si^2 \times \cB^2$. For a general function $F$ on $\fX_\Si^2 \times \cB^2$, define the rung $(F\bullet r)$ over $\fX_\Si$ by 
$$
(F\bullet r)({\sst y_1,y_2,y_3,y_4}) 
=  \smsum_{s'_1,s'_2\in \Si}\int_{\cB\times \cB} {\sst dx'_1 dx'_2 }\ 
F({\sst y_1,y_2;x'_1,x_2'})\ 
r({\sst (x_1',s_1'),(x_2',s_2'),y_3,y_4}) 
$$
if at least one of the arguments $y_1,\cdots,y_4$ lies in 
$\cB\times \Si \subset \fX_\Si$, and for 
$ \check\eta_1,\check\eta_2,\check\eta_3,\check\eta_4 
\in \check\cB\subset \fX_\Si$
$$
(F\bullet r)({\sst \check\eta_1,\check\eta_2,\check\eta_3,\check\eta_4})\,
{\sst (2\pi)^3}\de({\sst\check\eta_1+\check\eta_2+\check\eta_3+\check\eta_4}) 
=  \smsum_{s'_1,s'_2\in \Si}\int_{\cB\times \cB} {\sst dx'_1 dx'_2 }\ 
F({\sst \check\eta_1,\check\eta_2;x'_1,x_2'})\ 
r({\sst (x_1',s_1'),(x_2',s_2'),\check\eta_3,\check\eta_4}) 
$$
\Item ii) Let $\ell \ge 1\,$, $r_1,\cdots,r_{\ell+1}$ rungs over $\fX_\Si$  and 
$P_1,\cdots,P_\ell$  bubble propagators over $\cB$.
The ladder with rungs $r_1,\cdots,r_{\ell+1}$ and 
bubble propagators $P_1,\cdots,P_\ell$ is defined to be
$$
r_1\bullet P_1 \bullet r_2 \bullet P_2\bullet \cdots \bullet
 r_{\ell}\bullet P_\ell\bullet r_{\ell+1}
$$
If $r$ is a  rung over $\fX_\Si$ and $A,B$ are propagators over $\cB$, we define $L_\ell(r;A,B)$ as the ladder with   
$\ell+1$ rungs $r$ and  $\ell$ bubble propagators $\cC(A,B)$.

\Item iii) Definitions (i) and (ii) apply verbatim to the situation when
creation/annihilation indices are ignored, that is, when $\cB$ and $\fX_\Si$
are replaced by $\cB^\updownarrow$ and $\fY_\Si$, respectively.
}
\remark{\STM\remNPnoInternalPhi}{ The ladder 
$
r_1\bullet P_1 \bullet r_2 \bullet P_2\bullet \cdots \bullet
 r_{\ell}\bullet P_\ell\bullet r_{\ell+1}
$ depends only on
$r_1\big|_{\fX_\Si^2\times(\cB\times \Si)^2}$,
$r_2\big|_{(\cB\times \Si)^4}$, $\cdots$,
$r_\ell\big|_{(\cB\times\Si)^4}$ and
$r_{\ell+1}\big|_{(\cB\times \Si)^2\times\fX_\Si^2}$. 
If $r_1,\cdots, r_{\ell+1}$
are supported on $(\cB\times \Si)^4 \subset~\fX_\Si^4$ then Definitions \deNPdefsectbubbleprop.ii and \deftildeladders.ii agree. Also
Lemma \lemunordord.i carries over verbatim.

}

\vskip .3cm

The analog of Proposition \propparticleparticleladder\ is

\proposition{\STM\tildepropparticleparticleladder}{
Let $0<\La<\sfrac{\tau_2}{2M^j}$, where $\tau_2$ is the constant of 
Lemma \lemOSdiffpropbound\ of [FKTo3].
Let $u({\sst(\xi,s)},{\sst(\xi',s')})\in\cF_0(2,\Si_j)$ 
be an antisymmetric, spin independent, particle number conserving function whose
Fourier transforms obeys
 $|\check u(k)| \le \half |\imath k_0-e(k)|$ 
and such that 
$\ \v  u \v_{1,\Si_j} \le \La\cb_j\ $. 
Furthermore let $f \in \check\cF_{4,\Si_j}$. Then for all $\ell \ge 1$
$$\eqalign{
\V L_\ell \big(f;C^{(j)}_u,C^{(\ge j+1)}_u \big) \tV_{3,\Si_j}
& \le \big( \const\,\cb_j \big)^\ell\  \v f\tv_{3,\Si_j}^{\kern5pt \ell+1} \cr
\V V_{\rm pp} \Big( L_\ell \big(f;C^{(j)}_u,C^{(\ge j+1)}_u \big)^{\rm pp}
\Big) \tV_{3,\Si_j}
& \le \big( \const \,\fl_j^{\raise2pt\hbox{$\scriptscriptstyle{1/n_0}$}}\,
\cb_j \big)^\ell\   \v f\tv_{3,\Si_j}^{\kern5pt \ell+1}\cr
}$$
if the Fermi curve $F$ is strongly asymmetric in the sense of Definition
\defNPstrongasymm.
Here, $n_0$ is the constant of Definition
\defNPstrongasymm.
}

\prf
The first inequality is a direct consequence of Remark \remOSnaiveladderest\ of
[FKTo3] with $X=2\La M^j\cb_j$ and  $v'=u'=v=u$ followed by Corollary \corOSappMonoidIV.i
of [FKTo1] with $X=\tau_2,\ \mu=1$. If we set 
$\cC = \cC\big(C^{(j)}_u,C^{(\ge j+1)}_u\big)$,
then, by Lemma \lemunordord.i and Remark \remNPnoInternalPhi
$$
 L_\ell \big(f;C^{(j)}_u,C^{(\ge j+1)}_u \big)^{\rm pp}
= f^{\rm pp}\bullet {^{\rm pp}}\cC\bullet \cdots 
\bullet {^{\rm pp}}\cC\bullet f^{\rm pp} 
$$
Thus the second inequality follows from Theorem \:\theoremOSLadA\ of
[FKTo4]. 
\endproof

\remark{\STM\remtildepropparticleparticleladder}{
By Remarks \remNPsecdiffdecaynorm\  and \remNPnoInternalPhi, Proposition \propparticleparticleladder\ is a special case of Proposition \tildepropparticleparticleladder.
}

Definition \defcompLadder\ and Theorem \theoremcompLadder\ carry over almost verbatim to the present situation.

\definition{\STM\deftildecompLadder}{
Let $\vec F=\set{F^{(i)}}{i=2,3,\cdots}$ be a family of 
antisymmetric functions in $\check\cF_{4,\Si_i}$. Let
$\vec p=\big(p^{(2)},p^{(3)},\cdots \big)$ be a sequence of 
antisymmetric, spin independent, particle number conserving functions
$p^{(i)}({\sst(\xi,s)}, {\sst(\xi',s')}) \in \cF_0(2,\Si_i)$.
We define, recursively on 
$0~\le~j~<~\infty$, the iterated particle hole (or wrong way) ladders up
to scale $j$, denoted by $\,\cL^{(j)}(\vec p,\vec F)\,$, as
$$\eqalign{
\cL^{(0)}(\vec p,\vec F)&=0 \cr
\cL^{(j+1)}(\vec p,\vec F)&= \cL^{(j)}(\vec p,\vec F)_{\Si_j}
+ 2 \smsum_{\ell=1}^\infty {\sst (-1)^\ell(12)^{\ell+1}}
  L_\ell\big(w_j;C^{(j)}_{u_j}, C^{(\ge j+1)}_{u_j}\big)^{\rm ph}
\cr }$$
where $u_j=\smsum_{i=2}^{j-1}p^{(i)}_{\Si_j}$ and 
$w_j = \sum_{i=2}^{j}{F}_{\Si_j}^{(i)}
+\sfrac{1}{8} {\rm Ant\,}\Big( V_{\rm ph} 
 \big( \cL^{(j)}(\vec p,\vec F)\big)\Big)_{\Si_j}$. 
}

\theorem{\STM\theoremtildecompLadder}{
For every $\veps>0$ there are constants $\rho_0,\,\const$ such that the following
holds.  
Let $\vec F=\big(F^{(2)},F^{(3)},\cdots \big)$ be a sequence of 
antisymmetric, spin independent, particle number conserving functions
$F^{(i)}\in \check\cF_{4,\Si_i}$
and $\vec p=\big(p^{(2)},p^{(3)},\cdots \big)$ be a sequence of 
antisymmetric, spin independent, particle number conserving functions
$p^{(i)} \in \cF_0(2,\Si_i)$.
Assume that there is $\rho \le \rho_0$ such that for $i\ge 2$
$$
\v F^{(i)}\tv_{3,\Sigma_i} \le \sfrac{\rho}{M^{\veps i}}\cb_i\qquad\quad 
\v p^{(i)}\v_{1,\Sigma_i} \le \sfrac{\rho\,\fl_i}{M^i}\cb_i\qquad
\check p^{(i)}(0,\k)=0
$$
Then for all $j\ge 2$
$$
\V V_{\rm ph}\big( \cL^{(j)}(\vec p,\vec F)_{\Si_{j}} \big)\tV_{3,\Si_{j}} 
\le \const \rho^2 \,\cb_{j}
$$
}

\noindent
Theorem \theoremtildecompLadder\ follows from Theorem \:\theoremmodcompLadder\ below. Theorem \:\theoremmodcompLadder, in turn, is proven in [FKTl].

\remark{\STM\remtildecompLadder}{
With the same argument as in Remark \remtildepropparticleparticleladder,
Theorem \theoremcompLadder\ is a special case of Theorem \theoremtildecompLadder\ .

}

Theorem \theoremtildecompLadder\ is not general enough for controlling the
effect of a renormalization group step on the $\v\ \cdot\ \tv_{3,\Si_j}$ norms.
Consider an iterated particle--hole ladder  $\cL^{(j)}(\vec p,\vec f\,)$.
Integrating out subsequent scales can result in a propagator and a $\phi$ field being hooked to the $\psi$ legs of the ladder. See the 
$\psi+ C^{(j)}_{u,\Si_j} J\phi$ in Remark \remNPFnochangeI.iii. 
The resulting object is no longer an iterated 
particle--hole ladder. In \S\CHstep, this was harmless because ``external improvement'' with respect to the $\v\ \cdot\ \v_{3,\Si_j}$ norm 
(see Lemma \lemOSsectextimpr\ of [FKTo3]) 
led to bounds that were summable over scales. With the more sensitive norm
$\v\ \cdot\ \tv_{3,\Si_j}$, we have to control the ``shear'' from $\psi$ to 
$\phi$ fields in particle--hole ladders. Under a shear transformation, 
a Grassmann function $\cW(\phi,\psi)$ is mapped to $\cW(\phi,\psi+\hat B\phi)$.
\definition{\STM\defNPshear}{Let $\Si$ be a sectorization, $B(k)$ a function 
on $\bbbr\times\bbbr^2$ and $f\in\check\cF_{n;\Si}$.
\Item i)
The {\bf shear} of $f$ with respect to $B$ is the element 
$\sh(f,B)\in\check\cF_{n;\Si}$ defined by
$$\eqalign{
&\sh(f,B)\big|_{\vec i}(y_1,\cdots,y_n)\cr
&\hskip.5cm=\hskip-8pt
\sum_{\vec j\in\{0,1\}^n\atop j_p\ge i_p,\ 1\le p\le n}
\prod_{1\le\nu\le n\atop i_\nu=0,\ j_\nu=1}\hskip-10pt
\Big\{\smsum_{s_\nu\in\Si}
\int_{\cB}d\xi_\nu\  E_+(\check\et_\nu,\xi_\nu)B(k_\nu)\Big\}
f\big|_{\vec j}(z_1,\cdots,z_n)
\Big|_{z_\nu=(\xi_\nu,s_\nu)\hbox{$\sst {\rm\ if\ }i_\nu=0,\ j_\nu=1$}
   \atop \hskip-20pt z_\nu=y_\nu\hskip14.5pt\hbox{$\sst {\rm\ if\ }i_\nu=j_\nu$} }
}$$
where, for $i_\nu=0$, $y_\nu=\check\et_\nu=(k_\nu,\si_\nu,a_\nu)\in\check\cB$ 
 and $E_+$ was defined before Definition \defNPfourtrans.

\Item ii)
We use $Gr(\phi,\psi;f)$ to denote
 the Grassmann function with kernel $f$. That is,
$$
Gr(\phi,\psi;f) =\sum_{\vec i\in\{0,1\}^n}\
\int dy_1\cdots dy_n\ 
\hat f\big|_{\vec i}(y_1,\cdots,y_n)\prod_{p=1}^n
\cases{\phi(y_p)& if $i_p=0$\cr
       \psi(y_p)& if $i_p=1$\cr}
$$
where factors in the product are in the order specified by the index $p$,
$\hat f$ is the Fourier transform of $f$ with respect to its $\phi$ arguments
and $y_\nu$ runs over $\cB$ when $\nu=0$ and over $\cB\times\Si$ when $\nu=1$.

}

\noindent
The definition of shear has been chosen so that
$$
Gr(\phi,\psi;\sh(f,B))=Gr(\phi,\psi+\hat B\phi;f)
\EQN\eqnNPshear$$
where, with some abuse of notation, we set
$\ 
(\hat B\phi)(\xi,s)=\int_\cB d\xi'\  \hat B(\xi,\xi')\phi(\xi')$, 
for all $s\in\Si$, retaining the $\hat B$ defined in Notation  \notNPfourierTI.

\corollary{\STM\cortildecompLadder (to Theorem \theoremtildecompLadder)}{
For every $\veps>0$ and $c_B>0$ there are constants 
$\rho_0,\,\const$ such that the following holds.  
Let $\vec v=\big(v^{(2)},v^{(3)},\cdots \big)$ be a sequence of 
antisymmetric, spin independent, particle number conserving functions
$v^{(i)}\in \check\cF_{4,\Si_i}$
and $\vec p=\big(p^{(2)},p^{(3)},\cdots \big)$ be a sequence of 
antisymmetric, spin independent, particle number conserving functions
$p^{(i)} \in \cF_0(2,\Si_i)$.
Assume that there is $\rho \le \rho_0$ such that for $i\ge 2$
$$
\v v^{(i)}\tv_{3,\Sigma_i} \le \sfrac{\rho}{M^{\veps i}}\cb_i\qquad\quad 
\v p^{(i)}\v_{1,\Sigma_i} \le \sfrac{\rho\,\fl_i}{M^i}\cb_i\qquad
\check p^{(i)}(0,\k)=0
$$
Let $B(k)$ be a function obeying $\| B(k)\tnorm \le  c_B \cb_j$,
set $f^{(i)}=\sh\big(v^{(i)},\nu^{(\ge i)} B\,\big)\in\check\cF_{4,\Si_i}$
and let $\vec f=\big(f^{(2)},f^{(3)},\cdots \big)$.
\Item i)
For all $j\ge 2$,
$$
\V V_{\rm ph}\big( \cL^{(j)}(\vec p,\vec f\,)_{\Si_{j}} \big)\tV_{3,\Si_{j}} 
\le \const \rho^2 \,\cb_{j}
$$

\Item ii)
Let $B'(k)$ obey $\| B'(k)\tnorm \le  c'c_B \cb_j$,
set $f^{(i)}_s=\sh\big(v^{(i)},\nu^{(\ge i)}( B+sB')\,\big)\in\check\cF_{4,\Si_i}$
and let $\vec f_s=\big(f_s^{(2)},f_s^{(3)},\cdots \big)$.
For all $j\ge 2$ and all $c'>0$,
$$
\V \sfrac{d\hfill}{ds}V_{\rm ph}\big( \cL^{(j)}(\vec p,\vec f_s\,)_{\Si_{j}} \big)\big|_{s=0}\tV_{3,\Si_{j}} 
\le \const c'\rho^2 \,\cb_{j}
$$

}

In the proof, which follows Remark \:\remNPshearprime,
 we will use auxiliary external fields, named $\phi'$. We now 
extend the notation of Definitions \defNPdisjointOrd\ and  
\defNPdisjointfnspaces\ to include them.

\definition{\STM\defNPmoredisjointOrd}{
Let $\Si$ be a sectorization.
Set $\fX_{-1}=\fX_0 = \check \cB$, $\fX_1 = \cB \times \Si$ and
$\fX'_\Si=\fX_{-1}\dunion\fX_0\dunion\fX_1$. 
Let $\vec i = (i_1,\cdots,i_n) \in \{-1, 0,1\}^n$. 

\Item i) The inclusions of $\fX_{i_j},\ j=1,\cdots,n,\ $ in $\fX'_\Si$ induce an inclusion of 
$\fX_{i_1} \times\cdots\times \fX_{i_n}$ in ${\fX'}_\Si^n$. We identify 
$\fX_{i_1} \times\cdots\times \fX_{i_n}$ with its image in ${\fX'}_\Si^n$.

\Item ii) Set $m'(\vec i) = \#\set{1\le j\le n}{i_j=-1}$ and 
$m(\vec i) = \#\set{1\le j\le n}{i_j=0}$. 

\Item iii)
If $f$ is a function on $ \fX_{i_1} \times\cdots\times \fX_{i_n}$, then 
$\ord f$ is the function on 
$\check \cB^{m'(\vec i)}\times\check \cB^{m(\vec i)}\times (\cB \times \Si)^{n-m'(\vec i)-m(\vec i)}$ 
obtained from $f$ by permuting the arguments so that all $\fX_{-1}$ arguments
appear before all $\fX_{0}$ arguments and all $\fX_{0}$ arguments appear before
all $\fX_1$ arguments, while preserving the relative order of the 
$\fX_j$ arguments, $j=-1,0,1$,  and multiplying by the sign of the permutation that implements the reordering of the arguments. That is, 
$\ord f(x_1,\cdots,x_n)= \sgn\, \si\ f(x_{\si(1)},\cdots,x_{\si(n)})$
where the permutation $\si\in S_n$ is determined by
$\ 
\si(j)<\si(j')\hbox{ if } i_{j}<i_{j'}\hbox{ or } i_{j}=i_{j'},\ j<j'.
\ $
\Item iv)
Using the identification of part (i), 
$$
{\fX'}_\Si^n\ =\ 
\bigcup_{i_1,\cdots,i_n \in \{-1,0,1\}}\kern-3.45em\cdot\kern3.45em \fX_{i_1} \times\cdots\times \fX_{i_n}
$$
where, on the right hand side we have a disjoint union. If $f$ is a function on
${\fX'}_\Si^n$ and $\vec i = (i_1,\cdots,i_n) \in \{-1,0,1\}^n$, we denote 
by $f\big|_{\vec i}$ the restriction of $f$ to 
$\fX_{i_1} \times\cdots\times \fX_{i_n}$.
}

\definition{\STM\defNPmoredisjointfnspaces}{
Let $\Si$ be a sectorization and $m',m,n \ge 0$.
\Item i)
For $n\ge 1$, denote by  $\check\cF_{m',m}(n;\Si)$ the space
of all translation invariant, complex valued functions 
$\ 
f({\sst\check\eta_1,\cdots,\check\eta_{m'+m};\,(\xi_1,s_1),\cdots,(\xi_n,s_n)} )
\ $
on $\fX_{-1}^{m'} \times\fX_0^m \times  \big( \cB \times\Si \big)^n$ 
whose Fourier transform $\check f({\sst\check\eta_1,\cdots,\check\eta_m;
\,(\check\xi_1,s_1),\cdots,(\check\xi_n,s_n)} )$
vanishes unless 
$ k_i\in \tilde s_i$ for all $1\le j\le n$. 
Here, $\check\xi_i=(k_i,\si_i,a_i)$.
Also, let $\check\cF_{m',m}(0;\Si)$  be the space of all momentum conserving,
complex valued functions 
$\ 
f({\sst\check\eta_1,\cdots,\check\eta_{m'+m}} )
\ $
on $\fX_{-1}^{m'} \times\fX_0^m$.

\Item ii) We denote by $\check \cF'_{n;\Si}$ the set of functions 
on ${\fX'}_\Si^n$ with the property that for each 
$\vec i = (i_1,\cdots,i_n) \in \{-1,0,1\}^n$ with $m'(\vec i)+m(\vec i) <n$
$$
\ord \big(f\big|_{\vec i} \big) \in \check \cF_{m'(\vec i),m(\vec i)}
(n-m'(\vec i)-m(\vec i);\Si)
$$
and such that for each $\vec i = (i_1,\cdots,i_n) \in \{-1,0,1\}^n$ with $m'(\vec i)+m(\vec i)=n$
there is a function 
$g_{m',m}\in\check\cF_{m',m}(0;\Si)$ such that
$$
\ord \big(f\big|_{\vec i} \big)(\check\eta_1,\cdots,\check\eta_n) = 
(2\pi)^3\de(\check\eta_1+\cdots+\check\eta_n)\,g_{m',m}
(\check\eta_1,\cdots,\check\eta_n)
$$

\Item iii) There is a natural identification $\Pi: \check\cF_{m',m}(n;\Si)
\rightarrow \check\cF_{m'+m}(n;\Si)$ obtained by identifying 
$\fX_{-1}^{m'}\times\fX_0^m=\check\cB^{m'}\times\check\cB^m$ with 
$\fX_0^{m'+m}=\check\cB^{m'+m}$. Similarly, if 
$\vec i = (i_1,\cdots,i_n) \in \{-1, 0,1\}^n$ and $f$ is a function on 
$\fX_{i_1} \times\cdots\times \fX_{i_n}$ then the function $\Pi(f)$
on $\fX_{\pi(i_1)} \times\cdots\times \fX_{\pi(i_n)}$ where $\pi(-1)=\pi(0)=0$
and $\pi(1)=1$, is obtained by identifying $\fX_{-1}$ with $\fX_0$.
We extend the map to 
$\Pi: \check\cF'_{n;\Si}\rightarrow \check\cF_{n;\Si}$ by
$$
\Pi(f)\big|_{\vec i}=\sum_{\vec j\in\{-1,0,1\}^n\atop\pi(\vec j)=\vec i}
\Pi\big(f\big|_{\vec j}\big)\qquad\hbox{for all $\vec i\in\{0,1\}$}
$$

\Item iv)
For a function $f\in \check \cF_{m',m}(n;\Si)$ and a natural number $p$ we set
$$
\v f\tv_{p,\Si} =  \v \Pi(f) \tv_{p,\Si} 
$$
For a function $f\in \check \cF'_{n;\Si}$ and a natural number $p$ we set
$$
\v f\tv_{p,\Si} =  \sum_{\vec i\in\{-1,0,1\}^n}
\V \Pi\big(f\big|_{\vec i}\big)\tV_{p,\Si}
$$
}

\lemma{\STM\lemNPremoveprimes}{ For $\vec\ka\in\bbbc^n$, define
$S_{\vec\ka}:\check\cF'_{n,\Si}\rightarrow\check\cF'_{n,\Si}$ by
$$
\big(S_{\vec\ka}f\big)\big|_{\vec i}
=\Big(\smprod_{j=1}^n\ka_{j}^{1-i_j}\Big)f\big|_{\vec i}
\qquad\hbox{for all $\vec i\in\{-1,0,1\}^n$}
$$
Then, for all $f\in\check\cF'_{n,\Si}$
$$
\v \Pi(f) \tv_{p,\Si}\le \v f\tv_{p,\Si}
\le 3^n\sup_{\vec\ka\in\bbbc^n\atop |\ka_j|\le 1,\ 1\le j\le n}
\v \Pi(S_{\vec\ka}f) \tv_{p,\Si}
$$

}
\prf The first bound is just the triangle inequality.
$$
\v \Pi(f) \tv_{p,\Si}=
\sum_{\vec i\in\{0,1\}^n}\V \Pi(f)\big|_{\vec i} \tV_{p,\Si}
\le \sum_{\vec i\in\{0,1\}^n}\sum_{\vec j\in\{-1,0,1\}^n\atop\pi(\vec j)=\vec i}
\V \Pi\big(f\big|_{\vec j}\big) \tV_{p,\Si}
= \v f\tv_{p,\Si}
$$
For the second bound, we just use the Cauchy integral formula
$$
f\big|_{\vec i}=\Big(\smprod_{j=1}^n\sfrac{1}{(1-i_j)!}
\sfrac{d^{1-i_j}\hfill}{d\ka^{1-i_j}_j}\Big)
S_{\vec\ka}f\Big|_{\vec\ka=\vec 0}
=\Big(\smprod_{j=1}^n\oint_{|\ka_j|=1}
\sfrac{d\ka_j}{2\pi \imath}\sfrac{1}{\ka_j^{2-i_j}}\Big)
S_{\vec\ka}f
$$
to prove that, for each $\vec i\in\{-1,0,1\}^n$,
$$
\V f\big|_{\vec i} \tV_{p,\Si}
=\V \Pi\big(f\big|_{\vec i}\big) \tV_{p,\Si}
\le \sup_{\vec\ka\in\bbbc^n\atop |\ka_j|\le 1,\ 1\le j\le n}
\v \Pi(S_{\vec\ka}f) \tv_{p,\Si}
$$
and then sum over $\vec i\in\{-1,0,1\}^n$.
\endproof

Let $B(k)$ be a kernel which is the product of two other kernels $B_1(k)$ 
and $B_2(k)$. Then, for any Grassmann function $\cW(\phi,\psi)$,
$$
\cW(\phi,\psi+\hat B\phi)=\cW(\phi,\psi+\hat B_1\phi')\big|_{\phi'=\hat B_2\phi}
$$
Thus the shear transformation with respect to $B$
 may be written as the composition of another shear--like transformation 
with respect to $B_1$ and a ``scaling transformation'' with respect to $B_2$.
To make this precise, we have

\definition{\STM\defNPshearprime}{Let $\Si$ be a sectorization and
 $B(k)$ a function on $\bbbr\times\bbbr^2$. 
\Item i)
If $f\in\check\cF_{n;\Si}$, then the element $\sh'(f,B)\in\check\cF'_{n;\Si}$ 
is defined by
$$\eqalign{
&\sh'(f,B)\big|_{\vec i}(y_1,\cdots,y_n)\cr
&\hskip.5cm=
\prod_{1\le\nu\le n\atop i_\nu=-1}\Big\{\smsum_{s_\nu\in\Si}
\int_{\cB}d\xi_\nu\  E_+(\check\et'_\nu,\xi_\nu)B(k'_\nu)\Big\}
f\big|_{|\vec i|}(z_1,\cdots,z_n)
\Big|_{z_\nu=(\xi_\nu,s_\nu)\hbox{$\sst {\rm\ if\ }i_\nu=-1$}
   \atop z_\nu=y_\nu\hskip15pt\hbox{$\sst {\rm\ if\ }i_\nu\ne-1$} }
}$$
where $|\vec i|=(|i_1|,\cdots,|i_n|)$ and 
$y_\nu=\check\et'_\nu=(k'_\nu,\si_\nu,a_\nu)\in\check\cB$ when $i_\nu=-1$.

\Item ii)
If $f\in\check\cF'_{n;\Si}$, then the element $\sct(f,B)\in\check\cF'_{n;\Si}$ 
is defined by
$$\eqalign{
&\sct(f,B)\big|_{\vec i}(y_1,\cdots,y_n)
=
\Big\{\smprod_{1\le\nu\le n\atop i_\nu=-1}B(k'_\nu)\Big\}
f\big|_{\vec i}(y_1,\cdots,y_n)
}$$
where $y_\nu=(k'_\nu,\si_\nu,a_\nu)\in\check\cB$ if $i_\nu=-1$.

\Item iii) 
If $f\in\check\cF_{n;\Si}$, then the element 
$\Sct(f,B)\in\check\cF_{n;\Si}$ is defined by
$$
\Sct(f,B)\big|_{\vec i}(y_1,\cdots,y_n)
=
\Big\{\smprod_{1\le\nu\le n\atop i_\nu=0}B(k_\nu)\Big\}
f\big|_{\vec i}(y_1,\cdots,y_n)
$$
where $y_\nu=(k_\nu,\si_\nu,a_\nu)\in\check\cB$ if $i_\nu=0$.

}

\remark{\STM\remNPshearprime}{
\Item i) If $B_1(k)$ and $B_2(k)$ are  functions on $\bbbr\times\bbbr^2$
and $f\in\check\cF_{n;\Si}$, then
$$
\sh(f,B_1B_2)=\Pi\Big(\sct\big(\sh'(f,B_1),B_2\big)\Big)
$$
\Item ii) Let $f\in\check\cF_{n;\Si_i}$, $f'\in\check\cF'_{n;\Si_i}$
 and $B(k)$ be a function obeying $\| B(k)\tnorm \le c_B\cb_i$. 
Then, by repeated application of equation (\eqnNPderivampprelim) of [FKTo3],
with $j$ replaced by $i$, $X=0$, $X_B=1$,
 there is a constant $\abcst$, depending on $n$, such that
$$\eqalign{
\v\sh'(f,B)\tv_{p,\Si_i}
&\le\abcst\,\max\{1,c_B\}^n\cb_i\, \v f\tv_{p,\Si_i}\cr
\v\Sct(f,B)\tv_{p,\Si_i}
&\le\abcst\,\max\{1,c_B\}^n\cb_i\, \v f\tv_{p,\Si_i}\cr
\v\sct(f',B)\tv_{p,\Si_i}
&\le\abcst\,\max\{1,c_B\}^n\cb_i\, \v f'\tv_{p,\Si_i}\cr
}$$

}

\proof{of Corollary \cortildecompLadder}
i) Let $\tilde v^{(i)}=\sh'(v^{(i)},\nu^{(\ge i)})$. Then,  
by Remark \remNPshearprime.i, 
$$
f^{(i)}=\Pi\big(\sct(\tilde v^{(i)},B)\big)
$$
and, extending the definition of 
$\cL^{(j)}(\vec p,\vec F)$ to $F^{(i)}\in\check\cF'_{n;\Si}$ in the obvious
way (just replace $\check \cB$ by $\check \cB\dunion\check \cB$ in Definitions
\deftildeladders\ and \deftildecompLadder),
$$\eqalign{
V_{\rm ph}\big( \cL^{(j)}(\vec p,\vec f\,)_{\Si_j}\big)
&= V_{\rm ph}\Big( \cL^{(j)}\big(\vec p,
               \Pi(\sct(\vec{\tilde v},B))\big)_{\Si_j}\Big)\cr
&= \Pi\Big(\sct\Big(V_{\rm ph}\big( \cL^{(j)}\big(\vec p,\vec{\tilde v}\,)
                \big)_{\Si_j},B\Big)\Big)\cr
}$$
Hence, by Lemma \lemNPremoveprimes\ and Remark \remNPshearprime.ii,
$$\eqalign{
\V V_{\rm ph}\big( \cL^{(j)}(\vec p,\vec f\,)_{\Si_{j}} \big)\tV_{3,\Si_{j}} 
&\le \VV \sct\Big(V_{\rm ph}\big( \cL^{(j)}\big(\vec p,\vec{\tilde v}\,)
                \big)_{\Si_j},B\Big)\tVV_{3,\Si_{j}} \cr
&\le \abcst\,\cb_j\ 
\V V_{\rm ph}\big( \cL^{(j)}\big(\vec p,\vec{\tilde v}\,)
                \big)_{\Si_j}\tV_{3,\Si_{j}} \cr
&\le \abcst\,\cb_j\ \sup_{\vec\ka\in\bbbc^4\atop |\ka_\nu|\le 1,\ 1\le \nu\le 4}
\VV \Pi\Big(S_{\vec\ka}V_{\rm ph}\big( \cL^{(j)}\big(\vec p,\vec{\tilde v}\,)
                \big)_{\Si_j}\Big)\tVV_{3,\Si_{j}} \cr
}$$
Observe that 
$$
\Pi\Big(S_{\vec\ka}\,V_{\rm ph}\big( \cL^{(j)}\big(\vec p,\vec{\tilde v}\,)
                \big)_{\Si_j}\Big)
=V_{\rm ph}\Big( \cL^{(j)}\big(\vec p,\Pi(S_{\vec\ka}\vec{\tilde v}\,)\big)
                \Big)_{\Si_j}
$$
By Definition \defNPscales.ii, $\| \nu^{(\ge i)}(k)\tnorm \le \abcst\,\cb_i$.
Hence, by Lemma \lemNPremoveprimes\ and Remark \remNPshearprime.ii,
$$\eqalign{
\sup_{\vec\ka\in\bbbc^4\atop |\ka_\nu|\le 1,\ 1\le \nu\le 4}
\V \Pi\big(S_{\vec\ka}{\tilde v}^{(i)}\big)\tV_{3,\Si_i}
&\le \sup_{\vec\ka\in\bbbc^4\atop |\ka_\nu|\le 1,\ 1\le \nu\le 4}
\V S_{\vec\ka}{\tilde v}^{(i)}\tV_{3,\Si_i}\cr
&\le \V {\tilde v}^{(i)}\tV_{3,\Si_i}
\le \abcst\,\cb_{i}\V {v}^{(i)}\tV_{3,\Si_i}
\le \abcst\,\sfrac{\rho}{M^{\veps i}}\cb^2_{i}\cr
&\le\abcst\,\sfrac{\rho}{M^{\veps i}}\cb_{i}\cr
}$$
So Theorem \theoremtildecompLadder\  gives
$$\eqalign{
\V V_{\rm ph}\big( \cL^{(j)}(\vec p,\vec f\,)_{\Si_{j}} \big)\tV_{3,\Si_{j}}
&\le \abcst\,\cb_j\ \sup_{\vec\ka\in\bbbc^4\atop |\ka_\nu|\le 1,\ 1\le \nu\le 4}
\VV V_{\rm ph}\Big( \cL^{(j)}\big(\vec p,\Pi(S_{\vec\ka}\vec{\tilde v})\big)
                \Big)_{\Si_j}\tVV_{3,\Si_{j}} \cr
&\le \const \rho^2 \,\cb_{j}
}$$

\Item ii) Part (i), with $c_B$ replaced by $2c_B$ and $B$ replaced by
$B+sB'$, implies that, for all $s\in\bbbc$ obeying $|s|\le\sfrac{1}{c'}$,
$$
\V V_{\rm ph}\big( \cL^{(j)}(\vec p,\vec f_s\,)_{\Si_{j}} \big)\tV_{3,\Si_{j}}
\le \const \rho^2 \,\cb_{j}
$$
Since $V_{\rm ph}\big( \cL^{(j)}(\vec p,\vec f_s\,)_{\Si_{j}}$ is a
polynomial in $s$, the desired bound now follows from the Cauchy
integral formula
$$
\sfrac{d\hfill}{ds}V_{\rm ph}\big( \cL^{(j)}(\vec p,\vec f_s\,)_{\Si_{j}} \big)\big|_{s=0}
=\sfrac{1}{2\pi i}\oint_{|s|={1\over c'}}\hskip-15pt ds \hskip5pt
\sfrac{1}{s^2}
V_{\rm ph}\big( \cL^{(j)}(\vec p,\vec f_s\,)_{\Si_{j}}
$$

\vfill\eject

\chap{Recursion Step for Momentum Green's Functions}\PG\pgNPXV

This section provides the analog of \S\CHstep\ for the $\v\ \ \tv$--norms.
Recall, from \S\CHintroIII, that we assuming that the interaction $V$ 
satisfies the reality condition (\eqnNPreal) and is
bar/unbar exchange invariant in the sense of (\eqnNPphexchange).

\titleb{More Input and Output Data}\PG\pgNPXVa 
We now supplement the conditions on the input and output data
of Definitions \stepInputData\ and \stepOutputData\ in order get more
detailed information on the behaviour of the two-- and four--point 
functions. Recall from Theorem \theoremNPtildeinduction\ that 
$\half<\aleph<\aleph'<\sfrac{2}{3}$. We generalize the notation 
$\cb_j$ and $\fe_j(X)$ of Definition \defNPFancynormdomain\ to
$$\eqalign{
\cb_{i,j}
&=\sum_{|\bde|\le r\atop |\de_0|\le r_0}  M^{i\de_0} M^{j|\bde|}\,t^\de
+\sum_{|\bde|> r\atop {\rm or\ }|\de_0|> r_0}\infty\, t^\de
\in\fN_{d+1}\cr
\fe_{i,j}(X)&=\sfrac{\cb_{i,j}}{1-M^jX}\cr
}\EQN\eqnTNPcbfeij$$
so that we can track different degrees of smoothness in temporal and spatial
directions.

\definition{\STM\moreInputData (More Input Data)}{
 Let $\tilde\cD^{(j)}_{\rm in}$ be the set of 
interaction quadruples,  $(\cW,\cG,u,\vec p)$, that fulfill Definitions
\defNPintquad, \stepInputData\ and the following.
Let  $w(\phi,\psi;K)$ be the $\Si_j$--sectorized representative of $\cW(K)$
specified in (I1) and $w^a(\phi,\psi;K)$ its amputation in the sense of
Definition \defNPampGreen.
\item{($I^\sim 1$)}  
The Grassmann function $w^a(\phi,\psi;K)-w^a(0,\psi;K)$ vanishes unless 
$\hat\nu^{(<j)}\phi$ is nonzero and
$$\eqalign{
N^\sim_j\big(w^a(K),64\al,\|K\|_{1,\Si_j}\big) 
&\le \fe_j\big(\|K\|_{1,\Si_j}\big) \cr
N^\sim_j\big(\sfrac{d\hfill}{ds} w^a(K+sK')\big|_{s=0}
,64\al,\,\| K\|_{1,\Si_{j}}\big) 
   &\le M^j \,\fe_j\big(\|K\|_{1,\Si_j}\big) \,  \|K'\|_{1,\Si_j}
\cr
}$$
for all $ K\in\fK_{j}$ and all $ K'$.

\item{($I^\sim 2$)}  
There is a family $\vec v$ of  antisymmetric kernels
$$
v^{(i)}\in\check\cF_{4,\Si_i},\qquad 2\le i\le j-1
$$ 
(independent of $K$) and an antisymmetric kernel 
$\de f^{(j)}(K)\in\check\cF_{4,\Si_j}$ obeying
$$\eqalign{
\V v^{(i)}\tV_{3,\Si_i,\tilde\rho}&\le  \sfrac{\vi_i}{\al^7}\,\cb_i
\qquad\hbox{for all $2\le i\le j-1$}
\cr
\V \de f^{(j)}(0)\tV_{3,\Si_j,\tilde \rho}
&\le  \sfrac{\vi_j}{\al^7}\cb_j
\cr
\V \sfrac{d\hfill}{ds}{\de f}^{(j)}(K+sK')\big|_{s=0}\tV_{3,\Si_{j},\tilde\rho}
&\le \sfrac{1}{64\al^4\IB^2}\fe_j({\sst\|K\|_{1,\Si_j}})M^j\|K'\|_{1,\Si_j}\cr
}$$
for all $K\in\fK_{j}$,
from which the quartic parts of $w^a$ and $\cG^a$ are built as follows.
Let  $f^{(i)}=\sh\big(v^{(i)},C^{[i,j)}_{u(0)}(k)A(k)\big)\in\check\cF_{4,\Si_i},\ 
2\le i\le j-1$, where $\sh(f,B)$ was defined in Definition \defNPshear.
Then the kernel of the quartic part of $w^a+\cG^a$ is
$$
\de f^{(j)}(K)+\smsum_{i=2}^{j-1}f^{(i)}_{\Si_j}+
\sfrac{1}{8}{\rm Ant\,}
\Big( V_{\rm ph}\big( \cL^{(j)}(\vec p,\vec f)\big) \Big)_{\Si_j}
$$
where the particle--hole value $V_{\rm ph}$ was defined in 
Definition \deftildeParicleHoleDecomp. The kernel $v^{(i)}$ vanishes unless
all of its $\phi$ momenta are in the support of $\nu^{(<i)}$.
\item{($I^\sim 3$)}
Let $\int d\eta_1 d\eta_2\ G_{2}(\eta_1, \eta_2;K)\ \phi(\eta_1)\phi(\eta_2)$
be the part of $\cG(K)$ that is homogeneous of degree two. Then  
2$\check G_{2}(k;K)$ has the decomposition 
$$
2\check G_{2}(k;K)=  C^{(< j)}_{u(0)}(k)+ \sfrac{1}{(ik_0-e(\k))^2}
\sum_{i=2}^{j-1}\Big\{\de q^{(i)}(k;K)+\smsum_{\ell=i}^{j}q^{(i,\ell)}(k)\Big\}
$$
with $q^{(i,\ell)}(k)$, $i\le\ell\le j$  and $\de q^{(i)}(k;K)$
vanishing in the $(i+2)^{\rm nd}$ neighbourhood and when $\k$ is not in
the support of $U(\k)$, $\de q^{(i)}(k;0)=0$ for $2\le i\le j-1$ and
$$\eqalign{
\big\| q^{(i,\ell)}(k) \Tnorm &\le 
\la_0^{1-2\upsilon}\sfrac{\fl_\ell}{M^\ell}M^{\aleph'(\ell-i)}\cb_{i,\ell}\cr
\big\| \sfrac{d\hfill}{ds}\de q^{(i)}(K+sK')\big|_{s=0}\,\Tnorm
&\le M^{\aleph'(j-i)}\fe_{i+{1\over 2},j+{1\over 2}}\big(\|K\|_{1,\Si_j}\big) \,  \|K'\|_{1,\Si_j}\cr
}$$
Let $w^a_{1,1}(\et,(\xi,s);K)$ be the kernel of the part of $w^a(K)$ that
is of degree 1 in $\phi$ and degree 1 in $\psi$.
Then
$$
\v w_{1,1}^{a\sim}(K)\tv_{1,\Si_j,\tilde\rho}
\le \sfrac{1}{\al^6}\sfrac{\fl_j}{M^j}\fe_j({\sst \| K\|_{1,\Si_{j}}})
$$
\item{($I^\sim 4$)}   $\cW$, $\cG$,  the $\Si_j$--sectorized representative
$w(\phi,\psi;K)$ of $\cW(K)$ and  all of the $F^{(i)}$'s and $v^{(i)}$'s
are bar/unbar invariant in the sense of Definition \defOSsymmetries.B of
[FKTo2]. If $K$ is real, then $\cW$, $\cG$,  
the $\Si_j$--sectorized representative
$w(\phi,\psi;K)$ of $\cW(K)$ and  all of the $q^{(i,\ell)}$'s
are $k_0$--reversal real in the sense of Definition \defOSsymmetries.R of
[FKTo2].

}

\remark{\STM\remMorebarunbar}{
By Remark \remOSsymmetryConsequences.ii of [FKTo2], for any interaction 
quadruple $(\cW,\cG,u,\vec p)$, $u$ and every $p^{(i)}$ is bar/unbar invariant.

}

\definition{\STM\moreOutputData (More Output Data)}{
Let $\tilde\cD^{(j)}_{\rm out}$ be the set of interaction 
quadruples $(\cW,\cG,u,\vec p)$ that fulfill Definitions
\defNPintquad, \stepOutputData\ and the following.
Let  $w(\phi,\psi;K)$ be the $\Si_j$--sectorized representative of $\cW(K)$
specified in (O1) and $w^a(\phi,\psi;K)$ its amputation in the sense of
Definition \defNPampGreen.
\item{($O^\sim 1$)}  
The Grassmann function $w^a(\phi,\psi;K)-w^a(0,\psi;K)$  vanishes unless 
$\hat\nu^{(\le j)}\phi$ is nonzero and
$$\eqalign{
N^\sim_j\big(w^a(K),\al,\|K\|_{1,\Si_j}\big) 
&\le \fe_j\big(\|K\|_{1,\Si_j}\big) \cr
N^\sim_j\big(\sfrac{d\hfill}{ds} w^a(K+sK')\big|_{s=0}
,\al,\,\| K\|_{1,\Si_{j}}\big) 
   &\le M^j \,\fe_j\big(\|K\|_{1,\Si_j}\big) \,  \|K'\|_{1,\Si_j}
\cr
}$$
for all $ K\in\fK_{j}$ and all $ K'$.
\item{($O^\sim 2$)}  
There is a family $\vec v$ of  antisymmetric kernels
$$
v^{(i)}\in\check\cF_{4,\Si_i},\qquad 2\le i\le j
$$ 
(independent of $K$) and an antisymmetric kernel 
$\de f^{(j+1)}(K)\in\check\cF_{4,\Si_j}$ obeying
$$\eqalign{
\V v^{(i)}\tV_{3,\Si_i,\tilde\rho}&\le  \sfrac{\vi_i}{\al^7}\,\cb_i
\qquad\hbox{for all $2\le i\le j$}
\cr
\V \de f^{(j+1)}(0)\tV_{3,\Si_j,\tilde \rho}
&\le \sfrac{\vi_{j+1}}{\al^8}\cb_j
\cr
\V\sfrac{d\hfill}{ds}{\de f}^{(j+1)}(K+sK')\big|_{s=0}\tV_{3,\Si_{j},\tilde\rho}
&\le \sfrac{1}{\al^4\IB^2}\fe_j({\sst\|K\|_{1,\Si_j}})M^j\|K'\|_{1,\Si_j}\cr
}$$
for all $K\in\fK_{j}$,
from which the quartic parts of $w^a$ and $\cG^a$ are built as follows.
Let  $f^{(i)}=\sh\big(v^{(i)},C^{[i,j]}_{u(0)}(k)A(k)\big)\in\check\cF_{4,\Si_i},\ 
2\le i\le j$.
Then the kernel of the quartic part of $w^a+\cG^a$ is
$$
\de f^{(j+1)}(K)+\smsum_{i=2}^{j}f^{(i)}_{\Si_j}+
\sfrac{1}{8}{\rm Ant\,}
\Big( V_{\rm ph}\big( \cL^{(j+1)}(\vec p,\vec f)\big) \Big)
$$
 The kernel $v^{(i)}$ vanishes unless
all of its $\phi$ momenta are in the support of $\nu^{(<i)}$.
\item{($O^\sim 3$)}
Let $\int d\eta_1 d\eta_2\ G_{2}(\eta_1, \eta_2;K)\ \phi(\eta_1)\phi(\eta_2)$
be the part of $\cG(K)$ that is homogeneous of degree two. Then  
2$\check G_{2}(k;K)$ has the decomposition 
$$
2\check G_{2}(k;K)=  C^{(\le j)}_{u(0)}(k)+ \sfrac{1}{(ik_0-e(\k))^2}
\sum_{i=2}^{j}\Big\{\de q^{(i)}(k;K)+\smsum_{\ell=i}^{j}q^{(i,\ell)}(k)\Big\}
$$
with $q^{(i,\ell)}(k)$, $i\le\ell\le j$  and $\de q^{(i)}(k;K)$
vanishing in the $(i+2)^{\rm nd}$ neighbourhood  and when $\k$ is not in
the support of $U(\k)$,
$\de q^{(i)}(k;0)=0$ for $2\le i\le j$ and
$$\eqalign{
\big\| q^{(i,\ell)}(k) \Tnorm
&\le \la_0^{1-2\upsilon}\sfrac{\fl_\ell}{M^\ell}M^{\aleph'(\ell-i)}\cb_{i,\ell}\cr
\big\| \sfrac{d\hfill}{ds}\de q^{(i)}(K+sK')\big|_{s=0}\,\Tnorm
&\le M^{\aleph'(j-i)}\fe_{i+{1\over 2},j+{1\over 2}}\big(\|K\|_{1,\Si_j}\big) \,  \|K'\|_{1,\Si_j}
\qquad\hbox{for $i<j$}\cr
\big\| \sfrac{d\hfill}{ds}\de q^{(j)}(K+sK')\big|_{s=0}\,\Tnorm
&\le \sqrt{M^{\aleph'-\aleph}}
\fe_{j+{1\over 2}}\big(\|K\|_{1,\Si_j}\big) \,  \|K'\|_{1,\Si_j}
}$$
Let $w^a_{1,1}(\et,(\xi,s);K)$ be the kernel of the part of $w^a(K)$ that
is of degree 1 in $\phi$ and degree 1 in $\psi$.
Then
$$\eqalign{
\V w_{1,1}^{a\sim}(K)_{\Si_{j+1}}\tV_{1,\Si_{j+1},\tilde\rho}
&\le \sfrac{1}{\al^7} \sfrac{\fl_j}{M^j}\fe_j({\sst \| K\|_{1,\Si_{j}}})\cr
\V \sfrac{d\hfill}{ds}w^{a\sim}_{1,1}({\sst K+sK'})_{\Si_{j+1}}
                                       \big|_{s=0}\tV_{1,\Si_{j+1},\tilde\rho}
&\le\big(\sfrac{1}{\al}+\la_0^{\upsilon/8}\big)
      \sfrac{1}{\al^2\IB}\,\fe_j({\sst \| K\|_{1,\Si_{j}}})\,\| K'\|_{1,\Si_{j}}
}$$
\item{($O^\sim 4$)}   $\cW$, $\cG$,  the $\Si_j$--sectorized representative
$w(\phi,\psi;K)$ of $\cW(K)$ and  all of the $F^{(i)}$'s and $v^{(i)}$'s
are bar/unbar invariant. If $K$ is real, then
$\cW$, $\cG$,  the $\Si_j$--sectorized representative
$w(\phi,\psi;K)$ of $\cW(K)$ and  all of the $q^{(i,\ell)}$'s
are $k_0$--reversal real in the sense of Definition \defOSsymmetries.R of
[FKTo2].

}

\goodbreak
\titleb{Integrating Out a Scale}\PG\pgNPXVb

\theorem{\STM\thmTildeIntoOut}{
If $(\cW,\cG,u,\vec p)\in\tilde\cD^{(j)}_{\rm in}$ then 
$\Om_j(\cW,\cG,u,\vec p)\in\tilde\cD^{(j)}_{\rm out}$.

}

The rest of this subsection is devoted to the proof of this theorem.
Let 
$$
w=\sum_{m,n}\om_{m,n}(K)
$$ 
with
$$\eqalign{
\om_{m,n}(\phi,\psi;K) &= \smsum_{s_1,\cdots,s_n\in\Si_j}\
\int {\sst d\eta_1\cdots d\eta_m\,d\xi_1\cdots d\xi_n}\ 
w_{m,n}({\sst \eta_1,\cdots, \eta_m\,(\xi_1,s_1),\cdots ,(\xi_n,s_n)};K)\cr
& \hskip 5cm \phi({\sst \eta_1})\cdots \phi({\sst \eta_m})\
\psi({\sst (\xi_1,s_1)})\cdots \psi({\sst (\xi_n,s_n)\,})\cr
}$$
be the $\Si_j$--sectorized representative of $\cW$ specified in (I1).
Let
$\ 
(\cW',\cG',u,\vec p)=\Om_j(\cW,\cG,u,\vec p)
\ $ and choose the sectorized representative $w'$ of $\cW'$ as in Remark
\remNPFnochangeI.iii. Define $z$ by
$$\eqalign{
\lW z(\phi,\psi;K)\rW_{\psi,D_j(u;K)_{\Si_j}}
&=\Om_{C^{(j)}_{u,\Si_j}}
\big(\lW w^a-\om^a_{1,1}\rW_{\psi,C_j(u;K)_{\Si_j}}\big)
(\phi,\psi+\hat B\phi+\widehat{\de B}\phi )\cr
}\EQN\eqnNPzdef$$
where
$$\eqalign{B(k)&=\big(ik_0-e(\k)\big)C^{(j)}_{u(0)}(k)\cr
\de B(k;K)&=\big(ik_0-e(\k)\big)
\big\{C^{(j)}_{u(K)}(k)\big(1+\check w_{1,1}(k;K)\big)-C^{(j)}_{u(0)}(k)
\big\}\cr
}$$
and $\ \check w^a_{1,1}(k;K)\ $ is the Fourier transform, 
as in the last part of Definition \defNPfourtrans,
 of the function $(\et,\xi)\mapsto \sum_{s\in\Si_j}w^a_{1,1}(\et,(\xi,s);K)$.
As in Proposition \propOSfunctorialitySect\ of [FKTo3], with some abuse
of notation,
$$
(\hat B\phi )(\xi,s)=\int d\xi'\ \hat B(\xi,\xi')\phi(\xi')
$$
for all $s\in\Si_j$.
Recall that if, for each $s,s'\in\Si_j$, $c\big((\,\cdot\,,s),(\,\cdot\,,s')\big)$
is the Fourier transform, as in (\eqnNPcovFT) and (\eqnNPantisymmCov), 
of $\chi_s(k)C(k)\chi_{s'}(k)$, then 
$C_{\Si_j}\big((\xi,s),(\xi',s')\big)
=\sum_{t,t'\in\Si_j\atop{s\cap t\ne\emptyset\atop s'\cap t'\ne\emptyset}}
c\big((\xi,t),(\xi',t')\big)$. The integration and initial/final Wick ordering
covariances $C^{(j)}_{u,\Si_j}$, $C_j(u;K)_{\Si_j}$, $D_j(u;K)_{\Si_j}$
are constructed in this way from the $C^{(j)}_{u}(k)$, $C_j(u;K)(k)$, 
$D_j(u;K)(k)$ specified in Definition 
\defNPCovariances.

\lemma{\STM\lemNPMwprimez}{
$$\eqalign{
{w'}^a(\phi,\psi;K)&=z(\phi,\psi;K)+\om^a_{1,1}(\phi,\psi;K)-z(\phi,0;K)\cr
{\cG'}^a(\phi) &= \cG^a(\phi) +z(\phi,0;K)
-\half(J(1+\check w_{1,1})^{\hat{}}\hat A\phi)C^{(j)}_{u(K)}
J(1+\check w_{1,1})^{\hat{}}\hat A\phi\cr
}$$

}
\prf
In this proof, set $C=C^{(j)}_{u,\Si_j}$ and $C_j=C_j(u;K)_{\Si_j}$.
Define, for each $s\in\Si_j$,  $\ B_s(k)\ $ to be the Fourier transform, 
as in the last part of Definition \defNPfourtrans,
 of the function $(\et,\xi)\mapsto w_{1,1}(\et,(\xi,s);K)$. Observe that
$$
\sum_{s\in\Si_j}B_s(k)=\check w_{1,1}(k;K)
$$
By Lemma \lemOSphipsistruct\footnote{$^{(1)}$}{This is the only step in the proof that substantially uses bar/unbar invariance. Without it, the argument
$\hat B\phi+\widehat{\de B}\phi$ of (\eqnNPzdef)
would have a more general, but still controllable form.} of [FKTo2],
$$
\om_{1,1}(\phi,\psi)
=\sum_{s\in\Si_j}\int d\et d\xi\ \psi(\xi,s) (J\hat B_s)(\xi,\et)\phi(\et)
\EQN\eqnNPwoneoneBhat$$
Consequently, by Lemma \lemOSappGrassI\ of [FKTo2],
$$\eqalign{
&\Om_{C}\big(\lw w\rw_{\psi,C_j}\big)(\phi,\psi)
 = \log\sfrac{1}{Z} 
    \int e^{\lw w(\phi,\psi+\ze;K)\rw_{\psi,C_j}}\,d\mu_{C}(\ze)  \cr
&\hskip0.3in = \log\sfrac{1}{Z} 
    \int e^{\om_{1,1}(\phi,\psi)+\om_{1,1}(\phi,\ze)}\, e^{\lw (w-\om_{1,1})(\phi,\psi+\ze;K)\rw_{\psi,C_j}}\,d\mu_{C}(\ze)  \cr
&\hskip0.3in = \om_{1,1}(\phi,\psi)+\log\sfrac{1}{Z} 
    \int e^{\Si_s \ze(s) J\hat B_s\phi}\, e^{\lw (w-\om_{1,1})(\phi,\psi+\ze;K)\rw_{\psi,C_j}}\,d\mu_{C}(\ze)  \cr
&\hskip0.3in = \om_{1,1}(\phi,\psi)-\half\smsum_{s,s'\in\Si_j}
 (J\hat B_s\phi)C(s,s')(J\hat B_{s'}\phi)\cr
&\hskip1in+\log\sfrac{1}{Z} 
    \int e^{\lw (w-\om_{1,1})(\phi,\psi+\Si_sC(\,\cdot\,,s)J\hat B_s\phi+\ze;K)\rw_{\psi,C_j}}\,d\mu_{C}(\ze)  \cr
&\hskip0.3in = \om_{1,1}(\phi,\psi)-\half\smsum_{s,s'\in\Si_j}
 (J\hat B_s\phi)C(s,s')(J\hat B_{s'}\phi)\cr
&\hskip1in+\Om_{C}\big(\lw w-\om_{1,1}\rw_{\psi,C_j}\big) (\phi,\psi+\Si_sC(\,\cdot\,,s)J\hat B_s\phi ) \cr
&\hskip0.3in = \om_{1,1}(\phi,\psi)
-\half (J\widehat{\check w_{1,1}}\phi)C^{(j)}_u(J\widehat{\check w_{1,1}}\phi)
+\Om_{C}\big(\lw w-\om_{1,1}\rw_{\psi,C_j}\big) (\phi,
\psi+C^{(j)}_u J\widehat{\check w_{1,1}}\phi ) \cr
}$$
Therefore the Grassmann function $w''$ of Remark \remNPFnochangeI.iii
is determined by
$$\eqalign{
\lW w''(\phi,\psi;K)\rW_{\psi,D_{j}(u;K)_{\Si_j}}
&=\half\phi JC^{(j)}_{u}J\phi
+\Om_{C}(\lw w \rw_{\psi,C_j})(\phi,\psi+ C^{(j)}_u J\phi)\cr
&=-\half(J\phi)C^{(j)}_{u}J\phi
+\om_{1,1}(\phi,C^{(j)}_u J\phi)
-\half (J\widehat{\check w_{1,1}}\phi)C^{(j)}_u(J\widehat{\check w_{1,1}}\phi)
\cr
&\hskip.3in+\om_{1,1}(\phi,\psi)+\Om_{C}\big(\lw w-\om_{1,1}\rw_{\psi,C_j}\big) 
(\phi,\psi+ C^{(j)}_u J(1+\check w_{1,1})^{\hat{}}\phi ) \cr
}$$  
By (\eqnNPwoneoneBhat), the antisymmetry of $C^{(j)}_u$ and repeated use
of Lemma \lemOSjhat\ of [FKTo2] 
$$\eqalign{
\om_{1,1}(\phi,C^{(j)}_u J\phi)
&=\sum_s (C^{(j)}_uJ\phi)J\hat B_s\phi
=(C^{(j)}_uJ\phi)J\widehat{\check w_{1,1}}\phi\cr
&=- (J\phi)C^{(j)}_u(J\widehat{\check w_{1,1}}\phi)
=- (J\widehat{\check w_{1,1}}\phi)C^{(j)}_u(J\phi)
}$$
so that 
$$\eqalign{
\lW w''(\phi,\psi;K)\rW_{\psi,D_{j}(u;K)_{\Si_j}}
&=-\half(J(1+\check w_{1,1})^{\hat{}}\phi)C^{(j)}_{u}
J(1+\check w_{1,1})^{\hat{}}\phi
\cr
&\hskip.2in+\om_{1,1}(\phi,\psi)+\Om_{C}\big(\lw w-\om_{1,1}\rw_{\psi,C_j}\big) 
(\phi,\psi+ [C^{(j)}_u(k)(1+\check w_{1,1})]^{\hat{}}\phi ) \cr
}$$  
Amputating and applying parts (i) and (ii) of Lemma \lemOSjhat\   of [FKTo2],
$$
{w''}^a(\phi,\psi;K)
=-\half(J(1+\check w_{1,1})^{\hat{}}\hat A\phi)C^{(j)}_{u}
J(1+\check w_{1,1})^{\hat{}}\hat A\phi
+\om_{1,1}^a(\phi,\psi)+z(\phi,\psi;K)
$$  
The Lemma now follows by Remark \remNPFnochangeI.iii.
\endproof

\lemma{\STM\lemTNPbcbStruct}{
$$
(J(1+\check w_{1,1})^{\hat{}}\hat A\phi)C^{(j)}_{u(K)}
J(1+\check w_{1,1})^{\hat{}}\hat A\phi
=(J\hat A\phi)  C^{(j)}_{u(0)}(J\hat A\phi)
+\phi J\hat E^{(j)}\phi
$$
with
$$\eqalign{
E^{(j)}(k;K) &= -A(k)^2\big\{C^{(j)}_{u(K)}(k) \check w_{1,1}(k;K)
\big(2+\check w_{1,1}(k;K)\big)
+C^{(j)}_{u(K)}(k)- C^{(j)}_{u(0)}(k)\big\}\cr
}$$
}
\prf
Set
$$
B'(k;K)=1+\check w_{1,1}(k;K)
$$
By repeated use of Lemma \lemOSjhat\ of [FKTo2] and the facts that $J^2=-\bbbone$
and $J^t=-J$,
$$\deqalign{
 (J\hat B'\hat A\phi)C^{(j)}_{u(K)}(J\hat B'\hat A\phi)
&= -(J\widehat{B' A}J^2\phi)C^{(j)}_{u(K)}(J\widehat {B' A}\phi)
&= (\widehat{B' A}^tJ\phi)C^{(j)}_{u(K)}(J\widehat {B' A}\phi)\cr
&= (J\phi)\widehat{B' A} C^{(j)}_{u(K)}J\widehat {B' A}\phi
&= -\phi J\widehat{B' A} C^{(j)}_{u(K)}J\widehat {B' A}\phi\cr
}$$
Similarly,
$$\deqalign{
 (J\hat A\phi)  C^{(j)}_{u(0)}(J\hat A\phi)
&= -\phi J\hat A C^{(j)}_{u(0)}J\hat A\phi\cr
}$$
Subtracting and observing that
$$\eqalign{
& -A(k)^2\big\{C^{(j)}_{u(K)}(k)B'(k)^2- C^{(j)}_{u(0)}(k)\big\}\cr
&= -A(k)^2\big\{C^{(j)}_{u(K)}(k)\big(B'(k)^2-1\big)
+C^{(j)}_{u(K)}(k)- C^{(j)}_{u(0)}(k)\big\}\cr
&=E^{(j)}(k;K)
}$$
yields 
$$
(J\hat B'\hat A\phi)C^{(j)}_{u(K)}(J\hat B'\hat A\phi)-
(J\hat A\phi)  C^{(j)}_{u(0)}(J\hat A\phi)
=\phi J\hat E^{(j)}\phi
$$ 
\endproof

\lemma{\STM\lemNPBbnd}{
\Item i)
$$\eqalign{
\big\|C^{(j)}_{u(K)}(k) \Tnorm
&\le \const M^j\,\fe_j\big(\|K\|_{1,\Si_j}\big)\cr
\big\| \big(ik_0-e(\k)\big)C^{(j)}_{u(K)}(k) \Tnorm
&\le \const \fe_j\big(\|K\|_{1,\Si_j}\big)\cr
\big\|\sfrac{d\hfill}{ds} C^{(j)}_{u(K+sK')}(k)\big|_{s=0} \Tnorm 
&\le \const M^{2j}\,\fe_j\big(\|K\|_{1,\Si_j}\big)\,\| K'\|_{1,\Si_{j}}  \cr
\big\|\sfrac{d\hfill}{ds} (ik_0-e(\k))C^{(j)}_{u(K+sK')}(k)\big|_{s=0} \Tnorm 
&\le \const M^j\,\fe_j\big(\|K\|_{1,\Si_j}\big)\,\| K'\|_{1,\Si_{j}}  \cr
\big\|\sfrac{d\hfill}{ds} (ik_0-e(\k))^2C^{(j)}_{u(K+sK')}(k)\big|_{s=0} \Tnorm 
&\le \abcst\,\fe_{j+{1\over 2}}\big(\|K\|_{1,\Si_j}\big)\,\| K'\|_{1,\Si_{j}}\cr
}$$
\Item ii)
$$\eqalign{
\| B\tnorm 
&\le \const \cb_j\cr
\| \de B(K)\tnorm 
&\le  \const \fe_j\big(\|K\|_{1,\Si_j}\big)
\Big[M^j\| K\|_{1,\Si_{j}}  
+ \sfrac{\la_0^{1-2\upsilon}\fl_j}{\al^6}\Big]\cr
\big\| \sfrac{d\hfill}{ds}\de B(K+sK')\big|_{s=0}\Tnorm 
&\le \const M^j\,\fe_j\big(\|K\|_{1,\Si_j}\big)\|K'\|_{1,\Si_j}\cr
}$$
\Item iii)
$$\eqalign{
\| E^{(j)}(k;0)\tnorm 
&\le \const \sfrac{\la_0^{1-8\upsilon/7}}{\al^6}\sfrac{\fl_j}{M^j} \cb_j\cr
\big\| \sfrac{d\hfill}{ds} E^{(j)}(K+sK')\big|_{s=0}\Tnorm 
&\le \abcst\,\fe_{j+{1\over 2}}\big(\|K\|_{1,\Si_j}\big)\,
\| K'\|_{1,\Si_{j}}\cr
}$$

}
\prf i) Observe that
$$
(ik_0-e(\k))C^{(j)}_{u(K)}(k)
=(ik_0-e(\k)) \sfrac{\nu^{(j)}(k)}{ik_0-e(\k)-\check u(k;K)}
= \frac{\nu^{(j)}(k)}
    {1-\sfrac{\check u(k;K)}{ik_0-e(\k)}}
$$
and 
$$\eqalign{
\sfrac{d\hfill}{ds} (ik_0-e(\k))^2C^{(j)}_{u(K+sK')}(k)
&=(ik_0-e(\k))^2
\sfrac{d\hfill}{ds} \sfrac{\nu^{(j)}(k)}{ik_0-e(\k)-\check u(k;K+sK')}\cr
&=(ik_0-e(\k))^2
 \sfrac{\nu^{(j)}(k)}{[ik_0-e(\k)-\check u(k;K+sK')]^2}\sfrac{d\hfill}{ds}\check u(k;K+sK')\cr
&= \frac{\nu^{(j)}(k)\sfrac{d\hfill}{ds}\check u(k;K+sK')}
    {\big[1-\sfrac{\check u(k;K+sK')}{ik_0-e(\k)}\big]^2}\cr
}$$ 
Next use the fact 
that, on the support of $\nu^{(j)}$,
$\sfrac{1}{\sqrt{M}\,M^j}\le |ik_0-e(\k)|\le \sfrac{\sqrt{2M}}{M^j}$
to show that, if $f(k)$ vanishes except on the support of $\nu^{(\le j)}(k)$,
$$\eqalign{
\big\|\nu^{(j)}(k) \Tnorm&\le\abcst\,\cb_{j+{1\over 2}}\cr
\big\|\sfrac{f(k)}{ik_0-e(\k)}\Tnorm
&\le\abcst\,M^{j+{1\over 2}}\cb_{j+{1\over 2}}\|f(k)\tnorm\cr
}\EQN\eqnTNPjhalfbnds$$
Also use Lemma \lemNPpptyu\ and Lemma \lemOSNormMom\ of [FKTo3] to show that
$$\eqalign{
\big\|\check u(K)\Tnorm
&\le 2\v u(K)\v_{1,\Si_j}
\le\abcst\,\Big[\sfrac{\la_0^{1-\upsilon}}{M^{j-1}}+\| K\|_{1,\Si_{j}}
\Big]\fe_j({\sst \| K\|_{1,\Si_{j}}})\cr
&\le\abcst\,\sfrac{1}{M^j}\fe_j({\sst \| K\|_{1,\Si_{j}}})\cr
\big\|\sfrac{d\hfill}{ds}\check u(K+sK')\big|_{s=0} \Tnorm
&\le 2\V \sfrac{d\hfill}{ds} u(K+sK')\big|_{s=0} \V_{1,\Si_j} \cr
&\le\abcst\,\fe_j({\sst \| K\|_{1,\Si_{j}}})\,\| K'\|_{1,\Si_{j}} \cr
}\EQN\eqnTNPudubnds$$
Since $\Big|\sfrac{\check u(k;K)}{ik_0-e(\k)}\Big|\le \half$ on the support
of $\nu^{(j)}$, we have
$$\eqalign{
\big\| (ik_0-e(\k))C^{(j)}_{u(K)}(k)\Tnorm 
&\le \abcst\, \fe_{j+{1\over 2}}\big(\|K\|_{1,\Si_j}\big)\cr
\big\|\sfrac{d\hfill}{ds} (ik_0-e(\k))^2C^{(j)}_{u(K+sK')}(k)\big|_{s=0} \Tnorm 
&\le \abcst\, \fe_{j+{1\over 2}}\big(\|K\|_{1,\Si_j}\big)\,\| K'\|_{1,\Si_{j}}\cr
}$$
The remaining bounds follow from these, the second bound of (\eqnTNPjhalfbnds),
Corollary \corOSappMonoidIV\ of [FKTo1]
and the observation that 
$\fe_{j+{1\over 2}}({\sst \| K\|_{1,\Si_{j}}})
\le \const\fe_j({\sst \| K\|_{1,\Si_{j}}})$ for some
$M$, $r_0$ and $r$ dependent constant.
\Item ii)
By ($I^\sim 3$), Remark \remNPdiffdecay\ and ($I^\sim 1$),
$$\eqalign{
\big\| \big(ik_0-e(\k)\big)\check w_{1,1}(k;K) \Tnorm 
&\le  2\v w_{1,1}^{a\sim}(K)\tv_{1,\Si_j}
=2\la_0^{1-8\upsilon/7}\v w_{1,1}^{a\sim}(K)\tv_{1,\Si_j,\tilde\rho}\cr
&\le 2\sfrac{\la_0^{1-8\upsilon/7}}{\al^6}
\sfrac{\fl_j}{M^j}\fe_j({\sst \| K\|_{1,\Si_{j}}})\cr
\big\|\big(ik_0-e(\k)\big)\sfrac{d\hfill}{ds} \check w_{1,1}(k;K+sK')\big|_{s=0} \Tnorm 
&\le \sfrac{\la_0^{1-8\upsilon/7}}{\IB\al^2}\fe_j({\sst \| K\|_{1,\Si_{j}}})\,\| K'\|_{1,\Si_{j}} \cr
}\EQN\eqnTNPwoneonebnd$$
All three bounds follow by using Leibniz and Corollary \corOSappMonoidIV.ii
of [FKTo1] to combine (\eqnTNPwoneonebnd) with the bounds of part (i).
\Item iii)
For the first bound, just apply 
$\big\|A(k)C^{(j)}_{u(0)}(k)\Tnorm\le\const\cb_j$ from
part (i), the first bound of (\eqnTNPwoneonebnd)
and 
$$
\big\| \check w_{1,1}(k;K) \Tnorm 
\le \const\sfrac{\la_0^{1-8\upsilon/7}}{\al^6}
\fl_j\fe_j({\sst \| K\|_{1,\Si_{j}}})
\EQN\eqnTNPwoneonenonampbnd$$
which follows from (\eqnTNPwoneonebnd), (\eqnTNPjhalfbnds) and the fact that 
$\check w_{1,1}(k;K)$ vanishes except on the support of $\nu^{(>j)}$.

Now for the second bound.
By part (i),
$$
\big\|\sfrac{d\hfill}{ds} A(k)^2C^{(j)}_{u(K+sK')}(k)\big|_{s=0} \Tnorm 
\le \abcst\,\fe_{j+{1\over 2}}\big(\|K\|_{1,\Si_j}\big)\,\| K'\|_{1,\Si_{j}}
$$
By part (i), (\eqnTNPwoneonebnd) and (\eqnTNPwoneonenonampbnd),
$$\eqalign{
\big\|A(k)\check w_{1,1}({\sst k;K})
\big(2\!+\!\check w_{1,1}({\sst k;K})\big)
\sfrac{d\hfill}{ds} A(k)C^{(j)}_{u(K+sK')}(k)\big|_{s=0} \Tnorm
&\le\const\!\sfrac{\la_0^{1-{8\over 7}\upsilon}}{\al^6} \fl_j
\fe_j({\sst \|K\|_{1,\Si_j}})\| K'\|_{1,\Si_{j}}\cr
&\le \sfrac{\la_0^{1-2\upsilon}}{\al^6} \fl_j\,
\fe_j({\sst \|K\|_{1,\Si_j}})\,\| K'\|_{1,\Si_{j}}\cr
\big\|A(k)C^{(j)}_{u(K)}(k)
\big(1+\check w_{1,1}({\sst k;K})\big)
\sfrac{d\hfill}{ds} A(k)\check w_{1,1}({\sst k;K+sK'})\big|_{s=0} \Tnorm
&\le \const\sfrac{\la_0^{1-{8\over7} \upsilon}}{\al^2} \,
\fe_j({\sst \|K\|_{1,\Si_j}})\,\| K'\|_{1,\Si_{j}}\cr
&\le \sfrac{\la_0^{1-2\upsilon}}{\al^2} \,
\fe_j({\sst \|K\|_{1,\Si_j}})\,\| K'\|_{1,\Si_{j}}\cr
}$$\endproof

\proof{of Theorem \thmTildeIntoOut}
We have already proven in Theorem \thmIntoOut\ that 
$(\cW',\cG',u,\vec p)\in\cD^{(j)}_{\rm out}$.
We shall use $\om_{m,n},\ \fz_{m,n},\ \cdots$ to denote the part of 
$w,\ z,\ \cdots$ that is of degree $m$ in $\phi$ and degree $n$ in $\psi$
and $w_{m,n},\ z_{m,n},\ \cdots$ to denote the corresponding kernels.
We shall also use $\om_{n},\ \fz_{n},\ \cdots$ to denote the part of 
$w,\ z,\ \cdots$ that is of degree $n$ in $\phi$ and  $\psi$ combined
and $w_{n},\ z_{n},\ \cdots$ to denote the corresponding kernels.
Recall that 
$$\eqalign{
\lW z(\phi,\psi;K)\rW_{\psi,D_j(u;K)_{\Si_j}}
&=\Om_{C^{(j)}_{u,\Si_j}}
\big(\lW w^a-\om^a_{1,1}\rW_{\psi,C_j(u;K)_{\Si_j}}\big)
(\phi,\psi+\hat B\phi+\hat \de B\phi )\cr
}$$
and define $z''$ by
$$\eqalign{
\lW z''(\phi,\psi;K)\rW_{\psi,D_j(u;K)_{\Si_j}}
&=\Om_{C^{(j)}_{u,\Si_j}}
\big(\lW w^a-\om^a_{1,1}\rW_{\psi,C_j(u;K)_{\Si_j}}\big)(\phi,\psi )\cr
}$$
so that
$$
z(\phi,\psi;K)
          =z''(\phi,\psi+\hat B\phi+\widehat{\de B}(K)\phi;K)
\EQN\eqnTNPzzpp$$

\Item {\it Preparation for the verification of ($O^\sim1$), ($O^\sim2$) and ($O^\sim3$):} 

We now apply Theorems \thOSmomrengroupestimate\ and
\thOSmomrengroupdiffestimate\ of [FKTo3] to bound $z$. 
By ($I^\sim1$), parts (i) and (ii) of Lemma \lemNPpptyu\ and 
Lemma \lemNPBbnd.ii,
the hypotheses of Theorem \thOSmomrengroupestimate\ of [FKTo3], 
with $w=w^a-\om^a_{1,1}$, $B$ replaced by $B+\de B$, $w'=z$, $w''=z''$,
$\mu=\abcst$, $\La=\sfrac{\la_0^{1-\upsilon}}{M^{j-1}}$, 
$\ga=\la_0^{\upsilon/7}$ and $X=\big\| K\big\|_{1,\Si_{j}}$,
are fulfilled. Therefore,
$$\eqalign{
N_j^\sim(z-w^a+\om^a_{1,1},\al,\| K\|_{1,\Si_{j}}) 
&\le\const\big(\sfrac{1}{\al}+\la_0^{\upsilon/7}\big)\,
\sfrac{N^\sim_j(w^a,64\al,\| K\|_{1,\Si_{j}})}
{1-{\const \over\al}N^\sim_j(w^a,64\al,\| K\|_{1,\Si_{j}})}\cr
\v  z_{2,0} \tv _{1,\Si_j,\tilde\rho},\v  z_{1,1} \tv _{1,\Si_j,\tilde\rho}  
&\le  \sfrac{\const}{\al^8}\sfrac{ \fl_j}{M^j}\,
\sfrac{N_j^\sim(w^a,64\al,\| K\|_{1,\Si_{j}})^2}
           {1-{\const \over\al}N_j^\sim(w^a,64\al,\| K\|_{1,\Si_{j}})} \cr
}\EQN\eqnTNPodI$$
and
$$
\V z''_4-w^a_4 -
\sfrac{1}{4} 
\smsum_{\ell=1}^\infty {\sst (-1)^\ell(12)^{\ell+1}}
{\rm Ant\,}L_\ell(w^a_4;\,{\sst C^{(j)}_u,D_j})\tV_{3,\Si_j,\tilde\rho}
\le\sfrac{\const}{\al^{10}}\,\fl_j\,
\sfrac{N_j^\sim(w^a,64\al,\| K\|_{1,\Si_{j}})^2}
           {1-{\const \over\al}N_j^\sim(w^a,64\al,\| K\|_{1,\Si_{j}})} 
\EQN\eqnTNPodIfour$$
The hypotheses of Theorem \thOSmomrengroupdiffestimate\
of [FKTo3], with, in addition, $Y=\abcst\big\| K'\big\|_{1,\Si_{j}}$,
$Z=\const M^j\big\| K'\big\|_{1,\Si_{j}}$
and $\veps=\const M^j\,\| K'\|_{1,\Si_{j}}\big|_{t=0}$,
are fulfilled. Hence 
$$\eqalign{
N^\sim_j\big(&\sfrac{d\hfill}{ds}\big[z(K+sK')
             +\om_{1,1}^a(K+sK')-w^a(K+sK')\big]_{s=0}
               ,\al,\,\| K\|_{1,\Si_{j}}\big)\cr
\le\ \ &\ \const\Big\{\la_0^{\upsilon/7}+\sfrac{1}{\al^2}\,
     \sfrac{N^\sim_j(w^a,64\al,\,\| K\|_{1,\Si_{j}})}
     {1-{\const\over\al^2}N^\sim_j(w^a,64\al,\,\| K\|_{1,\Si_{j}})}\Big\}
     N^\sim_j\big(\,\sfrac{d\hfill}{ds}w^a(K+sK')\big|_{s=0}
,16\al,\,\| K\|_{1,\Si_{j}}\big)\cr
+&\ \const
     \sfrac{N^\sim_j(w^a,64\al,\,\| K\|_{1,\Si_{j}})}
      {1-{\const\over\al^2}N^\sim_j(w^a,64\al,\,\| K\|_{1,\Si_{j}})}
\Big\{\sfrac{1}{\al^2}N^\sim_j(w^a,64\al,\,\| K\|_{1,\Si_{j}})
+\la_0^{\upsilon/7}\Big\}M^j\| K'\|_{1,\Si_{j}}
}\EQN\eqnTNPodIb$$
By ($I^\sim 1$) and Corollary \corOSappMonoidIV.ii of [FKTo1],
$$
\sfrac{N^\sim_j(w^a,64\al,\,\| K\|_{1,\Si_{j}})}
           {1-{\const \over\al}N^\sim_j(w^a,64\al,\,\| K\|_{1,\Si_{j}})}
\le \sfrac{\fe_j(\| K\|_{1,\Si_{j}})}
{1-{\const \over\al}\fe_j(\| K\|_{1,\Si_{j}})}
\le \const \fe_j(\| K\|_{1,\Si_{j}})
$$
and
$$
\sfrac{N^\sim_j(w^a,64\al,\,\| K\|_{1,\Si_{j}})^2}
          {1-{\const \over\al}N^\sim_j(w^a,64\al,\,\| K\|_{1,\Si_{j}})}
\le \const \fe_j(\| K\|_{1,\Si_{j}})^2
\le \const \fe_j(\| K\|_{1,\Si_{j}})
$$
Therefore, by (\eqnTNPodI) and ($I^\sim 1$), recalling that $w^a$ vanishes when $\psi=0$,
$$\eqalignno{
N^\sim_j(z+\om^a_{1,1},\al,\,\| K\|_{1,\Si_{j}}) 
&\le N^\sim_j(w^a,\al,\,\| K\|_{1,\Si_{j}})
+\const\big(\sfrac{1}{\al}+\la_0^{\upsilon/7}\big)\,\fe_j(\| K\|_{1,\Si_{j}})\cr
&\le\sfrac{1}{64} N^\sim_j(w^a,64\al,\,\| K\|_{1,\Si_{j}})
+\half\,\fe_j(\| K\|_{1,\Si_{j}})\cr
&\le \fe_j(\| K\|_{1,\Si_{j}})
&\EQNO\eqnTNPodII}$$
and
$$
\v  z_{2,0} \tv _{1,\Si_j,\tilde\rho} , \v  z_{1,1} \tv _{1,\Si_j,\tilde\rho}  
\le  \sfrac{\const}{\al^8}\sfrac{ \fl_j}{M^j}\,
\fe_j(\| K\|_{1,\Si_{j}})
\le  \sfrac{1}{\al^7}\sfrac{ \fl_j}{M^j}\,
\fe_j(\| K\|_{1,\Si_{j}})
\EQN\eqnTNPodIII$$
and
$$
\V z''_4-w^a_4 -
\sfrac{1}{4} \smsum_{\ell=1}^\infty {\sst (-1)^\ell(12)^{\ell+1}}
{\rm Ant\,}L_\ell(w^a_4;\,{\sst C^{(j)}_u,D_j})\tV_{3,\Si_j,\tilde\rho}
\le\sfrac{\const}{\al^{10}}\,\fl_j\,
\fe_j(\| K\|_{1,\Si_{j}}) 
\EQN\eqnTNPodIfourB$$
Furthermore, by (\eqnTNPodIb), ($I^\sim 1$) and Corollary \corOSappMonoidIV.ii of [FKTo1],
$$\eqalign{
&N^\sim_j\big(\sfrac{d\hfill}{ds}\big[z(K+sK')+\om_{1,1}^a(K+sK')-w^a(K+sK')\big]_{s=0}
,\al,\,\| K\|_{1,\Si_{j}}\big) \cr
&\hskip2.5in\le\const\big(\sfrac{1}{\al^2}+\la_0^{\upsilon/7}\big)
         \,\fe_j(\| K\|_{1,\Si_{j}})\,M^j\| K'\|_{1,\Si_{j}}\cr
}\EQN\eqnTNPodIIba$$
and 
$$\eqalignno{
&N^\sim_j\big(\sfrac{d\hfill}{ds}\big[z(K+sK')+\om_{1,1}^a(K+sK')\big]_{s=0}
,\al,\,\| K\|_{1,\Si_{j}}\big) \cr
&\hskip0.3in\le N^\sim_j\big(\sfrac{d\hfill}{ds}w^a(K+sK')\big|_{s=0}
,\al,\,\| K\|_{1,\Si_{j}}\big)
+\const\big(\sfrac{1}{\al^2}+\la_0^{\upsilon/7}\big)
\,\fe_j(\| K\|_{1,\Si_{j}})\,M^j\| K'\|_{1,\Si_{j}}\cr
&\hskip0.3in\le\sfrac{1}{64} N^\sim_j\big(\sfrac{d\hfill}{ds}w^a(K+sK')\big|_{s=0}
,64\al,\,\| K\|_{1,\Si_{j}}\big)
+\half\,\fe_j(\| K\|_{1,\Si_{j}})\,M^j\| K'\|_{1,\Si_{j}}\cr
&\hskip0.3in\le M^j\fe_j(\| K\|_{1,\Si_{j}})\,\| K'\|_{1,\Si_{j}}
&\EQNO\eqnTNPodIIb}$$

\Item {\it Verification of ($O^\sim1$):} 
Observe that $B(k)$ and $\de B(k;K)$ vanish unless $k$ is in the support 
of $\nu^{(j)}$. Also, by ($I^\sim 1$), $w^a(\phi,\psi;K)-w^a(0,\psi;K)$ 
vanishes unless $\hat\nu^{(<j)}\phi$ is nonzero. Consequently, 
$z(\phi,\psi;K)-z(0,\psi;K)$, and hence
${w'}^a(\phi,\psi;K)-{w'}^a(0,\psi;K)$, vanishes unless 
$\hat\nu^{(\le j)}\phi$ is nonzero.
The bounds of ($O^\sim 1$)  follow from Lemma \lemNPMwprimez,
(\eqnTNPodII) and (\eqnTNPodIIb).

\goodbreak
\Item {\it Verification of ($O^\sim2$):} 

By Lemma \lemNPMwprimez,
$$\eqalign{
\om^{\prime a}_4(\phi,\psi;K)+\cG^{\prime a}_4(\phi;K)
&=\fz_4(\phi,\psi;K)+\cG^a_4(\phi;K)\cr
&=\fz_4(\phi,\psi;0)+\cG^a_4(\phi;0)+\de \ff'_0(\phi,\psi;K)\cr
}$$
where
$$
\de \ff'_0(\phi,\psi;K)
=\fz_4(\phi,\psi;K)+\cG^a_4(\phi;K)
-\fz_4(\phi,\psi;0)-\cG^a_4(\phi;0)
$$
As the corresponding kernel $\de f'_0(K)$ obeys
$$
\sfrac{d\hfill}{ds} \de f'_0(K+sK')
=\sfrac{d\hfill}{ds} \Big[z_4(K+sK')-w^a_4(K+sK')\Big]
+\sfrac{d\hfill}{ds} \de f^{(j)}(K+sK') 
$$
(\eqnTNPodIIba) and ($I^\sim 2$) give
$$\eqalign{
\V \sfrac{d\hfill}{ds} \de f'_0(K+sK')\big|_{s=0}\tV_{3,\Si_j,\tilde \rho}
&\le \Big[\sfrac{\const}{\al^4}\big(\al^2+\la_0^{\upsilon/7}\big)
                          +\sfrac{1}{64\al^4\IB^2}\Big]
\,\fe_j(\| K\|_{1,\Si_{j}})\,M^j\| K'\|_{1,\Si_{j}}\cr
}\EQN\eqnTNPdFz$$
By  (\eqnTNPzzpp),
$$\eqalign{
\fz_4(\phi,\psi;0)+\cG^a_4(\phi;0)
&=\fz''_4(\phi,\psi+\hat B\phi+\hat \de B(0)\phi;0)
+\cG^a_4(\phi;0)\cr
&=\fz''_4(\phi,\psi+\hat B\phi;0)+\cG^a_4(\phi;0)
+\de \ff'_1(\phi,\psi)\cr
}$$
where
$$
\de \ff'_1(\phi,\psi)
=\fz''_4(\phi,\psi+\hat B\phi+\hat \de B(0)\phi;0)
-\fz''_4(\phi,\psi+\hat B\phi;0)
$$
By Lemma \lemNPBbnd.ii\ and Theorem \thOSmomrengroupdiffestimate\ of [FKTo3],
with $w=w^a(0)-\om^a_{1,1}(0)$, $C_\ka=C^{(j)}_{u(0)}$, $D_\ka=D_j(u;0)$,
independent of $\ka$, $B_\ka=B+(\ka_0+\ka)\de B(0)$ where $0\le \ka_0\le 1$,
$\ga=\la_0^{\upsilon/7}$, $\mu=\abcst$, $\La=\sfrac{\la_0^{1-\up}}{M^{j-1}}$,
$c_B=\const$, $X=Y=\epsilon=0$ and
$Z=\sfrac{\la_0^{1-2\upsilon}\fl_j}{\al^6}$, the corresponding
kernel $\de f'_1$ is bounded by
$$\eqalign{
\V \de f'_1\tV_{3,\Si_{j},\tilde\rho}
&\le \const\la_0^{\upsilon/7}\sfrac{\la_0^{1-2\upsilon}\fl_j}{\al^6}
\sfrac{1}{\al^{4}\IB^2}\,\cb_j
}\EQN\eqnTNPdFpone$$
Next, let  
$$
\de f'_2
=\sh\Big(
z_4''(0)-w^a_4(0) -
\sfrac{1}{4} \smsum_{\ell=1}^\infty {\sst (-1)^\ell(12)^{\ell+1} }
{\rm Ant\,}L_\ell\big(w^a_4(0);C^{(j)}_{u(0)}, D_j(u(0);0)\big),B\Big)
$$
By (\eqnTNPodIfourB), Lemma \lemNPBbnd.ii\ and Lemma \lemOStildesourceterm\
of [FKTo3], with $c_B=\const$, $X=0$, $X_B=1$,
$$
\V \de f'_2\V_{3,\Si_j,\tilde\rho} 
 \le\sfrac{\const}{\al^{10}}\,\fl_j\,\cb_j 
\le\sfrac{1}{\al^9}\,\fl_j\,\cb_j
\EQN\eqnTNPdFptwo$$
Define 
$$
\de f'_3=
\sh\bigg(\sfrac{1}{4} \smsum_{\ell=1}^\infty {\sst (-1)^\ell(12)^{\ell+1}}
         {\rm Ant\,} V_{\rm pp} \Big(L_\ell\big(w^a_4(0);C^{(j)}_{u(0)}, 
                  C^{(\ge j+1)}_{u(0)}\big)^{\rm pp} \Big),B\bigg)
$$
and
$$
{\de f'}^{(j+1)}(K)=\de f'_0(K)+\de f'_1+\de f'_2+\de f'_3\qquad 
{\de \ff'}^{(j+1)}(\phi,\psi;K)=Gr\big(\phi,\psi;{\de f'}^{(j+1)}(K)\big)
$$
where $Gr(\phi,\psi,f)$ is the Grassmann function associated to the kernel $f$
as in Definition  \defNPshear.ii.
By Proposition \propparticleparticleladder, for $\ell \ge 1$
$$
\VV  V_{\rm pp} \Big(L_\ell\big(w_4^a(0);C^{(j)}_{u(0)}, 
                  C^{(\ge j+1)}_{u(0)}\big)^{\rm pp} \Big)
\VV_{3,\Si_j} 
\le \big(\const\vi_j\cb_j\big)^\ell\, \v w_4^a(0)\v_{3,\Si_{j}}^{\ell+1}
$$
so that
$$
\VV  V_{\rm pp} \Big(L_\ell\big(w_4^a(0);C^{(j)}_{u(0)}, 
                  C^{(\ge j+1)}_{u(0)}\big)^{\rm pp} \Big)
\VV_{3,\Si_j,\tilde\rho} 
\le \big(\const\la_0^{1-\upsilon}
\vi_j\cb_j\big)^\ell\, \v w_4^a(0)\v_{3,\Si_j,\tilde\rho}^{\ell+1}
$$

The hypotheses of this Proposition are fulfilled by parts (i) and (ii) of 
Lemma \lemNPpptyu. 
Observe that, by ($I^\sim 1$),
$$
\sfrac{M^{2j}}{\fl_j}\fe_j({\sst \| K\|_{1,\Si_{j}}})
(64\al)^4\big(\sfrac{\fl_j\IB}{M^j}\big)^2\ \sfrac{1}{\fl_j}\ 
\v w_{0,4}(K)\v_{3,\Si_j,\tilde\rho}
\le \fe_j({\sst \| K\|_{1,\Si_{j}}})
$$
so that
$$
\v w_{0,4}(K)\v_{3,\Si_j,\tilde\rho}\le\sfrac{1}{(64\al)^4\IB^2} 
\fe_j({\sst \| K\|_{1,\Si_{j}}})
$$
In particular,
$$
\v w_4^a(0)\v_{3,\Si_j,\tilde\rho}\le\sfrac{1}{(64\al)^4\IB^2} \cb_j
$$
Therefore,  by Lemma \lemOStildesourceterm\
of [FKTo3] and Corollary \corOSappMonoidIV.ii of [FKTo1],
$$\eqalignno{
\V \de f'_3\V_{3,\Si_j,\tilde\rho}
&\le \smsum_{\ell=1}^\infty \big(\const\la_0^{1-\upsilon}\vi_j\cb_j\big)^\ell\, 
       \v w_4^a(0)\v_{3,\Si_j,\tilde\rho}^{\ell+1}\cr
&\le \smsum_{\ell=1}^\infty \big(\const\la_0^{1-\upsilon}\vi_j\cb_j\big)^\ell\, 
       \big(\sfrac{1}{\al^4} 
\cb_j\big)^{\ell+1}\cr
&\le  \const \sfrac{\la_0^{1-\upsilon}}{\al^8}\vi_j\ \cb_j^3
\ \frac{1}{1-\const \sfrac{\la_0^{1-\upsilon}}{\al^4}
            \vi_j\,\cb_j^2 }\cr
&\le  \const \sfrac{\la_0^{1-\upsilon}}{\al^8} \vi_j\ \cb_j
&\EQNO\eqnTNPdFpthree\cr
}$$
Hence, by   (\eqnTNPdFpone), (\eqnTNPdFptwo) and (\eqnTNPdFpthree), 
$$
\V {\de f'}^{(j+1)}(0)\V_{3,\Si_j,\tilde\rho}
\le \Big\{
      \sfrac{\la_0^{1-2\upsilon}}{\al^{10}}\fl_j
     +\sfrac{\fl_j}{\al^9}
     + \sfrac{\const\la_0^{1-\upsilon}}{\al^8}\vi_j
     \Big\}\,\cb_j
\le \sfrac{\vi_{j+1}}{\al^8}\cb_j
\EQN\eqnTNPFjplusI$$
and by (\eqnTNPdFz),
$$\eqalign{
\V \sfrac{d\hfill}{ds} {\de f'}^{(j+1)}(K+sK')\big|_{s=0}\tV_{3,\Si_j,\tilde \rho}
\le \sfrac{1}{\al^4\IB^2}\,\fe_j(\| K\|_{1,\Si_{j}})\,M^j\| K'\|_{1,\Si_{j}}
}\EQN\eqnTNPFjplusIderiv$$

\vskip0.3cm
\goodbreak

Observe that $D_j(u(0);0) = C^{(\ge j+1)}_{u(0)}$ and, by Lemma
\lemtildeunordord,
$$\eqalign{
L_\ell\big(w_4^a(0);C^{(j)}_{u(0)},  C^{(\ge j+1)}_{u(0)}\big)
&= V_{\rm pp} \Big(L_\ell\big(w_4^a(0);C^{(j)}_{u(0)}, 
                  C^{(\ge j+1)}_{u(0)}\big)^{\rm pp} \Big)\cr
&\hskip3cm+V_{\rm ph} \Big(L_\ell\big(w_4^a(0);C^{(j)}_{u(0)}, 
                  C^{(\ge j+1)}_{u(0)}\big)^{\rm ph} \Big) \cr
}$$ 
so that the kernel of $\fz''_4(\phi,\psi+\hat B\phi;0)$ is 
$$\eqalign{
\sh\Big(w_4^a(0)+
\sfrac{1}{4} \smsum_{\ell=1}^\infty {\sst (-1)^\ell(12)^{\ell+1} }
{\rm Ant\,} V_{\rm ph}\Big(L_\ell\big(w_4^a(0);C^{(j)}_{u(0)}, 
                  C^{(\ge j+1)}_{u(0)}\big)^{\rm ph} \Big),B\Big)
+\de f'_2+\de f'_3\cr
}\EQN\eqnTNPodFIa$$

To verify ($O^\sim 2$), set ${v'}^{(i)}=v^{(i)}$ for all $2\le i\le j-1$, 
${v'}^{(j)}=\de f^{(j)}(0)$ and define  ${f'}^{(i)}$, $2\le i\le j$, by
$$\eqalign{
{f'}^{(i)}
&=\sh\big({v'}^{(i)},C^{[i,j]}_{u(0)}(k) A(k)\big)
=\cases{\sh\big(f^{(i)},B\big)
                                              & if $2\le i\le j-1$\cr
\sh\big(\de f^{(j)}(0),B\big)
                                              & if $i=j$\cr
}\cr
}$$
Observe that ${v'}^{(j)}$ vanishes unless
all of its $\phi$ momenta are in the support of $\nu^{(<j)}$, since the
same is true for $\de f^{(j)}(0)$ by ($I^\sim 1$) and ($I^\sim 2$).
Since $B(k)$ is supported on the $j^{\rm th}$ neighbourhood
$\ 
{f'}^{(i)}_{\Si_j}=\sh\big(f^{(i)}_{\Si_j},B\big)
\ $
for all $2\le i\le j-1$. Consequently, observing that the pure $\psi$ parts
of $f^{(i)}$ and ${f'}^{(i)}$ coincide and that only the pure $\psi$ parts
contribute to internal ladder vertices,
$$
\sh\Big({\rm Ant\,}
\Big( V_{\rm ph}\big( \cL^{(j)}(\vec p,\vec f)\big) \Big)_{\Si_j},B\Big)
={\rm Ant\,}
\Big( V_{\rm ph}\big( \cL^{(j)}(\vec p,\vec f')\big) \Big)_{\Si_j}
$$
By ($I^\sim 2$), the kernel of 
$
\om_4^a(\phi,\psi;0)+\cG_4^a(\phi;0)
$
is
$
\de f^{(j)}(0)+\smsum_{i=2}^{j-1}f^{(i)}_{\Si_j}+
\sfrac{1}{8}{\rm Ant\,}
\Big( V_{\rm ph}\big( \cL^{(j)}(\vec p,\vec f)\big) \Big)_{\Si_j}
$
so that the kernel of 
$
\om_4^a(\phi,\psi+\hat B \phi;0)+\cG_4^a(\phi;0)
$
is
$\smsum_{i=2}^{j}{f'}^{(i)}_{\Si_j}+
\sfrac{1}{8}{\rm Ant\,}
\Big( V_{\rm ph}\big( \cL^{(j)}(\vec p,\vec f')\big) \Big)_{\Si_j}
$.
Therefore, by  the Definition \defcompLadder\ of iterated particle--hole 
ladders,
$$\eqalign{
&\om^{\prime a}_4(\phi,\psi;K)+\cG^{\prime a}_4(\phi;K)\cr
&\hskip.1in=Gr\Big(\phi,\psi+\hat B\phi;w_4^a(0)
+\sfrac{1}{4} \smsum_{\ell=1}^\infty {\sst (-1)^\ell(12)^{\ell+1} }
{\rm Ant\,} V_{\rm ph}\Big(L_\ell\big(w_4^a(0);C^{(j)}_{u(0)}, 
                  C^{(\ge j+1)}_{u(0)}\big)^{\rm ph} \Big)\Big)\cr
&\hskip.4in+\cG^{a}_4(\phi;0)+\de{\ff'}^{(j+1)}(\phi,\psi;K)
\cr
&=Gr\Big(\phi,\psi;\sum_{i=2}^{j}{f'}_{\Si_j}^{(i)} +{\de f'}^{(j+1)}(K)
+\sfrac{1}{8} {\rm Ant\,}
\Big( V_{\rm ph} \big( \cL^{(j)}(\vec p,\vec f')\big)\Big)_{\Si_j}\Big)\cr
&\hskip.5in
+Gr\Big(\phi,\psi+\hat B\phi;\sfrac{1}{4}  \smsum_{\ell=1}^\infty {\sst (-1)^\ell(12)^{\ell+1}}
{\rm Ant\,} V_{\rm ph} \Big(L_\ell\big(w_4^a(0);C^{(j)}_{u(0)}, 
                  C^{(\ge j+1)}_{u(0)}\big)^{\rm ph} \Big)\Big)\cr
&=Gr\Big(\phi,\psi;\sum_{i=2}^{j}{f'}_{\Si_j}^{(i)}+{\de f'}^{(j+1)}(K)
+\sfrac{1}{8} {\rm Ant\,}\Big( V_{\rm ph} 
\big( \cL^{(j+1)}(\vec p,\vec f')\big)\Big)\Big)\cr
}$$
The estimates on $\V {\de f'}^{(j+1)}(K)\V_{3,\Si_{j},\tilde \rho}$ required 
for ($O^\sim2$) were proven in (\eqnTNPFjplusI) and (\eqnTNPFjplusIderiv).  
By ($I^\sim2$),
$$
\V {v'}^{(j)}\V_{3,\Si_j,\tilde \rho}
=\V \de f^{(j)}(0)\V_{3,\Si_j,\tilde \rho}\le 
\sfrac{\vi_j}{\al^7}\cb_j
$$
and
$\ \V {v'}^{(i)}\V_{3,\Si_i,\tilde \rho}
=\V v^{(i)}\V_{3,\Si_i,\tilde \rho}\le 
\sfrac{\vi_i}{\al^7}\cb_i
\ $ for $2\le i\le j-1$.

\vskip.6cm

\Item {\it Verification of ($O^\sim3$):} 
Let 
$$
2\check G_{2}(k;K)=  C^{(< j)}_{u(0)}(k)+ \sfrac{1}{(ik_0-e(\k))^2}
\sum_{i=2}^{j-1}\Big\{\de q^{(i)}(k;K)+\smsum_{\ell=i}^{j}q^{(i,\ell)}(k)\Big\}
$$
be the decomposition of ($I^\sim 3$) and set
$$\deqalign{
q^{\prime(i,\ell)}(k)&=q^{(i,\ell)}(k)
& 2\le i\le j-1,\ i\le \ell\le j \cr
q^{\prime(j,j)}(k)&=2\check z_{2,0}(k;0)- E^{(j)}(k;0) 
\hidewidth\cr
\de q^{\prime(i)}(k;K)&=\de q^{(i)}(k;K)
\hskip0.5in& 2\le i\le j-1\hskip1.5in \cr
\de q^{\prime(j)}(k;K)&=2\check z_{2,0}(k;K)-2\check z_{2,0}(k;0)
- E^{(j)}(k;K)+ E^{(j)}(k;0)
\hidewidth\cr
}$$
By Lemmas \lemNPMwprimez\ and  \lemTNPbcbStruct,
both conclusions of Lemma \lemOSjhat.i of [FKTo2]
and the fact that the transpose of $J$ is $-J$,
$$
2\check G'_{2}(k;K)=  C^{(\le j)}_{u(0)}(k)+ \sfrac{1}{(ik_0-e(\k))^2}
\sum_{i=2}^{j}\Big\{\de q^{\prime(i)}(k;K)
                     +\smsum_{\ell=i}^{j}q^{\prime(i,\ell)}(k)\Big\}
$$
By (\eqnTNPodIII), Lemma \lemOSNormMom\ of [FKTo3], Lemma \lemNPBbnd.iii\ 
and  (\eqnTNPodIIb),
$$\eqalign{
\big\| q^{(j,j)}(k) \Tnorm
&\le 4 \sfrac{\la_0^{1-9\upsilon/7}}{\al^7}\sfrac{ \fl_j}{M^j} \cb_j
+\const \sfrac{\la_0^{1-8\upsilon/7}}{\al^6}\sfrac{\fl_j}{M^j} \cb_j
\le\la_0^{1-2\upsilon}\sfrac{\fl_j}{M^j}\cb_{j}\cr
\big\| \sfrac{d\hfill}{ds}\de q^{\prime(j)}({\sst K+sK'})\big|_{s=0}\,\Tnorm
&\le\Big\{\sfrac{\const\la_0^{1-9\upsilon/7}}{\al^2}
              \fe_j\big({\sst \|K\|_{1,\Si_j}}\big)
          +\abcst\,\fe_{j+{1\over 2}}\big({\sst \|K\|_{1,\Si_j}}\big)\Big\}
 \| K'\|_{1,\Si_{j}}\cr
&\le\sqrt{M^{\aleph'-\aleph}}\fe_{j+{1\over 2}}\big(\|K\|_{1,\Si_j}\big) \,  \|K'\|_{1,\Si_j}\cr
}$$
if $M$ is large enough.

Recall that $\om^a_{1,1}(\phi,\psi)$ vanishes unless $\hat\nu^{(<j)}\phi$
is nonzero. Hence, by conservation of momentum,
$\big(w^a_{1,1}\big)_{\Si_{j+1}}$ vanishes
and, by (\eqnTNPodIII) and Proposition \propOSresectorI.ii of [FKTo4],
$$\eqalign{
\V \big(w'^{a\sim}_{1,1}\big)_{\Si_{j+1}}\tV_{1,\Si_{j+1},\tilde\rho}
&=\V \big(z_{1,1}^{\sim}\big)_{\Si_{j+1}}\tV_{1,\Si_{j+1},\tilde\rho}
\le\abcst\,\cb_j\V z_{1,1}^{\sim}\tV_{1,\Si_{j},\tilde\rho}
\le  \sfrac{\const}{\al^8}
\sfrac{\fl_j}{M^j}\fe_j({\sst \| K\|_{1,\Si_{j}}})\cr
&\le \sfrac{1}{\al^7}
\sfrac{\fl_j}{M^j}\fe_j({\sst \| K\|_{1,\Si_{j}}})\cr
}$$
and, by (\eqnTNPodIIba),
$$\eqalign{
\V \sfrac{d\hfill}{ds}w'^{a\sim}_{1,1}({\sst K+sK'})_{\Si_{j+1}}
                                       \big|_{s=0}\tV_{1,\Si_{j+1},\tilde\rho}
&=\V \sfrac{d\hfill}{ds}z_{1,1}^{\sim}({\sst K+sK'})_{\Si_{j+1}}
                                      \big|_{s=0}\tV_{1,\Si_{j+1},\tilde\rho}\cr
&\le\abcst\,\cb_j\V \sfrac{d\hfill}{ds}z_{1,1}^{\sim}({\sst K+sK'})
                                    \big|_{s=0}\tV_{1,\Si_{j},\tilde\rho}\cr
&\le\sfrac{\const}{\al^2\IB}\big(\sfrac{1}{\al^2}+\la_0^{\upsilon/7}\big)\,
    \fe_j(\| K\|_{1,\Si_{j}})\,\| K'\|_{1,\Si_{j}}\cr
&\le\big(\sfrac{1}{\al}+\la_0^{\upsilon/8}\big)\sfrac{1}{\al^2\IB}\,
    \fe_j(\| K\|_{1,\Si_{j}})\,\| K'\|_{1,\Si_{j}}\cr
}$$

\Item {\it Verification of ($O^\sim4$):} 
Apply Remark \remOSrengrppreserves\ of [FKTo2].

\endproof

\goodbreak
\titleb{Sector Refinement, ReWick ordering and Renormalization}\PG\pgNPXVc

\theorem{\STM\thmTildeOuttoIn}{
If $(\cW,\cG,u,\vec p)\in \tilde\cD^{(j)}_{\rm out}$ then  
$\cO_j(\cW,\cG,u,\vec p)\in \tilde\cD^{(j+1)}_{\rm in}$.

}

\noindent The rest of this subsection is devoted to the proof of this theorem.
Let
$
(\cW',\cG',u',{\vec p}^{\,\prime})=\cO_j(\cW,\cG,u,\vec p)
$. 

\lemma{\STM\lemNPmoreBbnd}{
$$\deqalign{
\Big\|A(k)^2\big[C^{(\le j)}_{u'(0)}(k)-C^{(\le j)}_{u(0)}(k)\big]\TNorm
&\le \const\, \sfrac{\la_0^{1-\upsilon}}{\al^5}\sfrac{\fl_{j+1}}{M^{j+1}}
\cb_{j,j+1}\hidewidth\cr
\Big\|A(k)\big[C^{(\le j)}_{u'(0)}(k)-C^{(\le j)}_{u(0)}(k)\big]\TNorm
&\le \const\, \sfrac{\la_0^{1-\upsilon}}{\al^5}\fl_j\cb_{j+1}
\hidewidth\cr
\Big\|A(k)\big[C^{[i,j]}_{u'(0)}(k)-C^{[i,j]}_{u(0)}(k)\big]\TNorm
&\le \const\, \sfrac{\la_0^{1-\upsilon}}{\al^5}\fl_j
\cb_{j+1}
&\qquad\hbox{for all $2\le i\le j$}\cr
\big\|A(k)C^{(\le j)}_{u(0)}(k)\Tnorm,\ 
\big\|A(k)C^{(\le j)}_{u'(0)}(k)\Tnorm
&\le \const\, \cb_{j+1}
\hidewidth\cr
\big\|A(k)C^{[i,j]}_{u(0)}(k)\Tnorm,\ 
\big\|A(k)C^{[i,j]}_{u'(0)}(k)\Tnorm
&\le \const\, \cb_{j+1}
&\qquad\hbox{for all $2\le i\le j$}\cr
}$$

}
\prf Recall from Definition \defNPintquad\ that 
$\check u'(k;0)-\check u(k;0)=\check p^{(j)}(k)$. Hence
$$\eqalign{
A(k)^2\big[C^{(\le j)}_{u'(0)}(k)-C^{(\le j)}_{u(0)}(k)\big]
&=\frac{[ik_0-e(\k)]^2[\check u'(k;0)-\check u(k;0)]\nu^{(\le j)}(k)}
{[ik_0-e(\k)-\check u(k;0)][ik_0-e(\k)-\check u'(k;0)]}\cr
&=\frac{\check p^{(j)}(k)\nu^{(\le j)}(k)}
{\big[1-\sfrac{\check u(k;0)}{ik_0-e(\k)}\big]
\big[1-\sfrac{\check u'(k;0)}{ik_0-e(\k)}\big]}\cr
}$$
By (\eqnTNPjhalfbnds), (\eqnTNPudubnds) and (\eqnNPpboundal),
$$\eqalign{
\big\|\nu^{(\le j)}(k) \Tnorm&\le\abcst\,\cb_{j+{1\over 2}}\cr
\| \check u(0)\tnorm\,,\ \|\check u'(0)\tnorm
&\le\abcst\,\sfrac{1}{M^j}\cb_j\cr
\big\| \check p^{(j)}(k)\Tnorm
&\le 2\sfrac{\la_0^{1-\upsilon}}{\al^5}\,\sfrac{\fl_j}{M^j}\cb_j\cr
}$$
Since $\Big|\sfrac{\check u(k;0)}{ik_0-e(\k)}\Big|\le \half$ and
 $\Big|\sfrac{u'(k;0)}{ik_0-e(\k)}\Big|\le \half$ on the support
of $\nu^{(\le j)}$, we have, using the second bound of (\eqnTNPjhalfbnds)
to control the denominator,
$$
\bigg\| \frac{\check p^{(j)}(k)\nu^{(\le j)}(k)}
{\big[1-\sfrac{\check u(k;0)}{ik_0-e(\k)}\big]
\big[1-\sfrac{\check u'(k;0)}{ik_0-e(\k)}\big]}\,\TNOrm
\le\abcst\, \sfrac{\la_0^{1-\upsilon}}{\al^5}\sfrac{\fl_j}{M^j}\cb_{j+{1\over2}}
\le\const\, \sfrac{\la_0^{1-\upsilon}}{\al^5}\sfrac{\fl_{j+1}}{M^{j+1}}
\cb_{j,j+1}
$$
The proof of the second and third bounds are virtually identical,  using
$$
\big\|\nu^{[i,j]}(k) \Tnorm\le\abcst\,\cb_{j+{1\over 2}}
$$
and with one additional use of the second bound of (\eqnTNPjhalfbnds).
The proof of the final two bounds uses
$$
(ik_0-e(\k))C^{I}_{u(0)}(k)
=(ik_0-e(\k)) \sfrac{\nu^{I}(k)}{ik_0-e(\k)-\check u(k;0)}
= \frac{\nu^{I}(k)}
    {1-\sfrac{\check u(k;0)}{ik_0-e(\k)}}
$$
with $I=[i,j]$ and $I=(\ge j)$,
and (\eqnTNPudubnds) and the corresponding properties with $u(0)$ replaced
by $u'(0)$.
\endproof

\proof {of Theorem \thmTildeOuttoIn}
Let 
$$\eqalign{
w&=\sum_{m,n} \smsum_{s_1,\cdots,s_n\in\Si_j}\
\int {\sst d\eta_1\cdots d\eta_m\,d\xi_1\cdots d\xi_n}\ 
w_{m,n}({\sst \eta_1,\cdots, \eta_m\,(\xi_1,s_1),\cdots ,(\xi_n,s_n)};K)\cr
& \hskip 5cm \phi({\sst \eta_1})\cdots \phi({\sst \eta_m})\
\psi({\sst (\xi_1,s_1)})\cdots \psi({\sst (\xi_n,s_n)\,})\cr
}$$
be the $\Si_j$--sectorized representative of $\cW$ specified in (O1).
Choose the $\Si_{j+1}$--sectorized representative $w'$ of $\cW'$ as in (\eqnNPwpwpp) where $w''$, the $\Si_{j}$--sectorized representative,
was defined in (\eqnNPtwwpp) and $\tilde w$ was defined in (\eqnNPtwrewick).
Recall that, by (\eqnNPgup),
$$
\cG'(\phi;K') = \cG(\phi;K(K'))  +\tilde w(\phi,0;K') -\tilde  w(0,0;K')
\EQN\eqnTNPgtogprime$$

\goodbreak
\Item {\it Preparation for the verification of ($I^\sim1$), ($I^\sim2$) and ($I^\sim3$):} 

Recall from  (\eqnNPwickwtw) that
$$\eqalign{
\tilde w(\phi,\psi;K')=\lw w(\phi,\psi;K(K'))\rw_{\psi,-E_{\Si_j}(K';q_0)}\cr
}$$
where $E_{\Si_j}(K';q)$ was defined just before (\eqnNPtwrewick).
We proved in Lemma \lemRWintbnd.iii that 
$|E(K';q_0)|\le\sfrac{\la_0^{1-\upsilon}\fl_j}{|ik_0-e(\k)|}$.
Hence, by Proposition \propOSmomcontrintboundsectors.ii of [FKTo3], 
$\sqrt{2\la_0^{1-\upsilon}\fl_j\IB_3\sfrac{\fl_j}{M^j}}
\le \sqrt{\la_0^{1-\upsilon}\fl_j}\sqrt{\sfrac{\fl_j\IB}{M^j}}$ is an 
integral bound for $E_{\Si_j}$  for the configuration $\v\,\cdot\,\tv_{p,\Si_j,\tilde\rho}$ of seminorms. Hence
by Corollary \corwicknorm.ii\ of [FKTr1],
$$\eqalign{
N^\sim_j\Big(\tilde w^a(K')-w^a\big(K(K')\big),\sfrac{\al}{2},\,X\Big) 
&\le \sfrac{8\la_0^{1-\upsilon}}{\al^2}\fl_j\
   N^\sim_j\big(w^a(K(K')),\al,\,X\big) \cr
}\EQN\eqnNPTreWickBnd
$$
for all $X\in\fN_{d+1}$. In particular
$$
N^\sim_j\big(\tilde w^a(K'),\sfrac{\al}{2},\,X\big)
\le \ \sfrac{3}{2}\, N^\sim_j\big(w^a(K(K')),\al,\,X\big)
\EQN\eqnNPTreWickBndbis$$
As in (\eqnNPderivwtildeest),
$$\eqalign{
&N^\sim_j\Big(\sfrac{d\hfill}{ds} \tilde w^a (K'+sK'')\big|_{s=0}
   -\sfrac{d\hfill}{ds} w^a({\sst K(K'+sK'')})\big|_{s=0}
               \, ,\,\sfrac{\al}{2}\,,\,{\sst \|K'\|_{1,\Si_{j+1}}} \Big)\cr
&\hskip2in\le \sfrac{\const}{(\al-1)^2}
      \,M^{j}\fe_{j+1}({\sst \|K'\|_{1,\Si_{j+1}}}) \|K''\|_{1,\Si_{j+1}}
}\EQN\eqnNPTderivwchangeest$$
and
$$
N^\sim_j\Big(\sfrac{d\hfill}{ds} \tilde w^a (K'+sK'')\big|_{s=0}
               \, ,\,\sfrac{\al}{2}\,,\,{\sst \|K'\|_{1,\Si_{j+1}}} \Big)
\le \abcst\,M^{j+\aleph}\fe_{j+1}({\sst \|K'\|_{1,\Si_{j+1}}}) \|K''\|_{1,\Si_{j+1}}
\EQN\eqnNPTderivwtildeest$$

\goodbreak
\Item {\it Verification of ($I^\sim3$):} 
Let $\tilde w_{2,0}$ be kernel of the part of $\tilde w$ that is of degree two 
in $\phi$ and degree zero in $\psi$. By  (\eqnTNPgtogprime) and ($O^\sim 3$),
$$\eqalign{
2\check G'_{2}(k;K')
&=  2\check G_{2}(k;K(K'))+2(\tilde w_{2,0})^{\check{}}(k;K')\cr
&= C^{(\le j)}_{u(0)}(k)+ \sfrac{1}{(ik_0-e(\k))^2}
\sum_{i=2}^{j}\Big\{
        \de q^{(i)}(k;K(K'))+\smsum_{\ell=i}^{j}q^{(i,\ell)}(k)\Big\}
+2(\tilde w_{2,0})^{\check{}}(k;K')\cr
&=C^{(< j+1)}_{u'(0)}(k)+ \sfrac{1}{(ik_0-e(\k))^2}
\sum_{i=2}^{j}\Big\{\de q^{\prime(i)}(k;K')
          +\smsum_{\ell=i}^{j+1}q^{\prime(i,\ell)}(k)\Big\}
\cr
}$$
with 
$$\deqalign{
q^{\prime(i,\ell)}(k)&=q^{(i,\ell)}(k)
& 2\le i\le j,\ i\le \ell\le j \cr
q^{\prime(i,j+1)}(k)&=\de q^{(i)}(k;K(0))
& 2\le i\le j-1 \cr
\de q^{\prime(i)}(k;K')&=\de q^{(i)}(k;K(K'))-\de q^{(i)}(k;K(0))
\hskip0.5in& 2\le i\le j-1\hskip0.5in \cr
q^{\prime(j,j+1)}(k)&=\de q^{(j)}(k;K(0))
+2(\tilde w^a_{2,0})^{\check{}}(k;0)
+[ik_0-e(\k)]^2\big[C^{(\le j)}_{u(0)}(k)-C^{(\le j)}_{u'(0)}(k)\big]
\hidewidth \cr
\de q^{\prime(j)}(k;K')&=\de q^{(j)}(k;K(K'))-\de q^{(j)}(k;K(0))
+2(\tilde w^a_{2,0})^{\check{}}(k;K')-2(\tilde w^a_{2,0})^{\check{}}(k;0)
\hidewidth \cr
}$$
By ($O^\sim 3$), Lemma \lemNPdeKbnd.i and Remark \remOSappMonoidIV\ of [FKTo1],
$$\eqalign{
\big\| \de q^{(i)}(k;K(0)) \Tnorm
&= \big\| \de q^{(i)}(k;\de K(0)) \Tnorm\cr
&\le M^{\aleph'(j-i)}\fe_{i+{1\over 2},j+{1\over 2}}
           \big(\|\de K(0)\|_{1,\Si_j}\big) \, \|\de K(0)\|_{1,\Si_j}
        \cases{1 & if $i<j$\cr
               \sqrt{M^{\aleph'-\aleph}}& if $i=j$\cr}\cr
&\le \abcst\,\sqrt{M^{\aleph'-\aleph}}M^{\aleph'(j-i)}\,  
           \sfrac{\la_0^{1-\upsilon}}{\al^6}\sfrac{\fl_j}{M^j}
           \cb_{i+{1\over 2},j+{1\over 2}}\cr
&\le M^{\aleph'(j+1-i)}\,          
   \sfrac{\la_0^{1-\upsilon}}{\al^5}\sfrac{\fl_{j+1}}{M^{j+1}}\cb_{i,j+1}\cr
}$$
if $\al$ is large enough.
This implies the desired bound on $q^{\prime(i,j+1)}$ for $i\le j-1$.
By Lemma \lemOSNormMom\ of [FKTo3], (\eqnNPTreWickBnd) (recall that
$w^a_{2,0}=0$), ($O^\sim 1$) and
Lemma \lemNPdeKbnd.iii,
$$\eqalign{
\big\| (\tilde w^a_{2,0})^{\check{}}(k;0)\,\Tnorm
&\le 2\,\v(\tilde w^a_{2,0})^{\check{}}(k;0)\tv_{1,\Si_j} 
\le \sfrac{\const\la_0^{1-9\upsilon/7}}{\al^2M^j}
\sfrac{8\la_0^{1-\upsilon}}{\al^2}\fl_j
\fe_j({\sst \|\de K(0)\|_{1,\Si_j}})
\le \sfrac{\la_0^{2-3\upsilon}}{\al^4}
\sfrac{\fl_j}{M^j}\cb_{j,j+1}
}$$
So, by Lemma \lemNPmoreBbnd,
$$
\big\|q^{\prime(j,j+1)} \Tnorm
\le \la_0^{1-\upsilon}\sfrac{\fl_{j+1}}{M^{j+1}}\cb_{j,j+1}
\Big[\sfrac{M^{\aleph'}}{\al^5} 
      + 2\sfrac{\la_0^{1-2\upsilon}}{\al^{4}}M^\aleph
      +\sfrac{\const}{\al^5}\Big]
\le \sfrac{\la_0^{1-\upsilon}}{\al^4}\sfrac{\fl_{j+1}}{M^{j+1}}\cb_{j,j+1}
$$
By (\eqnNPTderivwtildeest),
$$\eqalign{
\big\|\sfrac{d\hfill}{ds} (\tilde w_{2,0}^a)^{\check{}} (K'+sK'')\big|_{s=0}\,\Tnorm
&\le \sfrac{\const\la_0^{1-9\upsilon/7}}{\al^2M^j}
\,M^{j+\aleph}\fe_{j+1}({\sst \|K'\|_{1,\Si_{j+1}}}) \|K''\|_{1,\Si_{j+1}}\cr
&\le \sfrac{\la_0^{1-2\upsilon}}{\al^2}
\fe_{j+{1\over 2},j+1}({\sst \|K'\|_{1,\Si_{j+1}}}) \|K''\|_{1,\Si_{j+1}}\cr
}\EQN\eqnTNPderivwtildetwoest$$
and, by ($O^\sim 3$) and parts (ii) and (iii) of Lemma \lemNPdeKbnd
$$\eqalign{
&\big\| \sfrac{d\hfill}{ds}\de q^{(i)}(K({\sst K'+sK''}))\big|_{s=0}\,\Tnorm
=\Big\| \sfrac{d\hfill}{ds}\de q^{(i)}\big( K({\sst K'})
           +s\sfrac{d\hfill}{dx} K({\sst K'+xK''})\big|_{x=0}\big)
\big|_{s=0}\,\TNorm\cr
&\hskip.25in\le M^{\aleph'(j-i)}\fe_{i+{1\over 2},j+{1\over 2}}
   \big(\|K({\sst K'})\|_{1,\Si_j}\big) 
   \,  \Big\|\sfrac{d\hfill}{dx} K({\sst K'+xK''})\big|_{x=0}\Big\|_{1,\Si_j}
     \cases{1 & if $i<j$\cr
               \sqrt{M^{\aleph'-\aleph}}& if $i=j$\cr}\cr
&\hskip.25in\le \abcst\,M^{\aleph'(j-i)}\fe_{i+{1\over 2},j+{3\over 2}}({\sst \|K'\|_{1,\Si_{j+1}}}) 
\,  M^\aleph\fe_{0,j}({\sst \|K'\|_{1,\Si_{j+1}}})\|K''\|_{1,\Si_{j+1}}
\sqrt{M^{\aleph'-\aleph}}\cr
&\hskip.25in\le \half\,M^{\aleph'(j+1-i)}\fe_{i+{1\over 2},j+{3\over 2}}({\sst \|K'\|_{1,\Si_{j+1}}}) 
\, \|K''\|_{1,\Si_{j+1}}\cr
}$$
if $M$ is large enough. This and, when $i=j$, (\eqnTNPderivwtildetwoest) give 
the desired bound
on $\sfrac{d\hfill}{ds}\de q^{\prime(i)}({\sst K'+sK''})$.

As $w^{\prime a}_{1,1}(K')=\tilde w^a_{1,1}(K')_{\Si_{j+1}}$, we have, by
Proposition \propOSresectorI.ii of [FKTo4],  (\eqnNPTreWickBnd),
($O^\sim 1$), ($O^\sim 3$) and Lemma \lemNPdeKbnd.iii,
$$\eqalign{
&\V w^{\prime a\sim}_{1,1}(K')\tV_{1,\Si_{j+1},\tilde\rho}
\le \abcst\,\cb_j\,\V \tilde w^{a\sim}_{1,1}(K')- w^{a\sim}_{1,1}\big(K(K')\big)\tV_{1,\Si_j,\tilde\rho}
+\V  w^{a\sim}_{1,1}\big(K(K')\big)_{\Si_{j+1}}\tV_{1,\Si_{j+1},\tilde\rho}\cr
&\hskip.3in\le \abcst\,\cb_j\,\sfrac{1}{M^j\IB(\al/2)^2}
\sfrac{8\la_0^{1-\upsilon}}{\al^2}\fl_jN^\sim_j(w^a({\sst K(K')}),\al,\,
{\sst \| K(K')\|_{1,\Si_j}})
+\V  w^{a\sim}_{1,1}({\sst K(K')})_{\Si_{j+1}}\tV_{1,\Si_{j+1},\tilde\rho}\cr
&\hskip.3in\le \abcst\,\cb_j\,\sfrac{\la_0^{1-\upsilon}}{\al^4}
\sfrac{\fl_j}{M^j}\fe_{j}({\sst \| K(K')\|_{1,\Si_j}})
+\sfrac{1}{\al^7}
\sfrac{\fl_j}{M^j}\fe_{j}({\sst \| K(K')\|_{1,\Si_j}})\cr
&\hskip.3in\le \abcst\,\sfrac{\fl_j}{M^j}
\Big[\sfrac{\la_0^{1-\upsilon}}{\al^4}+
\sfrac{1}{\al^7}\Big]\fe_{j+1}({\sst \| K'\|_{1,\Si_{j+1}}})\cr
&\hskip.3in\le \sfrac{1}{\al^6}
\sfrac{\fl_{j+1}}{M^{j+1}}\fe_{j+1}({\sst \| K'\|_{1,\Si_{j+1}}})\cr
}\EQN\eqnTNPwponeoneest$$
The derivative is bounded similarly, using  (\eqnNPTderivwchangeest), 
($O^\sim 3$),  and parts (ii) and (iii) of Lemma \lemNPdeKbnd,
$$\eqalign{
&\V\sfrac{d\hfill}{ds}w^{\prime a\sim}_{1,1}(K'+sK'')\big|_{s=0}
         \tV_{1,\Si_{j+1},\tilde\rho}\cr
&\hskip.2in\le \abcst\,\cb_j\,\V \sfrac{d\hfill}{ds}
\big\{\tilde w^{a\sim}_{1,1}(K'+sK'')- w^{a\sim}_{1,1}\big(K(K'+sK'')\big)\big\}
\big|_{s=0}\tV_{1,\Si_j,\tilde\rho}\cr
&\hskip1.7in
+\V\sfrac{d\hfill}{ds}w^{a\sim}_{1,1}\big(K(K'+sK'')\big)_{\Si_{j+1}}\big|_{s=0}
\tV_{1,\Si_{j+1},\tilde\rho}\cr
&\hskip.2in\le \sfrac{\const}{(\al-1)^2\al^2\IB}
      \,\fe_{j+1}({\sst \|K'\|_{1,\Si_{j+1}}}) \|K''\|_{1,\Si_{j+1}}\cr
&\hskip1.7in+\big(\sfrac{1}{\al}+\la_0^{\upsilon/8}\big)
    \sfrac{1}{\al^2\IB}\fe_j({\sst \|K(K')\|_{1,\Si_{j}}})
   \big\|\sfrac{d\hfill}{dx}K(K'+xK'')\big|_{x=0}\big\|_{1,\Si_j}\cr
&\hskip.2in\le \sfrac{1}{\al^2\IB}\Big[\sfrac{\const}{(\al-1)^2}+
\abcst\,M^\aleph\big(\sfrac{1}{\al}+\la_0^{\upsilon/8}\big)
      \Big]
\fe_{j+1}({\sst \|K'\|_{1,\Si_{j+1}}}) \|K''\|_{1,\Si_{j+1}}\cr
}\EQN\eqnTNPwponeonederivest$$

\Item {\it Verification of ($I^\sim2$):} 

Observe that by (\eqnNPTreWickBnd), ($O^\sim1$) and Lemma \lemNPdeKbnd.iii
$$\eqalignno{
\sfrac{M^{2j}}{\fl_j}\,\sfrac{\al^4}{16}\,\Big(\sfrac{\fl_j \IB}{M^j} \Big)^2
\sfrac{1}{\fl_j}\,&
\V \tilde w^{a\sim}_4(K')-w^{a\sim}_4\big(K(K')\big)\tV_{3,\Si_j,\tilde\rho}\cr 
&\le \sfrac{8\la_0^{1-\upsilon}}{\al^2}\fl_j\
   N^\sim_j\big(w^a(K(K')),\al,\,\|K(K')\|_{1,\Si_j}\big) \cr
&\le \sfrac{8\la_0^{1-\upsilon}}{\al^2}\fl_j\
   \fe_j({\sst \|K(K')\|_{1,\Si_j}})\cr
&\le \abcst \sfrac{\la_0^{1-\upsilon}}{\al^2}\fl_j\,\fe_{j+1}({\sst
\|K'\|_{1,\Si_{j+1}}}) \cr
 }$$
so that, by Proposition \propOSresectorI.ii of [FKTo4] and Corollary 
\corOSappMonoidIV.ii of [FKTo1]
$$
\VV \Big( \tilde w^{a\sim}_4(K')-w^{a\sim}_4\big(K(K')\big)\Big)_{\Si_{j+1}}
              \tVV_{3,\Si_{j+1},\tilde\rho} 
\le \abcst \sfrac{\la_0^{1-\upsilon}}{\al^6}\fl_j\,
\fe_{j+1}({\sst\|K'\|_{1,\Si_{j+1}}}) 
\EQN\eqnTNPdifffourlegged$$
Set, for all $2\le i\le j$,
$\ {v'}^{(i)}=v^{(i)}$ and define 
$$
{f'}^{(i)}
   =\sh\Big({v'}^{(i)},C^{[i,j+1)}_{u'(0)}(k)A(k)\Big)
   =\sh\Big(f^{(i)},\big[C^{[i,j]}_{u'(0)}(k)-C^{[i,j]}_{u(0)}(k)\big]A(k)\Big)
$$  
Also set
$$\eqalign{
{\de f'}^{(j+1)}(K')&={\de f}^{(j+1)}(K(K'))_{\Si_{j+1}}
   +\smsum_{i=2}^{j}\big[f^{(i)}_{\Si_{j+1}}-f^{\prime(i)}_{\Si_{j+1}}\big]
   +\big( \tilde w^a_4(K')-w^a_4(K(K'))\big)_{\Si_{j+1}}\cr
&\hskip.5in +\sfrac{1}{8} {\rm Ant\,}\Big( V_{\rm ph} \big( \cL^{(j+1)}
({\vec p}^{\,\prime}, \vec f)\big) \Big)_{\Si_{j+1}}
-\sfrac{1}{8} {\rm Ant\,}\Big( V_{\rm ph} \big( \cL^{(j+1)}
({\vec p}^{\,\prime}, \vec f')\big) \Big)_{\Si_{j+1}}   
}$$
These definitions have been chosen so that
$$\eqalign{
w^{\prime a}_4(K')+G^{\prime a}_4(K')
&=\tilde w^a_{4}(K')_{\Si_{j+1}}+G_4^a(K(K'))\cr
&=\de f^{\prime(j+1)}(K')+\smsum_{i=2}^{j}f^{\prime(i)}_{\Si_{j+1}}
 +\sfrac{1}{8} {\rm Ant\,}\Big( V_{\rm ph} \big( \cL^{(j+1)}
({\vec p}^{\,\prime}, \vec f')\big) \Big)_{\Si_{j+1}}
}$$
Furthermore the required bounds
$$
\V {v'}^{(i)}\tV_{3,\Si_i,\tilde\rho}\le \sfrac{\vi_i}{\al^7}\cb_i
\qquad\qquad\hbox{for all }2\le i\le j
$$
are trivially satisfied, so it remains only to bound 
$\V {\de f'}^{(j+1)}(K')\tV_{3,\Si_{j+1},\tilde\rho}$.

Since $\check u'(k;0) -\check u(k;0)=\check p_j(k)$, $C^{[i,j]}_{u'(0)}(k)-C^{[i,j]}_{u(0)}(k)$ is supported in the 
$j^{\rm th}$ neighbourhood and 
$$
{f'}^{(i)}_{\Si_j}
   =\sh\Big(f^{(i)}_{\Si_j},\big[C^{[i,j]}_{u'(0)}(k)-C^{[i,j]}_{u(0)}(k)\big]               A(k)\Big)
$$  
Hence by Lemma \lemOStildesourceterm\ of [FKTo3], with $X=0$ and
$X_B=\const\sfrac{\la_0^{1-\upsilon}}{\al^5}\fl_j$ and Lemma \lemNPmoreBbnd,
followed by Proposition \propOSresectorI.ii of [FKTo4],
$$
\V {f'}^{(i)}_{\Si_j}-{f}^{(i)}_{\Si_j}\tV_{3,\Si_j,\tilde\rho}
\le \const\sfrac{\la_0^{1-\upsilon}}{\al^5}\fl_j\,\cb_j
\V {f}^{(i)}_{\Si_j}\tV_{3,\Si_j,\tilde\rho}
\le \const\sfrac{\la_0^{1-\upsilon}}{\al^5}\fl_j\,\cb_j
\V {f}^{(i)}\tV_{3,\Si_i,\tilde\rho}
\EQN\eqnTNPprefprime$$
By Lemma \lemOStildesourceterm\ of [FKTo3], with $j$ replaced by $i$,
$X=0$, $X_B=\const\cb_{j+1}$ and  Lemma \lemNPmoreBbnd, 
followed by ($O^\sim 2$),
$$\deqalign{
\V {f}^{(i)}\tV_{3,\Si_i,\tilde\rho}
&\le \const\cb_{j+1}\V {v}^{(i)}\tV_{3,\Si_i,\tilde\rho}
&\le \const\sfrac{\vi_i}{\al^7}\cb_{j+1}\cr
\V {f'}^{(i)}\tV_{3,\Si_i,\tilde\rho}
&\le \const\cb_{j+1}\V {v}^{(i)}\tV_{3,\Si_i,\tilde\rho}
&\le \const\sfrac{\vi_i}{\al^7}\cb_{j+1}\cr
}\EQN\eqnTNPfprime$$
so that, by Proposition \propOSresectorI.ii of [FKTo4] and 
Corollary \corOSappMonoidIV\  of [FKTo1],
$$
\V \smsum_{i=2}^{j}\big[f^{(i)}_{\Si_{j+1}}-f^{\prime(i)}_{\Si_{j+1}}\big]
\tV_{3,\Si_{j+1},\tilde\rho}
\le \const\smsum_{i=2}^{j}\sfrac{\la_0^{1-\upsilon}}{\al^{12}}\fl_j\vi_i\cb_{j+1}
\le \sfrac{\la_0^{1-\upsilon}}{\al^{11}}\fl_j\cb_{j+1}
\EQN\eqnTNPdiffffp$$
Also by Lemma \lemNPmoreBbnd\ and Corollary \cortildecompLadder.ii,
with $j$ replaced by $j+1$, $\rho=\la_0^{1-9\upsilon/7}$, 
$\veps=\sfrac{\aleph}{n_0}$, $B'=[C^{(\le j)}_{u'(0)}(k)
-C^{(\le j)}_{u(0)}(k)]A(k)$,
$B=C^{(\le j)}_{u(0)}(k)A(k)+sB'$, for $0\le s\le 1$,
$c_B=\const$ and $c'=\sfrac{\la_0^{1-\upsilon}}{\al^5}\fl_j$
$$
\V {\rm Ant\,}V_{\rm ph} \big( \cL^{(j+1)}
({\vec p}^{\,\prime}, \vec f)_{\Si_{j+1}}\big) 
-{\rm Ant\,} V_{\rm ph} \big( \cL^{(j+1)}
({\vec p}^{\,\prime}, \vec f')_{\Si_{j+1}}\big) \tV_{3,\Si_{j+1},\tilde\rho}
\le\const \sfrac{\la_0^{2-18\upsilon/7}}{\al^5}\fl_j\cb_{j+1}
\EQN\eqnTNPdiffladdladdp$$
By Definition \defcompLadder, $\cL^{(j+1)}({\vec p}, \vec f)$ depends only on 
$p^{(2)},\cdots, p^{(j-1)}$ and $f^{(2)},\cdots, f^{(j)}$. In particular
$ \cL^{(j+1)}({\vec p}^{\,\prime}, \vec f) =\cL^{(j+1)}({\vec p}, \vec f)$.

To bound $\V {\de f'}^{(j+1)}(0)\tV_{3,\Si_{j+1},\tilde\rho}$,
we first use Proposition \propOSresectorI.ii of [FKTo4], 
(\eqnTNPdifffourlegged), (\eqnTNPdiffffp) and (\eqnTNPdiffladdladdp)
to get the first line, then
($O^\sim 2$) to get the second line and  Lemma \lemNPdeKbnd\ and Corollary 
\corOSappMonoidIV.ii of [FKTo1] to get the third line.
$$\eqalignno{
&\V {\de f'}^{(j+1)}(0)\tV_{3,\Si_{j+1},\tilde\rho}
\le \abcst\, \cb_j \,\V \de f^{(j+1)}(K(0))\tV_{3,\Si_j,\tilde\rho} 
+ \abcst\, \sfrac{\la_0^{1-\upsilon}}{\al^5}\fl_j\,\cb_{j+1}  \cr
&\hskip.2in\le \abcst\,  \cb_j\Big\{
      \sfrac{\vi_{j+1}}{\al^8}
     +\sfrac{1}{\IB^2\al^4}  M^j\| K(0)\|_{1,\Si_j} 
    \Big\} \fe_j({\sst \|K(0)\|_{1,\Si_{j}}})
 + \abcst \sfrac{\la_0^{1-\upsilon}}{\al^5}\fl_j\,\cb_{j+1}\cr
&\hskip.2in\le \abcst\,\Big\{
        \sfrac{\vi_{j+1}}{\al^8}
     + \sfrac{\la_0^{1-\upsilon}}{\al^5}\,\fl_j
     \Big\}\cb_{j+1} \cr
&\hskip.2in\le \sfrac{\vi_{j+1}}{\al^7}\cb_{j+1}
&\EQNO\eqnTNPgoodFprimeest
}$$
since $\ \const\,\la_0^{1-\upsilon} \fl_j\le \sfrac{\vi_{j+1}}{2\al^2}$.
To bound the derivative of $ {\de f'}^{(j+1)}$,
we first use Proposition \propOSresectorI.ii of [FKTo4] to get the first line, 
then ($O^\sim 2$) and (\eqnNPTderivwchangeest) to get the second line 
and  Lemma \lemNPdeKbnd\ and Corollary \corOSappMonoidIV.ii of [FKTo1] to get the third line.
$$\eqalignno{
&\V \sfrac{d\hfill}{ds}{\de f'}^{(j+1)}(K'+sK'')
                 \big|_{s=0}\tV_{3,\Si_{j+1},\tilde\rho}\cr
&\hskip.2in\le \abcst\, \cb_j \,\V \sfrac{d\hfill}{ds}\de f^{(j+1)}(K(K'+sK''))
                 \big|_{s=0}\tV_{3,\Si_j,\tilde\rho} \cr
&\hskip1.5in+ \abcst\, \cb_j \,\V \sfrac{d\hfill}{ds} 
         \tilde w_4^{a\sim} (K'+sK'')\big|_{s=0}
   -\sfrac{d\hfill}{ds} w_4^{a\sim}({\sst K(K'+sK'')})\big|_{s=0}
                \tV_{3,\Si_j,\tilde\rho}  \cr
&\hskip.2in\le\abcst \sfrac{1}{\al^4\IB^2}\fe_j({\sst \|K(K')\|_{1,\Si_{j}}})
M^j\big\|\sfrac{d\hfill}{dx}K(K'+xK'')\big|_{x=0}\big\|_{1,\Si_j}
\cr&\hskip1.5in
+\sfrac{\const}{(\al-1)^2\al^4\IB^2}
      \,M^{j}\fe_{j+1}({\sst \|K'\|_{1,\Si_{j+1}}}) \|K''\|_{1,\Si_{j+1}}\cr
&\hskip.2in\le \sfrac{1}{\al^4\IB^2}\Big\{
      \abcst \sfrac{1}{M^{1-\aleph}}
      +\sfrac{\const}{(\al-1)^2}
     \Big\}\fe_{j+1}({\sst\|K'\|_{1,\Si_{j+1}}}) M^{j+1}\|K''\|_{1,\Si_{j+1}}\cr
&\hskip.2in\le \sfrac{1}{64\al^4\IB^2}\fe_{j+1}({\sst\|K'\|_{1,\Si_{j+1}}}) M^{j+1}\|K''\|_{1,\Si_{j+1}}
&\EQNO\eqnTNPgoodFprimederivest
}$$
if $M$ and $\al$ are large enough.

\goodbreak
\Item {\it Verification of ($I^\sim1$):} 

Let $\tilde \om_{n}$ and $\om'_{n}$ be the parts of $\tilde w$ and $w'$,
respectively, that are of degree $n$ in   $\phi$ and $\psi$ combined.
By  (\eqnTNPgoodFprimeest), (\eqnTNPgoodFprimederivest), 
(\eqnTNPfprime), Proposition \propOSresectorI.ii of 
[FKTo4] and Corollary \cortildecompLadder.i
(with $\rho =\la_0^{1-9\upsilon/7}$, 
$\veps =\sfrac{\aleph}{\scriptscriptstyle n_0}$, $B=C^{(\le j)}_{u'(0)}(k)A(k)$,
$c_B=\const$) 
$$\eqalign{
&\V w'^{a\sim}_{4}(K')\tV_{3,\Si_{j+1},\tilde\rho}\cr
 &\le \V \de f^{\prime(j+1)}(K')\tV_{3,\Si_{j+1},\tilde\rho}
 +\smsum_{i=2}^{j} \V f^{\prime(i)}_{\Si_{j+1}}\tV_{3,\Si_{j+1},\tilde\rho} 
 +\sfrac{1}{8} \VV {\rm Ant\,}\Big( V_{\rm ph} \big( \cL^{(j+1)}
({\vec p}^{\,\prime}, \vec f')\big) \Big)_{\Si_{j+1}} \tVV_{3,\Si_{j+1},\tilde\rho} \cr
& \le  \sfrac{1}{\al^4\IB^2}
      \big(\sfrac{\abcst}{M^{1-\aleph}}\!+\!\sfrac{\const}{(\al-1)^2}\big) 
      {\sst M^{j+1}\|K'\|_{1,\Si_{j+1}}}\fe_{j+1}({\sst\|K'\|_{1,\Si_{j+1}}}) 
     \!+\! \const\!\smsum_{i=2}^{j+1}\! \sfrac{\vi_{i}}{\al^7}\cb_{j+1}
            \!+\!\const \la_0^{1-2\upsilon}\cb_{j+1}\cr
& \le \sfrac{1}{\al^4\IB^2}\Big\{
      \sfrac{\const}{\al^3}
        +\sfrac{\abcst}{M^{1-\aleph}} +\sfrac{\const}{(\al-1)^2}
      \Big\}\fe_{j+1}({\sst\|K'\|_{1,\Si_{j+1}}}) \cr
}$$
Consequently, by Proposition \propOSthreetoonenorm \ of [FKTo4]
$$
\V w'^{a\sim}_4(K')\tV_{1,\Si_{j+1},\tilde\rho} 
\le \sfrac{1}{\al^4\IB^2\fl_{j+1}}\Big\{
      \sfrac{\const}{\al^3}
        +\sfrac{\abcst}{M^{1-\aleph}} +\sfrac{\const}{(\al-1)^2}
      \Big\}\fe_{j+1}({\sst\|K'\|_{1,\Si_{j+1}}}) 
$$
Therefore, by Corollary \corOSappMonoidIV\ of [FKTo1]  
$$\eqalignno{
N^\sim_{j+1}&(\om'^a_4(K'),64\al,\,\|K'\|_{1,\Si_{j+1}})  \cr
&\le 2^{24}\,\al^{4}\,\IB^2\,\,\fl_{j+1}\,
\fe_{j+1}({\sst \|K'\|_{1,\Si_{j+1}}})
\Big( \V w'^{a\sim}_4(K')\tV_{1,\Si_{j+1},\tilde\rho}
+\V w'^{a\sim}_4(K')\tV_{2,\Si_{j+1},\tilde\rho}\cr
&\hskip2in
+\sfrac{1}{\fl_{j+1}}\V w'^{a\sim}_4(K')\tV_{3,\Si_{j+1},\tilde\rho}
+\sfrac{1}{\fl_{j+1}}\V w'^{a\sim}_4(K')\tV_{4,\Si_{j+1},\tilde\rho} \Big) \cr
&\le 2^{25}\,\al^{4}\,\IB^2\,\,\fl_{j+1}\,
\fe_{j+1}({\sst \|K'\|_{1,\Si_{j+1}}})
\Big( \V w'^{a\sim}_4(K')\tV_{1,\Si_{j+1},\tilde\rho} + 
\sfrac{1}{\fl_{j+1}}\V w'^{a\sim}_4(K')\tV_{3,\Si_{j+1},\tilde\rho} \Big) \cr
& \le \Big\{\sfrac{\const}{\al^3}
        +\sfrac{\abcst}{M^{1-\aleph}} +\sfrac{\const}{(\al-1)^2}
      \Big\}\fe_{j+1}({\sst\|K'\|_{1,\Si_{j+1}}})    \cr
& \le \sfrac{1}{3} \fe_{j+1}({\sst\|K'\|_{1,\Si_{j+1}}}) 
&\EQNO\eqnTNPfourlegest
}$$
if $M$ and $\al$ are large enough. We have used that
$\v\ \cdot\ \tv_{2,\Si_{j+1},\tilde\rho}\le \v\ \cdot\ \tv_{1,\Si_{j+1},\tilde\rho}$ and 
$\v\ \cdot\ \tv_{4,\Si_{j+1},\tilde\rho}\le \v\ \cdot\ \tv_{3,\Si_{j+1},\tilde\rho}$.
Similarly, since $\om'^a_{2,0}(K')=\om'^a_{0,2}(K')=0$, (\eqnTNPwponeoneest)
implies
$$\eqalignno{
N^\sim_{j+1}(\om'^a_2(K'),64\al,\,\|K'\|_{1,\Si_{j+1}})  
&\le 2^{12}\,\al^{2}\,\IB\,\,M^{j+1}\,
\fe_{j+1}({\sst \|K'\|_{1,\Si_{j+1}}})
 \V w'^{a\sim}_{1,1}(K')\tV_{1,\Si_{j+1},\tilde\rho}  \cr
& \le \sfrac{\const}{\al^4}
\fl_{j+1}\fe_{j+1}({\sst \| K'\|_{1,\Si_{j+1}}}) ^2  \cr
& \le \sfrac{1}{3} \fe_{j+1}({\sst\|K'\|_{1,\Si_{j+1}}}) 
&\EQNO\eqnTNPtwolegest
}$$
By (\eqnNPtwwpp), (\eqnNPTreWickBndbis), ($O^\sim1$) and 
Lemma \lemNPdeKbnd.iii, 
$$\eqalign{
N^\sim_j( w''^a(K'),\sfrac{\al}{2},0)
&\le \sfrac{3}{2}
\, N^\sim_j(w^a(K(K')),\al,0) \cr
&\le \sfrac{3}{2}
\, N^\sim_j(w^a(K(K')),\al,\,\|K(K')\|_{1,\Si_j}) \cr
&\le \sfrac{3}{2}\,\fe_j(\sst{\|K(K')\|_{1,\Si_j}})  \cr
&\le  \abcst\,\fe_{j+1}({\sst\|K'\|_{1,\Si_{j+1}}})\cr
}$$
so that, by Corollary \corOStildeirrelevantresect\ of [FKTo4] and 
Corollary \corOSappMonoidIV\ of [FKTo1],
$$\eqalign{
&N^\sim_{j+1}(w'^a(K')-\om'^a_4(K')-\om'^a_2(K'),64\al,\|K'\|_{1,\Si_{j+1}})\cr
&\hskip.7in
\le \sfrac{1}{M^{(1-\aleph)/8}}\,\fe_{j+1}({\sst\|K'\|_{1,\Si_{j+1}}})\,
N^\sim_j\big( w''^a(K')-\om''^a_4(K')-\om''^a_2(K'),
                 \sfrac{\al}{2},\|K'\|_{1,\Si_{j+1}}\big)\cr
&\hskip.7in
\le \sfrac{1}{M^{(1-\aleph)/8}}\,\fe_{j+1}({\sst\|K'\|_{1,\Si_{j+1}}})^2\,
N^\sim_j\big( w''^a(K'),\sfrac{\al}{2},0\big)\cr
&\hskip.71in
\le \abcst\,\sfrac{1}{M^{(1-\aleph)/8}}
\,\fe_{j+1}({\sst\|K'\|_{1,\Si_{j+1}}})^3\cr
&\hskip.7in \le \sfrac{1}{3} \fe_{j+1}({\sst\|K'\|_{1,\Si_{j+1}}}) 
}\EQN\eqnTNPnonfourlegest$$
Combining (\eqnTNPfourlegest), (\eqnTNPtwolegest) and (\eqnTNPnonfourlegest), 
we get
$$
N^\sim_{j+1}(w'(K'),64\al,\|K'\|_{1,\Si_{j+1}})
\le \fe_{j+1}({\sst\|K'\|_{1,\Si_{j+1}}}) 
$$

\vskip.25in\noindent
By (\eqnNPtwwpp) and (\eqnNPTderivwtildeest),
$$
N^\sim_j\Big(\sfrac{d\hfill}{ds} w''^a (K'+sK'')\big|_{s=0}
               \, ,\,\sfrac{\al}{2}\,,\,{\sst \|K'\|_{1,\Si_{j+1}}} \Big)
\le \abcst\,M^{j+\aleph}\fe_{j+1}({\sst \|K'\|_{1,\Si_{j+1}}}) \|K''\|_{1,\Si_{j+1}}
$$
Therefore by Corollary \corOStildeirrelevantresect\ of [FKTo4] and 
Corollary \corOSappMonoidIV\ of [FKTo1],
$$\eqalign{
&N^\sim_{j+1}\Big(\sfrac{d\hfill}{ds} w'^a (K'+sK'')\big|_{s=0}
-\sfrac{d\hfill}{ds} \om'^a_2 (K'+sK'')\big|_{s=0}
,64\al,\|K'\|_{1,\Si_{j+1}}\Big)\cr
&\hskip1in
\le \abcst\,\fe_{j+1}({\sst\|K'\|_{1,\Si_{j+1}}})\,
N^\sim_j\Big(\sfrac{d\hfill}{ds} w'' (K'+sK'')\big|_{s=0},\sfrac{\al}{2},\|K'\|_{1,\Si_{j+1}}\Big)\cr
&\hskip1in
\le \abcst\,M^{j+\aleph}
\,\fe_{j+1}({\sst\|K'\|_{1,\Si_{j+1}}})^2\|K''\|_{1,\Si_{j+1}}\cr
&\hskip1in \le \half M^{j+1}
\,\fe_{j+1}({\sst\|K'\|_{1,\Si_{j+1}}})\|K''\|_{1,\Si_{j+1}}
}$$
By (\eqnTNPwponeonederivest),
$$\eqalign{
&N^\sim_{j+1}\Big(\sfrac{d\hfill}{ds} \om'^a_2 (K'+sK'')\big|_{s=0}
,64\al,\|K'\|_{1,\Si_{j+1}}\Big)\cr
&\hskip1in\le 64^2M^{j+1}\Big[\sfrac{\const}{(\al-1)^2}+
\abcst\,M^\aleph\big(\sfrac{1}{\al}+\la_0^{\upsilon/8}\big)
      \Big]
\fe_{j+1}({\sst \|K'\|_{1,\Si_{j+1}}})^2 \|K''\|_{1,\Si_{j+1}}\cr
&\hskip1in\le \half M^{j+1}
\,\fe_{j+1}({\sst\|K'\|_{1,\Si_{j+1}}})\|K''\|_{1,\Si_{j+1}}\cr
}$$
so 
$$
N^\sim_{j+1}\Big(\sfrac{d\hfill}{ds} w'^a (K'+sK'')\big|_{s=0}
,64\al,\|K'\|_{1,\Si_{j+1}}\Big)
 \le  M^{j+1}
\,\fe_{j+1}({\sst\|K'\|_{1,\Si_{j+1}}})\|K''\|_{1,\Si_{j+1}}
$$
as desired. That  $w'^a(\phi,\psi;K)-w'^a(0,\psi;K)$ vanishes unless 
$\hat\nu^{(<j+1)}\phi$ is nonzero is inherited from the same property of $w^a$.

\Item {\it Verification of ($I^\sim4$):} 
Bar/unbar invariance and $k_0$--reversal reality are inherited 
from the corresponding quantities in
$\tilde\cD^{(j)}_{\rm out}$ by Remarks  \remOSgrassmannsymmetries\ 
and \remOSrengrppreserves\ of [FKTo2].

\endproof
\vskip.25in
\proof{of Theorem \theoremNPtildeinduction}

\Item {\it Initialization at $j=j_0$.} 
As at the beginning of \S\CHrecurs, let 
$\tilde w(\phi,\psi;K)$ be the $\Si_{j_0}$--sectorized representative for 
$\ 
\tilde\Om_{C_{u(K)}^{(\le j_0)}}\big(\cV{\sst (\psi)}\big) (\phi,\psi) 
 -\half\phi JC^{(\le j_0)}_{u_{j_0}(K)}J\phi
\ $ 
chosen in Theorems \thmNPsetupinduction\ and \thmNPTsetupinduction\ and set
$$\eqalign{
w(\phi,\psi;K) &=\tilde w(\phi,\psi;K) -\tilde w(\phi,0;K)\cr
}$$
\noindent Set
\item{$\circ$} $v^{(2)}=\cdots=v^{(j_0)}=0$
\item{$\circ$} $f^{(2)}=\cdots=f^{(j_0)}=0$ 
\item{$\circ$} $q^{(i,\ell)}=0$ for all $2\le i\le\ell\le j_0$ 
                     except $i=\ell=j_0$
\item{$\circ$} $\de q^{(2)}=\cdots=\de q^{(j_0-1)}=0$ 

\noindent and define 
$$\eqalign{
\de f^{(j_0+1)}(K)&=\hbox{the kernel of the part of $\tilde w(\phi,\psi;K)$ 
that is quartic in $(\phi,\psi)$ }\cr
q^{(j_0,j_0)}&=2\tilde w^a_{2,0}(k;0)\cr
\de q^{(j_0)}(K)&=2\tilde w^a_{2,0}(k;K)-2\tilde w^a_{2,0}(k;0)
+\big[C^{(\le j_0)}_{u_{j_0}(K)}(k)-C^{(\le j_0)}_{u_{j_0}(0)}(k)\big]
           \big(ik_0-e(\k)\big)^2\cr
&=2\tilde w^a_{2,0}(k;K)-2\tilde w^a_{2,0}(k;0)
-\sfrac{U(\k)-\nu^{(>j_0)}(k)}{1+\check K(\k)/(ik_0-e(\k))}\check K(\k)\cr
}$$
By Theorem \thmNPTsetupinduction, with $w^a$ replace by $\tilde w^a$
$$\eqalign{
N^\sim_{j_0}\big(\tilde w^a(K),\al,\|K\|_{1,\Si_{j_0}}\big) 
&\le \const\,\al^4\la_0^\upsilon\,\fe_{j_0}\big(\|K\|_{1,\Si_{j_0}}\big) \cr
N^\sim_{j_0}\big(\sfrac{d\hfill}{ds}\tilde w^a(K+sK')\big|_{s=0}
,\al,\,\| K\|_{1,\Si_{j_0}}\big) 
   &\le \const\,\al^4\la_0^\upsilon\,\fe_{j_0}\big(\|K\|_{1,\Si_{j_0}}\big) \,  \|K'\|_{1,\Si_{j_0}}
\cr
}$$
If $\la_0$ is sufficiently small, depending on $M$, 
and $\al<\sfrac{1}{\la_0^{\up/10}}$ these bounds and 
Lemma \lemNPBbnd.i (with $C^{(\le j_0)}$ replacing $C^{(j)}$)
imply the bounds on
\item{$\circ$} $w^a(K)$ and $\sfrac{d\hfill}{ds} w^a(K+sK')$
imposed by ($O^\sim 1$)
\vskip.05in
\item{$\circ$} $\V \de f^{(j+1)}(0)\tV_{3,\Si_j,\tilde \rho}$ 
and $\V\sfrac{d\hfill}{ds}{\de f}^{(j+1)}(K+sK')\big|_{s=0}
                       \tV_{3,\Si_{j},\tilde\rho}$
imposed by ($O^\sim 2$)
\vskip.05in
\item{$\circ$} $\big\| q^{(j_0,j_0)}(k) \Tnorm$ and
$\big\| \sfrac{d\hfill}{ds}\de q^{(j_0)}(K+sK')\big|_{s=0}\,\Tnorm$ 
imposed by ($O^\sim 3$)
\vskip.05in
\item{$\circ$} $\V w_{1,1}^{a\sim}(K)_{\Si_{j_0+1}}\tV_{1,\Si_{j_0+1},\tilde\rho}$
and $\V \sfrac{d\hfill}{ds}w^{a\sim}_{1,1}({\sst K+sK'})_{\Si_{j_0+1}}
                                       \big|_{s=0}\tV_{1,\Si_{j_0+1},\tilde\rho}$
imposed by ($O^\sim 3$)

\noindent The support properties required by ($O^\sim 1$) and ($O^\sim 3$)
follow from the conclusion in Theorem \thmNPTsetupinduction\  that 
$\tilde w^a(\phi,\psi,K)-\tilde w^a(0,\psi,K)$ vanishes unless 
$\hat\nu^{(\le j_0)}\phi$ is nonzero. Finally ($O^\sim 4$)
follows from Remark \remOSrengrppreserves\ of [FKTo2] and
the requirement, stated in the Introduction, \S\CHintroIII,
to this part, that $V$ satisfy the reality condition (\eqnNPreal) and
be bar/unbar exchange invariant.

\Item {\it Recursive step $j-1\rightarrow j$.} 
In the proof of Theorem \theoremNPinduction\ in \S\CHrecurs, we constructed
$\cG^\rg$ so that $\big(\cW_j,\cG^\rg_{j},u_{j},({\sst p^{(2)},\cdots,p^{(j-1)}})\big)\in\cD^{(j)}_{\rm out}$. 
By Theorems \thmTildeIntoOut\ and \thmTildeOuttoIn, we have, in addition,
that
$\big(\cW_j,\cG^\rg_{j},u_{j},({\sst p^{(2)},\cdots,p^{(j-1)}})\big)\in
\tilde\cD^{(j)}_{\rm out}$. Let 
$$
2\check G^\rg_{j,2}(k)=  C^{(\le j)}_{u(0)}(k)+ \sfrac{1}{[ik_0-e(\k)]^2}
\sum_{i=2}^{j}\smsum_{\ell=i}^{j}q^{(i,\ell)}(k)
$$
be the decomposition of ($O^\sim 3$), but with $K=0$. By ($O^\sim 3$),
$$
\sup_k\big|\rD^\de q^{(i,\ell)}(k) \big|
\le\de!\la_0^{1-2\upsilon}\sfrac{\fl_\ell}{M^\ell}M^{\aleph'(\ell-i)}
M^{\de_0 i}M^{|\bde|\ell}
$$
By ($O^\sim 4$),  $q^{(i,\ell)}(-k_0,\k)=\overline{q^{(i,\ell)}(k_0,\k)}$
for all of the required $i,\ell$.
\endproof

\vskip.25in
\proof{of Theorem \theoremNPmainthIII}
By ($O^\sim2$), the kernel of the quartic part of $\cG^\rg_j(0)$,
amputated by $ik_0-e(\k)$ rather than  $\sfrac{1}{\check G_2(k)}
=ik_0-e(\k)-\Si(k)$, is
$$
P_\phi\Big[\de f^{(j+1)}(0)+\smsum_{i=2}^{j}\big(f^{(i)}_j\big)_{\Si_j}+
\sfrac{1}{8}{\rm Ant\,}
\Big( V_{\rm ph}\big( \cL^{(j+1)}(\vec p,\vec f_j)\big) \Big)\Big]
$$
where the $i^{\rm th}$ component of $\vec f_j$ is $f^{(i)}_j=\sh\big(v^{(i)},C^{[i,j]}_{u_j}(k)A(k)\big)$,
$u_j=\smsum_{i=2}^{j-1}p^{(i)}_{\Si_j}$ and $P_\phi$ is projection onto
the pure $\phi$ part.
Recall from Definition \defcompLadder\ that the ladders 
$\cL^{(n)}(\vec p,\vec f\,)$ were defined inductively
by
$$\eqalign{
\cL^{(0)}(\vec p,\vec f\,)&=0 \cr
\cL^{(n+1)}(\vec p,\vec f\,)&= \cL^{(n)}(\vec p,\vec f\,)_{\Si_n}
+ 2 \smsum_{\ell=1}^\infty {\sst (-1)^\ell(12)^{\ell+1}}
  L_\ell\big(w_n;C^{(n)}_{u_n}, C^{(\ge n+1)}_{u_n}\big)^{\rm ph}
\cr }\EQN\eqnGFitphl$$
where $w_n = \sum_{i=2}^{n}{f}_{\Si_n}^{(i)}
+\sfrac{1}{8} {\rm Ant\,}\Big( V_{\rm ph} 
 \big( \cL^{(n)}(\vec p,\vec f\,)\big)\Big)_{\Si_n}$. By the construction of Theorem \theoremNPinduction, the quartic part of $\cG$,
again amputated by $ik_0-e(\k)$ rather than  $ik_0-e(\k)-\Si(k)$, is 
$$
\smsum_{i=2}^{\infty}P_\phi f^{(i)}
+\lim_{j\rightarrow\infty}\sfrac{1}{8}{\rm Ant\,}
\Big( V_{\rm ph}\big( \cL^{(j+1)}(\vec p,\vec f\,)\big) \Big)
$$
with $f^{(i)}=\sh\big(v^{(i)},C^{(\ge i)}_{P}(k)A(k)\big)$
and $P(k)=\sum_{i=2}^\infty \check p^{{i}}(k)$ as in Lemma 
\lemNPgtwoamp. The sum  makes sense even though different terms
have different sectorization scales because $\phi$ arguments are 
not involved in sectorization. We shall prove convergence shortly.

The quartic part of $\cG$,
correctly amputated by $ik_0-e(\k)-\Si(k)$, is 
$$
\smsum_{i=2}^{\infty}P_\phi F^{(i)}
+\lim_{j\rightarrow\infty}\sfrac{1}{8}{\rm Ant\,}
\Big( V_{\rm ph}\big( \cL^{(j+1)}(\vec p,\vec F)\big) \Big)
$$
where 
$$\eqalign{
F^{(i)}&=\Sct\big(f^{(i)},\sfrac{ik_0-e(\k)-\Si(k)}{ik_0-e(\k)}\big)\cr
&=\Sct\Big(\sh\big(v^{(i)},\nu^{(\ge i)}\sfrac{ik_0-e(\k)}{ik_0-e(\k)-P(k)}\big)
\ \ ,\ \ \sfrac{ik_0-e(\k)-P(k)}{ik_0-e(\k)}A_2(k)\Big)\cr
&=\Sct\Big(\sh\Big\{\Sct\big(v^{(i)},\sfrac{ik_0-e(\k)-P(k)}{ik_0-e(\k)}\big)\ ,
\ \nu^{(\ge i)}\Big\}\ \ ,\ \ A_2(k)\Big)\cr
&=\Sct\Big(\sh\Big\{\Sct\big(v^{(i)},A_1(k)\big)\ ,
\ \nu^{(\ge i)}\Big\}\ \ ,\ \ A_2(k)\Big)\cr
}$$
since, by ($O^\sim 2$), $v^{(i)}$ vanishes unless its $\phi$ momenta 
are in the support of $\nu^{(<i)}$ and the factor $\nu^{(\le i)}$ 
in the definition of $A_1$ (in Lemma \lemNPampcorrect) is identically 
one on that support. Set
$$\eqalign{
\tilde v^{(i)}
&=\sh\Big(\Sct\big(v^{(i)},A_1(k)\big)\ ,
\ \nu^{(\ge i)}\Big)\cr
}$$
In this notation, the quartic part of $\cG$,
correctly amputated by $ik_0-e(\k)-\Si(k)$, is 
$$
\Sct\Big(\smsum_{i=2}^{\infty}P_\phi\Big[\tilde v^{(i)}
+\lim_{j\rightarrow\infty}\sfrac{1}{8}{\rm Ant\,}
\Big( V_{\rm ph}\big( \cL^{(j+1)}(\vec p,\vec{\tilde v})\big) \Big)\Big]\ ,\ A_2(k)\Big)
$$
We have proven in Lemma \lemNPampcorrect.ii that $A_2(k)$ is $C^{1/2}$.
So it suffices to prove that  
$\smsum\limits_{i=2}^{\infty}\!P_\phi\tilde v^{(i)}$ and
$\lim\limits_{j\rightarrow\infty}\sfrac{1}{8}{\rm Ant\,}
\Big( V_{\rm ph}\big( \cL^{(j+1)}(\vec p,\vec{\tilde v})\big)\!\Big)$
 have the continuity properties specified in the statement of the theorem.

By $(O^\sim 2)$, Lemma \lemNPampcorrect.i and Remark \remNPshearprime.ii, 
$$
\V \tilde v^{(i)}\tV_{3,\Si_i}
\le \const\la_0^{1-11\upsilon/7}\fl_i^{1/n_0} \,\cb_{i}
\EQN\eqnGFtvbnd$$
Define, for $f\in\check\cF_{4,\Si}$,
$$\eqalign{
\tn f\trn &=\sup
\big\{\ 
|g({\sst(k_1,\si_1,1),(k_2,\si_2,0),(k_3,\si_3,1),(k_4,\si_4,0)})|
\ \big|\ \si_i\in\{\uparrow,\downarrow\},\ k_1+k_3=k_2+k_4\ \big\}\cr
}$$
where $g$ is the function on $\check \cB_4$ such that
$$
f\big|_{(0,\cdots, 0)}(\check\eta_1,\cdots,\check\eta_n) = 
(2\pi)^3\de(\check\eta_1+\cdots+\check\eta_n)\,
g(\check\eta_1,\cdots,\check\eta_n)
$$
By Lemma \lemOSNormMom\ of [FKTo3],
$$\eqalign{
\tn\tilde v^{(i)}\trn&\le  \const\la_0^{1-11\upsilon/7}\fl_i^{1/n_0}\cr 
}$$
and, if $\rD$ is dd--operator of type $(4,0)$ with $|\de(\rD)|=1$, 
in the sense of Definition \defNPdiffdecay, 
$$
\tn\rD\tilde v^{(i)}\trn\le \const\la_0^{1-11\upsilon/7}\fl_i^{1/n_0} M^i
$$
By Lemma \lemNPhoelder, with $\al=\sfrac{\aleph}{n_0}$ and $\be=1-\sfrac{\aleph}{n_0}$,
$\smsum\limits_{i=2}^{\infty}P_\phi\tilde v^{(i)}$ is $C^{\aleph/n_0}$. 

\vskip.3cm
The remaining estimates use cancellations between scales. For this reason, we 
want to replace the $j$--dependent functions 
$\check u_j=\smsum_{i=2}^{j-1}\check p^{(i)}(k)$ in the recursive 
definition (\eqnGFitphl) of iterated particle hole ladders by one 
$j$--independent function $P(k)=\smsum_{i=2}^\infty\check p^{(i)}(k)$.
As in Definition \defmodcompLadder, we define, recursively on 
$0\le n <\infty$, the compound particle hole  ladder up
to scale $n$ as
$$\eqalign{
\cL^{(0)}_P(\vec f\,)&=0 \cr
\cL^{(n+1)}_P(\vec f\,)&= \cL^{(n)}_{P}(\vec f\,)_{\Si_n}
+ 2 \smsum_{\ell=1}^\infty {\sst (-1)^\ell(12)^{\ell+1}}
  L_\ell\big(w_{P,n};C^{(n)}_P, C^{(\ge n+1)}_P\big) ^{\rm ph}
\cr }\EQN\eqnGFcmpphl$$
where
$w_{P,n} = \sum_{i=2}^n{f}_{\Si_n}^{(i)}
+\sfrac{1}{8}{\rm Ant\,}\Big( 
          V_{\rm ph}\big(\cL^{(n)}_P(\vec f\,)\big)
    \Big)_{\Si_n}$.
For $j\ge 1$ set 
$$\eqalign{
\tilde w_j&= \sum_{i=2}^j\tilde v_{\Si_j}^{(i)}
+\sfrac{1}{8} {\rm Ant\,}\Big( V_{\rm ph} 
 \big( \cL^{(j)}(\vec p,\vec{\tilde v})\big)\Big)_{\Si_j}\cr
\tilde v^{\prime\,(j+1)} &= \tilde v^{(j+1)}
-\sfrac{1}{4} \smsum_{\ell=1}^\infty \! {\sst (-1)^\ell(12)^{\ell+1}}
{\rm Ant\,}\, V_{\rm ph} 
\Big( L_\ell ({\sst \tilde w_j;\,C^{(j)}_P, C^{(\ge j+1)}_P })^{\rm ph}\!-\! 
L_\ell ({\sst \tilde w_j;\,C^{(j)}_{u_j}, C^{(\ge j+1)}_{u_j}})^{\rm ph}
 \Big)_{\Si_{j+1}}\cr
}$$
By (\eqnmodcompLadderII), with the replacements 
$F^{(i)}\rightarrow \tilde v^{(i)}$,
$F^{\prime\,(i)}\rightarrow \tilde v^{\prime\,(i)}$,
$w_j\rightarrow \tilde w_j$,
$v_j\rightarrow u_j$ and
$v\rightarrow P$,
$$
\tilde w_j = \smsum_{i=2}^{j}{\tilde v}_{\Si_j}^{\prime\,(i)}
+\sfrac{1}{8} {\rm Ant\,}\Big( V_{\rm ph} 
 \big( \cL^{(j)}_P(\vec{\tilde v}^{\,\prime})\big)\Big)_{\Si_j}
$$
for all $j\ge 2$. Set
$$
\de\cL^{(n)}=2 \smsum_{\ell=1}^\infty {\sst (-1)^\ell(12)^{\ell+1}}
  \Big(L_\ell\big(\tilde w_n;C^{(n)}_{u_n}, C^{(\ge n+1)}_{u_n}\big)
      -L_\ell\big(\tilde w_n;C^{(n)}_P, C^{(\ge n+1)}_P\big) \Big)^{\rm ph}
$$
Subtracting (\eqnGFcmpphl)  from  (\eqnGFitphl)
$$
 \cL^{(n+1)}(\vec p,\vec{\tilde v})-\cL^{(n+1)}_P(\vec{\tilde v}^{\,\prime})
=\cL^{(n)}(\vec p,\vec{\tilde v})_{\Si_n}
      -\cL^{(n)}_P(\vec{\tilde v}^{\,\prime})_{\Si_n}
+\de\cL^{(n)}
$$
Since $ \cL^{(1)}(\vec p,\vec{\tilde v})=\cL^{(1)}_v(\vec{\tilde v}^{\,\prime})=0 $
$$
 \cL^{(n+1)}(\vec p,\vec{\tilde v})-\cL^{(n+1)}_P(\vec{\tilde v}^{\,\prime})
=\sum_{i=1}^n\de\cL^{(i)}_{\Si_n}
$$

By (\eqnGFtvbnd), the hypotheses of Theorem \theoremtildecompLadder\ 
with $\vec F=\vec{\tilde v}$, $\rho=\const\la_0^{1-11\upsilon/7}$ 
and $\veps = \sfrac{\aleph}{n_0}$ are satisfied. Hence, 
by (\eqnmodcompLadderVIII), 
$$
\V V_{\rm ph}\big( \de \cL^{(i)}\big)\tV_{3,\Si_i}
\le\const\la_0^{3-33\upsilon/7}\fl_i\cb_i
$$
By Lemma \lemOSNormMom\ of [FKTo3],
$$
\TN V_{\rm ph}\big( \de \cL^{(i)}\big)\TRN
\le  \const\la_0^{3-33\upsilon/7}\fl_i\qquad\hbox{ and }\qquad 
\TN\rD V_{\rm ph}\big( \de \cL^{(i)}\big)\TRN
\le \const\la_0^{3-33\upsilon/7}\fl_i M^i
$$
for all  dd--operators, $\rD$, with $|\de(\rD)|=1$.
By Lemma \lemNPhoelder, with $\al=\aleph$ and $\be=1-\aleph$,
$\smsum\limits_{i=2}^{\infty}P_\phi V_{\rm ph}\big( \de \cL^{(i)}\big)$ 
is $C^{\aleph}$. 
Thus 
$\smsum\limits_{i=2}^{\infty}P_\phi\Big[\tilde v^{(i)}+\sfrac{1}{8}{\rm Ant}
\big(V_{\rm ph}\big(\de\cL^{(i)}\big)\big)\Big]$ is $C^{\aleph/n_0}$. 
We choose $N_{\si_1,\si_2,\si_3,\si_4}$ to be
$\Sct\big(\ \cdot\ ,\ A_2(k)\big)$ applied to this sum and 
$L_{\si_1,\si_2,\si_3,\si_4}(q_1,q_2,t)$ to be
$\Sct\big(\ \cdot\ ,\ A_2(k)\big)$ applied to
$\lim\limits_{n\rightarrow\infty}P_\phi 
 V_{\rm ph}\big(\cL^{(n+1)}_P(\vec{\tilde v}^{\,\prime})\big)$
and evaluated at $k_1=q_1+\sfrac{t}{2}$, $k_2=q_1-\sfrac{t}{2}$, 
$k_3=q_2+\sfrac{t}{2}$ and $k_4=q_2+\sfrac{t}{2}$.
By [FKTl, Theorem \thmLADmodcompLaddercont]
the limit exists pointwise for all $t\ne 0$ and is continuous there.
The same theorem provides the existence of continuous extensions of 
$L_{\si_1,\si_2,\si_3,\si_4}(q_1,q_2,(0,\t))$ 
and $L_{\si_1,\si_2,\si_3,\si_4}(q_1,q_2,(t_0,\0))$ to $t=0$.
\endproof

\vfill\eject

\appendix{\APappHoelder}{H\"older Continuity of Limits}\PG\pgNPC

\lemma{\STM\lemNPhoelder}{
Let $\al,\ \be,\ C_0,\ C_1>0$ and $M>1$. There is a constant $C'$such that if
$$
f(t)=\sum_{j=0}^\infty f_j(t)
$$
with
$$
\sup_t\big|f_j(t)\big|\le \sfrac{C_0}{M^{\al j}}\qquad
\sup_t\big|f'_j(t))\big|\le C_1 M^{\be j}
$$
then
$$
\big|f(t)-f(t')\big|
\le C'\,C_0^{\be\over\al+\be}\,C_1^{\al\over\al+\be}\, |t-t'|^{\al/(\al+\be)}
$$

}
\prf
Since
$$
\big|f_j(t)-f_j(t')\big|
\le 2\min\Big\{C_1M^{\be j}|t-t'|, \sfrac{C_0}{M^{\al j}}\Big\}
$$
we have
$$\eqalign{
\big|f(t)-f(t')\big|
&\le\sum_{j=0}^\infty \big|f_j(t)-f_j(t')\big|\cr
&\le\hskip.3in 2\hskip-.3in
   \sum_{0\le j<\infty\atop {1\over M^{(\al+\be)j}}\ge {C_1\over C_0}|t-t'|}
               C_1 M^{\be j}|t-t'|
\hskip.3in +\ \ 2\hskip-.3in
   \sum_{0\le j<\infty\atop {1\over M^{(\al+\be)j}}\le {C_1\over C_0}|t-t'|} 
                \sfrac{C_0}{M^{\al j}}\cr
&\le 2C_1\sfrac{M^\be}{M^\be-1}
        \big(\sfrac{C_1}{C_0}|t-t'|\big)^{-\be/(\al+\be)}|t-t'|
+2C_0\sfrac{1}{1-M^{-\al}}
        \big(\sfrac{C_1}{C_0}|t-t'|\big)^{\al/(\al+\be)}
\cr
&= C'\,C_0^{\be\over\al+\be}\,C_1^{\al\over\al+\be}\, |t-t'|^{\al/(\al+\be)}
}$$
with
$$
C'=2\big\{\sfrac{M^\be}{M^\be-1}+\sfrac{M^\al}{M^\al-1}\big\}
$$

\endproof

\vfill\eject

\appendix{\APappPhladders}{Another Description of}
\null\vskip-.6in
\centerline{\tafontt Particle--Hole Ladders}\PG\pgNPD
\vskip.3in

The estimates on iterated particle hole ladders in [FKTl] use cancellations between scales. For this reason, we want to replace the $j$--dependent functions $u_j$ in the recursive Definition \deftildecompLadder\ of iterated particle hole ladders by one $j$--independent function.

\definition{\STM\defmodcompLadder}{
Let $\vec F=\big(F^{(2)},F^{(3)},\cdots \big)$ be a sequence of 
antisymmetric, spin independent, particle number conserving functions
$F^{(i)}\in \check\cF_{4,\Si_i}$
and $v(k)$ a function on $\bbbr\times\bbbr^2$ such that
 $|v(k)| \le \half|\imath k_0 -e(\k)|$.
We define, recursively on 
$0\le j <\infty$, the compound particle hole (or wrong way) ladders up
to scale $j$, denoted by $\,\cL^{(j)}=\cL^{(j)}_v(\vec F)\,$, as
$$\eqalign{
\cL^{(0)}&=0 \cr
\cL^{(j+1)}&= \cL^{(j)}_{\Si_j}
+ 2 \smsum_{\ell=1}^\infty {\sst (-1)^\ell(12)^{\ell+1}}
  L_\ell\big(w;C^{(j)}_{v}, C^{(\ge j+1)}_{v}\big) ^{\rm ph}
\cr }$$
where
$w = \sum_{i=2}^{j}{F}_{\Si_j}^{(i)}
+\sfrac{1}{8} {\rm Ant\,}\Big( V_{\rm ph}  \big( \cL^{(j)}\big)\Big)_{\Si_j}$.
}

In [FKTl, Remark \remLADmodcompLadder] we prove

\theorem{\STM\theoremmodcompLadder}{
For every $\veps>0$ there are constants $\tilde\rho_0,\const$ such that the following
holds.  
Let $\vec F=\big(F^{(2)},F^{(3)},\cdots \big)$ be a sequence of 
antisymmetric, spin independent, particle number conserving functions
$F^{(i)}\in \check\cF_{4,\Si_i}$
and $\vec p=\big(p^{(2)},p^{(3)},\cdots \big)$ be a sequence of 
antisymmetric, spin independent, particle number conserving functions
$p^{(i)} \in \cF_0(2,\Si_i)$.
Assume that there is $\rho \le \tilde\rho_0$ such that for $i\ge 2$
$$
\v F^{(i)}\tv_{3,\Sigma_i} \le \sfrac{\rho}{M^{\veps i}}\cb_i\qquad\quad 
\v p^{(i)}\v_{1,\Sigma_i} \le \sfrac{\rho\,\fl_i}{M^i}\cb_i\qquad
\check p^{(i)}(0,\k)=0
$$
Set 
$\,v(k) =\smsum_{i=2}^\infty \check p^{(i)}(k)\,$. Then for all $j\ge 1$
$$
\V V_{\rm ph}\big( \cL^{(j+1)}_v(\vec F) \big)\tV_{3,\Si_j} 
\le \const \rho^2 \,\cb_j
$$
}

\proof{that Theorem \theoremtildecompLadder\ follows from Theorem \theoremmodcompLadder}
Let $\vec F=\big(F^{(2)},F^{(3)},\cdots \big)$ and 
$\vec p=\big(p^{(2)},p^{(3)},\cdots \big)$ 
be as in the hypotheses of Theorem \theoremtildecompLadder. Because $\check p^{(i)}$ 
vanishes at $k_0=0$, Lemma \lemOSNormMom\ of [FKTo3] implies
that
$$
\big|\check p^{(i)}(k)\big|
\le 2|k_0|\sfrac{\partial\hfill}{\partial t_{(1,0,0)}}
                \v p^{(i)}\v_{1,\Si_i}\Big|_{t=0}
\le 2|\imath k_0-e(\k)|\sfrac{\rho\fl_i}{M^i}M^i
= 2 \rho\,\fl_i|\imath k_0-e(\k)|
\EQN\eqnmodcompLadderI
$$

Set $w_j = \sum_{i=2}^{j}{F}_{\Si_j}^{(i)}
+\sfrac{1}{8} {\rm Ant\,}\Big( V_{\rm ph} 
 \big( \cL^{(j)}(\vec p,\vec F\big)\Big)_{\Si_j}$ and  $v_j(k)=\smsum_{i=2}^{j-1}\check p^{(i)}(k)$.
By definition
$$
\cL^{(j+1)}(\vec p,\vec F)= \cL^{(j)}(\vec p,\vec F)_{\Si_j}
+ 2 \smsum_{\ell=1}^\infty {\sst (-1)^\ell(12)^{\ell+1}}
  \Big(L_\ell\big(w_j;C^{(j)}_{v_j}, C^{(\ge j+1)}_{v_j}\big) \Big)^{\rm ph}
\EQN\eqnmodcompLadderIIa$$
For $j\ge 1$ set 
$$
F^{\prime\,(j+1)} = F^{(j+1)}
-\sfrac{1}{4} \smsum_{\ell=1}^\infty  {\sst (-1)^\ell(12)^{\ell+1}}
{\rm Ant\,}\, V_{\rm ph} 
\Big( L_\ell ({\sst w_j;\,C^{(j)}_{v}, C^{(\ge j+1)}_{v} })^{\rm ph}  - 
L_\ell ({\sst w_j;\,C^{(j)}_{v_j}, C^{(\ge j+1)}_{v_j}})^{\rm ph}
 \Big)_{\Si_{j+1}}
$$
By induction ones sees that for $j\ge 2$
$$
w_j = \smsum_{i=2}^{j}{F}_{\Si_j}^{\prime\,(i)}
+\sfrac{1}{8} {\rm Ant\,}\Big( V_{\rm ph} 
 \big( \cL^{(j)}_v(\vec F')\big)\Big)_{\Si_j}
\EQN\eqnmodcompLadderII$$

Since $\check p^{(i)}(k)$ is supported on the $i^{\rm th}$ 
extended neighbourhood (defined in Definition \defNPscales)
and $\nu^{(j)}$ is supported on the $j^{\rm th}$ shell, 
$\,C^{(j)}_{v}=C^{(j)}_{v_j+\check p^{(j)}+\check p^{(j+1)}}$. By Proposition \propOSresectorI.iii of [FKTo4], 
$$
\V p^{(j)}+ p^{(j+1)}_{\Si_j} \V_{1,\Si_j} 
\le \const\,\big( \v p^{(j)} \v_{1,\Si_j}+  
\sfrac{\fl_j}{\fl_{j+1}}\cb_j\, \V p^{(j+1)} \V_{1,\Si_{j+1}} \big)
\le \const\, \sfrac{\rho\,\fl_j} {M^j}\cb_{j+1}
$$
It follows from (\eqnmodcompLadderI) that
$$
|v(k) - v_j(k)| \le  \abcst\,\rho\, \fl_j\, |\imath k_0 - e(\k)|
$$
By Corollary \:\corOSresectorvanishkzero\  of [FKTo4],
$$
\VV \smsum_{i=2}^{j-1}p^{(i)}_{\Si_j} \VV_{1,\Si_j},\ 
\VV \smsum_{i=2}^{j+1}p^{(i)}_{\Si_j}  \VV_{1,\Si_j} 
\le \const\sfrac{\rho}{M^j}\cb_{j+1}
$$
We apply Proposition \:\propOSNaiveLadder.ii of [FKTo3] 
with $u=\smsum\limits_{i=2}^{j+1}p^{(i)}_{\Si_j}$,
$v=\smsum\limits_{i=2}^{\infty}p^{(i)}_{\Si_j}$, 
$u'=v'=\smsum\limits_{i=2}^{j-1}p^{(i)}_{\Si_j}$,
$\veps = \const\, \rho\,\fl_j$ , $X=\tau_2\, \cb_{j+1}$
(where $\tau_2$ was defined in Lemma \lemOSdiffpropbound\ of [FKTo3]), $f=w_j$
and get, that for all $\ell\ge 1$
$$\eqalign{
&\VV  V_{\rm ph} 
\Big(   L_\ell\big(w_j;C^{(j)}_{v}, C^{(\ge j+1)}_{v}\big)^{\rm ph}  - 
       L_\ell\big(w_j;C^{(j)}_{v_j}, C^{(\ge j+1)}_{v_j}\big)^{\rm ph} 
\Big) \tVV_{3,\Si_j} 
\le \rho\,\fl_j\,
\big( \const \,\cb_{j+1}\big)^\ell\,
\v w_j\tv_{3,\Si_j}^{\ \ \ell+1}  \cr
}\EQN\eqnmodcompLadderIIIa$$
Proposition \propOSresectorI.ii of [FKTo4] implies that also
$$\eqalign{
&\VV  {\rm Ant\,}\, V_{\rm ph} 
\Big( L_\ell\big(w_j;C^{(j)}_{v}, C^{(\ge j+1)}_{v}\big)^{\rm ph}  - 
L_\ell\big(w_j;C^{(j)}_{v_j}, C^{(\ge j+1)}_{v_j}\big)^{\rm ph} 
\Big)_{\Si_{j+1}} \tVV_{3,\Si_{j+1}} \cr
& \hskip 8cm \le \rho\,\fl_j\,
\big( \const \,\cb_{j+1}\big)^\ell\,
\v w_j\tv_{3,\Si_j}^{\ \ \ell+1}  \cr
}\EQN\eqnmodcompLadderIII$$

We now prove by induction that, if $\rho_0$ is small enough, then for all $j\ge 2$
$$\eqalign{
\V F^{\prime\,(j)}\tV_{3,\Si_j} 
 &\le 2 \,\rho\,\max \big\{\fl_j,\,\sfrac{1}{M^{\veps j}} \big\} \cb_j\cr
\V F^{\prime\,(j)}-F^{(j)}\tV_{3,\Si_j} 
 &\le \const\,\rho^3\, \fl_{j-1} \cb_j\cr
}\EQN\eqnmodcompLadderIV$$
The induction beginning is trivial since $F^{\prime\,(2)}=F^{(2)}$. 
For the induction step, assume that 
(\eqnmodcompLadderIV) holds for $j'\le j$. Then, using Proposition \propOSresectorI.ii of [FKTo4], 
$$
\V \smsum_{i=2}^{j}{F}_{\Si_j}^{\prime\,(i)} \tV_{3,\Si_j}
\ \le\ \abcst\,\rho\,\smsum_{i=2}^{j}
\max \big\{\fl_i,\,\sfrac{1}{M^{\veps i}} \big\} \cb_j
\ \le\ \abcst \,\rho\,\cb_j
$$
Since $\cL^{(j)}_v(\vec F')$ only depends on $F^{\prime\,(2)},\cdots,F^{\prime\,(j-1)}$, Theorem \theoremmodcompLadder\ applies whenever $\rho \le \sfrac{1}{2}\tilde \rho_0$, and
$$
\V  V_{\rm ph} \big( \cL^{(j)}_v(\vec F')\big)\tV_{3,\Si_{j-1}}
\le \cst{}{0} \rho^2 \cb_{j-1}
\EQN\eqnmodcompLadderV$$
with a $j$--independent constant $\cst{}{0}$. 
Therefore, by (\eqnmodcompLadderII)
$$
\v w_j \tv_{3,\Si_j} \le \abcst(1+\cst{}{0} \rho)\, \rho\,\cb_j
\EQN\eqnmodcompLadderVI$$
Hence, by the Definition of $F^{\prime\,(j+1)}$ and (\eqnmodcompLadderIII)
$$\eqalign{
\V F^{\prime\,(j+1)} - F^{(j+1)} \tV_{3,\Si_{j+1}} 
&\le \const\,\rho^2\,\fl_j\,\cb_j\,\smsum_{\ell=1}^\infty
\big( \const\,\rho \,\cb_j \,\cb_{j+1}\big)^\ell \cr
&\le \cst{'}{}\,\rho^3\,\fl_j\,\cb_{j+1} \cr
}$$
Therefore 
$$
\V F^{\prime\,(j+1)}\tV_{3,\Si_{j+1}} 
\ \le \ \V F^{(j+1)}\tV_{3,\Si_{j+1}} + \const''\,\rho^3\,\fl_{j+1}\,\cb_{j+1} 
 \le 2\,\rho\,\max \big\{\fl_{j+1},\,\sfrac{1}{M^{\veps (j+1)}} \big\} \cb_{j+1}
$$
and (\eqnmodcompLadderIV) is proven for all $j$.

Set
$$
\de\cL^{(j)}=2 \smsum_{\ell=1}^\infty {\sst (-1)^\ell(12)^{\ell+1}}
  \Big(L_\ell\big(w_j;C^{(j)}_{v_j}, C^{(\ge j+1)}_{v_j}\big)
      -L_\ell\big(w_j;C^{(j)}_{v}, C^{(\ge j+1)}_{v}\big) \Big)^{\rm ph}
$$
Subtracting Definition \defmodcompLadder\  from  (\eqnmodcompLadderIIa)
$$
 \cL^{(j+1)}(\vec p,\vec F)-\cL^{(j+1)}_v(\vec F')
=\cL^{(j)}(\vec p,\vec F)_{\Si_j}-\cL^{(j)}_v(\vec F')_{\Si_j}
+\de\cL^{(j)}
$$
Since $ \cL^{(1)}(\vec p,\vec F)=\cL^{(1)}_v(\vec F')=0 $
$$
 \cL^{(j+1)}(\vec p,\vec F)-\cL^{(j+1)}_v(\vec F')
=\sum_{i=1}^j\de\cL^{(i)}_{\Si_j}
\EQN\eqnmodcompLadderVII$$
By (\eqnmodcompLadderIIIa) and (\eqnmodcompLadderVI)
$$\eqalign{
\V V_{\rm ph}\big( \de \cL^{(j)}\big)\tV_{3,\Si_j}
&\le\sum_{\ell=1}^\infty  \rho\,\fl_j\, \big( \const \,\cb_{j+1}\big)^\ell\,
\v w_j\tv_{3,\Si_j}^{\ \ \ell+1}\cr
&\le\sum_{\ell=1}^\infty  \abcst\,\rho^2\cb_j\,\fl_j\,
\big( \const\rho\,\cb_j\cb_{j+1}\big)^\ell\cr
&\le\const\rho^3\fl_j\cb_j\cr
}\EQN\eqnmodcompLadderVIII$$
Hence by Proposition \propOSresectorI.ii of [FKTo4] 
$$\eqalign{
\V V_{\rm ph}\big( \cL^{(j)}(\vec p,\vec F)-\cL^{(j)}_v(\vec F')\big)
\tV_{3,\Si_{j-1}}
&\le\sum_{i=1}^{j-1}\V  V_{\rm ph} 
\big(\de\cL^{(i)}\big)_{\Si_{j-1}}\tV_{3,\Si_{j-1}}\cr
&\le\sum_{i=1}^{j-1}\abcst\,\cb_{j-2}\V  V_{\rm ph} 
\big(\de\cL^{(i)}\big)\tV_{3,\Si_i}\cr
&\le\sum_{i=1}^{j-1}\const\rho^3\fl_i\cb_i\cb_{j-2}\cr
&\le\const\rho^3\cb_{j-1}\cr
}$$
As pointed out in (\eqnmodcompLadderV) , (\eqnmodcompLadderIV) implies that
$$
\V  V_{\rm ph} \big( \cL^{(j)}_v(\vec F')\big) \tV_{3,\Si_{j-1}}
\le \cst{}{0} \rho^2 \cb_{j-1}
$$
for all $j\ge 2$ and Theorem \theoremtildecompLadder\ follows
by yet another application of Proposition \propOSresectorI.ii of [FKTo4].
\endproof

The inductive Definition \deftildecompLadder\ of iterated particle--hole ladders is suited to direct application in the renormalization group analysis of \S\CHstep\ and \S\CHtildestep. 
Its relation to Definition \defmodcompLadder\ of compound particle hole ladders has been exhibited above.
In [FKTl] we use a more conceptual description of particle--hole ladders than that of Definition \defmodcompLadder. Corollary
\:\corcompLadder\ below and Remark \remLADmodcompLadder\ of [FKTl] show
 that the description of  particle--hole ladders in
Definition \defmodcompLadder\ agrees with that of Definition 
\defLADcompoundphladder\ in [FKTl].
Therefore, Theorem \theoremmodcompLadder\ follows from Theorem 
\thmLADmodcompLadder\ in [FKTl].

\noindent
A compound particle--hole ladder of scale $j$ 
may have a rung which is a particle--hole ladder of a scale $i < j$ , and this
rung may be  ``perpendicular'' to the direction of the big ladder. 

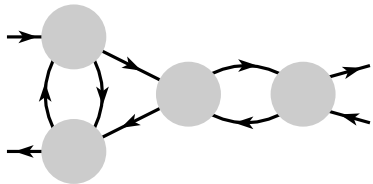
\centerline{\figplace{cmpPHladder}{0 in}{0 in}}

\noindent
To make this concept precise,  we define 
\definition{\STM\defInversym}{ Let $\fX$ be a measure space and
$F$ a four legged kernel on $\fX$. The associated flipped kernel is
$$
F^f({\sst x_1,x_2,x_3,x_4}) = - F({\sst x_1,x_3,x_2,x_4})
$$
The kernel $F$ is called inversion 
symmetric if
$$
F({\sst x_4,x_3,x_2,x_1}) = F({\sst x_1,x_2,x_3,x_4})
$$
}

\lemma{\STM\lemAntInvsym}{
Let $L$ be an inversion symmetric four legged kernel over
$\fY^\updownarrow_\Si$. Then
$$
\big( {\rm Ant\,} V_{\rm ph}(L) \big)^{\rm ph}  
=\sfrac{1}{3}(L+L^f)
$$
}
The proof is left to the reader.

\example{\STM\exInversym}{
\Item {i)} The propagators $\cC\big(C^{(j)}_u, C^{(\ge j+1)}_u\big)$ over
$\cB$ and $\cC\big(C^{(j)}_u, C^{(\ge j+1)}_u\big)^{\rm ph}$ over
$\cB^\updownarrow$ are inversion symmetric.
\Item {ii)} If $f$ is an antisymmetric kernel over $\fX_\Si$ then its particle
hole reduction $f^{\rm ph}$ is inversion symmetric.
\Item {iii)} If $f'_1,\,f'_2$ are inversion symmetric four-legged kernels 
over $\fY^\updownarrow_\Si$ and $P$ is an inversion symmetric bubble propagator over $\cB^\updownarrow$
then 
$$
(f'_1 \bullet P\bullet f'_2)({\sst z'_4,z'_3,z'_2,z'_1})
 = (f'_2\bullet P\bullet f'_1)({\sst z'_1,z'_2,z'_3,z'_4})
$$ 
\Item{iv)} If $f'$ is an inversion symmetric four-legged kernel over 
$\fY^\updownarrow_\Si$, $P$ an inversion symmetric bubble propagator over $\cB^\updownarrow$
and $\ell\ge 1$, then $(f'\bullet P)^\ell\bullet f'$ is also 
inversion symmetric.
\Item{v)} One proves by induction on $j$ that the compound particle hole ladders
$ \cL^{(j)}(\vec p,\vec F)$ are inversion symmetric. 
}

\proposition{\STM\propPHladsimp}{ Let $\Si$ be a sectorization.
Let $f$ be a particle number conserving, antisymmetric four legged 
kernel over $\fX_\Si$, $L$ an inversion symmetric four legged kernel over
$\fY^\updownarrow_\Si$  and $P$ a particle number conserving, inversion symmetric 
bubble propagator over $\cB$ that obeys 
$P(\xi_1,\xi_2,\xi_3,\xi_4) =P(\xi_2,\xi_1,\xi_4,\xi_3)$. Set 
$w=f + \sfrac{1}{8} {\rm Ant\,}\big(V_{\rm ph}(L)\big)$ .
Then for $\ell \ge1$
$$
{\sst(12)^{\ell+1}} \big( [ w\bullet P ]^\ell \bullet w \big)^{\rm ph}  
= \half\big[ \big({\sst 24}\,f^{\rm ph} + L + L^f \big) 
       \bullet {^{\rm ph}\!}P\big]^\ell
   \hskip-3pt \bullet \hskip-2pt
   \big({\sst 24}\,f^{\rm ph} + L + L^f \big) 
$$

}

\prf
By Lemma \lemunordord.i\ and Lemma \lemAntInvsym
$$\eqalign{
{\sst(12)^{\ell+1}} \big( [ w\bullet P ]^\ell \bullet w &\big)^{\rm ph} 
= \half(24)^{\ell+1} 
\big(f+\sfrac{1}{8} {\rm Ant\,} V_{\rm ph} L \big)^{\rm ph}
\bullet {^{\rm ph}\!}P\bullet \cdots {^{\rm ph}\!}P \bullet
\big(f+\sfrac{1}{8} {\rm Ant\,} V_{\rm ph} L \big)^{\rm ph} \cr
&= \half \big( 24 f^{\rm ph} +3({\rm Ant\,} V_{\rm ph} L)^{\rm ph} \big)
\bullet {^{\rm ph}\!}P\bullet \cdots {^{\rm ph}\!}P \bullet
\big( 24 f^{\rm ph} +3({\rm Ant\,} V_{\rm ph} L)^{\rm ph} \big) \cr
&= \half \big( 24 f^{\rm ph} +L+L^f \big)
\bullet {^{\rm ph}\!}P\bullet \cdots {^{\rm ph}\!}P \bullet
\big( 24 f^{\rm ph} +L+L^f \big) \cr
}$$
\endproof

\corollary{\STM\corcompLadder}{
Let $\vec F=\big(F^{(2)},F^{(3)},\cdots \big)$ be a sequence of 
antisymmetric, spin independent, particle number conserving functions
$F^{(i)}\in \check\cF_{4,\Si_i}$
and $v(k)$ a function on $\bbbr\times\bbbr^2$ such that
 $|v(k)| \le \half|\imath k_0 -e(\k)|$. As in Definition \defmodcompLadder,
let $\,\cL^{(j)}=\cL^{(j)}_v(\vec F)\,$ be the compound particle hole ladder up
to scale $j$.
Then for $j\ge 0$
$$
\cL^{(j+1)}= \cL^{(j)}_{\Si_j}
+ \smsum_{\ell=1}^\infty {\sst (-1)^\ell}
\big( 24 F +\cL^{(j)}_{\Si_j}+\cL^{(j)\,f}_{\Si_j} \big)
\bullet \cC\bullet \cdots \cC
\bullet
\big( 24 F +\cL^{(j)}_{\Si_j}+\cL^{(j)\,f}_{\Si_j} \big)
$$
where  $F = \sum_{i=2}^{j}{F}_{\Si_j}^{(i)\,{\rm ph}}$
and $\cC = {^{\rm ph}}\cC\big(C^{(j)}_v, C^{(\ge j+1)}_v \big)$.
 }

\prf 
Since $\cL^{(j)}$ and $\cC$ are inversion symmetric by Example \exInversym\
and $P=\cC\big(C^{(j)}_v, C^{(\ge j+1)}_v \big)$ obeys 
$P(\xi_1,\xi_2,\xi_3,\xi_4) =P(\xi_2,\xi_1,\xi_4,\xi_3)$,
the Corollary follows from Definition \defmodcompLadder\ and Proposition
\propPHladsimp.
\endproof

\vfill\eject

\titleb{References}\PG\pgNPIIIref

\item{[FKTf1]} J. Feldman, H. Kn\"orrer, E. Trubowitz, 
{\bf A Two Dimensional Fermi Liquid, Part 1: Overview}, preprint.
\smallskip%
\item{[FKTf2]} J. Feldman, H. Kn\"orrer, E. Trubowitz, 
{\bf A Two Dimensional Fermi Liquid, Part 2: Convergence}, preprint.
\smallskip%
\item{[FKTo1]} J. Feldman, H. Kn\"orrer, E. Trubowitz, 
{\bf Single Scale Analysis of Many Fermion Systems, Part 1: Insulators}, preprint.
\smallskip%
\item{[FKTo2]} J. Feldman, H. Kn\"orrer, E. Trubowitz, 
{\bf Single Scale Analysis of Many Fermion Systems, Part 2: The First Scale}, preprint.
\smallskip%
\item{[FKTo3]} J. Feldman, H. Kn\"orrer, E. Trubowitz, 
{\bf Single Scale Analysis of Many Fermion Systems, Part 3: Sectorized Norms}, preprint.
\smallskip%
\item{[FKTo4]} J. Feldman, H. Kn\"orrer, E. Trubowitz, 
{\bf Single Scale Analysis of Many Fermion Systems, Part 4: Sector Counting}, preprint.
\smallskip%
\item{[FKTr1]} J. Feldman, H. Kn\"orrer, E. Trubowitz, 
{\bf Convergence of Perturbation Expansions in Fermionic Models, Part 1: Nonperturbative Bounds}, preprint.
\smallskip%
\item{[FST3]} J.\ Feldman, M.\ Salmhofer, and E.\ Trubowitz,
{\bf  Regularity of Interacting Nonspherical Fermi Surfaces: The Full 
Self--Energy}, Communications on Pure and Applied Mathematics, {\bf LII},
 273-324  (1999).

\vfill\eject

\hoffset=-0.2in
\titlea{Notation}\PG\pgNPIIInot
\vfil
\titleb{Norms}
\centerline{
\vbox{\offinterlineskip
\hrule
\halign{\vrule#&
         \strut\hskip0.05in\hfil#\hfil&
         \hskip0.05in\vrule#\hskip0.05in&
          #\hfil\hfil&
         \hskip0.05in\vrule#\hskip0.05in&
          #\hfil\hfil&
           \hskip0.05in\vrule#\cr
height2pt&\omit&&\omit&&\omit&\cr
&Norm&&Characteristics&&Reference&\cr
height2pt&\omit&&\omit&&\omit&\cr
\noalign{\hrule}
height2pt&\omit&&\omit&&\omit&\cr
&$\tn\ \cdot\ \tn_{1,\infty}$&&no derivatives, external positions, acts on 
functions&&Definition \defNPSymmNorm&\cr
height4pt&\omit&&\omit&&\omit&\cr
&$\|\ \cdot\ \|_{1,\infty}$&&derivatives, external positions, acts on functions&&Definition \defNPSymmNorm&\cr
height4pt&\omit&&\omit&&\omit&\cr
&$\|\ \cdot\ \tnorm$&&derivatives, external momenta, acts on functions
&&Definition \defNPdiffdecaynorm&\cr
height4pt&\omit&&\omit&&\omit&\cr
&$\tn\ \cdot\ \tn_{\infty}$&&no derivatives, external positions, acts on 
functions&&Definition \defNPsectGrnorm&\cr
height4pt&\omit&&\omit&&\omit&\cr
&$\v \ \cdot\ \v_{p,\Si}$&&derivatives, external positions, 
all but $p$ sectors summed
&&Definition \defNPsectnorm&\cr
height4pt&\omit&&\omit&&\omit&\cr
&$\tn \ \cdot\ \tn_{1,\Si}$&&no derivatives, all but $1$ sector summed
&&(\eqnOVtriponesi)&\cr
height4pt&\omit&&\omit&&\omit&\cr
&$\tn \ \cdot\ \tn_{3,\Si}$&&no derivatives, all but $3$ sectors summed
&& (\eqnOVtripthreesi) &\cr
height4pt&\omit&&\omit&&\omit&\cr
&$\v \ \cdot\ \tv_{p,\Si}$&&derivatives, external momenta, all but
&&\omit&\cr
&\omit&& \hskip.5in($p-$\# external momenta) sectors summed
&&Definition \defNPsectdiffdecaynorm&\cr
height4pt&\omit&&\omit&&\omit&\cr
&$\|\ \cdot\ \|_{1,\Si}$&& like $\v \ \cdot\ \v_{1,\Si}$, but for functions
on $\big(\bbbr^2\times\Si\big)^2$
&&[Def'n \defOSzerosectorext, FKTo4]&\cr
height4pt&\omit&&\omit&&\omit&\cr
&$\v \varphi \v_j$&&$\rho_{m;n}\cases{
\v \varphi \v_{1,\Si_j} + \sfrac{1}{\fl_j}\,\v \varphi \v_{3,\Si_j}
+ \sfrac{1}{\fl_j^2}\,\v \varphi \v_{5,\Si_j} 
    & if $m=0$ \cr
\sfrac{\fl_j}{M^{2j}}\,\v \varphi \v_{1,\Si_j} & if $m\ne0$} $
&&Definition \defNPsectnorm&\cr
height4pt&\omit&&\omit&&\omit&\cr
&$N_j(w,\al,X)$&&$\sfrac{M^{2j}}{\fl_j}\,\fe_j(X) 
\smsum_{m,n\ge 0}\,
\al^{n}\,\big(\sfrac{\fl_j\IB}{M^j}\big)^{n/2} \,\v w_{m,n}\v_j$
&&Definition \defNPsectGrnorm&\cr
height4pt&\omit&&\omit&&\omit&\cr
&$N(\cG)$&&$
\smsum_{m> 0}\,\sfrac{1}{\la_0^{(1-\upsilon)\max\{m-2,2\}/2}}
\,\tn G_m\tn_\infty$
&&Definition \defNPsectGrnorm&\cr
height4pt&\omit&&\omit&&\omit&\cr
&$\v f \tv_j$&&$\tilde\rho_{m;n}\cases{
\smsum_{p=1}^6\sfrac{1}{\fl_j^{[(p-1)/2]}}\v f\tv_{p,\Si_j}& if $m\ne 0$\cr
\noalign{\vskip.05in}
\v f \tv  _{1,\Si_j} + \sfrac{1}{\fl_j}\,\v f \tv_{3,\Si_j}
+ \sfrac{1}{\fl_j^2}\,\v f \tv_{5,\Si_j}& if $m= 0$\cr
}$
&&Definition \defNPmomscalednorms&\cr
height4pt&\omit&&\omit&&\omit&\cr
&$\v f \tv_{p,\Si_j,\tilde\rho}$&&$\tilde\rho_{m;n} \v f \tv_{p,\Si_j} $
&&Definition \defNPmomscalednorms&\cr
height4pt&\omit&&\omit&&\omit&\cr
& && &&Definition \defNPdisjointfnspaces&\cr
height4pt&\omit&&\omit&&\omit&\cr
&$N_j^\sim(w,\al,X)$&&$\sfrac{M^{2j}}{\fl_j}\,\fe_j(X) 
\smsum_{m,n\ge 0}\,
\al^{m+n}\,\big(\sfrac{\fl_j\,\IB}{M^j}\big)^{(m+n)/2} \,
\v w^\sim_{m,n}\tv_j $
&&Definition \defNPmomscalednorms&\cr
height4pt&\omit&&\omit&&\omit&\cr
}\hrule}}
\vfill\goodbreak
\titleb{Spaces}
\centerline{
\vbox{\offinterlineskip
\hrule
\halign{\vrule#&
         \strut\hskip0.05in\hfil#\hfil&
         \hskip0.05in\vrule#\hskip0.05in&
          #\hfil\hfil&
         \hskip0.05in\vrule#\hskip0.05in&
          #\hfil\hfil&
           \hskip0.05in\vrule#\cr
height2pt&\omit&&\omit&&\omit&\cr
&Not'n&&Description&&Reference&\cr
height2pt&\omit&&\omit&&\omit&\cr
\noalign{\hrule}
height2pt&\omit&&\omit&&\omit&\cr
&$\cE$&&counterterm space&&Definition \defNPCTMSpace&\cr
height2pt&\omit&&\omit&&\omit&\cr
&$\fK_j$&&space of future counterterms for scale $j$
&&Definition \defNPCTmSpace&\cr
height2pt&\omit&&\omit&&\omit&\cr
&$\cB$&&$\bbbr \times \bbbr^d \times \{\uparrow, \downarrow\}\times\{0,1\}$
 viewed as position space&&before Def \defbubbleprop&\cr
height2pt&\omit&&\omit&&\omit&\cr
&$\check \cB$&&$\bbbr\times\bbbr^d\times\{\uparrow,\downarrow\}\times\{0,1\}$ 
viewed as momentum space&&beginning of \S\CHnewsectors&\cr
height2pt&\omit&&\omit&&\omit&\cr
&$\check \cB_m$&&$\set{(\check \eta_1,\cdots,\check \eta_m)\in \check \cB^m}
{\check \eta_1+\cdots+\check \eta_m=0} $ 
&&Definition \defNPdiffdecay&\cr
height2pt&\omit&&\omit&&\omit&\cr
&$\fX_\Si$&&$=\fX_0\dunion\fX_1=\check \cB\dunion(\cB\times\Si)$
&&Definition \defmoredisjointunions&\cr
height2pt&\omit&&\omit&&\omit&\cr
&$\fX'_\Si$&&$=\fX_{-1}\dunion\fX_0\dunion\fX_1
=\check \cB\dunion\check \cB\dunion(\cB\times\Si)$
&&Definition \defNPmoredisjointOrd&\cr
height2pt&\omit&&\omit&&\omit&\cr
&$\cB^\updownarrow$&&$\bbbr \times \bbbr^d \times \{\uparrow, \downarrow\}$
 viewed as position space&&Definition \NPsomespaces&\cr
height2pt&\omit&&\omit&&\omit&\cr
&$\check\cB^\updownarrow$&&$\bbbr\times\bbbr^d\times\{\uparrow,\downarrow\}$ 
viewed
as momentum space&&Definition \defmoredisjointunions&\cr
height2pt&\omit&&\omit&&\omit&\cr
&$\fY^\updownarrow_\Si$&&$\check \cB^\updownarrow\dunion(\cB^\updownarrow\times\Si)$&&Definition \defmoredisjointunions&\cr
height2pt&\omit&&\omit&&\omit&\cr
&$\cF_m(n;\Si)$&&functions on $\cB^m \times  \big( \cB \times\Si \big)^n$,
internal momenta in sectors&&Definition \defNPsectrepr.ii&\cr
height2pt&\omit&&\omit&&\omit&\cr
&$\check\cF_m(n;\Si)$&&functions on $\check\cB^m \times  \big( \cB \times\Si \big)^n$,
internal momenta in sectors&&Definition \defNPsectcheckcF&\cr
height2pt&\omit&&\omit&&\omit&\cr
&$\check \cF_{n;\Si}$&&functions on $\fX_\Si^n$ that reorder to 
$\check \cF_{m}(n-m;\Si)$'s&&Definition \defNPdisjointfnspaces&\cr
height2pt&\omit&&\omit&&\omit&\cr
&$\check\cF_{m',m}(n;\Si)$&&fns on 
$\fX^{m'}_{-1}\times\fX_0^m \times  \big( \cB \times\Si \big)^n$,
internal momenta in sectors&&Definition \defNPmoredisjointfnspaces&\cr
height2pt&\omit&&\omit&&\omit&\cr
&$\check \cF'_{n;\Si}$&&functions on ${\fX'}_\Si^n$ that reorder to 
$\check \cF_{m,m'}(n-m-m';\Si)$'s&&Definition \defNPmoredisjointfnspaces&\cr
height2pt&\omit&&\omit&&\omit&\cr
&$\cD_{\rm in}^{(j,\form)}$&&formal input data for scale $j$
&&Definition \defNPinputData&\cr
height2pt&\omit&&\omit&&\omit&\cr
&$\cD_{\rm out}^{(j,\form)}$&&formal output data for scale $j$
&&Definition \defNPoutputData&\cr
height2pt&\omit&&\omit&&\omit&\cr
&$\cD^{(j)}_{\rm in}$&&input data for scale $j$&&Definition \stepInputData&\cr
height2pt&\omit&&\omit&&\omit&\cr
&$\cD^{(j)}_{\rm out}$&&output data for scale $j$
&&Definition \stepOutputData&\cr
height2pt&\omit&&\omit&&\omit&\cr
&$\tilde\cD^{(j)}_{\rm in}$&& more input data for scale $j$
&&Definition \moreInputData&\cr
height2pt&\omit&&\omit&&\omit&\cr
&$\tilde\cD^{(j)}_{\rm out}$&&more output data for scale $j$
&&Definition \moreOutputData&\cr
height4pt&\omit&&\omit&&\omit&\cr
}\hrule}}

\vfil
\goodbreak
\titleb{Other Notation}
\centerline{
\vbox{\offinterlineskip
\hrule
\halign{\vrule#&
         \strut\hskip0.05in\hfil#\hfil&
         \hskip0.05in\vrule#\hskip0.05in&
          #\hfil\hfil&
         \hskip0.05in\vrule#\hskip0.05in&
          #\hfil\hfil&
           \hskip0.05in\vrule#\cr
height2pt&\omit&&\omit&&\omit&\cr
&Not'n&&Description&&Reference&\cr
height2pt&\omit&&\omit&&\omit&\cr
\noalign{\hrule}
height2pt&\omit&&\omit&&\omit&\cr
&$r_0$&&number of $k_0$ derivatives tracked&&following (\eqnNPinteraction)&\cr
height2pt&\omit&&\omit&&\omit&\cr
&$r$&&number of $\k$ derivatives tracked&&following (\eqnNPinteraction)&\cr
height2pt&\omit&&\omit&&\omit&\cr
&$M$&&scale parameter, $M>1$&&before Definition \defNPscales&\cr
height2pt&\omit&&\omit&&\omit&\cr
&$\const$&&generic constant, independent of scale&& &\cr
height2pt&\omit&&\omit&&\omit&\cr
&$\abcst$&&generic constant, independent of scale and $M$&& &\cr
height2pt&\omit&&\omit&&\omit&\cr
&$\nu^{(j)}(k)$&&$j^{\rm th}$ scale function&&Definition \defNPscales&\cr
height2pt&\omit&&\omit&&\omit&\cr
&$\nu^{(\ge j)}(k)$&&$\smsum_{i\ge j}\nu^{(j)}(k)$&&Definition \defNPscales&\cr
height2pt&\omit&&\omit&&\omit&\cr
&$n_0$&&degree of asymmetry&&Definition \defNPstrongasymm&\cr
height2pt&\omit&&\omit&&\omit&\cr
&$J$&&particle/hole swap operator&&(\eqnNPjdef)&\cr
height2pt&\omit&&\omit&&\omit&\cr
&$\Om_S(\cW)(\phi,\psi)$
&&$\log\sfrac{1}{Z} \int  e^{\cW(\phi,\psi+\ze)}\,d\mu_{S}(\ze)$
&&Definition \defNPrengroupmap&\cr
height2pt&\omit&&\omit&&\omit&\cr
&$\tilde \Om_C(\cW)(\phi,\psi)$
&&$\log \sfrac{1}{Z}\int e^{\phi J\ze}\,e^{\cW(\phi,\psi +\ze)} d\mu_C(\ze)$
&&Definition \defNPrengroupmap&\cr
height2pt&\omit&&\omit&&\omit&\cr
&$\aleph$&&$\half<\aleph< \sfrac{2}{3}$&&following Definition \defNPsectrepr&\cr
height2pt&\omit&&\omit&&\omit&\cr
&$\aleph'$&&$\aleph<\aleph'< \sfrac{2}{3}$&&Theorem  \theoremNPtildeinduction&\cr
height2pt&\omit&&\omit&&\omit&\cr
&$\la_0$&&maximum allowed ``coupling constant''&&Theorem \eqnNPnbndA&\cr
height2pt&\omit&&\omit&&\omit&\cr
&$\upsilon$&&$0<\upsilon< \sfrac{1}{4}$, power of $\la_0$ eaten by bounds&&Definition \defNPrhomn&\cr
height2pt&\omit&&\omit&&\omit&\cr
&$\rho_{m;n}(\la)$&&$\la^{-(1-\upsilon)\max\{m+n-2,2\}/2}$&&Definition \defNPrhomn&\cr
height2pt&\omit&&\omit&&\omit&\cr
&$\rho_{m;n}$&&$\rho_{m;n}(\la_0)\big\{\hbox{1 if $m=0$\ ;\ $\root{4}\of{\fl_j
M^j}$ if $m>0$}$&&Definition \defNPsectnorm.ii&\cr
height2pt&\omit&&\omit&&\omit&\cr
&$\tilde\rho_{m;n}$&&$\rho_{m;n}(\la_0)\la_0^{m\upsilon/7}$&&Theorem 
\thmNPTfirststep&\cr
height2pt&\omit&&\omit&&\omit&\cr
&$\fl_j$&&$=\sfrac{1}{M^{\aleph j}}$ = 
  length of sectors of scale $j$&&following Definition \defNPsectrepr&\cr
height2pt&\omit&&\omit&&\omit&\cr
&$\Si_j$&&the sectorization at scale $j$ of length $\fl_j$&&following Definition \defNPsectrepr&\cr
height2pt&\omit&&\omit&&\omit&\cr
&$\IB$&&$j$--independent constant&&Definitions \defNPsectGrnorm, \defNPmomscalednorms&\cr
height2pt&\omit&&\omit&&\omit&\cr
&$\cb_j$&& $
=\sum_{|\bde|\le r\atop |\de_0|\le r_0}  M^{j|\de|}\,t^\de
+\sum_{|\bde|> r\atop {\rm or\ }|\de_0|> r_0}\infty\, t^\de
\in\fN_{d+1}
$&&Definition \defNPFancynormdomain&\cr
height2pt&\omit&&\omit&&\omit&\cr
&$\cb_{i,j}$&& $
=\sum_{|\bde|\le r\atop |\de_0|\le r_0}  M^{i\de_0} M^{j|\bde|}\,t^\de
+\sum_{|\bde|> r\atop {\rm or\ }|\de_0|> r_0}\infty\, t^\de
\in\fN_{d+1}
$&&(\eqnTNPcbfeij)&\cr
height2pt&\omit&&\omit&&\omit&\cr
&$\fe_j(X)$&& $= \sfrac{\cb_j}{1-M^j X}$&&Definition \defNPFancynormdomain&\cr
height2pt&\omit&&\omit&&\omit&\cr
&$\fe_{i,j}(X)$&& $=\sfrac{\cb_{i,j}}{1-M^jX}$&&(\eqnTNPcbfeij)&\cr
height2pt&\omit&&\omit&&\omit&\cr
&$f_{\rm ext}$&& extends $f(\x,\x')$ to 
$f_{\rm ext}\big((x_0,\x,\si,a),(x_0',\x',\si',a')\big)$&&
[Definition \defOSzeroext, FKTo4]&\cr
height2pt&\omit&&\omit&&\omit&\cr
&$*$&& convolution&&Definition \defNPconvol&\cr
height2pt&\omit&&\omit&&\omit&\cr
&$\bullet$&& ladder convolution&&Definition \deNPdefsectbubbleprop,&\cr
height2pt&\omit&&\omit&&\omit&\cr
& &&  &&Definition \deftildeladders&\cr
height2pt&\omit&&\omit&&\omit&\cr
&$\sh(\ \cdot\ ,B)$&&maps kernel of $\cW(\phi,\psi)$ to kernel of $\cW(\phi,\psi+B\phi)$&&Definition  \defNPshear&\cr
height2pt&\omit&&\omit&&\omit&\cr
&$\sh'(\ \cdot\ ,B)$&&maps kernel of $\cW(\phi,\psi)$ to kernel of $\cW(\phi,\psi+B\phi')$&&Definition  \defNPshearprime&\cr
height2pt&\omit&&\omit&&\omit&\cr
&$\sct(\ \cdot\ ,B)$&&maps kernel of $\cW(\phi',\phi,\psi)$ to kernel of $\cW(B\phi',\phi,\psi)$&&Definition \defNPshearprime&\cr
height2pt&\omit&&\omit&&\omit&\cr
&$\Pi$&&maps kernel of $\cW(\phi',\phi,\psi)$ to kernel of $\cW(\phi,\phi,\psi)$&&Definition \defNPmoredisjointfnspaces&\cr
height2pt&\omit&&\omit&&\omit&\cr
&$\hat\mu$&&Fourier transform&&Notation \notNPfourierTI&\cr
height4pt&\omit&&\omit&&\omit&\cr
}\hrule}}

\end

%% file: jfvmacros.tex

\def\ifundefined#1{\expandafter\ifx\csname#1\endcsname\relax}
\ifundefined{ftmagnification}  \def\ftmagnification{1200} \fi
\ifundefined{spacingNumerator}  \def\spacingNumerator{5} \fi
\ifundefined{spacingDenominator}  \def\spacingDenominator{4} \fi


\magnification\ftmagnification
\tolerance=10000
\hsize=17truecm\vsize=23truecm

\parindent=40pt
\mathsurround=0pt
     \multiply\baselineskip by \spacingNumerator
     \divide \baselineskip by \spacingDenominator 

%
%
\def\today{\ifcase\month\or January\or February\or March\or April\or
     May\or June\or July\or August\or September\or October\or November\or
     December\fi\space\number\day, \number\year}
%
%
\def\dst{\displaystyle}
\def\sst{\scriptstyle}
\def\tst{\textstyle}
%
%
\def\frac#1#2{\dst {#1\over#2}}     
\def\sfrac#1#2{{\tst{#1\over#2}}}   

\def\deqalign#1{\vcenter{\openup1\jot \mathsurround=0pt \ialign{
                \strut\hfil$\displaystyle{##}$&&$\displaystyle{{}##}$\hfil
                \crcr
                #1\crcr}}}         

\def\meqalign#1{\vcenter{\openup1\jot \mathsurround=0pt \ialign{
                &\strut\hfil$\displaystyle{##}$&$\displaystyle{{}##}$\hfil&
                \quad$##$\crcr
                #1\crcr}}}         

%
%
\def\al{\alpha}
\def\be{\beta}
\def\ga{\gamma}
\def\de{\delta}
\def\ep{\epsilon}
\def\ze{\zeta}
\def\et{\eta}

\def\ka{\kappa}
\def\la{\lambda}

\def\si{\sigma}

\def\up{\upsilon}

\def\om{\omega}

\def\La{\Lambda}
\def\Si{\Sigma}

\def\Om{\Omega}   
%
%
\def\pmb#1{\setbox0=\hbox{#1}       
     \kern-.025em\copy0\kern-\wd0
     \kern.05em\copy0\kern-\wd0
     \kern-.025em\box0}             
\def\0{{\bf 0}}

\def\k{{\bf k}}

\def\t{{\bf t}}

\def\x{{\bf x}}

\def\p{{\bf p}}

\def\cB{{\cal B}}
\def\cE{{\cal E}}
\def\cF{{\cal F}}
\def\cG{{\cal G}}

\def\cL{{\cal L}}
\def\cO{{\cal O}}

%
%
\font\tenfrak                 = eufm10
\font\sevenfrak               = eufm7
\font\fivefrak                = eufb5
\newfam\frakfam
     \textfont\frakfam=\tenfrak
     \scriptfont\frakfam=\sevenfrak   
     \scriptscriptfont\frakfam=\fivefrak
\def\frak{\fam\frakfam\tenfrak}
\font \tensans                = cmss10
\font \fivesans               = cmss10 at 5pt
\font \sevensans              = cmss10 at 7pt
\newfam\sansfam
     \textfont\sansfam=\tensans
     \scriptfont\sansfam=\sevensans
     \scriptscriptfont\sansfam=\fivesans
\def\sans{\fam\sansfam\tensans}
%
%
\def\bbbr{{\rm I\!R}}  
\def\bbbn{{\rm I\!N}}

\def\bbbone{{\mathchoice {\rm 1\mskip-4mu l} {\rm 1\mskip-4mu l}    
{\rm 1\mskip-4.5mu l} {\rm 1\mskip-5mu l}}}
\def\bbbc{{\mathchoice {\setbox0=\hbox{$\displaystyle\rm C$}\hbox{\hbox 
to0pt{\kern0.4\wd0\vrule height0.9\ht0\hss}\box0}}
{\setbox0=\hbox{$\textstyle\rm C$}\hbox{\hbox
to0pt{\kern0.4\wd0\vrule height0.9\ht0\hss}\box0}}
{\setbox0=\hbox{$\scriptstyle\rm C$}\hbox{\hbox
to0pt{\kern0.4\wd0\vrule height0.9\ht0\hss}\box0}}
{\setbox0=\hbox{$\scriptscriptstyle\rm C$}\hbox{\hbox
to0pt{\kern0.4\wd0\vrule height0.9\ht0\hss}\box0}}}}
\def\bbbq{{\mathchoice {\setbox0=\hbox{$\displaystyle\rm               
Q$}\hbox{\raise
0.15\ht0\hbox to0pt{\kern0.4\wd0\vrule height0.8\ht0\hss}\box0}}
{\setbox0=\hbox{$\textstyle\rm Q$}\hbox{\raise
0.15\ht0\hbox to0pt{\kern0.4\wd0\vrule height0.8\ht0\hss}\box0}}
{\setbox0=\hbox{$\scriptstyle\rm Q$}\hbox{\raise
0.15\ht0\hbox to0pt{\kern0.4\wd0\vrule height0.7\ht0\hss}\box0}}
{\setbox0=\hbox{$\scriptscriptstyle\rm Q$}\hbox{\raise
0.15\ht0\hbox to0pt{\kern0.4\wd0\vrule height0.7\ht0\hss}\box0}}}}
\def\bbbz{{\mathchoice {\hbox{$\sans\textstyle Z\kern-0.4em Z$}}       
{\hbox{$\sans\textstyle Z\kern-0.4em Z$}}
{\hbox{$\sans\scriptstyle Z\kern-0.3em Z$}}
{\hbox{$\sans\scriptscriptstyle Z\kern-0.2em Z$}}}}
%
%
\def\const{{\rm const}\,}
\def\sgn{{\rm sgn}}
\def\half{\sfrac{1}{2}}

\def\optbar#1{\vbox{\ialign{##\crcr\hfil${\scriptscriptstyle(}\mkern -1mu
         \vrule height 1.2pt width 3pt depth -.8pt
         {\scriptscriptstyle)}$\hfil\crcr
          \noalign{\kern-1pt\nointerlineskip}$\hfil\displaystyle{#1}\hfil$\crcr}}}
\def\<{\left<}
\def\>{\right>}

\def\smprod{\mathop{\textstyle\prod}}
\def\smsum{\mathop{\textstyle\sum}}
\def\set#1#2{\big\{ \ #1\ \big|\ #2\ \big\}}
\def\eval#1{\big|\lower4pt\hbox{$\displaystyle\sst #1$}}
%
%
\font \tafontt                = cmbx10 scaled\magstep2
\font \tbfontt                = cmbx10 scaled\magstep1
\def\titlea#1{\centerline{\tafontt #1 }\vskip.5truein}
\def\titleb#1{\removelastskip\vskip.3truein%
\noindent{\tbfontt #1 }\vskip.25truein}

%
%
\def\newenvironment#1#2#3#4{\long\def#1##1##2{%
\removelastskip\penalty-100\vskip\baselineskip%
\noindent{#3#2\if!##1!.\else\unskip\ \ignorespaces
##1\unskip\fi\ }{#4\ignorespaces##2\vskip\baselineskip}}}
\newenvironment\lemma{Lemma}{\bf}{\it}
\newenvironment\proposition{Proposition}{\bf}{\it}
\newenvironment\theorem{Theorem}{\bf}{\it}
\newenvironment\corollary{Corollary}{\bf}{\it}
\newenvironment\example{Example}{\bf}{\rm}
\newenvironment\problem{Problem}{\bf}{\rm}
\newenvironment\definition{Definition}{\bf}{\rm}
\newenvironment\remark{Remark}{\bf}{\rm}
\newenvironment\hypothesis{Hypothesis}{\bf}{\it}
\newenvironment\convention{Convention}{\bf}{\it}

\def\Item{\vskip.1in\noindent}

%
%
\long\def\proof#1{\removelastskip\penalty-100\vskip\baselineskip\noindent{\bf
            Proof\if!#1!\else\ \ignorespaces#1\fi:\ }\ \ \ignorespaces}
\long\def\prf{\removelastskip\penalty-100\vskip\baselineskip\noindent{\bf
            Proof:\ }\ \ \ignorespaces}
\def\endproof{\hfill\vrule height .6em width .6em depth 0pt\goodbreak\vskip.25in }

\ifundefined{warnForwardRef}  \def\warnForwardRef{n} \fi
\newcount\chapno
\newcount\sectno
\newcount\equano
\newcount\theono
\newcount\probno

\def\IgNoRe#1{}

\chapno=0
\sectno=0
\equano=0
\theono=0
\probno=0
\def\eqhead{}
\def\frefwarning{\if\warnForwardRef y\immediate\write16{   Forward reference on line \the\inputlineno}\fi}
\def\qqqrefwarning{\immediate\write16{   ??? reference on line \the\inputlineno}}

\def\chap#1{\equano=0\sectno=0\theono=0\probno=0\global\advance\chapno by 1%
\def\eqhead{\ifcase\chapno\or I\or II\or III\or IV\or V\or VI\or VII\or
VIII\or IX\or X\or XI\or XII\or XIII\or XIV\or XV\or XVI\or XVII\or XVIII\or
XIX\or XX\or XXI\or XXII\or XXIII\or XXIV\or XXV\or XXVI\or XXVII\or XXVIII\or XXIX\or XXX\or XXXI\or XXXII\or XXXIII\or XXXIV\or XXXV\or XXXVI\or XXXVII\or XXXVIII\or XXXIX\fi.}%
\titlea{\eqhead \hglue 5pt #1}%
}

\def\sect#1{\global\advance\sectno by 1%
\titleb{\eqhead\number\sectno  \hglue 5pt #1}%
}%

\def\appendix#1#2{\equano=0\sectno=0\theono=0\probno=0\def\eqhead{#1.}
\titlea{Appendix #1: #2}%
}

\def\:#1{\def\temp{\expandafter\IgNoRe\string#1}%
\expandafter\ifx\csname\temp\endcsname\relax%
\expandafter\gdef#1{\qqqrefwarning ???}\fi#1}

\def\Eqn{{\hbox{\global\advance\equano by 1}}%
\eqno ({\rm \eqhead\number\equano})}%

\def\Eqno{{\hbox{\global\advance\equano by 1}}%
 ({\rm \eqhead\number\equano})}%

\def\EQN#1{\Eqn\edef\Zwi{\eqhead\number\equano}%
\global\let #1=\Zwi
}

\def\EQNO#1{\Eqno\edef\Zwi{\eqhead\number\equano}%
\global\let #1=\Zwi
}

\def\STM#1{{\global\advance \theono by 1}%
\eqhead\number\theono
\edef\Zwi{\eqhead\number\theono }
\global\let#1=\Zwi
}

\def\PRB#1{{\global\advance \probno by 1}%
\eqhead\number\probno
\edef\Zwi{\eqhead\number\probno }
\global\let#1=\Zwi
}

\def\PG#1{\def\Zwi{\number\pageno }
\global\let#1=\Zwi
}

\def\Stm{{\global\advance \theono by 1}%
\eqhead\number\theono
}

\def\Prb{{\global\advance \probno by 1}%
\eqhead\number\probno
}

\def\EDEF#1#2{
\def\tEmP{#1}\expandafter\gdef\tEmP{#2}
}



\def\suffix{ps}
\newcount\system
\global\system=3   

\def\ifundefined#1{\expandafter\ifx\csname#1\endcsname\relax}
\ifundefined{figdir}\def\figdir{}\fi
%
%
\newcount\firstline
\newdimen\pswidth  \newdimen\xleft
\newdimen\psheight \newdimen\ytop \newdimen\ybot
\newcount\justx \newcount\justy
\global\justx=0 \global\justy=0
\newdimen\vpos \newtoks\labeL 
\newread\labeLfile \newdimen\xcoord \newdimen\ycoord
\newif\ifdoit 
\newbox\labox
\newdimen\xdvikwid 
\newdimen\xdvikht
\newdimen\pspoints
\newdimen\rwi
\pspoints=1bp
\newcount\temp
\def\readdim#1{\global\read\labeLfile to \temp
\global #1=\temp pt}
%
%
%
%
\def\figcrop#1{\par
\openin\labeLfile=\figdir#1.lbl                                              
\global\read\labeLfile to\firstline\message{#1}               
\global\read\labeLfile to\temp
\readdim{\ybot}
\readdim{\xleft}
\readdim{\ytop}
\global\read\labeLfile to\justx
\global\read\labeLfile to\justy
\global\read\labeLfile to\labeL
\readdim{\pswidth}
\global\advance\pswidth by -\xleft
\readdim{\psheight}
\global\advance\ybot by -\psheight
\global\advance\psheight by -\ytop
\global\read\labeLfile to\justx
\global\read\labeLfile to\justy
\global\read\labeLfile to\labeL
\vbox to\psheight{\vfill
\ifnum\system=1
\fi 
\ifnum\system=2
\fi 
\ifnum\system=3
                                                 \fi         
\ifnum\system=4
\fi         
\ifnum\system=1
\hbox to \pswidth{\kern-\xleft\special{postscriptfile \figdir#1.\suffix }\hfil}\fi
\ifnum\system=2
\hbox to \pswidth{\kern-\xleft\special{ps: plotfile \figdir#1.\suffix }\hfil}\fi
\ifnum\system=3
\hbox to \pswidth{\kern-\xleft\includegraphics{\figdir#1.\suffix}\hfil}\fi
\ifnum\system=4
\hbox to \pswidth{\kern-\xleft\includegraphics{\figdir#1.\suffix}\hfil}\fi
\ifnum\system=5
\hbox to \pswidth{\kern-\xleft\includegraphics{\figdir#1.\suffix}\hfil}\fi 
\ifnum\system=6
   \xdvikwid=\pswidth
   \xdvikht=\psheight
   {\global\divide\xdvikwid by \pspoints}
   {\global\divide\xdvikht by \pspoints}
   \rwi=\xdvikwid
    {\global\multiply\rwi by 10}
\hbox to \pswidth{\kern-\xleft\includegraphics{\figdir#1.\suffix\space}\hfil}\fi                   
\vskip -\baselineskip
\vskip -\ybot 
\vskip-\psheight %
\hbox to\pswidth  {\hss}%
\parindent=0pt\offinterlineskip                                       
\vpos=0 pt%
\loop\readdim{\xcoord}                                 
\ifdim \xcoord < -999pt \doitfalse\else\doittrue\fi                        
\ifdoit \advance \xcoord by -\xleft
\readdim{\ycoord}
\advance \ycoord by -\ytop                              
\global\read\labeLfile to\justx                                       
\global\read\labeLfile to\justy                                       
\global\read\labeLfile to\labeL
\global\setbox\labox=\hbox{\labeL\hskip-0.3em}%
\advance\vpos by-\ycoord                                              
\vskip-\vpos \vpos=\ycoord                                         
\hbox to\pswidth{\hskip\xcoord %
\hbox to 0pt{\ifnum\justx>0\hss\fi%
\vbox to0pt{%
\ifnum\justy<2\vss\fi%
\copy\labox\kern0pt%
\ifnum\justy>0\vss\fi}%
\ifnum\justx<2\hss\fi}%
\hss}%
\repeat%
\advance\vpos by-\psheight%
\vskip-\vpos %
}\closein\labeLfile}
%
%
%
\def\figplace#1#2#3{
\openin\labeLfile=\figdir#1.lbl
\ifeof \labeLfile
       \immediate\write16{***Can't find \figdir#1.lbl; Skipping it.***}
\else  \closein\labeLfile
       \null\hskip#2\raise #3 \hbox{\figcrop{#1}}
\fi
}
%
%
%
%


\font\tenscript                 = pzcmi at 10pt
\font\sevenscript               = pzcmi at 7pt
\font\fivescript                = pzcmi at 5pt
\newfam\scriptfam
     \textfont\scriptfam=\tenscript
     \scriptfont\scriptfam=\sevenscript   
     \scriptscriptfont\scriptfam=\fivescript
\def\script{\fam\scriptfam\tenscript}

    \newenvironment\notation{Notation}{\bf}{\rm}

          \def\stoday{\number\day\space\ifcase\month\or Jan\or Feb\or 
                      Mar\or Apr\or May\or Jun\or Jul\or Aug\or Sep\or 
                      Oct\or Nov\or Dec\fi, \number\year}

         \def\squiggle{\raise2pt\hbox{${\scriptstyle\sim}$}}

    \def\form{{\ssst\script form}}

    \def\dunion{\cup\kern-0.7em\cdot\kern0.45em}
    \def\cb{{\frak c}}

    \def\IB{{\rm\sst B}}
    \def\vi{\fl^{\raise2pt\hbox{$\scriptscriptstyle{1/n_0}$}}}

    \def\ord{{\rm Ord}\,}

    \def\rg{{\rm rg}}
    \def\Im{{\rm Im\,}}
    \def\sh{{\rm shear}}
    \def\sct{{\rm sct}'}
    \def\Sct{{\rm sct}}

    \def\cst#1#2{{\rm const}^{#1}_{#2}\,}
    \def\abcst{{\sst const}}

     \def\veps{\varepsilon}
     \def\ssst{\scriptscriptstyle}
     \def\bde{{\mathchoice{\pmb{$\de$}}{\pmb{$\de$}}
                              {\pmb{$\sst\de$}}{\pmb{$\ssst\de$}}}}
    \def\jbar{{\mathchoice
                   {{\smash{\lower1ex\hbox{$\mathchar'26$}}\mkern-9mu j}}
                   {{\smash{\lower1ex\hbox{$\mathchar'26$}}\mkern-9mu j}}
                   {{\smash{\lower1.2ex\hbox{$\mathchar'26$}}\mkern-10.2mu j}}
                   {{\smash{\lower1.2ex\hbox{$\mathchar'26$}}\mkern-10.2mu j}}}}

    \def\cC{{\cal C}}
    \def\cD{{\cal D}}
    
    \def\cL{{\cal L}}

    \def\cV{{\cal V}}
    \def\cW{{\cal W}}

    \def\fe{{\frak e}}
    \def\ff{{\frak f}}
    \def\fl{{\frak l}}
    
    \def\fN{{\frak N}}
    \def\fK{{\frak K}}

    \def\fX{{\frak X}}
    \def\fY{{\frak Y}}
    
    \def\fz{{\frak z}}
    
    \def\rD{{\rm D}}

    \def\v{\pmb{$\vert$}}
    \def\V{\pmb{$\big\vert$}}
    \def\VV{\pmb{$\Big\vert$}}

     \def\tv{\kern8pt\tilde{\kern-8pt\pmb{$\vert$}}}
     \def\tV{\kern8pt\tilde{\kern-8pt\pmb{$\big\vert$}}}
     \def\tVV{\kern8pt\tilde{\kern-8pt\pmb{$\Big\vert$}}}

    \def\tn{|\kern-1pt|\kern-1pt|}
    \def\TN{\big|\kern-1.5pt\big|\kern-1.5pt\big|}
    \def\TTN{\Big|\kern-2pt\Big|\kern-2pt\Big|}

    \def\trn{|\kern-1pt|\kern-1pt|^{\,\tilde{\,}}}
    \def\TRN{\big|\kern-1.5pt\big|\kern-1.5pt\big|^{\,\tilde{\,}}}
    \def\TTRN{\Big|\kern-2pt\Big|\kern-2pt\Big|^{\,\tilde{\,}}}

     \def\tnorm{\kern8pt\tilde{\kern-8pt\|}}
     \def\Tnorm{\kern8pt\tilde{\kern-8pt\big\|}}
     \def\TNorm{\kern8pt\tilde{\kern-8pt\Big\|}}
     \def\TNOrm{\kern8pt\tilde{\kern-8pt\bigg\|}}

    \def\rw{\mathclose{:}}
    \def\lw{\mathopen{:}}
    \def\lW{\mathopen{{\tst{\hbox{.}\atop\raise 2.5pt\hbox{.}}}}}
    \def\rW{\mathclose{{\tst{{.}\atop\raise 2.5pt\hbox{.}}}}}
    \def\lww{\mathopen{{\tst{\raise 1pt\hbox{.}\atop\raise 1pt\hbox{.}}}}}
    \def\rww{\mathclose{{\tst{\raise 1pt\hbox{.}\atop\raise 1pt\hbox{.}}}}}

   \font\sixrm=cmr6   \font\eightrm=cmr8  
   \font\sixi=cmmi6   \font\eighti=cmmi8  
  \font\sixsy=cmsy6  \font\eightsy=cmsy8 
  \font\sixbf=cmbx6  \font\eightbf=cmbx8 
                     \font\eightit=cmti8 
                     \font\eightsl=cmsl8 
                     \font\eighttt=cmtt8 

\font\eightfrak=eufm7 at 8pt

\def\eightpoint{\def\rm{\fam0\eightrm}
 \textfont0=\eightrm \scriptfont0=\sixrm \scriptscriptfont0=\fiverm
 \textfont1=\eighti \scriptfont1=\sixi \scriptscriptfont1=\fivei
 \textfont2=\eightsy \scriptfont2=\sixsy \scriptscriptfont2=\fivesy
 \textfont3=\tenex \scriptfont3=\tenex \scriptscriptfont3=\tenex
 \textfont\itfam=\eightit \def\it{\fam\itfam\eightit}%
 \textfont\slfam=\eightsl \def\sl{\fam\slfam\eightsl}%
 \textfont\ttfam=\eighttt \def\tt{\fam\ttfam\eighttt}%
 \textfont\frakfam=\eightfrak \def\frak{\fam\frakfam\tenfrak}%
 \textfont\bffam=\eightbf \scriptfont\bffam=\sixbf
 \scriptscriptfont\bffam=\fivebf \def\bf{\fam\bffam\eightbf}%
 \normalbaselineskip=9pt
 \setbox\strutbox=\hbox{\vrule height7pt depth2pt width0pt}%
 \let\sc=\sixrm \let\big=\eightbig \normalbaselines\rm}
\catcode`@=11
\def\footnote#1{\edef\@sf{\spacefactor\the\spacefactor}#1\@sf
     \insert\footins\bgroup\eightpoint
     \interlinepenalty100 \let\par=\endgraf
     \leftskip=0pt \rightskip=0pt
     \splittopskip=10pt plus 1pt minus 1pt \floatingpenalty=20000
     \smallskip\item{#1}\bgroup\strut\aftergroup\@foot\let\next}
\skip\footins=12pt plus 2pt minus 4pt
\dimen\footins=30pc
\catcode`@=12


  \IgNoRe{PG}
  \IgNoRe{STM Assertion }
  \IgNoRe{PG}
  \IgNoRe{PG}
  \IgNoRe{STM Assertion }
  \IgNoRe{PG}
  \IgNoRe{STM Assertion }
  \IgNoRe{STM Assertion }
  \IgNoRe{EQN}
  \IgNoRe{STM Assertion }
  \IgNoRe{STM Assertion }
  \IgNoRe{PG}
  \IgNoRe{STM Assertion }
  \IgNoRe{STM Assertion }
  \IgNoRe{EQN}
  \IgNoRe{STM Assertion }
  \IgNoRe{STM Assertion }
  \IgNoRe{STM Assertion }
  \IgNoRe{STM Assertion }
  \IgNoRe{STM Assertion }
  \IgNoRe{STM Assertion }
  \IgNoRe{PG}
  \IgNoRe{STM Assertion }
  \IgNoRe{STM Assertion }
  \IgNoRe{STM Assertion }
  \IgNoRe{STM Assertion }
  \IgNoRe{STM Assertion }
  \IgNoRe{STM Assertion }
  \IgNoRe{STM Assertion }
  \IgNoRe{STM Assertion }
  \IgNoRe{STM Assertion }
  \IgNoRe{STM Assertion }
  \IgNoRe{STM Assertion }
  \IgNoRe{STM Assertion }
  \IgNoRe{PG}
  \IgNoRe{EQN}
  \IgNoRe{STM Assertion }
  \IgNoRe{STM Assertion }
  \IgNoRe{STM Assertion }
  \IgNoRe{PG}
  \IgNoRe{STM Assertion }
  \IgNoRe{STM Assertion }
 \def\corwicknorm{\frefwarning II.32} \IgNoRe{STM Assertion }
  \IgNoRe{STM Assertion }
  \IgNoRe{EQN}
  \IgNoRe{EQN}
  \IgNoRe{STM Assertion }
  \IgNoRe{PG}
  \IgNoRe{PG}
  \IgNoRe{STM Assertion }
  \IgNoRe{EQN}
  \IgNoRe{STM Assertion }
  \IgNoRe{STM Assertion }
  \IgNoRe{STM Assertion }
  \IgNoRe{EQN}
  \IgNoRe{STM Assertion }
  \IgNoRe{STM Assertion }
  \IgNoRe{STM Assertion }
  \IgNoRe{PG}
  \IgNoRe{STM Assertion }
  \IgNoRe{STM Assertion }
  \IgNoRe{PG}
  \IgNoRe{STM Assertion }
  \IgNoRe{STM Assertion }
  \IgNoRe{STM Assertion }
  \IgNoRe{PG}
  \IgNoRe{STM Assertion }
  \IgNoRe{STM Assertion }
  \IgNoRe{STM Assertion }
  \IgNoRe{STM Assertion }
  \IgNoRe{STM Assertion }
  \IgNoRe{STM Assertion }
  \IgNoRe{STM Assertion }
  \IgNoRe{STM Assertion }
  \IgNoRe{PG}
  \IgNoRe{STM Assertion }
  \IgNoRe{STM Assertion }
  \IgNoRe{STM Assertion }
  \IgNoRe{STM Assertion }
  \IgNoRe{STM Assertion }
  \IgNoRe{STM Assertion }
  \IgNoRe{STM Assertion }
  \IgNoRe{STM Assertion }
  \IgNoRe{PG}
  \IgNoRe{STM Assertion }
  \IgNoRe{STM Assertion }
  \IgNoRe{PG}
  \IgNoRe{PG}
  \IgNoRe{STM Assertion }
  \IgNoRe{STM Assertion }
  \IgNoRe{PG}
  \IgNoRe{PG}
  \IgNoRe{STM Assertion }
  \IgNoRe{STM Assertion }
  \IgNoRe{EQN}
  \IgNoRe{STM Assertion }
  \IgNoRe{PG}
  \IgNoRe{STM Assertion }
  \IgNoRe{STM Assertion }
  \IgNoRe{STM Assertion }
  \IgNoRe{PG}
  \IgNoRe{STM Assertion }
  \IgNoRe{STM Assertion }
  \IgNoRe{STM Assertion }
  \IgNoRe{EQN}
  \IgNoRe{STM Assertion }
  \IgNoRe{PG}
  \IgNoRe{EQN}
  \IgNoRe{STM Assertion }
  \IgNoRe{STM Assertion }
  \IgNoRe{STM Assertion }
  \IgNoRe{STM Assertion }
  \IgNoRe{EQN}
  \IgNoRe{EQN}
  \IgNoRe{PG}
  \IgNoRe{PG}
  \IgNoRe{STM Assertion }
  \IgNoRe{STM Assertion }
  \IgNoRe{STM Assertion }
  \IgNoRe{STM Assertion }
  \IgNoRe{STM Assertion }
  \IgNoRe{STM Assertion }
  \IgNoRe{STM Assertion }
  \IgNoRe{STM Assertion }
  \IgNoRe{PG}
  \IgNoRe{STM Assertion }
  \IgNoRe{STM Assertion }
  \IgNoRe{STM Assertion }
  \IgNoRe{STM Assertion }
  \IgNoRe{STM Assertion }
  \IgNoRe{STM Assertion }
  \IgNoRe{STM Assertion }
  \IgNoRe{STM Assertion }
  \IgNoRe{STM Assertion }
  \IgNoRe{STM Assertion }
  \IgNoRe{PG}
  \IgNoRe{STM Assertion }
  \IgNoRe{STM Assertion }
  \IgNoRe{STM Assertion }
  \IgNoRe{PG}
  \IgNoRe{STM Assertion }
  \IgNoRe{STM Assertion }
  \IgNoRe{STM Assertion }
  \IgNoRe{STM Assertion }
  \IgNoRe{PG}
  \IgNoRe{STM Assertion }
  \IgNoRe{STM Assertion }
  \IgNoRe{PG}
  \IgNoRe{STM Assertion }
  \IgNoRe{PG}
  \IgNoRe{STM Assertion }
  \IgNoRe{PG}
  \IgNoRe{STM Assertion }
  \IgNoRe{STM Assertion }
  \IgNoRe{STM Assertion }
  \IgNoRe{STM Assertion }
  \IgNoRe{PG}
  \IgNoRe{PG}
  \IgNoRe{STM Assertion }
  \IgNoRe{STM Assertion }
  \IgNoRe{EQN}
  \IgNoRe{EQN}
  \IgNoRe{STM Assertion }
  \IgNoRe{STM Assertion }
  \IgNoRe{STM Assertion }
  \IgNoRe{STM Assertion }
  \IgNoRe{PG}
  \IgNoRe{STM Assertion }
  \IgNoRe{STM Assertion }
  \IgNoRe{STM Assertion }
  \IgNoRe{STM Assertion }
  \IgNoRe{STM Assertion }
  \IgNoRe{STM Assertion }
  \IgNoRe{STM Assertion }
  \IgNoRe{PG}
  \IgNoRe{STM Assertion }
  \IgNoRe{EQN}
  \IgNoRe{EQN}
  \IgNoRe{PG}
  \IgNoRe{STM Assertion }
  \IgNoRe{EQN}
  \IgNoRe{STM Assertion }
  \IgNoRe{STM Assertion }
  \IgNoRe{STM Assertion }
  \IgNoRe{PG}
  \IgNoRe{STM Assertion }
  \IgNoRe{EQN}
  \IgNoRe{STM Assertion }
  \IgNoRe{PG}
  \IgNoRe{PG}


  \IgNoRe{PG}
  \IgNoRe{PG}
  \IgNoRe{STM Assertion }
  \IgNoRe{EQN}
  \IgNoRe{STM Assertion }
  \IgNoRe{PG}
  \IgNoRe{STM Assertion }
  \IgNoRe{EQN}
  \IgNoRe{STM Assertion }
  \IgNoRe{EQN}
  \IgNoRe{STM Assertion }
  \IgNoRe{STM Assertion }
  \IgNoRe{STM Assertion }
  \IgNoRe{STM Assertion }
  \IgNoRe{PG}
  \IgNoRe{STM Assertion }
  \IgNoRe{STM Assertion }
  \IgNoRe{STM Assertion }
  \IgNoRe{STM Assertion }
  \IgNoRe{PG}
  \IgNoRe{STM Assertion }
  \IgNoRe{STM Assertion }
  \IgNoRe{EQN}
  \IgNoRe{STM Assertion }
  \IgNoRe{STM Assertion }
  \IgNoRe{STM Assertion }
  \IgNoRe{PG}
  \IgNoRe{PG}
  \IgNoRe{STM Assertion }
  \IgNoRe{EQN}
 \def\defLADcompoundphladder{\frefwarning I.19} \IgNoRe{STM Assertion }
 \def\thmLADmodcompLadder{\frefwarning I.20} \IgNoRe{STM Assertion }
  \IgNoRe{PG}
 \def\remLADmodcompLadder{\frefwarning I.21} \IgNoRe{STM Assertion }
 \def\thmLADmodcompLaddercont{\frefwarning I.22} \IgNoRe{STM Assertion }

  \IgNoRe{STM Assertion }
  \IgNoRe{STM Assertion }
  \IgNoRe{PG}
  \IgNoRe{PG}
  \IgNoRe{STM Assertion }
  \IgNoRe{STM Assertion }
  \IgNoRe{STM Assertion }
  \IgNoRe{STM Assertion }
  \IgNoRe{STM Assertion }
  \IgNoRe{STM Assertion }
  \IgNoRe{PG}
  \IgNoRe{STM Assertion }
  \IgNoRe{STM Assertion }
  \IgNoRe{STM Assertion }
  \IgNoRe{STM Assertion }
  \IgNoRe{STM Assertion }
  \IgNoRe{STM Assertion }
  \IgNoRe{EQN}
  \IgNoRe{PG}
  \IgNoRe{STM Assertion }
  \IgNoRe{STM Assertion }
  \IgNoRe{STM Assertion }
  \IgNoRe{EQN}
  \IgNoRe{STM Assertion }
  \IgNoRe{STM Assertion }
  \IgNoRe{STM Assertion }
  \IgNoRe{PG}
  \IgNoRe{STM Assertion }
  \IgNoRe{STM Assertion }
  \IgNoRe{STM Assertion }
  \IgNoRe{STM Assertion }
  \IgNoRe{EQN}
  \IgNoRe{EQN}
  \IgNoRe{EQN}
  \IgNoRe{EQN}
  \IgNoRe{STM Assertion }
  \IgNoRe{STM Assertion }
  \IgNoRe{EQN}
  \IgNoRe{STM Assertion }
  \IgNoRe{EQN}
  \IgNoRe{EQN}
  \IgNoRe{EQN}
  \IgNoRe{EQN}
  \IgNoRe{EQN}
  \IgNoRe{PG}
  \IgNoRe{STM Assertion }
  \IgNoRe{STM Assertion }
  \IgNoRe{EQN}

  \IgNoRe{EQN}
  \IgNoRe{EQN}
  \IgNoRe{PG}
  \IgNoRe{EQN}
  \IgNoRe{EQN}
  \IgNoRe{STM Assertion }
  \IgNoRe{STM Assertion }
  \IgNoRe{EQN}
  \IgNoRe{EQN}
  \IgNoRe{EQN}
  \IgNoRe{EQN}
  \IgNoRe{STM Assertion }
  \IgNoRe{STM Assertion }
  \IgNoRe{STM Assertion }
  \IgNoRe{STM Assertion }
  \IgNoRe{STM Assertion }
  \IgNoRe{STM Assertion }
  \IgNoRe{STM Assertion }
  \IgNoRe{STM Assertion }
  \IgNoRe{STM Assertion }
  \IgNoRe{STM Assertion }
  \IgNoRe{STM Assertion }
  \IgNoRe{STM Assertion }
  \IgNoRe{EQN}
  \IgNoRe{EQN}
  \IgNoRe{EQN}
  \IgNoRe{EQN}
  \IgNoRe{STM Assertion }
  \IgNoRe{EQN}
  \IgNoRe{STM Assertion }
  \IgNoRe{EQN}
  \IgNoRe{EQN}
  \IgNoRe{EQN}
  \IgNoRe{EQN}
  \IgNoRe{EQN}
  \IgNoRe{STM Assertion }
  \IgNoRe{STM Assertion }
  \IgNoRe{STM Assertion }
  \IgNoRe{STM Assertion }
  \IgNoRe{STM Assertion }
  \IgNoRe{EQN}
  \IgNoRe{EQN}
  \IgNoRe{STM Assertion }
  \IgNoRe{STM Assertion }
  \IgNoRe{EQN}
  \IgNoRe{EQN}
  \IgNoRe{STM Assertion }
  \IgNoRe{EQN}
  \IgNoRe{STM Assertion }
  \IgNoRe{STM Assertion }
  \IgNoRe{EQN}
  \IgNoRe{STM Assertion }
  \IgNoRe{EQN}
  \IgNoRe{EQN}
  \IgNoRe{EQN}
  \IgNoRe{EQN}
  \IgNoRe{STM Assertion }
  \IgNoRe{PG}
  \IgNoRe{STM Assertion }
  \IgNoRe{STM Assertion }
  \IgNoRe{PG}
  \IgNoRe{STM Assertion }

  \IgNoRe{PG}
  \IgNoRe{STM Assertion }
  \IgNoRe{EQN}
  \IgNoRe{EQN}
  \IgNoRe{EQN}
  \IgNoRe{STM Assertion }
  \IgNoRe{STM Assertion }
  \IgNoRe{STM Assertion }
  \IgNoRe{STM Assertion }
  \IgNoRe{EQN}
  \IgNoRe{STM Assertion }
  \IgNoRe{EQN}
  \IgNoRe{EQN}
  \IgNoRe{EQN}
  \IgNoRe{STM Assertion }
  \IgNoRe{STM Assertion }
  \IgNoRe{STM Assertion }
  \IgNoRe{EQN}

  \IgNoRe{STM Assertion }
  \IgNoRe{PG}
  \IgNoRe{STM Assertion }
  \IgNoRe{EQN}
  \IgNoRe{EQN}
  \IgNoRe{STM Assertion }
  \IgNoRe{EQN}

  \IgNoRe{STM Assertion }
  \IgNoRe{EQN}
  \IgNoRe{EQN}
  \IgNoRe{PG}
  \IgNoRe{EQN}
  \IgNoRe{EQN}
  \IgNoRe{EQN}
  \IgNoRe{EQN}
  \IgNoRe{EQN}
  \IgNoRe{STM Assertion }
  \IgNoRe{EQN}
  \IgNoRe{EQN}
  \IgNoRe{EQN}
  \IgNoRe{EQN}
  \IgNoRe{EQN}
  \IgNoRe{STM Assertion }
  \IgNoRe{EQN}
  \IgNoRe{EQN}
  \IgNoRe{EQN}
  \IgNoRe{EQN}
  \IgNoRe{EQN}
  \IgNoRe{EQN}
  \IgNoRe{EQN}
  \IgNoRe{EQN}
  \IgNoRe{STM Assertion }

  \IgNoRe{STM Assertion }
  \IgNoRe{PG}
  \IgNoRe{EQN}
  \IgNoRe{EQN}
  \IgNoRe{STM Assertion }
  \IgNoRe{EQN}
  \IgNoRe{EQN}
  \IgNoRe{STM Assertion }
  \IgNoRe{EQN}
  \IgNoRe{STM Assertion }
  \IgNoRe{EQN}
  \IgNoRe{EQN}
  \IgNoRe{PG}
  \IgNoRe{PG}


  \IgNoRe{PG}
  \IgNoRe{EQN}
  \IgNoRe{STM Assertion }
  \IgNoRe{PG}
  \IgNoRe{STM Assertion }
  \IgNoRe{STM Assertion }
  \IgNoRe{STM Assertion }
  \IgNoRe{STM Assertion }
  \IgNoRe{STM Assertion }
  \IgNoRe{STM Assertion }
  \IgNoRe{EQN}
  \IgNoRe{STM Assertion }
  \IgNoRe{STM Assertion }
  \IgNoRe{STM Assertion }
  \IgNoRe{STM Assertion }
  \IgNoRe{STM Assertion }
  \IgNoRe{STM Assertion }
  \IgNoRe{STM Assertion }
  \IgNoRe{STM Assertion }
  \IgNoRe{PG}
  \IgNoRe{STM Assertion }
  \IgNoRe{STM Assertion }
  \IgNoRe{STM Assertion }
  \IgNoRe{STM Assertion }
  \IgNoRe{STM Assertion }
  \IgNoRe{STM Assertion }
  \IgNoRe{STM Assertion }
  \IgNoRe{STM Assertion }
  \IgNoRe{EQN}
  \IgNoRe{PG}
  \IgNoRe{PG}
  \IgNoRe{STM Assertion }
  \IgNoRe{STM Assertion }
  \IgNoRe{STM Assertion }
  \IgNoRe{STM Assertion }
  \IgNoRe{STM Assertion }
  \IgNoRe{STM Assertion }
  \IgNoRe{PG}
  \IgNoRe{EQN}
  \IgNoRe{EQN}
  \IgNoRe{EQN}
  \IgNoRe{EQN}
  \IgNoRe{EQN}
  \IgNoRe{EQN}
  \IgNoRe{STM Assertion }
  \IgNoRe{STM Assertion }
  \IgNoRe{STM Assertion }
  \IgNoRe{STM Assertion }
  \IgNoRe{PG}
  \IgNoRe{STM Assertion }
  \IgNoRe{EQN}
  \IgNoRe{STM Assertion }
  \IgNoRe{STM Assertion }
  \IgNoRe{STM Assertion }
  \IgNoRe{STM Assertion }
  \IgNoRe{PG}
  \IgNoRe{STM Assertion }
 \def\corOSappMonoidIV{\frefwarning A.5} \IgNoRe{STM Assertion }
 \def\remOSappMonoidIV{\frefwarning A.6} \IgNoRe{STM Assertion }
  \IgNoRe{STM Assertion }
  \IgNoRe{PG}
  \IgNoRe{EQN}
  \IgNoRe{EQN}
  \IgNoRe{PG}
  \IgNoRe{STM Assertion }
  \IgNoRe{STM Assertion }
 \def\lemOStworengrpmaps{\frefwarning VII.3} \IgNoRe{STM Assertion }
  \IgNoRe{EQN}
  \IgNoRe{PG}
  \IgNoRe{STM Assertion }
  \IgNoRe{STM Assertion }
  \IgNoRe{EQN}
  \IgNoRe{STM Assertion }
  \IgNoRe{STM Assertion }
  \IgNoRe{STM Assertion }
  \IgNoRe{STM Assertion }
  \IgNoRe{PG}
  \IgNoRe{STM Assertion }
  \IgNoRe{STM Assertion }
  \IgNoRe{STM Assertion }
  \IgNoRe{STM Assertion }
  \IgNoRe{STM Assertion }
 \def\remOSthmV{\frefwarning VIII.7} \IgNoRe{STM Assertion }
  \IgNoRe{STM Assertion }
  \IgNoRe{STM Assertion }
  \IgNoRe{PG}
  \IgNoRe{STM Assertion }
  \IgNoRe{STM Assertion }
  \IgNoRe{STM Assertion }
 \def\lemOSjhat{\frefwarning IX.5} \IgNoRe{STM Assertion }
  \IgNoRe{STM Assertion }
  \IgNoRe{EQN}
  \IgNoRe{STM Assertion }
  \IgNoRe{STM Assertion }
  \IgNoRe{PG}
  \IgNoRe{STM Assertion }
  \IgNoRe{STM Assertion }
  \IgNoRe{STM Assertion }
  \IgNoRe{STM Assertion }
  \IgNoRe{STM Assertion }
  \IgNoRe{STM Assertion }
  \IgNoRe{STM Assertion }
  \IgNoRe{STM Assertion }
  \IgNoRe{STM Assertion }
  \IgNoRe{EQN}
 \def\thmOSTfirststep{\frefwarning X.12} \IgNoRe{STM Assertion }
 \def\defOSsymmetries{\frefwarning B.1} \IgNoRe{STM Assertion }
  \IgNoRe{PG}
 \def\remOSgrassmannsymmetries{\frefwarning B.2} \IgNoRe{STM Assertion }
 \def\remOSsymmetryConsequences{\frefwarning B.3} \IgNoRe{STM Assertion }
  \IgNoRe{STM Assertion }
 \def\remOSrengrppreserves{\frefwarning B.5} \IgNoRe{STM Assertion }
 \def\lemOSphipsistruct{\frefwarning B.6} \IgNoRe{STM Assertion }
 \def\lemOSappGrassI{\frefwarning C.1} \IgNoRe{STM Assertion }
  \IgNoRe{STM Assertion }
  \IgNoRe{PG}
  \IgNoRe{STM Assertion }
  \IgNoRe{PG}
  \IgNoRe{PG}
  \IgNoRe{STM Assertion }
  \IgNoRe{STM Assertion }
  \IgNoRe{PG}
  \IgNoRe{STM Assertion }
  \IgNoRe{STM Assertion }
  \IgNoRe{STM Assertion }
  \IgNoRe{STM Assertion }
  \IgNoRe{STM Assertion }
 \def\propOSfunctorialitySect{\frefwarning XII.8} \IgNoRe{STM Assertion }
  \IgNoRe{STM Assertion }
  \IgNoRe{STM Assertion }
  \IgNoRe{STM Assertion }
 \def\lemOSNormMom{\frefwarning XII.12} \IgNoRe{STM Assertion }
  \IgNoRe{STM Assertion }
  \IgNoRe{STM Assertion }
  \IgNoRe{STM Assertion }
  \IgNoRe{STM Assertion }
  \IgNoRe{STM Assertion }
  \IgNoRe{STM Assertion }
  \IgNoRe{EQN}
 \def\lemOSsectextimpr{\frefwarning XII.19} \IgNoRe{STM Assertion }
  \IgNoRe{STM Assertion }
  \IgNoRe{PG}
  \IgNoRe{STM Assertion }
  \IgNoRe{EQN}
  \IgNoRe{EQN}
  \IgNoRe{STM Assertion }
  \IgNoRe{EQN}
  \IgNoRe{EQN}
  \IgNoRe{STM Assertion }
  \IgNoRe{EQN}
  \IgNoRe{STM Assertion }
  \IgNoRe{EQN}
 \def\lemOSdiffpropbound{\frefwarning XIII.6} \IgNoRe{STM Assertion }
  \IgNoRe{STM Assertion }
  \IgNoRe{STM Assertion }
  \IgNoRe{STM Assertion }
  \IgNoRe{PG}
  \IgNoRe{STM Assertion }
  \IgNoRe{STM Assertion }
  \IgNoRe{STM Assertion }
  \IgNoRe{STM Assertion }
  \IgNoRe{EQN}
  \IgNoRe{EQN}
  \IgNoRe{STM Assertion }
  \IgNoRe{STM Assertion }
  \IgNoRe{PG}
  \IgNoRe{STM Assertion }
  \IgNoRe{STM Assertion }
  \IgNoRe{STM Assertion }
  \IgNoRe{EQN}
  \IgNoRe{STM Assertion }
  \IgNoRe{EQN}
  \IgNoRe{STM Assertion }
  \IgNoRe{STM Assertion }
  \IgNoRe{EQN}
  \IgNoRe{STM Assertion }
  \IgNoRe{EQN}
  \IgNoRe{EQN}
  \IgNoRe{EQN}
  \IgNoRe{EQN}
  \IgNoRe{STM Assertion }
  \IgNoRe{STM Assertion }
  \IgNoRe{STM Assertion }
  \IgNoRe{STM Assertion }
  \IgNoRe{PG}
  \IgNoRe{STM Assertion }
  \IgNoRe{STM Assertion }
  \IgNoRe{STM Assertion }
  \IgNoRe{STM Assertion }
  \IgNoRe{STM Assertion }
 \def\propOSmomcontrintboundsectors{\frefwarning XVI.8} \IgNoRe{STM Assertion }
  \IgNoRe{STM Assertion }
  \IgNoRe{STM Assertion }
  \IgNoRe{STM Assertion }
  \IgNoRe{STM Assertion }
  \IgNoRe{EQN}
  \IgNoRe{EQN}
  \IgNoRe{STM Assertion }
  \IgNoRe{PG}
  \IgNoRe{STM Assertion }
 \def\thOSmomrengroupestimate{\frefwarning XVII.3} \IgNoRe{STM Assertion }
  \IgNoRe{EQN}
  \IgNoRe{STM Assertion }
 \def\lemOStildesourceterm{\frefwarning XVII.5} \IgNoRe{STM Assertion }
 \def\eqnNPderivampprelim{\frefwarning XVII.3} \IgNoRe{EQN}
  \IgNoRe{EQN}
  \IgNoRe{EQN}
  \IgNoRe{EQN}
  \IgNoRe{EQN}
 \def\thOSmomrengroupdiffestimate{\frefwarning XVII.6} \IgNoRe{STM Assertion }
  \IgNoRe{STM Assertion }
  \IgNoRe{STM Assertion }
  \IgNoRe{EQN}
  \IgNoRe{EQN}
  \IgNoRe{EQN}
  \IgNoRe{EQN}
  \IgNoRe{EQN}
  \IgNoRe{EQN}
  \IgNoRe{STM Assertion }
  \IgNoRe{STM Assertion }
  \IgNoRe{STM Assertion }
  \IgNoRe{PG}
  \IgNoRe{STM Assertion }
  \IgNoRe{STM Assertion }
  \IgNoRe{EQN}
  \IgNoRe{EQN}
  \IgNoRe{STM Assertion }
  \IgNoRe{EQN}
  \IgNoRe{EQN}
  \IgNoRe{STM Assertion }
  \IgNoRe{EQN}
  \IgNoRe{EQN}
 \def\propOSNaiveLadder{\frefwarning D.7} \IgNoRe{STM Assertion }
 \def\remOSnaiveladderest{\frefwarning D.8} \IgNoRe{STM Assertion }
  \IgNoRe{PG}
  \IgNoRe{STM Assertion }
  \IgNoRe{STM Assertion }
  \IgNoRe{STM Assertion }
  \IgNoRe{PG}
  \IgNoRe{STM Assertion }
 \def\propOSthreetoonenorm{\frefwarning XIX.1} \IgNoRe{STM Assertion }
  \IgNoRe{STM Assertion }
  \IgNoRe{STM Assertion }
  \IgNoRe{PG}
 \def\propOSresectorI{\frefwarning XIX.4} \IgNoRe{STM Assertion }
  \IgNoRe{STM Assertion }
  \IgNoRe{STM Assertion }
  \IgNoRe{STM Assertion }
  \IgNoRe{STM Assertion }
 \def\corOStildeirrelevantresect{\frefwarning XIX.9} \IgNoRe{STM Assertion }
  \IgNoRe{STM Assertion }
  \IgNoRe{STM Assertion }
  \IgNoRe{STM Assertion }
 \def\corOSresectorvanishkzero{\frefwarning XIX.13} \IgNoRe{STM Assertion }
 \def\defOScreateSectoriz{\frefwarning XIX.14} \IgNoRe{STM Assertion }
 \def\propOScreateSectoriz{\frefwarning XIX.15} \IgNoRe{STM Assertion }
  \IgNoRe{STM Assertion }
  \IgNoRe{STM Assertion }
  \IgNoRe{PG}
  \IgNoRe{STM Assertion }
  \IgNoRe{PG}
  \IgNoRe{STM Assertion }
  \IgNoRe{STM Assertion }
  \IgNoRe{PG}
  \IgNoRe{STM Assertion }
  \IgNoRe{STM Assertion }
  \IgNoRe{EQN}
  \IgNoRe{PG}
  \IgNoRe{EQN}
  \IgNoRe{STM Assertion }
  \IgNoRe{STM Assertion }
  \IgNoRe{PG}
  \IgNoRe{EQN}
  \IgNoRe{EQN}
  \IgNoRe{EQN}
  \IgNoRe{EQN}
  \IgNoRe{EQN}
  \IgNoRe{STM Assertion }
  \IgNoRe{STM Assertion }
  \IgNoRe{EQN}
  \IgNoRe{STM Assertion }
  \IgNoRe{PG}
  \IgNoRe{PG}
  \IgNoRe{PG}
  \IgNoRe{STM Assertion }
  \IgNoRe{EQN}
  \IgNoRe{STM Assertion }
  \IgNoRe{PG}
  \IgNoRe{STM Assertion }
  \IgNoRe{EQN}
  \IgNoRe{STM Assertion }
  \IgNoRe{STM Assertion }
  \IgNoRe{PG}
  \IgNoRe{EQN}
  \IgNoRe{EQN}
  \IgNoRe{EQN}
  \IgNoRe{STM Assertion }
  \IgNoRe{STM Assertion }
  \IgNoRe{STM Assertion }
  \IgNoRe{EQN}
  \IgNoRe{STM Assertion }
  \IgNoRe{STM Assertion }
 \def\theoremOSLadA{\frefwarning XXII.8} \IgNoRe{STM Assertion }
 \def\defOSzeroext{\frefwarning E.1} \IgNoRe{STM Assertion }
  \IgNoRe{STM Assertion }
 \def\defOSzerosectorext{\frefwarning E.3} \IgNoRe{STM Assertion }
  \IgNoRe{PG}
  \IgNoRe{STM Assertion }
  \IgNoRe{STM Assertion }
  \IgNoRe{STM Assertion }
  \IgNoRe{STM Assertion }
  \IgNoRe{STM Assertion }
  \IgNoRe{STM Assertion }
  \IgNoRe{STM Assertion }
  \IgNoRe{PG}
  \IgNoRe{PG}


\newcount\CHAPNO
\newcount\APPNO
\CHAPNO=0
\APPNO=1
\def\advCHAPNO{\advance\CHAPNO by 1}
\def\advAPPNO{\advance\APPNO by 1}

\def\caproman#1{\ifcase#1\or I\or II\or III\or IV\or V\or VI\or VII\or
VIII\or IX\or X\or XI\or XII\or XIII\or XIV\or XV\or XVI\or XVII\or XVIII\or
XIX\or XX\or XXI\or XXII\or XXIII\or XXIV\or XXV\or XXVI\or XXVII\or XXVIII\or XXIX\or XXX\or XXXI\or XXXII\or XXXIII\or XXXIV\or XXXV\or XXXVI\or XXXVII\or XXXVIII\or XXXIX\fi}%

\def\capletter#1{\ifcase#1\or A\or B\or C\or D\or E\or F\or G\or
H\or I\or J\or K\or L\or M\or N\or O\or P\or Q\or R\or
S\or T\or U\or V\or W\or X\or Y\or Z\fi}%

\newcount\cHintroI \cHintroI=\CHAPNO \advCHAPNO 
                              \edef\CHintroI{\caproman\CHAPNO}
\newcount\cHnorms  \cHnorms=\CHAPNO \advCHAPNO 
                              
\newcount\cHproprengrp \cHproprengrp=\CHAPNO \advCHAPNO 
                              
\newcount\cHcovbounds  \cHcovbounds=\CHAPNO \advCHAPNO 
                              
\newcount\cHinsulator \cHinsulator=\CHAPNO \advCHAPNO

 \advAPPNO

\newcount\cHintroII \cHintroII=\CHAPNO \advCHAPNO 
                              \edef\CHintroII{\caproman\CHAPNO}
\newcount\cHamputate \cHamputate=\CHAPNO \advCHAPNO
                              
\newcount\cHscales \cHscales=\CHAPNO \advCHAPNO
                              
\newcount\cHfourier \cHfourier=\CHAPNO \advCHAPNO
                              
\newcount\cHmomentum \cHmomentum=\CHAPNO \advCHAPNO

 \advAPPNO
 \advAPPNO

\newcount\cHintroIII \cHintroIII=\CHAPNO \advCHAPNO
                              \edef\CHintroIII{\caproman\CHAPNO}
\newcount\cHsectors \cHsectors=\CHAPNO \advCHAPNO
                              
\newcount\cHsecpropbounds \cHsecpropbounds=\CHAPNO \advCHAPNO
                              
\newcount\cHladdersNotn  \cHladdersNotn=\CHAPNO \advCHAPNO
                              
\newcount\cHestren  \cHestren=\CHAPNO \advCHAPNO
                              
\newcount\cHsecmomnorm \cHsecmomnorm=\CHAPNO \advCHAPNO
                              \edef\CHsecmomnorm{\caproman\CHAPNO}
\newcount\cHmomestren \cHmomestren=\CHAPNO \advCHAPNO

 \advAPPNO

\newcount\cHintroIV  \cHintroIV=\CHAPNO \advCHAPNO
                              
\newcount\cHcomparison   \cHcomparison=\CHAPNO \advCHAPNO
                              
\newcount\cHsumsmom  \cHsumsmom=\CHAPNO \advCHAPNO
                              
\newcount\cHsectorsmom   \cHsectorsmom=\CHAPNO \advCHAPNO
                              
\newcount\cHppladsect    \cHppladsect=\CHAPNO \advCHAPNO

 \advAPPNO


 \def\defNPCTMSpace{\frefwarning I.1} \IgNoRe{STM Assertion }
  \IgNoRe{PG}
 \def\eqnNPreal{\frefwarning I.1} \IgNoRe{EQN}
 \def\eqnNPphexchange{\frefwarning I.2} \IgNoRe{EQN}
 \def\eqnNPinteraction{\frefwarning I.3} \IgNoRe{EQN}
  \IgNoRe{EQN}
  \IgNoRe{EQN}
 \def\defNPscales{\frefwarning I.2} \IgNoRe{STM Assertion }
  \IgNoRe{STM Assertion }
 \def\theoremNPmainthI{\frefwarning I.4} \IgNoRe{STM Assertion }
 \def\theoremNPmainthII{\frefwarning I.5} \IgNoRe{STM Assertion }
  \IgNoRe{STM Assertion }
 \def\theoremNPmainthIII{\frefwarning I.7} \IgNoRe{STM Assertion }
  \IgNoRe{STM Assertion }
  \IgNoRe{STM Assertion }
 \def\defNPstrongasymm{\frefwarning I.10} \IgNoRe{STM Assertion }
  \IgNoRe{STM Assertion }
  \IgNoRe{STM Assertion }
  \IgNoRe{PG}
  \IgNoRe{PG}
  \IgNoRe{PG}
  \IgNoRe{PG}
  \IgNoRe{EQN}
  \IgNoRe{EQN}
  \IgNoRe{EQN}
  \IgNoRe{EQN}
  \IgNoRe{PG}
  \IgNoRe{PG}
  \IgNoRe{PG}
  \IgNoRe{EQN}
  \IgNoRe{PG}
  \IgNoRe{PG}
 \def\eqnOVtriponesi{\frefwarning II.6} \IgNoRe{EQN}
  \IgNoRe{EQN}
  \IgNoRe{EQN}
  \IgNoRe{EQN}
  \IgNoRe{EQN}
  \IgNoRe{EQN}
  \IgNoRe{EQN}
  \IgNoRe{EQN}
 \def\eqnOVtripthreesi{\frefwarning II.14} \IgNoRe{EQN}
  \IgNoRe{EQN}
  \IgNoRe{PG}
  \IgNoRe{PG}
 \def\eqnNPcovFT{\frefwarning III.1} \IgNoRe{EQN}
 \def\eqnNPantisymmCov{\frefwarning III.2} \IgNoRe{EQN}
 \def\eqnNPjdef{\frefwarning III.3} \IgNoRe{EQN}
 \def\defNPrengroupmap{\frefwarning III.1} \IgNoRe{STM Assertion }
  \IgNoRe{PG}
  \IgNoRe{EQN}
  \IgNoRe{EQN}
  \IgNoRe{EQN}
  \IgNoRe{EQN}
  \IgNoRe{STM Assertion }
  \IgNoRe{STM Assertion }
  \IgNoRe{STM Assertion }
  \IgNoRe{STM Assertion }
  \IgNoRe{STM Assertion }
  \IgNoRe{STM Assertion }
 \def\defNPinputData{\frefwarning III.8} \IgNoRe{STM Assertion }
 \def\defNPoutputData{\frefwarning III.9} \IgNoRe{STM Assertion }
  \IgNoRe{STM Assertion }
  \IgNoRe{EQN}
  \IgNoRe{EQN}
  \IgNoRe{EQN}
  \IgNoRe{EQN}
  \IgNoRe{STM Assertion }
  \IgNoRe{EQN}
  \IgNoRe{STM Assertion }
  \IgNoRe{EQN}
  \IgNoRe{STM Assertion }
  \IgNoRe{PG}
  \IgNoRe{EQN}
  \IgNoRe{STM Assertion }
  \IgNoRe{STM Assertion }
  \IgNoRe{EQN}
  \IgNoRe{PG}
  \IgNoRe{PG}
 \def\defNPdecayop{\frefwarning V.1} \IgNoRe{STM Assertion }
 \def\defNPFancynormdomain{\frefwarning V.2} \IgNoRe{STM Assertion }
  \IgNoRe{PG}
 \def\defNPSymmNorm{\frefwarning V.3} \IgNoRe{STM Assertion }
 \def\notNPfourierTI{\frefwarning V.4} \IgNoRe{STM Assertion }
  \IgNoRe{STM Assertion }
 \def\defNPrhomn{\frefwarning V.6} \IgNoRe{STM Assertion }
  \IgNoRe{STM Assertion }
 \def\thmNPfirststep{\frefwarning V.8} \IgNoRe{STM Assertion }
  \IgNoRe{STM Assertion }
 \def\defNPfourtrans{\frefwarning VI.1} \IgNoRe{STM Assertion }
  \IgNoRe{PG}
  \IgNoRe{STM Assertion }
 \def\defNPsectrepr{\frefwarning VI.3} \IgNoRe{STM Assertion }
 \def\defNPresector{\frefwarning VI.4} \IgNoRe{STM Assertion }
  \IgNoRe{STM Assertion }
  \IgNoRe{EQN}
 \def\defNPsectnorm{\frefwarning VI.6} \IgNoRe{STM Assertion }
 \def\defNPsectGrnorm{\frefwarning VI.7} \IgNoRe{STM Assertion }
  \IgNoRe{STM Assertion }
  \IgNoRe{EQN}
 \def\defNPCTmSpace{\frefwarning VI.9} \IgNoRe{STM Assertion }
  \IgNoRe{STM Assertion }
 \def\defNPCovariances{\frefwarning VI.11} \IgNoRe{STM Assertion }
 \def\thmNPsetupinduction{\frefwarning VI.12} \IgNoRe{STM Assertion }
  \IgNoRe{PG}
 \def\defbubbleprop{\frefwarning VII.1} \IgNoRe{STM Assertion }
 \def\deNPdefsectbubbleprop{\frefwarning VII.2} \IgNoRe{STM Assertion }
 \def\NPsomespaces{\frefwarning VII.3} \IgNoRe{STM Assertion }
 \def\defParicleHoleDecomp{\frefwarning VII.4} \IgNoRe{STM Assertion }
 \def\lemunordord{\frefwarning VII.5} \IgNoRe{STM Assertion }
 \def\propparticleparticleladder{\frefwarning VII.6} \IgNoRe{STM Assertion }
 \def\defcompLadder{\frefwarning VII.7} \IgNoRe{STM Assertion }
 \def\theoremcompLadder{\frefwarning VII.8} \IgNoRe{STM Assertion }
 \def\defNPintquad{\frefwarning VIII.1} \IgNoRe{STM Assertion }
 \def\eqnNPpbound{\frefwarning VIII.1} \IgNoRe{EQN}
  \IgNoRe{EQN}
  \IgNoRe{EQN}
  \IgNoRe{STM Assertion }
  \IgNoRe{PG}
  \IgNoRe{STM Assertion }
  \IgNoRe{STM Assertion }
 \def\theoremNPinduction{\frefwarning VIII.5} \IgNoRe{STM Assertion }
  \IgNoRe{EQN}
 \def\eqnNPnbndA{\frefwarning VIII.5} \IgNoRe{EQN}
  \IgNoRe{EQN}
 \def\defNPconvol{\frefwarning VIII.6} \IgNoRe{STM Assertion }
 \def\lemNPpptyu{\frefwarning VIII.7} \IgNoRe{STM Assertion }
  \IgNoRe{EQN}
  \IgNoRe{EQN}
  \IgNoRe{EQN}
 \def\stepInputData{\frefwarning IX.1} \IgNoRe{STM Assertion }
  \IgNoRe{PG}
  \IgNoRe{PG}
 \def\stepOutputData{\frefwarning IX.2} \IgNoRe{STM Assertion }
  \IgNoRe{STM Assertion }
  \IgNoRe{PG}
  \IgNoRe{STM Assertion }
 \def\thmIntoOut{\frefwarning IX.5} \IgNoRe{STM Assertion }
  \IgNoRe{EQN}
  \IgNoRe{EQN}
  \IgNoRe{EQN}
  \IgNoRe{EQN}
  \IgNoRe{EQN}
  \IgNoRe{EQN}
  \IgNoRe{EQN}
  \IgNoRe{EQN}
  \IgNoRe{EQN}
  \IgNoRe{EQN}
  \IgNoRe{EQN}
  \IgNoRe{EQN}
  \IgNoRe{EQN}
  \IgNoRe{EQN}
 \def\remNPFnochangeI{\frefwarning IX.6} \IgNoRe{STM Assertion }
  \IgNoRe{EQN}
  \IgNoRe{PG}
 \def\eqnNPtwrewick{\frefwarning IX.16} \IgNoRe{EQN}
  \IgNoRe{STM Assertion }
  \IgNoRe{EQN}
 \def\lemNPdeKbnd{\frefwarning IX.8} \IgNoRe{STM Assertion }
  \IgNoRe{STM Assertion }
 \def\eqnNPtwwpp{\frefwarning IX.18} \IgNoRe{EQN}
 \def\eqnNPgup{\frefwarning IX.19} \IgNoRe{EQN}
  \IgNoRe{EQN}
  \IgNoRe{STM Assertion }
 \def\eqnNPpboundal{\frefwarning IX.21} \IgNoRe{EQN}
  \IgNoRe{EQN}
  \IgNoRe{EQN}
  \IgNoRe{EQN}
  \IgNoRe{EQN}
  \IgNoRe{EQN}
  \IgNoRe{EQN}
 \def\eqnNPwpwpp{\frefwarning IX.28} \IgNoRe{EQN}
 \def\eqnNPwickwtw{\frefwarning IX.29} \IgNoRe{EQN}
  \IgNoRe{EQN}
  \IgNoRe{EQN}
  \IgNoRe{EQN}
  \IgNoRe{EQN}
  \IgNoRe{EQN}
 \def\eqnNPderivwtildeest{\frefwarning IX.35} \IgNoRe{EQN}
  \IgNoRe{EQN}
  \IgNoRe{EQN}
  \IgNoRe{EQN}
  \IgNoRe{EQN}
  \IgNoRe{EQN}
  \IgNoRe{EQN}
  \IgNoRe{STM Assertion }
  \IgNoRe{PG}
  \IgNoRe{PG}
  \IgNoRe{STM Assertion }
  \IgNoRe{PG}
  \IgNoRe{EQN}
  \IgNoRe{EQN}
  \IgNoRe{EQN}
  \IgNoRe{EQN}
 \def\lemRWintbnd{\frefwarning B.1} \IgNoRe{STM Assertion }
  \IgNoRe{EQN}
  \IgNoRe{PG}
  \IgNoRe{EQN}
  \IgNoRe{EQN}
  \IgNoRe{EQN}
  \IgNoRe{EQN}
  \IgNoRe{EQN}
  \IgNoRe{EQN}
  \IgNoRe{EQN}
  \IgNoRe{EQN}
  \IgNoRe{EQN}
  \IgNoRe{EQN}
  \IgNoRe{EQN}
  \IgNoRe{EQN}
  \IgNoRe{STM Assertion }
  \IgNoRe{STM Assertion }
  \IgNoRe{EQN}
  \IgNoRe{EQN}
  \IgNoRe{PG}
 \def\pgNPXI{\frefwarning 1} \IgNoRe{PG}
 \def\theoremNPtildeinduction{\frefwarning XII.1} \IgNoRe{STM Assertion }
 \def\eqnNPqderiv{\frefwarning XII.1} \IgNoRe{EQN}
 \def\lemNPgtwoamp{\frefwarning XII.2} \IgNoRe{STM Assertion }
 \def\pgNPXII{\frefwarning 2} \IgNoRe{PG}
 \def\eqnNPpderiv{\frefwarning XII.2} \IgNoRe{EQN}
 \def\eqnTNPQkbnd{\frefwarning XII.3} \IgNoRe{EQN}
 \def\eqnTNPQkbndb{\frefwarning XII.4} \IgNoRe{EQN}
 \def\lemNPgtwoft{\frefwarning XII.3} \IgNoRe{STM Assertion }
 \def\lemTNPabstractjump{\frefwarning XII.4} \IgNoRe{STM Assertion }
 \def\eqnLPone{\frefwarning XII.5} \IgNoRe{EQN}
 \def\eqnLPtwo{\frefwarning XII.6} \IgNoRe{EQN}
 \def\eqnLPthree{\frefwarning XII.7} \IgNoRe{EQN}
 \def\eqnLPthreep{\frefwarning XII.8} \IgNoRe{EQN}
 \def\eqnLPfour{\frefwarning XII.9} \IgNoRe{EQN}
 \def\lemNPpsebnd{\frefwarning XII.5} \IgNoRe{STM Assertion }
 \def\eqnTNPtildeqbnds{\frefwarning XII.10} \IgNoRe{EQN}
 \def\eqnTNPtildeQbnds{\frefwarning XII.11} \IgNoRe{EQN}
 \def\eqnTNPtildeQmQbnds{\frefwarning XII.12} \IgNoRe{EQN}
 \def\eqnNPexplicitdsigmadk{\frefwarning XII.13} \IgNoRe{EQN}
 \def\remNPpseregularity{\frefwarning XII.6} \IgNoRe{STM Assertion }
 \def\lemNPampcorrect{\frefwarning XII.7} \IgNoRe{STM Assertion }
 \def\eqnTNPAthreebnds{\frefwarning XII.14} \IgNoRe{EQN}
 \def\defNPtildefourtrans{\frefwarning XIII.1} \IgNoRe{STM Assertion }
 \def\defNPphicheckfourtrans{\frefwarning XIII.2} \IgNoRe{STM Assertion }
 \def\remNPphicheckfourtrans{\frefwarning XIII.3} \IgNoRe{STM Assertion }
 \def\defNPtransinv{\frefwarning XIII.4} \IgNoRe{STM Assertion }
 \def\pgNPXIII{\frefwarning 16} \IgNoRe{PG}
 \def\defNPdiffdecay{\frefwarning XIII.5} \IgNoRe{STM Assertion }
 \def\remNPdiffdecay{\frefwarning XIII.6} \IgNoRe{STM Assertion }
 \def\defNPdiffdecaynorm{\frefwarning XIII.7} \IgNoRe{STM Assertion }
 \def\remNPdiffdecaynorm{\frefwarning XIII.8} \IgNoRe{STM Assertion }
 \def\defNPampGreen{\frefwarning XIII.9} \IgNoRe{STM Assertion }
 \def\remNPampGreen{\frefwarning XIII.10} \IgNoRe{STM Assertion }
 \def\thmNPTfirststep{\frefwarning XIII.11} \IgNoRe{STM Assertion }
 \def\defNPsectdiffdecaynorm{\frefwarning XIII.12} \IgNoRe{STM Assertion }
 \def\remNPsecdiffdecaynorm{\frefwarning XIII.13} \IgNoRe{STM Assertion }
 \def\defNPsectcheckcF{\frefwarning XIII.14} \IgNoRe{STM Assertion }
 \def\defNPmomscalednorms{\frefwarning XIII.15} \IgNoRe{STM Assertion }
 \def\remNPmomscalednorms{\frefwarning XIII.16} \IgNoRe{STM Assertion }
 \def\thmNPTsetupinduction{\frefwarning XIII.17} \IgNoRe{STM Assertion }
 \def\defmoredisjointunions{\frefwarning XIV.1} \IgNoRe{STM Assertion }
 \def\deftildeParicleHoleDecomp{\frefwarning XIV.2} \IgNoRe{STM Assertion }
 \def\pgNPXIV{\frefwarning 23} \IgNoRe{PG}
 \def\lemtildeunordord{\frefwarning XIV.3} \IgNoRe{STM Assertion }
 \def\defNPdisjointOrd{\frefwarning XIV.4} \IgNoRe{STM Assertion }
 \def\remNPbigdisjointunion{\frefwarning XIV.5} \IgNoRe{STM Assertion }
 \def\defNPdisjointfnspaces{\frefwarning XIV.6} \IgNoRe{STM Assertion }
 \def\deftildeladders{\frefwarning XIV.7} \IgNoRe{STM Assertion }
 \def\remNPnoInternalPhi{\frefwarning XIV.8} \IgNoRe{STM Assertion }
 \def\tildepropparticleparticleladder{\frefwarning XIV.9} \IgNoRe{STM Assertion }
 \def\remtildepropparticleparticleladder{\frefwarning XIV.10} \IgNoRe{STM Assertion }
 \def\deftildecompLadder{\frefwarning XIV.11} \IgNoRe{STM Assertion }
 \def\theoremtildecompLadder{\frefwarning XIV.12} \IgNoRe{STM Assertion }
 \def\remtildecompLadder{\frefwarning XIV.13} \IgNoRe{STM Assertion }
 \def\defNPshear{\frefwarning XIV.14} \IgNoRe{STM Assertion }
 \def\eqnNPshear{\frefwarning XIV.1} \IgNoRe{EQN}
 \def\cortildecompLadder{\frefwarning XIV.15} \IgNoRe{STM Assertion }
 \def\defNPmoredisjointOrd{\frefwarning XIV.16} \IgNoRe{STM Assertion }
 \def\defNPmoredisjointfnspaces{\frefwarning XIV.17} \IgNoRe{STM Assertion }
 \def\lemNPremoveprimes{\frefwarning XIV.18} \IgNoRe{STM Assertion }
 \def\defNPshearprime{\frefwarning XIV.19} \IgNoRe{STM Assertion }
 \def\remNPshearprime{\frefwarning XIV.20} \IgNoRe{STM Assertion }
 \def\eqnTNPcbfeij{\frefwarning XV.1} \IgNoRe{EQN}
 \def\moreInputData{\frefwarning XV.1} \IgNoRe{STM Assertion }
 \def\pgNPXV{\frefwarning 34} \IgNoRe{PG}
 \def\pgNPXVa{\frefwarning 34} \IgNoRe{PG}
 \def\remMorebarunbar{\frefwarning XV.2} \IgNoRe{STM Assertion }
 \def\moreOutputData{\frefwarning XV.3} \IgNoRe{STM Assertion }
 \def\thmTildeIntoOut{\frefwarning XV.4} \IgNoRe{STM Assertion }
 \def\eqnNPzdef{\frefwarning XV.2} \IgNoRe{EQN}
 \def\lemNPMwprimez{\frefwarning XV.5} \IgNoRe{STM Assertion }
 \def\pgNPXVb{\frefwarning 37} \IgNoRe{PG}
 \def\eqnNPwoneoneBhat{\frefwarning XV.3} \IgNoRe{EQN}
 \def\lemTNPbcbStruct{\frefwarning XV.6} \IgNoRe{STM Assertion }
 \def\lemNPBbnd{\frefwarning XV.7} \IgNoRe{STM Assertion }
 \def\eqnTNPjhalfbnds{\frefwarning XV.4} \IgNoRe{EQN}
 \def\eqnTNPudubnds{\frefwarning XV.5} \IgNoRe{EQN}
 \def\eqnTNPwoneonebnd{\frefwarning XV.6} \IgNoRe{EQN}
 \def\eqnTNPwoneonenonampbnd{\frefwarning XV.7} \IgNoRe{EQN}
 \def\eqnTNPzzpp{\frefwarning XV.8} \IgNoRe{EQN}
 \def\eqnTNPodI{\frefwarning XV.9} \IgNoRe{EQN}
 \def\eqnTNPodIfour{\frefwarning XV.10} \IgNoRe{EQN}
 \def\eqnTNPodIb{\frefwarning XV.11} \IgNoRe{EQN}
 \def\eqnTNPodII{\frefwarning XV.12} \IgNoRe{EQN}
 \def\eqnTNPodIII{\frefwarning XV.13} \IgNoRe{EQN}
 \def\eqnTNPodIfourB{\frefwarning XV.14} \IgNoRe{EQN}
 \def\eqnTNPodIIba{\frefwarning XV.15} \IgNoRe{EQN}
 \def\eqnTNPodIIb{\frefwarning XV.16} \IgNoRe{EQN}
 \def\eqnTNPdFz{\frefwarning XV.17} \IgNoRe{EQN}
 \def\eqnTNPdFpone{\frefwarning XV.18} \IgNoRe{EQN}
 \def\eqnTNPdFptwo{\frefwarning XV.19} \IgNoRe{EQN}
 \def\eqnTNPdFpthree{\frefwarning XV.20} \IgNoRe{EQN}
 \def\eqnTNPFjplusI{\frefwarning XV.21} \IgNoRe{EQN}
 \def\eqnTNPFjplusIderiv{\frefwarning XV.22} \IgNoRe{EQN}
 \def\eqnTNPodFIa{\frefwarning XV.23} \IgNoRe{EQN}
 \def\thmTildeOuttoIn{\frefwarning XV.8} \IgNoRe{STM Assertion }
 \def\lemNPmoreBbnd{\frefwarning XV.9} \IgNoRe{STM Assertion }
 \def\pgNPXVc{\frefwarning 48} \IgNoRe{PG}
 \def\eqnTNPgtogprime{\frefwarning XV.24} \IgNoRe{EQN}
 \def\eqnNPTreWickBnd{\frefwarning XV.25} \IgNoRe{EQN}
 \def\eqnNPTreWickBndbis{\frefwarning XV.26} \IgNoRe{EQN}
 \def\eqnNPTderivwchangeest{\frefwarning XV.27} \IgNoRe{EQN}
 \def\eqnNPTderivwtildeest{\frefwarning XV.28} \IgNoRe{EQN}
 \def\eqnTNPderivwtildetwoest{\frefwarning XV.29} \IgNoRe{EQN}
 \def\eqnTNPwponeoneest{\frefwarning XV.30} \IgNoRe{EQN}
 \def\eqnTNPwponeonederivest{\frefwarning XV.31} \IgNoRe{EQN}
 \def\eqnTNPdifffourlegged{\frefwarning XV.32} \IgNoRe{EQN}
 \def\eqnTNPprefprime{\frefwarning XV.33} \IgNoRe{EQN}
 \def\eqnTNPfprime{\frefwarning XV.34} \IgNoRe{EQN}
 \def\eqnTNPdiffffp{\frefwarning XV.35} \IgNoRe{EQN}
 \def\eqnTNPdiffladdladdp{\frefwarning XV.36} \IgNoRe{EQN}
 \def\eqnTNPgoodFprimeest{\frefwarning XV.37} \IgNoRe{EQN}
 \def\eqnTNPgoodFprimederivest{\frefwarning XV.38} \IgNoRe{EQN}
 \def\eqnTNPfourlegest{\frefwarning XV.39} \IgNoRe{EQN}
 \def\eqnTNPtwolegest{\frefwarning XV.40} \IgNoRe{EQN}
 \def\eqnTNPnonfourlegest{\frefwarning XV.41} \IgNoRe{EQN}
 \def\eqnGFitphl{\frefwarning XV.42} \IgNoRe{EQN}
 \def\eqnGFtvbnd{\frefwarning XV.43} \IgNoRe{EQN}
 \def\eqnGFcmpphl{\frefwarning XV.44} \IgNoRe{EQN}
 \def\lemNPhoelder{\frefwarning C.1} \IgNoRe{STM Assertion }
 \def\pgNPC{\frefwarning 61} \IgNoRe{PG}
 \def\defmodcompLadder{\frefwarning D.1} \IgNoRe{STM Assertion }
 \def\theoremmodcompLadder{\frefwarning D.2} \IgNoRe{STM Assertion }
 \def\eqnmodcompLadderI{\frefwarning D.1} \IgNoRe{EQN}
 \def\eqnmodcompLadderIIa{\frefwarning D.2} \IgNoRe{EQN}
 \def\pgNPD{\frefwarning 62} \IgNoRe{PG}
 \def\eqnmodcompLadderII{\frefwarning D.3} \IgNoRe{EQN}
 \def\eqnmodcompLadderIIIa{\frefwarning D.4} \IgNoRe{EQN}
 \def\eqnmodcompLadderIII{\frefwarning D.5} \IgNoRe{EQN}
 \def\eqnmodcompLadderIV{\frefwarning D.6} \IgNoRe{EQN}
 \def\eqnmodcompLadderV{\frefwarning D.7} \IgNoRe{EQN}
 \def\eqnmodcompLadderVI{\frefwarning D.8} \IgNoRe{EQN}
 \def\eqnmodcompLadderVII{\frefwarning D.9} \IgNoRe{EQN}
 \def\eqnmodcompLadderVIII{\frefwarning D.10} \IgNoRe{EQN}
 \def\defInversym{\frefwarning D.3} \IgNoRe{STM Assertion }
 \def\lemAntInvsym{\frefwarning D.4} \IgNoRe{STM Assertion }
 \def\exInversym{\frefwarning D.5} \IgNoRe{STM Assertion }
 \def\propPHladsimp{\frefwarning D.6} \IgNoRe{STM Assertion }
 \def\corcompLadder{\frefwarning D.7} \IgNoRe{STM Assertion }
 \def\pgNPIIIref{\frefwarning 68} \IgNoRe{PG}
  \IgNoRe{PG}
  \IgNoRe{PG}
 \def\pgNPIIInot{\frefwarning 69} \IgNoRe{PG}


\newcount\CHAPNO
\newcount\APPNO
\CHAPNO=0
\APPNO=1
\def\advCHAPNO{\advance\CHAPNO by 1}
\def\advAPPNO{\advance\APPNO by 1}

\def\caproman#1{\ifcase#1\or I\or II\or III\or IV\or V\or VI\or VII\or
VIII\or IX\or X\or XI\or XII\or XIII\or XIV\or XV\or XVI\or XVII\or XVIII\or
XIX\or XX\or XXI\or XXII\or XXIII\or XXIV\or XXV\or XXVI\or XXVII\or XXVIII\or XXIX\or XXX\or XXXI\or XXXII\or XXXIII\or XXXIV\or XXXV\or XXXVI\or XXXVII\or XXXVIII\or XXXIX\fi}%

\def\capletter#1{\ifcase#1\or A\or B\or C\or D\or E\or F\or G\or
H\or I\or J\or K\or L\or M\or N\or O\or P\or Q\or R\or
S\or T\or U\or V\or W\or X\or Y\or Z\fi}%

\newcount\cHintroI \cHintroI=\CHAPNO \advCHAPNO 
                              \edef\CHintroI{\caproman\CHAPNO}         
\newcount\cHintroOverview  \cHintroOverview=\CHAPNO \advCHAPNO 
\newcount\cHrenmap \cHrenmap=\CHAPNO \advCHAPNO 
                              \edef\CHrenmap{\caproman\CHAPNO}         

\edef\APappModelComp{\capletter\APPNO} \advAPPNO

\newcount\cHintroII \cHintroII=\CHAPNO \advCHAPNO 
                              \edef\CHintroII{\caproman\CHAPNO}
\newcount\cHfirstscale \cHfirstscale=\CHAPNO \advCHAPNO
                              
\newcount\cHnewsectors \cHnewsectors=\CHAPNO \advCHAPNO
                              \edef\CHnewsectors{\caproman\CHAPNO}
\newcount\cHphladders \cHphladders=\CHAPNO \advCHAPNO
                              \edef\CHphladders{\caproman\CHAPNO}
\newcount\cHfinitescale \cHfinitescale=\CHAPNO \advCHAPNO
                              
\newcount\cHstep \cHstep=\CHAPNO \advCHAPNO
                              \edef\CHstep{\caproman\CHAPNO}
\newcount\cHrecurs \cHrecurs=\CHAPNO \advCHAPNO
                              \edef\CHrecurs{\caproman\CHAPNO}
\edef\APappRewick{\capletter\APPNO} \advAPPNO

\newcount\cHintroIII \cHintroIII=\CHAPNO \advCHAPNO
                              \edef\CHintroIII{\caproman\CHAPNO}
\newcount\cHtildefinitescale \cHtildefinitescale=\CHAPNO \advCHAPNO
                              
\newcount\cHtildenewsectors \cHtildenewsectors=\CHAPNO \advCHAPNO
                              \edef\CHtildenewsectors{\caproman\CHAPNO}
\newcount\cHtildephladders \cHtildephladders=\CHAPNO \advCHAPNO
                              
\newcount\cHtildestep  \cHtildestep=\CHAPNO \advCHAPNO
                              \edef\CHtildestep{\caproman\CHAPNO}

\edef\APappHoelder{\capletter\APPNO} \advAPPNO
\edef\APappPhladders{\capletter\APPNO} \advAPPNO